\documentclass[fleqn,12pt,twoside]{article}
\usepackage{bm}
\usepackage{myespcrc}
\usepackage{graphicx}
\usepackage[figuresright]{rotating}
\usepackage[abbr]{harvard}
\usepackage{makeidx}
\makeindex
\def\v#1{{\bf #1}}
\def\tensor#1{{\mathbf #1}}

\newcommand{\UGE}{UGe$_2$~}
\newcommand{\UBE}{UBe$_{13}$~}
\newcommand{\UTB}{U$_{1-x}$Th$_x$Be$_{13}$~}
\newcommand{\UPT}{UPt$_3$~}
\newcommand{\UPD}{UPd$_2$Al$_3~$~}
\newcommand{\URU}{URu$_2$Si$_2$~}
\newcommand{\UND}{UNi$_2$Al$_3~$~}

\newcommand{\la}{\langle}
\newcommand{\ra}{\rangle}

\newcommand{\ua}{\uparrow}
\newcommand{\da}{\downarrow}

\newcommand{\vk}{\bf k\rm}
\newcommand{\vd}{\bf d\rm}
\newcommand{\ve}{\bf e\rm}
\newcommand{\vq}{\bf q\rm}
\newcommand{\vQ}{\bf Q\rm}
\newcommand{\vv}{\bf v\rm}
\newcommand{\vH}{\bf H\rm}
\newcommand{\vcr}{\bf r\rm}

\newcommand{\RBC}{RNi$_2$B$_2$C~}
\newcommand{\YBC}{YNi$_2$B$_2$C~}
\newcommand{\LuBC}{LuNi$_2$B$_2$C~}
\newcommand{\DyBC}{DyNi$_2$B$_2$C~}
\newcommand{\HoBC}{HoNi$_2$B$_2$C~}
\newcommand{\ErBC}{ErNi$_2$B$_2$C~}
\newcommand{\TmBC}{TmNi$_2$B$_2$C~}

\newcommand{\PRS}{PrOs$_4$Sb$_{12}$~}
\newcommand{\RES}{RT$_4$X$_{12}$~}

\newcommand{\De}{$\Delta$(\bf k\rm)~}

\newcommand{\boldeta}{{\bm \eta}}
\newcommand{\boldsigma}{{\bm \sigma}}

\author{P. THALMEIER\\
Max-Planck-Institut f\"ur Chemische Physik fester Stoffe, 01187
Dresden\\[0.5cm]
G. ZWICKNAGL\\
Institut f\"ur Mathematische Physik, Technische Universit\"at
Braunschweig,\\ 38106 Braunschweig}

\title{Unconventional Superconductivity and Magnetism in Lanthanide
and Actinide Intermetallic Compounds\\[0.5cm]}

\begin{document}
\bibliographystyle{mykluwer}
\maketitle
\tableofcontents
\newpage


\begin{table}[hbtp]
{\em List of acronyms}\\[0.5cm]

\begin{tabular}{ll}
AF     &   antiferromagnetic\\
AFQ    &   antiferroquadrupolar\\ 
ANNNI  &   anisotropic next nearest neighbor Ising\\
ARPES  &   angle resolved photoemission spectroscopy\\
BIS    &   Bremsstrahlung isochromat spectroscopy\\
BZ     &   Brillouin zone\\
CDW    &   charge density wave\\
CEF    &   crystalline electric field\\
dHvA   &   de Haas-van Alphen\\
DLRO   &   diagonal long range order\\
DOS    &   density of states\\
e-e    &   electron-electron\\
e-h    &   electron-hole\\
e-p    &   electron-phonon\\
FFLO   &   Fulde-Ferrell-Larkin-Ovchinnikov\\
FL     &   Fermi liquid\\
FLEX   &   fluctuation exchange\\
FM     &   ferromagnet\\
FS     &   Fermi surface\\
GL     &   Ginzburg-Landau\\
HF     &   heavy fermion\\
IC     &   incommensurate\\
INS    &   inelastic neutron scattering\\
LDA    &   local density approximation\\
LSDA   &   local spin density approximation\\
mf     &   mean field\\
MFS    &   magnetic Fermi surface\\
nFl    &   non-Fermi liquid\\
n.n.   &   next neighbor\\ 
n.n.n. &   next nearest neighbor\\
NCA    &   non-crossing approximation\\
NMR    &   nuclear magnetic resonance\\
OAF    &   orbital antiferromagnet\\
ODLRO  &   off-diagonal long range order\\
OP     &   order parameter\\
PES    &   photoemission spectroscopy\\
QCP    &   quantum critical point\\
qp     &   quasi particle\\
RPA    &   random phase approximation\\
RKKY   &   Ruderman-Kittel-Kasuya-Yoshida\\
SBF    &   symmetry breaking field\\
SC     &   superconductivity\\
SDW    &   spin density wave\\
SN     &   spin nematic\\
s.o.   &   spin orbit\\
WFM    &   weak ferromagnetism

\end{tabular}
\end{table}
\newpage
\twocolumn
\begin{table}[pthb]
{\em List of symbols}\\[0.5cm]
\begin{tabular}{lp{5.5cm}}
$\alpha$(T)    &   ultrasonic attenuation\\
$\alpha_0$, $\beta$, $\gamma_0$& 
                   Landau free energy parameters\\
\v B, \vH, \v M&   magnetic induction, magnetic field and magnetisation\\
c, c$^\dagger$ &   conduction electron operators\\
C(T)           &   specific heat\\
$\delta$       &   CEF splitting energy\\
$\Delta(\v k)$ &   gap function (SC, SDW etc.; singlet, triplet etc.)\\
\v d(\v k)     &   vector of triplet gap functions\\
$\epsilon_{\vk}$&  quasiparticle energy in the normal state\\
E$_{\v k}$    &   quasiparticle energy in the ordered state (SC, SDW etc.)\\
E$_F$          &   Fermi energy\\
f(\vk)         &   form factor of gap functions\\
f(E$_{\v k}$)  &   Fermi function\\   
f$_L$, f$_{GL}$&   Landau and Ginzburg-Landau free energy density\\
$\gamma$       &   linear specific heat (Sommerfeld) coefficient\\
$\Gamma$       &   quasiparticle scattering rate\\
$|\Gamma\ra$   &   CEF state\\
${\bm \eta}$   &   SC vector order parameter\\
H$_{c2}$       &   upper critical field of the superconductor\\
I              &   residual on-site quasiparticle repulsion or contact
                   exchange\\
J(\v q)        &   effective interaction between quasiparticles or
                   local moments\\
\v J           &   total angular momentum of f- shell\\
$\kappa$(T)    &   thermal conductivity\\
\vk, \vq       &   conduction electron wave vectors\\
k$_F$          &   Fermi wave number\\
$\lambda$      &   electron-phonon interaction\\
$\mu$, $\mu_B$ &   magnetic moment, Bohr magneton\\
m              &   free electron mass\\
m$_b$          &   conduction electron band mass\\
m$^*$          &   effective quasiparticle mass\\
\v M$_{\v Q}$  &   staggered magnetization\\
\end{tabular}
\end{table}
\begin{table}[pthb]
\begin{tabular}{lp{5.5cm}}
N(E)           &   quasiparticle DOS\\
N$_n$          &   normal state quasiparticle DOS at E$_F$\\ 
p              &   pressure\\
\vQ            &   nesting or ordering wave vector\\
$\rho$(T)      &   electrical resistivity\\
$\rho_{\v Q}$  &   conduction electron charge density\\
\v s$_{\v Q}$  &   conduction electron spin density\\
${\bm \sigma}$ &   vector of Pauli matrices\\
$\theta, \phi$ &   polar angles of \v H\\
$\vartheta, \varphi$ &   
                   polar angles of \v k\\
t              &   nearest neighbor hopping matrix element\\
t'             &   next nearest neighbor hopping matrix element\\
T              &   temperature\\
T$^*$          &   characteristic or `Kondo' temperature of HF compound\\
T$_c$          &   superconducting transition temperature\\
T$_N$          &   N\'eel temperature\\
T$_1^{-1}$     &   NMR relaxation rate\\
U              &   bare on-site Coulomb interaction\\
u($\omega$)    &   local CEF susceptibility\\
v$_F$          &   Fermi velocity\\
\vv$_s$(\vcr)  &   superfluid velocity field\\
V$_l$          &   effective quasiparticle interaction (l = angular
                   momentum channel)\\
$\tilde{v}$    &   average Fermi velocity in uniaxial crystal\\
$\chi_n$, $\chi_s$(T)&   
                   static homogeneous (\vq~= 0) conduction
                   electron spin susceptibility in normal and SC state\\
$\chi$(\vq,$\omega$)& 
                   magnetic susceptibilities for conduction 
                   electrons or localised moments\\  
x              &   impurity or dopant concentration \\
Y(T)           &   Yoshida function\\
$\omega$       &   frequency\\
$\omega_E$(\vq)&   magnetic exciton dispersion\\[15cm]
\end{tabular}
\end{table}
\onecolumn
\newpage


\section{Introduction}

The Fermi liquid state in metals has two common
instabilities. Superconductivity (SC) is due to pair formation of electrons
and (spin-, charge-) density waves (CDW, SDW) are formed by pairing
electrons and holes. Theoretically it has been
suspected quite early that pair-wave functions other than s-wave as in
conventional superconductors and density waves may
exist. However it has taken surprisingly long to identify such
'unconventional' condensed pair states in real materials. 

Now there is an
abundance of non-s-wave superconductors which are frequently
associated with anisotropic `nodal' gap functions where the
quasiparticle excitations of the SC state vanishes on points or lines
on the Fermi surface. This leads to low temperature `power law'
behaviour in many physical quantities. Heavy fermion (HF) lanthanide
and actinide superconducting compounds were the first that were
supposed to have unconventional SC pair states mediated by low energy
spin fluctuations. High T$_c$ cuprates which exhibit d-wave
superconductivity are the most  prominent and important examples. But
nodal SC have also been found among the ruthenates and organic salts. 

Conventional CDW and SDW states are ubiquitous in metals with the 
commensurate SDW or antiferromagnetic (AF) order being the most common.
Compounds with confirmed unconventional density waves sofar are rather
scarce and with certainty have only been found in organic metals and
perhaps in uranium HF compounds and the `pseudogap' phase of
underdoped cuprates. This may in part be due to the difficulty
detecting such `hidden' order parameters which leave no signature in
standard neutron or x-ray diffraction experiments. 

In this review article we will summarize the knowledge on a class of 
unconventional superconductors and their competition and coexistence
with magnetism and hidden order phases that has accumulated over the
last decade. To be comprehensive we will restrict ourselves
exclusively to intermetallic lanthanide (4f) and actinide (5f)
systems. Except for occasional remarks we will leave out completely
the cuprate, cobaltate and ruthenate superconductors which are beyond
the scope of our work on intermetallic compounds. The oxide
superconductors are mostly close to a Mott insulator transition and
the underlying microscopic physics is very different from the
intermetallic f-electron compounds. Although on the phenomenological
level of SC and density wave order parameter classification and
investigation strong similarities exist. Likewise we do not discuss
unconventional organic superconductors. The
focus here is on stoichiometric 4f- and 5f-compounds where the
lanthanide or actinide atoms occupy regular sublattices. With the
exception of rare earth borocarbides the compounds reviewed are
heavy fermion metals to a varying degree.  
In our review we do not want to
provide an exhausting compilation of the physical properties in
this whole class of materials. We rather focus on a few important
compounds where each displays an important aspect of unconventional SC
and its relation to magnetism or hidden order that will be discussed
in detail in its physical and theoretical implications.

The complex low temperature phase diagrams of the HF metals results
from the partially filled f-shells of
the lanthanide and actinide ions which preserve atomic-like character.
Occupying the states according to Hund's rules leads to magnetic moments.
In a crystal their rotational degeneracy is partly lifted by the
crystalline electric field (CEF) and the hybridisation with broad
conduction bands corresponding to outer shell electrons.
As a result a large number of low-energy excitations appears. In the
ideal Fermi liquid case these
excitations correspond to heavy quasiparticles whose effective mass
m$^*$ is by orders of magnitude larger than the free electron mass
m. The corresponding quasiparticle band width T$^*$ is of the order
meV for true HF metals. The mass enhancement is reflected in large
increase of the specific heat $\gamma$-value, the Pauli
susceptibility and the T$^2$-coefficient of electrical
resistivity. However close to a quantum critical point (QCP) which
signifies the onset of density wave instabilities low temperature
anomalies in these quantities appear which are characteristic
signatures of a non-Fermi liquid state. 

At this point, we emphasize that the Landau theory of Fermi liquids
does not make assumptions concerning the microscopic nature of the
ground state and the low-lying excitations, it merely gives a counting
prescription. It does not address the question how they emerge in
an interacting electron system, this requires a microscopic
treatment. Generally speaking heavy quasiparticles arise from the
lifting of local degeneracies which result as a consequence
of strong local correlations. Various mechanisms for heavy mass
generation have been suggested.
While in Ce compounds with well localised 4f$^1$ states the Kondo 
mechanism is appropriate, and there is increasing evidence
that the dual, i.e. partly localised and partly itinerant, nature of
5f-states is responsible for
the mass renormalization in the U compounds. In both scenarios it is
assumed that the heavy quasiparticles predominantly have f-character.

Residual interactions among the quasiparticles lead to the pairing
instabilities. Two candidates have been identified in HF compounds :
Firstly pairing interactions via exchange of enhanced overdamped 
spin fluctuations of itinerant quasiparticles, presumably at an
antiferromagnetic wave vector. This
model is invoked for Ce-compunds, especially when SC appears near a
quantum critical point. It may also contain some truth for the more
itinerant U-HF compounds. However we now know for sure that in U-compounds with partly localised 5f-electrons a different mechanism is
at work: pairing mediated by the exchange of propagating internal
excitations ('magnetic excitons') of the localised 5f subsystem. A
variant of this mechanism may also be appropriate for the recently 
discovered Pr-skutterudite HF superconductor with quadrupolar
instead of magnetic excitons involved. 

A major topic of our review is the critical discussion of existing
evidence for unconventional SC order parameters as witnessed
frequently by the presence of nodal gap functions. Previously the
identification of gap symmetries has been an elaborate guess work
mostly built on indirect evidence from low temperature `power
laws'. This situation has dramatically improved in recent years with
the advent of genuine angular resolved magnetothermal and magnetotransport
experiments in the vortex phase.  They exploit the Doppler shift of SC
quasiparticle energies which leads to field-angle oscillations in specific
heat and thermal conductivity. Their analysis may lead to an
unambiguous determination of nodal positions of the gap
functions which facilitates a strong restriction and sometimes unique
determination of the possible gap symmetry.

Superconductivity in HF-compounds frequently coexists with
(spin-) density waves. In some cases they may be of the unconventional
(hidden order) type. We discuss the competition and coexistence
behaviour of these order parameters and related physical
effects. Since both order parameters appear in the itinerant
quasiparticle system this is a subtle interplay of Fermi surface
geometry, pairing potentials and gap structures which can only be
schematically understood in simplified toy models.

Coexistence behaviour is much simpler in the relatively new class of rare earth
borocarbide superconductors which have comparatively high
T$_c$'s. Although as electron-phonon superconductors they are set
aside from our main focus on HF
systems, they have been included in this review for various reasons:
Firstly magnetism and SC superconductivity are carried by different species of
electrons which only interact weakly through  contact exchange
interaction leading to a small effect of the local moment
molecular field on the SC conduction electrons. This can be nicely
treated by perturbation theory and allows a much better understanding
of coexistence behaviour as compared to the HF systems. Also they
provide the first example of homogeneous coexistence of SC and
ferromagnetism for all temperatures below T$_c$. 
Secondly the nonmagnetic rare earth borocarbides have extremely large gap
anisotropy ratios $\geq 10^{2}$ which means
that they are essentialy nodal superconductors. In fact nonmagnetic
borocarbides are the first which have been identified to have a SC gap
with point nodes that is of fully symmetric s+g wave type. Surely the standard
electron-phonon mechanism has to be supplemented by something else,
perhaps anisotropic Coulomb interactions to achieve this `quasi-unconventional' behaviour in borocarbides.

Considering the wide range of phenomena and questions involved we have
to be quite selective in our review. We will concentrate on the
exciting developments in 4f/5f intermetallic materials of the past
decade referring to the literature for the earlier work. A general
introduction to strongly correlated electron systems is given in
the textbooks of \citeasnoun{FuldeBook} and
\citeasnoun{Fazekas99}. Various aspects of heavy fermion physics are
described in the previous review articles
by \citeasnoun{Stewart84}, \citeasnoun{Ott87}, \citeasnoun{Ott87a},
\citeasnoun{Fulde88a} \citeasnoun{Grewe91}, \citeasnoun{Thalmeier91a}
and \citeasnoun{Zwicknagl92} as well as in the monographs by
\citeasnoun{KuramotoBook} and \citeasnoun{HewsonBook}. Reviews which
focus on theories of unconventional HF
superconductors are \citeasnoun{Sigrist91}, \citeasnoun{Sauls94} and
\citeasnoun{Joynt02} and the monograph by
\citeasnoun{MineevBook}. Reviews discussing the rare earth borocarbide
magnetic superconductors are given in \citeasnoun{Hilscher99} and
\citeasnoun{Mueller01}. Earlier reviews of coexistence of
superconductivity and localised magnetism can be found in
\citeasnoun{Fulde82a}, \citeasnoun{Fulde90} and \citeasnoun{Fischer90}.

Since we will not
discuss oxide or organic unconventional superconductors we would like
to refer to a few review articles where parallel developments,
especially on pairing mechanism and gap function symmetries are
disussed for these classes of materials: \citeasnoun{vanHarlingen95},
\citeasnoun{Scalapino95}, \citeasnoun{Izyumov99},
\citeasnoun{Moriya00}, \citeasnoun{Yanase03} and \citeasnoun{Tsuei00}
for cuprates, \citeasnoun{Mackenzie03} for ruthenates and
\citeasnoun{Maki96} and \citeasnoun{Lang03} for unconventional organic
superconductors.
 
 The present review focuses on chosen topics of high current interest.
We therefore select specific materials which allow us to discuss
physical effects and theoretical concepts in detail. 

The article is organized as follows: In sect. 2 we review the basic
theoretical concepts and experimental techniques to identify pair
condensate order parameters , especially we
discuss the important new tool of angular resolved methods for SC gap
investigation. Section 3 deals
with the Ce-based heavy fermion compounds. The central question there
is the competition between magnetism and Fermi liquid
states which may become superconducting. 
While superconducting ground states
seem to occur rather rarely in Ce-based heavy fermion systems they
are common in their U-based counterparts discussed in
sect. 4. The major challenge in this context is the investigation of the
unconventional and often exotic pairing mechanism and the identification
of the symmetry of superconducting order parameters which may coexist
with (hidden) long-range order. 
The low-temperature
phases of rare earth borocarbides are reviewed in sect. 5. Of particular
interest are the unusual magnetic ordering phenomena which may coexist
with very anisotropic electron-phonon superconductivity. 
Section 6 is devoted to the newly discovered superconducting rare
earth skutterudite cage compounds which may exhibit a quadrupolar
pairing mechanism. We conclude in sect. 7 by giving a summary and an outlook.

\section{Theory and Techniques}

\label{Sect:Theory}

The theoretical understanding of superconductivity and magnetism in
the Heavy Fermion systems is still in the state of rather schematic or
illustrative models without real predictive power. The difficulty
arises on two levels. Firstly the normal state quasiparticles
themselves can sofar be described only within effective one particle
renormalised band pictures with empirical input parameters. For some
compounds like UBe$_{13}$ and Ce-compounds close to the quantum
critical point the SC transition even takes place in a
state without well defined quasiparticles as witnessed by the
observation of non-Fermi liquid (nFl) behaviour. Secondly the effective pairing
interactions can only be described in an oversimplified way as in the spin
fluctuation models and its variants. They commonly neglect the
internal orbital structure of f-electron compounds due to the
intra-atomic spin orbit coupling and CEF potential. Attempts to
include these terms have not been carried very far.
Nevertheless it is important to understand these qualitative
theories. We first discuss normal state quasiparticles, namely Kondo
lattice models for Ce-compounds vs. dual 5f-electrons models for the
U-compounds. Starting from this basis the renormalised band theory provides
a way to describe the heavy quasiparticle bands within a Fermi liquid
approach. From there approximate models for the effective pairing
interactions may be obtained by standard many body techniques.
The symmetry classification of pairing order parameters is an important
step to understand the nodal structure of the gap and solve the gap
equations. Models for coexistence and competition of SC and density wave order
parameters frequently observed in HF metals will also be discussed.
Finally we make a survey of theoretical ideas and experimental methods
to identify the nodal structure of SC and density type order
parameters which is the most important ingredient to understand the physics
of HF superconductors.

\subsection{Heavy quasiparticles in Ce, U compounds and their interactions}

A prerequisite for a microscopic theory of superconductivity in heavy-fermion
compounds is a description of the normal state and of the 
low-energy excitations at low temperatures. We begin by reviewing the
physical processes which lead to the high density of low-energy
excitations reflected in the strongly enhanced specific heat. The
corrresponding microscopic many-body problems, i.~e.~, the Kondo model 
and the dual model, can be solved for (effective) impurities but not
for extended systems. The energy dispersion of the coherent
quasiparticle states in a periodic lattice, however, can be calculated 
from an effective single-particle Hamiltonian where the effective and
not necessarily local potential is devised to account for the relevant 
many-body effects. The residual interaction among the quasiparticles
eventually leads to the instability of the normal Fermi liquid
phase. We focus on Cooper pair formation induced by electron-electron 
interactions. Of particular importance are spin-fluctuation models
which were adopted in the majority of papers during the past decade.

\subsubsection{Kondo lattice \index{Kondo lattice} model for Ce-compounds}

\begin{figure}[h t b]
\includegraphics[width=.7\columnwidth,angle=0,clip]
{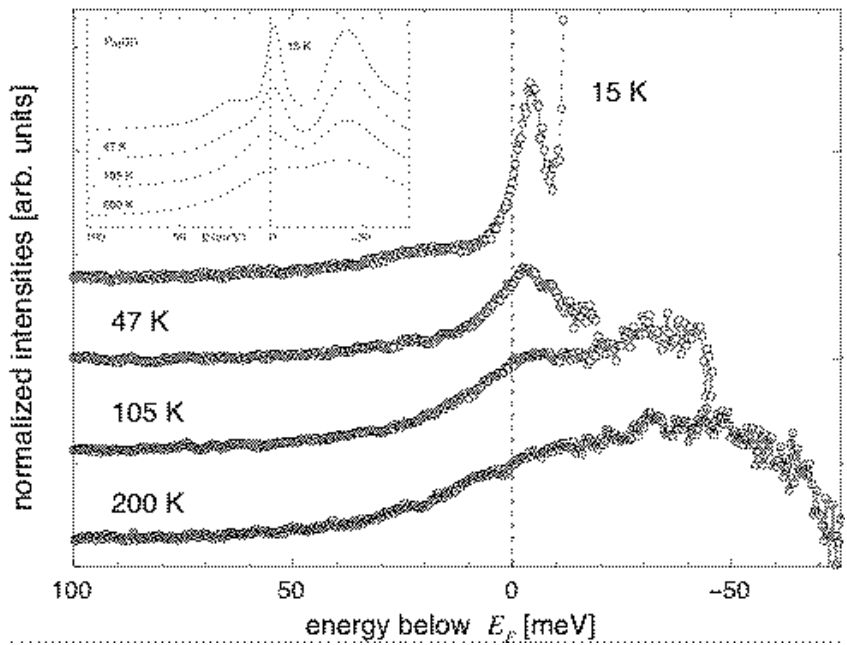}
\caption{PES spectra at various temperatures showing the apperance of the low
temperature Kondo resonance in \index{CeCu$_2$Si$_2$}
CeCu$_2$Si$_2$. The inset shows 4f-spectral density $\rho_{4f}$(E) of
the impurity Anderson model calculated within NCA (Reinert et al., 2001).}
\label{fig:CeCu2Si2KondoResonance}
\end{figure}

The similarities in the behavior of Ce-based heavy-fermion systems to
that of dilute magnetic alloys have led to the assumption that these
systems are `Kondo lattices' where the observed anomalous behavior can
be explained in terms of periodically repeated resonant Kondo scattering. 
This ansatz provides a microscopic model for the formation of a singlet 
ground state and the existence of heavy quasiparticles \index{heavy
fermions}. In addition, it
 explains why there is no magnetic pairbreaking associated with the
presence of the f-electrons. An extensive review is given by
\citeasnoun{HewsonBook}. The Kondo picture for the 
Ce-based heavy-fermion compounds is supported by the fact that both 
the thermodynamic properties at low temperatures (e. g., the specific
heat, the magnetic susceptibility) as well as the temperature-dependence
of the spectroscopic data can be reproduced by an Anderson model with the same
parameters \cite{Gunnarsson83}. The most direct evidence, however,
comes from photoelectron spectroscopy. The
characteristic features of a Kondo system can be summarized as
follows \cite{Allen92,Malterre96}: At high temperatures,
the combined PES/BIS spectra from 
photoemission and inverse photoemission exhibit two distinct
peaks below and above the Fermi energy. These two features correspond
to the valence transitions f\( ^{n}\rightarrow  \) f\( ^{n\pm 1} \), 
respectively. 
The changes in the occupation of the Ce 4f-shells are associated with
energies of order eV. The high-temperature state can be modelled by
weakly correlated conduction electrons which are weakly coupled to
local f-moments. The f-derived low-energy excitations are those of a
system of local moments. 
The direct manifestation of the low-energy scale is the appearance of a sharp
peak in the f-spectral density near the Fermi energy when the
temperature T is smaller than a \index{Kondo temperature}
characteristic (`Kondo') temperature T$^*$. In Ce systems, this 
many-body feature, i.~e., the  `Abrikosov-Suhl' or `Kondo'
resonance, is centered at E\( _{F}\)+kT$^*$ slightly above the 
Fermi edge E\( _{F} \) with T$^*\simeq$ W$\exp(\pi$E$_f$/N$_f\Delta)$ up
to a constant of order unity. Here W is the conduction band width,
E$_f<$ 0 is the f-level position below the Fermi level
E$_F\equiv$ 0, N$_f$ the f-level degeneracy and
$\Delta=\pi|V|^2$N(E$_F$) is the hybridization \index{hybridisation}
or charge fluctuation
width which can be estimated from the width of the transition 4f\(
^{1}\rightarrow \) 4f\( ^{0} \) (V = hybridization matrix element,
N(E$_F$)= conduction electron DOS). This energy scale characterises the
dynamical screening of the impurity spin by conduction electron spin
fluctuations within a ``screening cloud'' that extends to distances $\sim$
$\hbar$v$_F/T^*$ away from the impurity. 
The evolution of the Kondo resonance with temperature
was recently observed by high-resolution photoemission experiments. 
The spectra displayed in fig.~ \ref{fig:CeCu2Si2KondoResonance}
provide direct evidence for the presence of a Kondo resonance
\index{Kondo resonance} in the lattice.
%
%
The resonance is a genuine many-body feature reflecting the small
admixture of f$^0$-configurations to the ground state and the
low-lying excitations which are mainly built from
f$^1$-configurations. By hybridization with conduction states via
transitions f\( ^{1}\leftrightarrow  \)f\( ^{0} \), the
local magnetic degeneracies of the singly occupied 4f-shells are
lifted. The characteristic energy kT$^* \simeq$ 1 - 10 meV which can 
be surprisingly close to
the value of the corresponding dilute system sets the scale for the anomalous
low-temperature behavior. The width and the overall weight of the
resonance are of the order $\pi$kT\( ^{*} \)/N\( _{f} \) and 
\( \pi  \)kT\( ^{*} \)/N\(_{f}\)\( \Delta  \) = 1-n$_f$ $\ll$ 1
respectively. Here n$_f$ denotes the occupation of the f$^1$ state 
which in the Kondo limit $\Delta\ll|E_f|$ is slightly less than one due to
the finite hybridisation matrix element V with conduction electrons. Impurity
model calculations based on the NCA approximation for the Anderson
model indicate that the Kondo resonance states exist as long as
n$_f\geq$ 0.85. For smaller n$_f$ due to larger V one enters the mixed
valent regime.
At sufficiently low temperatures T\( \ll  \)T\( ^{*} \),
the contribution of the narrow resonance peak to the thermodynamic and 
transport properties can be described in terms of a Landau theory with 
heavy fermionic quasiparticles as suggested by the renormalization
group calculations for magnetic impurities immersed in a metallic host
\cite{Wilson75}. Based on the corresponding effective Hamiltonian 
\citeasnoun{Nozieres74} introduced a narrow resonant phase
shift to account for the impurity contribution to the low-energy
properties.

The novel feature observed in stoichiometric Ce-compounds is the
formation of narrow coherent bands of low-energy excitations
\cite{Garnier97,Garnier98,Zwicknagl99}. Following this line of
thought, heavy fermions \index{heavy fermions}  arise from a
decoherence-coherence
crossover. The strong local correlations in Kondo lattices lead to an 
observable many-body effect, i. e., the change with temperature of 
the volume of the Fermi surface. At high temperatures, the \( f
\)-degrees of freedom appear as localized magnetic moments, and the 
Fermi surface contains only the itinerant conduction electrons. 
At low temperatures, however, the \( f \) degrees
of freedom are now tied into itinerant fermionic quasiparticle
excitations and accordingly, have
to be included in the Fermi volume following Luttinger's
theorem. Consequently the Fermi surface is strongly modified. This
scenario \cite{Zwicknagl93a} 
was confirmed experimentally by measurements of the 
de Haas-van Alphen (dHvA) effect \cite{Lonzarich88,Aoki93,Tautz95}.

Competition between the formation of (local) Kondo singlets and the lifting
of degeneracies by long-range magnetic order is clearly evident in 
many Ce-based heavy fermion compounds. In the high-temperature regime 
the moments of the Ce 4f-shells are coupled by the RKKY interaction
which can favor parallel as well
as antiparallel orientation of the moments at neighboring sites. In
the majority of cases, there is a tendency towards antiferromagnetic 
alignment although also ferromagnetic HF systems are
known. Model calculations for two Kondo
impurities \cite{Jayprakash81,Jayprakash82,Jones87} showed that 
antiferromagnetic correlations between the magnetic sites
weaken the Kondo singlet formation reducing the characteristic energy scale
kT\( ^{*} \) to rather small values. Since the latter plays the
role of an effective Fermi energy for the heavy quasiparticles it is
not surprising that the Fermi liquid description fails to be valid in 
systems with sufficiently strong antiferromagnetic correlations.

For Ce-based systems, the natural tuning parameter is the
hybridization which measures the coupling between the strongly
correlated f-electrons and the weakly correlated conduction
states. This coupling, which can be increased by applying hydrostatic
pressure or chemical pressure via proper element substitution, affects
the f-electron system in two different ways. First,
it is responsible for the indirect exchange interaction building up
magnetic correlations between the moments at different sites. On the
other hand, it leads to the formation of local singlets via the Kondo
effect. The energy gain due to magnetic ordering follows a power law
for weak hybridization whereas the Kondo temperature depends
exponentially on the latter. Based on these considerations, 
\citeasnoun{Doniach77}
suggested that for sufficiently weak hybridization the f-derived
magnetic moments should order. With increasing hybridization, however, 
the magnetic ordering temperature should be suppressed and a Fermi
liquid ground
state characterized by Kondo-type correlations forms above a quantum
critical point (QCP) \index{quantum critical point}. At the QCP itself non-Fermi liquid 
behaviour is expected in the temperature dependence of physical
quantities. The resulting schematic ``Doniach'' phase diagram 
has been widely used to understand qualitatively the variation with
pressure of the anomalous low-temperature properties in heavy-fermion
systems. 
It seems that in the majority of Ce-based
heavy-fermion superconductors the superconducting phase develops in
the vicinity of a quantum critical point. 

\subsubsection{Dual model \index{dual model} for U-compounds}

The Kondo picture, however, does not apply in the case of the actinide 
compounds. The difficulties with this model have been discussed in 
\cite{CoxBook}. The difference between the Ce-based heavy-fermion 
compounds and their U-counterparts can be seen directly from the 
photoemission spectra \cite{Allen92}. In U-based heavy-fermion
compounds, the fingerprint character of the transitions 
f\( ^{n}\rightarrow  \) f\( ^{n\pm 1} \)
is lost. Instead of the well-defined f-derived peaks familiar for the 
Ce systems, we encounter a rather broad f-derived feature. This fact 
shows that the f-valence in the actinide heavy-fermion systems is not 
close to integer value as it is the case in Ce-based compounds. In fact, the 
f-valence of the U ions has been discussed rather controversially. 

There is growing evidence that actinide ions may have localized as
well as delocalized \( 5f \) electrons. This picture which was
suggested by susceptibility measurements  
\cite{Schoenes96} is supported by a great variety of experiments
including, e.~g., photoemission and neutron inelastic scattering 
experiments on \index{UPd$_2$Al$_3$} UPd$_{2}$Al$_{3}$ 
\cite{Takahashi95,Metoki98,Bernhoeft98} as well
as muon spin relaxation measurement in UGe$ _{2} $ \cite{Yaouanc02}.
The assumption is further supported
by quantum chemical calculations on uranocene U(C$ _{8} $H$ _{8} $)$ _{2} $
\cite{Liu97} which exhibit a number of low-lying excitations which are due to
intra-shell rearrangements of 5$ f $ electrons. There is clear
evidence that the presence of localized $5f$-states is
even responsible for the attractive interactions leading to 
superconductivity \cite{Sato01}. In addition the dual model should allow 
for a rather natural description of heavy fermion superconductivity 
coexisting with 5f-derived magnetism.

The above-mentioned observations form the basis of the dual model 
which assumes the coexistence of both itinerant and localized 
5$f$ electrons. The
former hybridize with the conduction states and form energy bands
while the latter form multiplets to reduce the local Coulomb
repulsion. The two subsystems interact which leads to the mass
enhancement \index{mass enhancement} of the delocalized
quasiparticles. The situation resembles 
that in Pr metal where a mass enhancement of the conduction electrons
by a factor of 5 results from virtual crystal field
(CEF) excitations of localized 4\( f^{2} \) electrons \cite{White81}.
The underlying hypothesis is supported by a
number of experiments. Detailed Fermi surface studies for UGa$_3$
\cite{Biasini02} show that the experimental data cannot be explained
by assuming all 5f-electrons to be itinerant nor by treating them as
fully localized. The basic assumptions underlying the dual model were recently
confirmed by measurements of the optical conductivity \cite{Dressel02} 
which show the evolution of the high effective mass at low
temperatures. The formation of the heavy quasiparticles \index{heavy
fermions} is also observed in ARPES \cite{Denlinger01}. 

\begin{figure}[h t b]
\begin{center}
\includegraphics[width=.7\columnwidth,angle=0,clip]{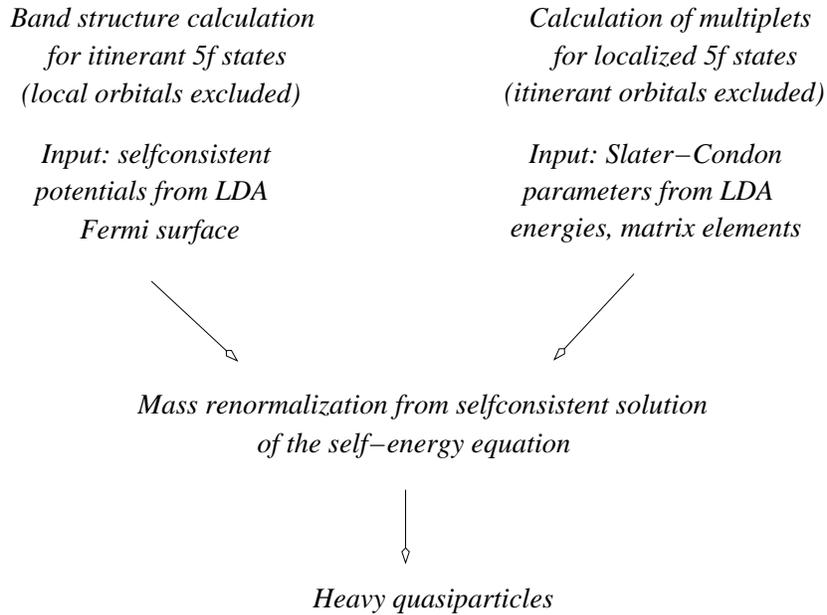}
\end{center}
\caption{%
Calculational scheme for heavy quasiparticles in U-based heavy-fermion 
compounds
}
\label{fig:DualModelScheme}
\end{figure} 

The dual model provides a microscopic theory for the heavy
quasiparticles in U compounds. The method reproduces the dHvA data in 
UPt$_3$ \cite{Zwicknagl02} and UPd$_2$Al$_3$ \cite{Zwicknagl03}. 
The calculation of the heavy bands proceeds in three steps 
schematically summarized in fig.~\ref{fig:DualModelScheme}. 
First, the band structure \index{band structure} is determined  by solving
the Dirac equation for the self-consistent LDA potentials 
but excluding two U 5$ f $~ (j=$\frac{5}{2}$) states from forming
bands. The choice of the itinerant and localized orbitals depends upon 
the symmetry of the crystal and the hybridization strengths.
The localized 5$ f $ orbitals are accounted for in the 
self-consistent density and accordingly in the potential seen by 
the conduction electrons. The 
intrinsic bandwidth of the itinerant U 5$ f $ j=$ \frac{5}{2} $
electrons is taken from the LDA
calculation while the position of the corresponding
band center is chosen such that the density distribution of the conduction
states as obtained within LDA remains unchanged. For \index{UPt$_3$}
UPt$_3$ and \index{UPd$_2$Al$_3$} UPd$_2$Al$_3$, the f-occupancy per U
atom for the delocalized 5f-electrons amounts to n$ _{\textrm{f}} $
= 0.65 indicating that we are dealing with a mixed valent situation. 
The calculations yield the dHvA frequencies which can be directly
compared with experimental data. In the second step, the localized 
U 5$ f $ states are calculated by diagonalizing the Coulomb matrix in
the restricted subspace of the localized 5$f$ states.
Assuming the jj-coupling scheme, the
Coulomb matrix elements are calculated from the radial functions of
the ab-initio band structure potentials. In a crystal, the degeneracies of the 
ground-state may be lifted by a CEF. This is in fact the case for
UPd$_2$Al$_3$ where the resulting singlet ground state \index{singlet
CEF state} of the localized 5$f^2$
is given by $|g\ra=(1/\sqrt{2})(|J_z=3\ra +|J_z=-3\ra)$ in the J = 4 subspace.
The coupling between the localized and delocalized 5$f$ electrons is
directly obtained from the expectation values of
the Coulomb interaction H$ _{\textrm{C}} $ in the 5$ f^{3} $
states. Finally, the renormalization of the effective masses
which results from the coupling between the two $ 5f $ subsystems is
determined. The enhancement factor is calculated from the self-consistent
solution of the self-energy equation \cite{White81} with the input
taken from the ab-initio electronic structure calculations for the
delocalized and the localized 5$f$ electrons. 

The coexistence of itinerant and localized 5f states is a consequence
of the interplay between hybridization with the conduction electrons
and local Coulomb correlations. This ''partial localization'' of the
5f states is found in many actinide intermetallic compounds. The
underlying microscopic mechanism is an area of active current
research  \cite{Lundin00,Soederlind00}. LDA calculations
show that the hopping matrix elements for different $5f$ orbitals
vary. But it is of interest to understand why only the largest one of 
them is important and why the other ones are suppressed.

Partial localization may arise from the competition between hopping
and angular correlations. This can be seen by exact diagonalization of 
small clusters
which model the U sites in heavy fermion compounds \cite{Efremov03}. 
We keep only the degrees of freedom of the 5f shells the conduction 
states being 
accounted for by (effective) anisotropic
intersite hopping. The Hamiltonian reads
\begin{equation}
H  =  - \sum _{\langle nm\rangle, j_{z}}\, t_{j_{z}}\left( 
  c_{j_{z}}^{\dagger}(n)\,c_{j_{z}}(m)+h.c.\right) +
  H_{\mathrm{C}}
\label{eq:TwoSiteHamiltonian}
\end{equation}
where the first sum is over neighboring sites $\langle nm\rangle$. 
Furthermore $c_{j_{z}}^{\dagger}(n)$ ($c_{j_{z}}(n)$), creates (annihilates) an
electron at site $n$ in the $5f\ j = 5/2$ state with $j_z=-5/2,\dots,5/2$. We will
consider two and three sites models. Since the relevant correlations
are local the results
for these small clusters are qualitatively similar to those of
four-site models. The effective hopping between sites results from
the hybridisation \index{hybridisation} of the 5$f$ states with the
orbitals of the ligands
and depends generally on the crystal structure. Rather than trying to
exhaust all possible different lattice symmetries, we shall concentrate
here on the special case that hopping conserves $j_z$. While this is
certainly an idealization, it allows us to concentrate on our main
interest, i.~e., a study of
the influence of atomic correlations on the renormalization of
hybridization matrix elements.
The parameters $t_{j_z}(=t_{-j_z})$ are chosen 
in accordance with density-functional calculations for bulk material
which use $jj_z$ basis states. 
The local Coulomb interactions \index{Coulomb interaction} can be
written in the form
\begin{equation}
\label{eqCoulomb}
  H_{\mathrm{C}}= \frac{1}{2}
  \sum_{n} 
  \sum_{j_{z1},\dots,j_{z4}} 
  U_{j_{z1}j_{z2},j_{z3}j_{z4}}
  c_{j_{z1}}^{\dagger }(n)\,c_{j_{z2}}^{\dagger }(n)\,c_{j_{z3}}(n)\,
  c_{j_{z4}}(n)
\end{equation}
with Coulomb matrix elements 
\begin{equation}
  U_{j_{z1}j_{z2},j_{z3}j_{z4}}
  = \delta_{j_{z1}+j_{z2},j_{z3}+j_{z4}} \
  \sum_J {\textstyle \langle \frac{5}{2}\,\frac{5}{2}\,j_{z1}\,j_{z2} | 
                     JJ_{z} \rangle }
  U_{J}  {\textstyle \langle JJ_{z}| \frac{5}{2}\,\frac{5}{2}
           \,j_{z3}\,j_{z4} \rangle }\quad.
\end{equation}
Here $J$ denotes the total angular momentum of two electrons and 
$J_{z}=j_{z1}+j_{z2}=j_{z3}+j_{z4} $. The sum is restricted 
by the antisymmetry of the Clebsch-Gordan coefficients 
\( \langle \frac{5}{2}\,\frac{5}{2}\,j_{z1}\,j_{z2}| JJ_{z}\rangle  \)
to even values $ J=0,2,4 $. 
We use in the actual calculations \( U_{J} \) values 
which are determined from LDA wavefunctions for 
UPt\( _{3} \) \cite{Zwicknagl02}, i.~e.,~$U_{J=4}$ = 17.21 eV, 
$U_{J=2}$ = 18.28 eV, and $U_{J=0}$ = 21.00 eV.
We expect $U_{J=4}<U_{J=2}<U_{J=0}$
always to hold for Coulomb interactions, independently of the
chemical environment. In contrast,
the relative order of the hopping matrix elements
will vary strongly from one compound to the next. The average Coulomb
repulsion of about 20\,eV is irrelevant for the low-energy physics of
the model. It simply restricts the relevant configurations to states such that
each site is occupied either by 2 or 3 f electrons.
The low-energy sector is exclusively determined by the differences of the 
$U_J$ values, which are of the order of 1\,eV and thus slightly larger
than typical bare $f$~- band widths. The latter are obtained, e.g., from LDA
calculations for metallic uranium compounds like UPt$_3$. Restricting
the model to $f^2$ and $f^3$ configurations is equivalent to let
the various $U_J\to \infty$ while their differences remain
finite. To mimic the situation in the U-based heavy-fermion compounds
we consider the intermediate valence regime. Note that in the absence
of a magnetic field all states of the two-site model with five electrons  
will be at least doubly degenerate because of Kramers' degeneracy.

The Hamiltonian eq.~(\ref{eq:TwoSiteHamiltonian}) conserves 
 ${\cal J}_z=\sum_n J_z(n)$ where ${\cal J}_z$ is the z-component of the total 
angular momentum of the system and the
$J_z(n)$ refer to angular momentum projections
on individual sites. We shall therefore characterize the
eigenstates by their ${\cal J}_z$ value. Since 
$t_{j_z}=t=\mbox{\rm const}$ the system is rotationally invariant. Then
$\mbox{\boldmath $\bf \cal J$}^2$ provides an additional good quantum number. 
Strong on-site correlations result in a considerable enhancement of anisotropies
in the bare hopping matrix elements. This can lead to 
a localization of electrons in orbitals with relatively weak hybridization.
The latter is effectively reduced to zero in those cases.

In order to quantify the degree of localization or, alternatively, of the 
reduction of hopping of a given $j_z$ orbital by local correlations,
we calculate the ratio of the $j_z$- projected kinetic energy $T_{j_z}$  
and the bare matrix element $t_{j_z}$ and obtain
\begin{equation} 
\label{eqDefDeltaTJ}
    \frac{T_{j_z}}{t_{j_z}}
= 
  \sum_{\langle nm \rangle, \pm} \langle\Psi_{\mathrm gs} | 
  (  c^\dagger_{\pm j_z}(n)\,  c_{\pm j_z}(m) + h.c. ) 
|\Psi_{\mathrm gs} \rangle 
  \nonumber 
  \, . 
\end{equation}
The ground-state wavefunction $| \Psi_{\mathrm gs} \rangle$ contains
the strong on-site correlations. A small ratio of $T_{j_z}/t_{j_z}$
indicates partial suppression of hopping for electrons in the
$\pm j_z$ orbitals. Two kinds of correlations may contribute to that
process. The first one is based on the reduction of charge
fluctuations to atomic $f^2$ and $f^3$ configurations. This is a result
for large values of $U_J$ and can be studied by setting all $U_J$ equal
to a value much larger than the different $t_{j_z}$. The second one is due
to differences in the $U_J$ values, i.~e.~, $U_{J=4}<U_{J=2}<U_{J=0}$.
The differences in the $U_J$ values are the basis of Hund's
rules. \index{Hund's rules}
Hopping counteracts Hund's rule correlations and vice versa.
What we want to stress is the fact that those correlations can lead to a 
complete suppression of hopping channels except for the dominant 
one which shows only little influence.

\begin{figure}[h t b]
\includegraphics[width=.5\columnwidth,angle=270,clip]{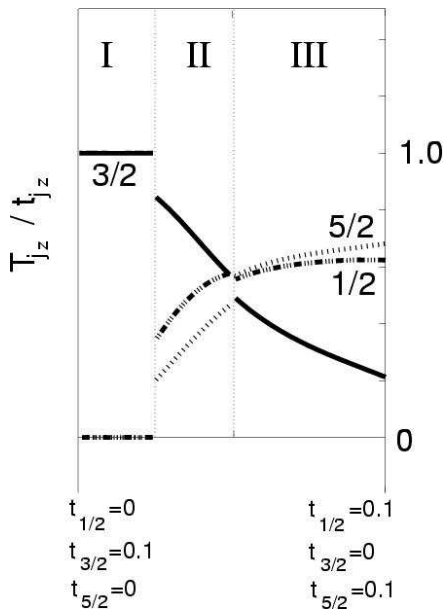}
\caption{Values  $T_{j_z}/t_{j_z}$ for the two-site cluster along the
line connecting linearly the points written below the figure. 
Regions with ${\cal J}_z=15/2$, 5/2 and 3/2, are labeled with I, II
and III respectively (Efremov et al. 2003).}
\label{fighop1}
\end{figure} 

Results for the ratios $T_{j_z}/t_{j_z}$ are shown in 
fig.~\ref{fighop1} for a two-site model. As the relevant correlations
are local the general results qualitatively agree with those found for 
a three-site cluster and four-site clusters \cite{Pollmann03}. 
We can distinguish three different regimes with ${\cal J}_z=15/2$, 5/2 and 3/2,
labeled I, II and III respectively. One observes that
in region I only the dominant hybridization of the $j_z=3/2$ orbital
survives while that of the $j_z=1/2$ and $j_z=5/2$
orbitals is completely suppressed. On the other hand in regions II and
III the correlation
effects on different orbitals are are not very different. These
findings demonstrate that in particular Hund's rule correlations   
strongly enhance anisotropies in the hopping. For a certain 
range of parameters this may result in a complete suppression of the 
effective hopping except for the largest one, which remains almost 
unaffected. This provides a microscopic justification of partial
localization of 5$f$ electrons which is observed in a number of
experiments on U compounds and which is the basis for further model
calculations described later.

\subsubsection{Fermi-liquid \index{Fermi liquid} state and heavy
quasiparticles: \index{heavy fermions} renormalized band theory
\index{band structure}}

Neither the Kondo model for Ce-based heavy-fermion systems nor the
dual model for their U-based counterparts can be solved for extended
systems described by lattice models. The excitation spectra of Ce metal in the 
$\alpha$- and $\gamma$-phase were recently calculated applying the dynamical 
mean field theory \cite{Held01}. The method yields a low-energy
resonance but the experimentally observed change of the f-valence at
the transition is not reproduced. A fully microscopic description is
not available to describe experiments where inter-site
effects become strongly manifest. Typical lattice effects are the
formation of coherent heavy quasiparticle bands whose Fermi surfaces
were observed experimentally. 

The Landau Fermi liquid (FL) \index{Fermi liquid} theory assumes a
one-to-one correspondence between
the states of the complex interacting system and those of a gas of independent
fermions which may move in an external potential,
\citename{Landau56} \citeyear{Landau56,Landau57,Landau58},
\citeasnoun{Abrikosov75}. The single-particle orbitals
and energies are determined from an effective Hamiltonian. The characteristic
properties of a system are reflected in an effective and not necessarily local
potential \( V_{eff} \) which describes the field of the nuclei and the modifications
arising from the presence of the other electrons. The essential many-body aspects
of the problem are then contained in the prescription for constructing the effective
potentials which have to be determined specifically for the problem under consideration.

The quasiparticle energies reflect the interaction among the fermions and therefore
may be altered when the overall configuration is changed. A characteristic feature
of interacting Fermi liquids is that the energy dispersion 
\( \epsilon_{\sigma }({\bf k}) \) depends on how many other
quasiparticles are present, 

\begin{equation}
\label{eq:evonk}
\epsilon_{\sigma }({\bf k})=E({\bf k})+
\sum _{{\bf k}'\sigma '}f_{\sigma \sigma '}({\bf k},{\bf k}')
\delta n_{\sigma '}({\bf k}')\quad .
\end{equation}

Here \( E({\bf k}) \) denotes the energy dispersion of a dilute gas of quasiparticles.
In systems with strong correlations it reflects the interactions and hence cannot
be calculated from the overlap of single-electron wave functions. The interactions
among the quasiparticles are characterized by the matrix \( f_{\sigma
\sigma '}({\bf k},{\bf k}') \). The deviations from the equilibrium
distribution are given by \( \delta n_{\sigma }({\bf k}) \).

Interaction effects must be accounted for in considering those situations where
the quasiparticle distribution function deviates from that of the equilibrium
case. In the phenomenological Landau FL theory the characteristic
properties of the quasiparticles, which can hardly be calculated
microscopically, are expressed in terms of parameters which are
determined from experiment.
Examples are the effective potentials, the interaction potential and the scattering
amplitudes. An important property of the quasiparticles is that they can be
considered as 'rigid' with respect to low-energy and long-wavelength
perturbations. Only such processes can be described within this
theoretical framework.

The energy dispersion \( E({\bf k}) \) of a dilute gas of noninteracting quasiparticles
is parametrized by the Fermi wave vector \( {\bf k}_{F} \)
\index{Fermi wave vector} and the Fermi velocity \index{Fermi
velocity} \( {\bf v}_{F} \)
\begin{equation}
\label{eq:Qtdis}
E({\bf k})={\bf v}_{F}(\hat{\bf k})\cdot ({\bf k}-{\bf k}_{F})
\end{equation}
 where \( \hat{\bf k} \) denotes the direction on the Fermi surface. The key idea
of the renormalized band method is to determine the quasiparticle states by
computing the band structure for a given effective potential. Coherence effects
which result from the periodicity of the lattice are then automatically accounted
for. The quantities to be parametrized are the effective potentials which include
the many-body effects. The parametrization of the quasiparticles is supplemented
by information from conventional band structure calculations as they are performed
for ``ordinary'' metals with weakly correlated electrons. The periodic potential
leads to multiple-scattering processes involving scattering off the individual
centers as well as the propagation between the centers which mainly depends
on the lattice structure and is therefore determined by geometry. The characteristic
properties of a given material enter through the information about single center
scattering which can be expressed in terms of properly chosen set of phase
shifts \( \{\eta _{\nu }^{i}(E)\} \) \index{phase shifts}
specifying the change in phase of a wave incident on site i with energy E and
symmetry \( \nu  \) with respect to the scattering center. Within the scattering
formulation of the band structure problem the values of the phase shifts at
the Fermi energy \( \{\eta _{\nu }^{i}(E_{F})\} \) together with their derivatives
\( \left\{ \left(d\eta _{\nu }^{i}/dE\right)_{E_{F}}\right\}  \)
determine the Fermi wave vectors \( {\bf k}_{F} \) and the Fermi
velocity \( {\bf v}_{F} \).

\begin{figure}[h t b]
\begin{center}
\includegraphics[width=.7\columnwidth,angle=0,clip]{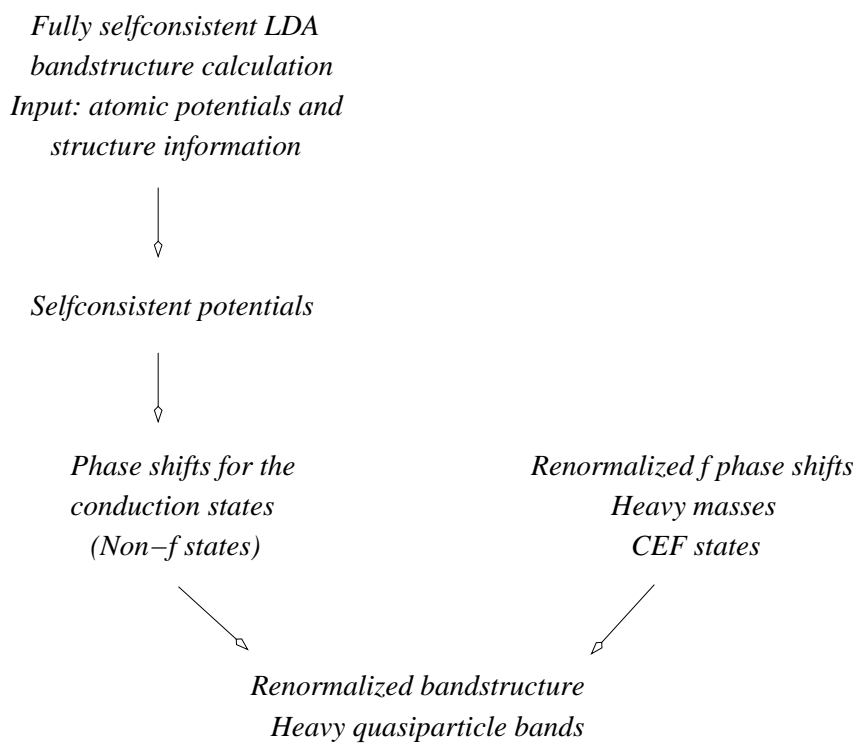}
\end{center}
\caption{
Schematic summary of renormalized band calculation for metals with
strongly correlated electrons
}\label{fig:RenBandScheme}
\end{figure} 

The calculation of realistic quasiparticle bands proceeds in several steps as
schematically summarized in fig.~\ref{fig:RenBandScheme}. The first 
step is a standard LDA band structure calculation \index{band
structure} by means of which the effective single-particle
potentials are self-consistently generated. The calculation starts, like any
other ab-initio calculation, from atomic potentials and structure information.
In this step, no adjustable parameters are introduced. The effective potentials
and hence the phase shifts of the conduction states are determined from first
principles to the same level as in the case of `ordinary' metals. The f-phase
shifts \index{phase shifts} at the lanthanide and actinide sites, on the other hand, are
described by a resonance type expression
\begin{equation}
\label{eq:efren}
\tilde{\eta }_{f}\simeq \arctan \frac{\tilde{\Delta }_{f}}{E-\tilde{\epsilon }_{f}}
\end{equation}
which renormalizes the effective quasiparticle mass.
One of the two remaining free parameters \( \widetilde{\epsilon }_{f} \)and
\( \tilde{\Delta }_{f} \) is eliminated by imposing the condition that the
charge distribution is not significantly altered as compared to the LDA calculation
by introducing the renormalization. The renormalized band method devised to
calculate the quasiparticles in heavy-fermion compounds thus is essentially
a one-parameter theory. We mention that spin-orbit and CEF splittings
can be accounted for in a straight-forward manner \cite{Zwicknagl92}.

\subsubsection{Quasiparticle interactions and spin fluctuation theory
\index{spin fluctuation}}

The low-energy excitations of heavy-fermion systems are described in terms of
quasiparticle bands which yield the high density of states and
specific heat $\gamma$- value. The preceding sections focussed on the case of a dilute
gas of quasiparticles whose energy dispersion can be explicitly calculated
by means of the renormalized band method. The many-body effects, however, lead
to deviations from the picture of independent fermions which fill the rigid
bands of the dilute gas of quasiparticles. An important consequence of
the quasiparticle interactions is the instability of the normal Fermi
liquid \index{Fermi liquid} with respect to charge or spin density waves
or superconductivity.  In heavy-fermion compounds,  quasiparticle
interactions are strongly evident in the electronic compressibility  
whose values are comparable to those of a normal metal and do not
reflect the enhancement of the specific heat. This experimental
fact indicates that there must be a strong  repulsion between two 
quasiparticles at the same lattice site.

The influence of the quasiparticle interactions on observable
quantities is usually described in terms of a
small set of interaction parameters. According to Landau, the 
compressibility $\kappa$ and susceptibility $\chi_s$ are given by
\begin{equation}
\frac{\kappa}{\kappa_0}=\frac{ \frac{m^*}{m}}{ 1+F_0^s} 
\qquad \mbox{and} \qquad
\frac{\chi_s}{\chi_s^0}=\frac{ \frac{m^*}{m}}{ 1+F_0^a} 
\end{equation}
where $\kappa_0$ denotes the compressibility of independent fermions
of mass $m$. From 
$\frac{\kappa}{\kappa_0} \simeq 1$ follows the order of magnitude estimate
$F_0^s \simeq \frac{m^*}{m} \simeq 10^2-10^3$ for the spin-independent
isotropic part of the quasiparticle interactions. The spin-dependent 
isotropic part of the interaction, on the other hand,
is reflected in the enhancement of the spin susceptibility $\chi_s$
over the independent quasiparticle value $\chi_s^0$.
From the observed value $\chi_s/\chi_s^0 \simeq m^*/m$ 
we infer that the enhanced spin susceptibility 
simply reflects the high density of quasiparticle states and that the
corresponding Landau FL parameter $F_0^a$ \index{Fermi liquid
parameters} plays only a minor role in
heavy-fermion compounds. Although these results are not sufficient to
specify the effective interaction in an anisotropic Fermi liquid they 
nevertheless impose important constraints on theoretical models.

\begin{figure}[h t b]
\begin{center}
\includegraphics[clip,width=12cm,angle=0]{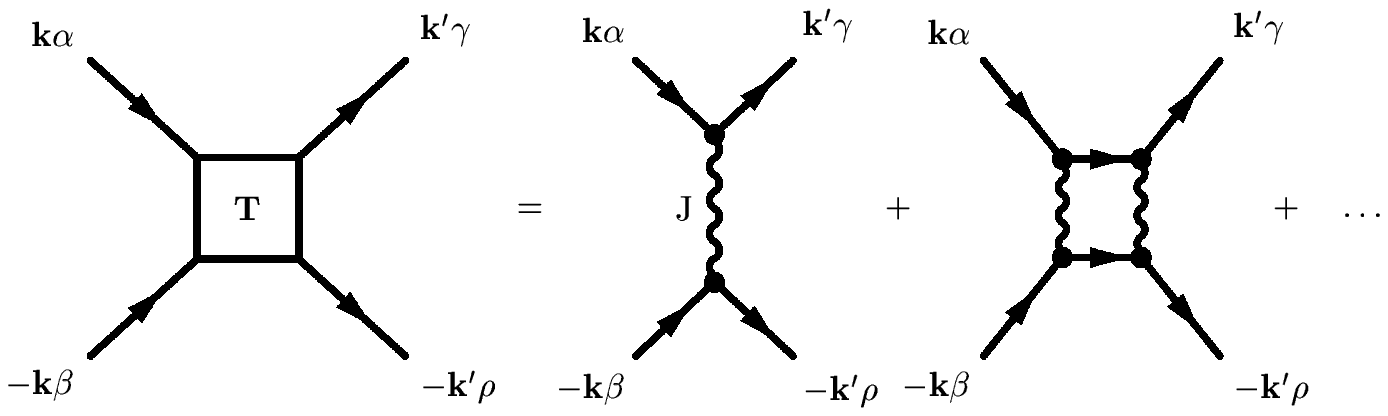}
\caption{
Left: two-particle (e-e) scattering matrix \( T_{\alpha \beta ;\gamma \rho }
\left({\bf k}_{1}{\bf k}_{2};{\bf k}_{3}{\bf k}_{4}\right)  \) for
quasiparticle scattering with initial momenta 
\( {\bf k}_{3} {\bf k}_{4} \) and spins \( \gamma \rho  \) to final
momenta \( {\bf k}_{1} {\bf k}_{2} \) and spins \( \alpha \beta
\).  Right: SC transition corresponds to singular pair scattering with
opposite momenta  \( {\bf k}_{3}=-{\bf k}_{4}={\bf k}' \) and
\( {\bf k}_{1}=-{\bf k} _{2}={\bf k}  \). It is approximated by ladder
diagrams involving the irreducible (with respect to e-e scattering)
four-point vertex for the effective interaction (wiggly line) which is
the central quantity in microscopic models of SC, see also
fig.~\ref{fig:RPADiagrams} and eqs.~(\ref{HEFFSF}),(\ref{EFFINT}).}
\label{fig:PairScattering}
\end{center}
\end{figure} 

The superconducting transition in a Fermi liquid is determined by a 
singularity in the scattering matrix
of two quasiparticles with opposite momenta at the Fermi surface.
In Landau FL theory a two-particle scattering matrix $T_{\alpha
\beta;\gamma \rho}({\bf k}_1 {\bf k}_2;{\bf k}_3 {\bf k}_4)$ is
introduced which describes the scattering of two quasiparticles with
momenta ${\bf k}_3 {\bf k}_4$ and spins $\gamma \rho$ into states with
momenta ${\bf k}_2 {\bf k}_1$ and spins $\alpha \beta$ on the Fermi
surface (see fig.~\ref{fig:PairScattering}). 
Assuming rotational invariance in spin space the general scattering matrix
$T_{\alpha,\beta; \gamma,\rho}$ can be expressed in terms of two
scalar amplitudes. Conventional choices are either the singlet and triplet
amplitudes in the particle-particle channel 
($\gamma \rho \rightarrow \alpha \beta$), $T_s$ and $T_t$ or alternatively, 
the singlet and triplet amplitudes \index{singlet pairing}
\index{triplet pairing} in one of the particle-hole channels
\index{electron-electron pairing} \index{electron-hole pairing}, $T^{(s)}$ and $T^{(a)}$,
\begin{eqnarray}
T_ {\alpha \beta; \gamma \delta} & = &
 -\frac{1}{2} \left(\sigma_2\right)_{\alpha \beta}  
\left(\sigma_2\right)_{\gamma \rho} T_s
+\frac{1}{2} \left(\bm{\sigma}\sigma_2\right)_{\alpha \beta} \cdot 
\left({\bm \sigma}\sigma_2\right)_{\gamma \rho} T_t \nonumber\\
  & = & \delta_{\alpha \beta} \delta_{\gamma \rho} T^{(s)}+
 {\bm \sigma}_{\alpha \beta}\cdot {\bm \sigma}_{\gamma \rho} T^{(a)}
\end{eqnarray}
where $\sigma_\mu; \mu=1-3$ denote the Pauli matrices. The two sets
of scalar amplitudes are related by
\begin{equation}
\label{CROSS}
T_s=T^{(s)}-3T^{(a)} \quad ; \quad T_t=T^{(s)}+T^{(a)}\nonumber\\
T^{(s)}=\frac{1}{4}(T_s+3T_t) \quad; T^{(a)}=\frac{1}{4}(T_t-T_s) 
\end{equation}
The symmetries can be obtained directly from the symmetries of the
two-particle Green's function. The Landau FL parameters \index{Fermi
liquid parameters} F$^s_0$ and
F$^a_0$ are determined by the isotropic Fermi surface averages
A$_0^{s,a}$ = $\la T^{s,a}(\v k_1\v k_2;\v k_1\v k_2)\ra_{FS}$ of 
the two particle forward-scattering amplitude by
F$_0^{s,a}$ = A$_0^{s,a}$[1-A$_0^{s,a}$]$^{-1}$.
As mentioned above, the transition to the
superconducting state is caused by a singularity in the scattering of
a pair with opposite momenta ${\bf k}_3=-{\bf k}_4={\bf k}'$ into pair states with 
${\bf k}_1=-{\bf k}_2={\bf k}$. 

The Fermi liquid  approach which attempts to construct phenomenological
models for the scattering amplitude
has been reviewed \cite{Fulde88a,Zwicknagl92}. In the present review
we shall rather concentrate
on microscopic models for the effective pairing interaction.
 
Microscopic theories of superconductivity focus on the attractive
interaction which results from the motion of the (heavy) quasiparticle
through a polarizable medium. The theories can be divided into two
major groups depending on whether the polarizable medium is distinct
from the particles which are participating in the attraction or not. 
The BCS theory \cite{Bardeen57,Schrieffer64} belongs to the former class
where Cooper-pair formation is due to quasiparticle-phonon
interactions which is appropriate for common metals. 
The spin fluctuation models assume that
pairing is mediated by overdamped low lying magnetic excitations with a
prefered wave vector in the itinerant electron system.
The dual model for U-based heavy-fermion systems suggests a novel
mechanism for the Cooper-pair formation, i.e., the exchange of weakly damped
propagating magnetic excitons. In this case, the polarizable medium is
provided by the CEF-split multiplets of the localized 5f-electrons. A 
model calculation for UPd$_2$Al$_3$ is presented in sect.~(\ref{Sect:UPd2Al3}).
where we also compare theoretical predictions with experimental
observations. 

The majority of theoretical calculations for heavy-fermion systems  
adopts spin-fluctuation models for the quasiparticle
attraction which belong to the second group. Early
theoretical efforts \cite{Anderson84} emphasized an analogy with
superfluid $^3$He. Even though a quantitative
microscopic theory of suprafluidity is still lacking for $^3$He it is
generally accepted that spin fluctuations qualitatively explain
observed features in $^3$He, in particular odd-parity spin-triplet
pairing \cite{AndersonBook,Vollhardtbook}. 
The influence of spin fluctuations on the Cooper pair formation in
metals was studied first by \citeasnoun{Berk66} and
\citename{Layzer71} \citeyear{Layzer71,Layzer74}. It has found its most
prominent applications in HF systems \cite{Miyake86} and especially in the
theory of high-T$_c$ superconductors \cite{Monthoux91,Monthoux94}.
The calculations proceed in close analogy to the strong-coupling
theory of phonon-mediated superconductivity. At this point we have to add that
the theory of spin fluctuation-mediated superconductivity does not
have a natural small parameter which would allow for a systematic
asymptotic expansion. In particular, there is no a priori
justification for including only selected contributions in a
diagrammatic expansion for the scattering matrix. 

\begin{figure}[h t b]
\begin{center}
\includegraphics[clip,width=4cm,angle=90]{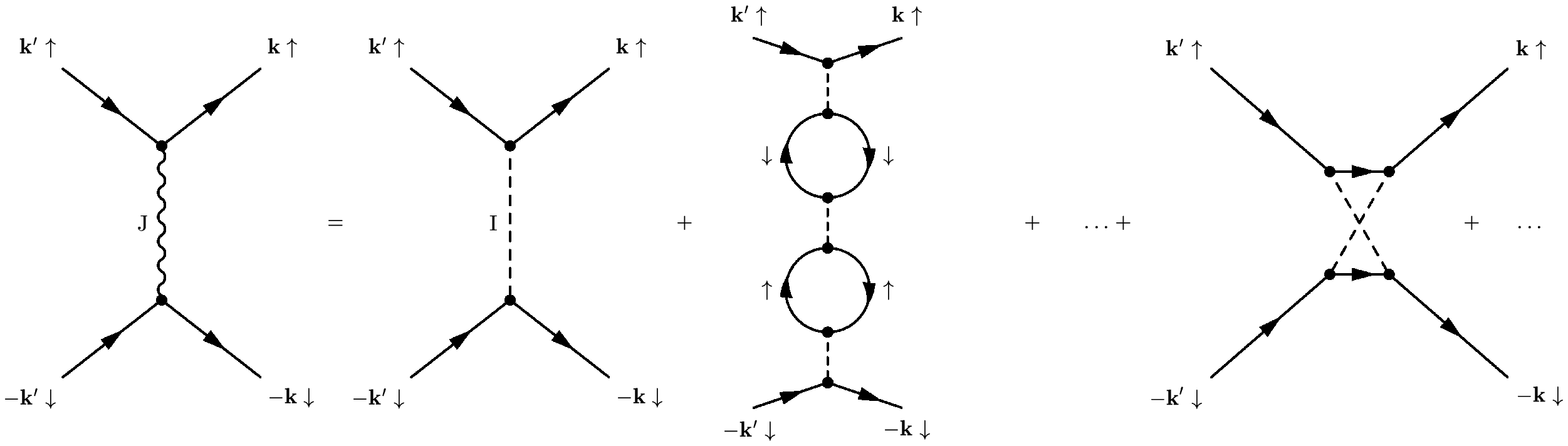}
\end{center}
\caption{
RPA diagrams for the effective interaction $J(\v k-\v k')$
\index{effective interaction} between
quasiparticles with opposite spins. The bare instantaneous interaction
is denoted by I (dashed line) as in eq.~(\ref{HUBB}). Contribution from polarisation
(bubble) diagrams contains odd number and from (maximally crossed)
exchange diagrams contain any number of interaction lines (I). 
For equal quasiparticle spins the RPA yields only a sum of bubble
diagrams with an even number of interaction lines.}  
\label{fig:RPADiagrams}
\end{figure}

The pair scattering for two quasiparticles with opposite momenta at
the Fermi surface is evaluated adopting the ladder approximation
displayed in fig.~\ref{fig:PairScattering}. The
problem is therefore reduced to finding the four-point vertex which is
irreducible with respect to particle-particle scattering. The basic
assumption is that the important structure in the scattering
amplitudes comes from exchange of collective modes in the two
particle-hole channels. 
The central quantity of these theories is the dynamic magnetic
susceptibility $\chi({\bf q},\omega)$ which can be determined - in
principle - by inelastic neutron scattering. Close to a magnetic
instability, the susceptibility diverges for $\omega \rightarrow 0$ at
some wave vector ${\bf Q}$. This behaviour results from a singularity
in the particle-hole scattering matrix which accounts for the dynamic
effects of the induced spin polarizations. The strong particle-hole
correlations affect the pair scattering amplitude in the
particle-particle channel. 
In explicit calculations, the effective interaction associated with
$\chi({\bf q},\omega)$ originates in a residual on-site quasiparticle
repulsion (I) described by the single band Hubbard model
\index{Hubbard model} for the heavy
quasiparticle bands $\epsilon_{\vk}$.
\begin{equation}
\label{HUBB}
H_I=\sum_{\v k,s}\epsilon_{\vk}n_{\vk} + 
I\sum_in_{i\uparrow}n_{i\downarrow}
\end{equation}
This is also the starting point in the theory of high-T$_c$ SC,
\index{high-T$_c$} albeit with bare Coulomb repulsion \index{Coulomb
interaction} and 2D quasiparticle band. In the
fluctuation exchange (FLEX) approach \cite{Pao94,Dahm95} the effective pair
interactions and gap equations are obtained and solved selfconsistently in the
context of a strong coupling approach. For the complicated 3D Fermi
surfaces of HF materials this is too difficult and one is restricted to
nonretarded model calculations. In this context the effective
interaction is obtained within the random phase approximation (RPA)
for pairs of quasiparticles with opposite and equal spins from the diagrams
displayed in fig.~\ref{fig:RPADiagrams}. To conserve
rotational invariance in spin space the maximally crossed
particle exchange contributions in fig.~\ref{fig:RPADiagrams}
have to be included. These terms were
studied by \citeasnoun{Berk66} who considered the influence of
spin-fluctuations on singlet pairing. The case of triplet scattering
 was discussed by \citeasnoun{Nakajima73} for $^3$He. He showed that
sufficiently close to a magnetic instability, the complete set of 
diagrams yields an effective spin-dependent interaction which is 
rotationally invariant in spin space \index{effective interaction}
\begin{equation}
\label{HEFFSF}
H_{eff}=\frac{1}{2} \sum_{{\bf k},{\bf k}' }J({\bf k}-{\bf k}')\  
{\bm \sigma}_{\alpha \gamma}\cdot {\bm \sigma}_{\beta \rho}\ 
c_{{\bf k}\alpha}^{\dagger}\  c_{-{\bf k}\beta}^{\dagger}\ 
c_{-{\bf k}'\rho}\  c_{{\bf k}'\gamma}
\end{equation}
where
\begin{eqnarray}
\label{EFFINT}
J(\v q)=-\frac{1}{2}\frac{I}{1-I\chi_0(\v q)}
\simeq -\frac{1}{2}I^2\chi(\v q) 
\qquad \mbox{and} \qquad 
\chi(\v q)=\frac{\chi_0(\v q)}{1-I\chi_0(\v q)}
\end{eqnarray} 
Here $\chi_0({\bf q})$ is the static susceptibility of 
the non-interacting quasiparticles and $\chi({\bf q})$ the RPA
susceptibility of the interacting system. The approximation is valid
for the enhanced spin fluctuation regime I$\chi_0({\bf q})\leq$ 1. In
this regime the effective pairing is therefore
completely determined by the collective static magnetic susceptibility
$\chi({\bf q})$. 

Assuming a spherical FS the appropriate basis function for the SC
order parameter are spherical harmonics of angular momentum l. The
interaction in the l-wave channel is given by (P$_l$= Legendre polynomial)
\begin{eqnarray}
\label{ANGINT}
V_l=a_l2\int_{0}^{1}dx xP_l(1-2x^2)[-J(2k_Fx)]
\end{eqnarray}
with x=1-(q/2k$_F$)$^2$ and a$_l$=3 (l even) or a$_l$ =-1 (l odd). 
If the effective interaction -J(\v q) strongly peaks for small q (x
$\simeq$ 1), i.e. for FM spin fluctuations, the integrand is positive
in this region and hence V$_l<$ 0 for odd l and  V$_l>$ 0 for even l.

Assuming that in HF systems the quasiparticle interaction is mediated 
by FM spin fluctuations one would expect odd-parity SC order
parameters. For UPt$_3$ however, the odd parity state was originally
claimed to be inconsistent with the power law behavior observed 
in thermodynamic and transport
properties at low temperatures. This statement was based on the 
assumption that the superconducting phase should be characterized 
by one of the symmetry-adapted order parameters which were derived 
in the limit of strongly coupled orbital and
spin moment of the Cooper pair \cite{Volovik85,Ueda85,Sigrist91}. We shall 
comment on this subtle point in the next section. 

The discovery of antiferromagnetic spin fluctuations in UPt$_3$
\cite{Aeppli87} prompted \citeasnoun{Miyake86} to study the
nature of pairing due to antiferromagnetic spin fluctuations. The
authors consider pairing in a single-band model with the effective electron
interaction of eq.~(\ref{HEFFSF}) where the nonretarded interaction
$J({\bf q})$ should have a maximum at an AF wave vector \v Q, e.g.
\vQ =($\frac{1}{2},\frac{1}{2},\frac{1}{2}$) or \v Q
=($\frac{1}{2},\frac{1}{2},0$). In the vicinity of its maximum at \v Q, J(\v q)
is approximated by
\begin{equation}
-J({\bf q})=J_0-J_1 \gamma_{\bf q}
\end{equation}
where $J_0$, $J_1$ are positive constants and the function
$\gamma_{\bf q}$ has a minimum at  ${\bf q}={\bf Q}$.
Examining this model for a parabolic single band 
\citeasnoun{Miyake86} demonstrated that the resulting pairing
interaction favors anisotropic even-parity SC order parameter in
cubic symmetry (V$_l <$ 0 for even l $>0$). 
This turned out to be an influential result. Because of the
general presence
of AF spin fluctuations in HF metals it has lead to the attitude to expect
generally singlet (even parity) SC in these compounds.

There are, however, counter examples like \UPT and \UND with AF spin
fluctuations and experimentally identified triplet order parameter. The
prediction of only even
parity unconventional pair states therefore seems to be an artefact of
the simplifications inherent in the model. One obvious deficiency is
the assumption of cubic symmetry and basis functions adjusted for a
spherical Fermi surface not appropriate for these compounds.
More sophisticated versions of the model incorporating experimental
data for the magnetic fluctuations via $\chi(\v q, \omega$) and
realistic band structures for the quasiparticles failed to reproduce the
symmetry of the superconducting order parameter in UPt$_3$. In
particular, the calculations do not produce a multicomponent E$_1$ or
E$_2$ order parameter as stable solution. This 
failure is discussed in detail in \cite{Heffner96}. Also the
generalizations of standard spin-fluctuation theory which account for
orbital effects \cite{Norman94} do not resolve the difficulties
encountered in real materials.

Aspects of strong coupling effects within spin-fluctutation theories,
especially the pair breaking role of low frequency bosons have been discussed by
\citeasnoun{Millis88}. More recently magnetically mediated superconductivity in
materials close to a
magnetic instability was investigated in a series of papers by 
\citename{Monthoux99} \citeyear{Monthoux99,Monthoux01,Monthoux02}. 
A phenomenological form for the retarded generalized magnetic 
susceptibility $\chi({\bf q},\omega)$ defining the effective 
quasiparticle interaction is adopted \index{effective interaction}
\begin{equation}
\chi({\bf q},\omega) = \frac{ \chi_0 \kappa_0^2}{\kappa^2+{\hat q}^2
- i \frac{\omega}{\eta({\hat q})}}
\end{equation}
where $\kappa$ and $\kappa_0$ are the inverse correlation lengths (in 
units of the inverse lattice constant) with and without strong
magnetic correlations, respectively. The functions ${\hat q}^2$ and 
$\eta({\hat q})$ are parametrized so as to directly compare ferromagnetic and
antiferromagnetic spin fluctuations. The instability of the normal
state is determined by solving the linearized Eliashberg
equations which yield the transition temperatures $T_c$ and the mass
renormalization. The questions addressed include the influence of
dimensionality on the robustness of magnetic pairing and the relative
stability of d-wave versus p-wave pairing.

Pairing instabilities may also appear in the electron-hole (Peierls-)
rather than electron-electron (Cooper-) channel leading to CDW/SDW
type instabilities described by T$^{(s)}$ and T$^{(a)}$ in
eq.~(\ref{CROSS}). They are strongly favored if the Fermi surface
shows the nesting property $\epsilon_{\v k+\v Q}$= -$\epsilon_{\v k}$
characteristic for flat FS portions connected by the nesting vector \v
Q as shown later in fig.~\ref{FIGOPDpair}. In this case the on-site interaction term
in the Hamiltonian eq.~(\ref{HUBB}) may be truncated in momentum space
because the e-h scattering is dominated by processes with momentum
transfer \v Q. In mf approximation the effective CDW/SDW
pairing Hamiltonian is given in eq.~(\ref{EHPAIR}). To obtain the full
variety of electron-hole pair states one needs more general
microscopic interactions, replacing the on-site Hubbard model of 
eq.~(\ref{HUBB}) by an extended Hubbard model which include inter-site
Coulomb repulsion and exchange \cite{Gulacsi87,Schulz89,Ozaki92}.

\subsection{Order parameters \index{order parameter} and their
coexistence \index{coexistence} in strongly correlated electron systems}

\label{Sect:OrderParametersinSCES}

Degenerate interacting Fermi systems are prone to
instabilities due to pair condensation. The Pauli
principle requires that states have to be filled up to the Fermi energy
E$_F$ leading to a large kinetic energy. Rearranging the occupation of
noninteracting states around the Fermi level may reduce the
interaction energy considerably. The large on-site Coulomb interactions will
achieve this by reducing  double occupancies of opposite spin
states below the Hartree-Fock level. This leads to a strongly correlated
electronic ground state which is ideally 
approached at low temperatures below a characteristic temperature
scale  T$^*$ without breaking of spatial or internal symmetries.
Excitations from this state may be described within the Landau
Fermi liquid (FL) picture with quasiparticles that have a strongly enhanced
effective electron mass m$^*\gg$ m. Eventually however this state
will become unstable against formation of electron (Cooper-) pairs or
electron- hole (Peierls-) pairs due to the residual screened
interactions between the quasiparticles. In the former case gauge symmetry is
broken leading to a superconducting (SC) state, in the latter spatial
symmetries and possibly spin rotational and time reversal symmetry are
broken leading to charge-density wave (CDW) \index{charge density
wave} or spin-density wave (SDW) \index{spin density wave}
states. Which state is more favorable
depends on the momentum and energy dependence of residual quasiparticle
interactions and on the geometric properties of the Fermi surface and
usually cannot be predicted with confidence for real materials. The
effective Hamiltonian obtained previously describing the low energy
pairing correlations  of quasiparticles is most frequently used for
studying possible SC
states. The condensation into pairs is described by a gap function \De which
characterises both the type of broken symmetry
state and its new quasiparticle excitations. Its experimental
determination is therefore of central
importance. Unfortunately this is a difficult task except in the
isotropic case where \De is a constant. In this section we
outline the possible type of pair states and gap functions and their symmetry
classification. We also discuss simple models of coexistence of SC and
CDW/SDW type order parameters based on 2D FS models with
nesting properties. 

\subsubsection{Order parameter classification}

Many physical properties of superconducting materials are directly  
determined by the symmetry of the SC order parameter. The possible
types of order parameters are restricted   
by crystal symmetry. This fact provides a classification scheme for   
different superconducting states and, in addition, allows one to      
construct the superconducting classes by means of group theory.

Superfluids are characterized by off-diagonal long range order (ODLRO)
which leads to nonvanishing correlations in the two-particle density
matrix for large separations of particles at points ${\bf r}_1$,${\bf
r}_2$ and ${\bf r}_1'$,${\bf r}_2'$,
\begin{equation} \Bigl\langle{\bf r}_1s_1;
{\bf r}_2s_2\Bigm|\rho^{(2)}\Bigm|
{\bf r}_1's_1';
{\bf r}_2's_2'\Bigr\rangle\rightarrow         
  {\psi_{s_1,s_2}^*({\bf r}_1,{\bf r}_2)}         
{\psi_{s_1',s_2'}({\bf r}_1',{\bf r}_2')}
\label{eq:order_param}
\end{equation}
Here, ${\bf r}_i$,       
$s_i$ denote the fermion positions and spins, respectively. 
In systems with strong spin-orbit (s.o.) interactions, the indices $s_i$     
refer to pseudo spins associated with Kramers degeneracy of conduction
bands. The ordered phase is characterized by the order parameter
$\psi$ which is a complex pseudo-wave function always connected with a
spontaneous breaking of U(1) gauge symmetry. In general, it depends on
the center-of mass and relative
coordinates, ${\bf R}={1 \over 2}({\bf r}_1+{\bf r}_2)$,     
${\bf r}={\bf r}_1-{\bf r}_2$ and the pseudo 
spins s$_1$ and s$_2$, respectively. We shall    
restrict ourselves to homogeneous systems neglecting the dependence on       
the center-of-mass variable of ${\bf R}$. Then the order
parameter is a function of the relative coordinate ${\bf r}$ only. Performing
a Fourier transformation with respect to ${\bf r}$ with
$\psi_{s_1s_2}({\bf r}) \rightarrow \psi_{s_1s_2}({\bf k})$ and
restricting the wave vector to the Fermi surface 
${\bf k}= k_F({\hat {\bf k}}){\hat {\bf k}}$ yields a pairing amplitude
$\psi_{s_1s_2}({\hat {\bf k}})$  which depends upon the
direction ${\hat {\bf k}}$ on the Fermi surface. In microscopic
theories the \index{gap function} gap function $\Delta_{s_1s_2}$ which also
determines the excitation spectrum is commonly used as order parameter. It is
given by
\begin{equation}
\Delta_{s_1s_2}(\hat{\v k})=-\sum_{\hat{\v k}' s_3 s_4}
V_{s_1s_2s_3s_4}(\hat{\v k},\hat{\vk}')\psi_{s_3s_4}(\hat{\v k}')
\end{equation}
where V is the effective pairing interaction. \index{effective
interaction} For the spin fluctuation
model we obtained V$_{s_1s_2s_3s_4}$(\v k,\v k')= 
-J(\v k-\v k')$\boldsigma_{s_1s_4}\boldsigma_{s_2s_3}$ in
eq.(~\ref{HEFFSF}). Concerning the symmetry classification of
the superconducting phases, one may use the gap function $\Delta_{s_1 s_2}$
or the pair amplitude $\psi_{s_1s_2}$ as order parameter due to their
identical transformation properties.
                                                  
The fundamental property of 
 $\psi_{s_1s_2}$ or $\Delta_{s_1s_2}$ is its behaviour as a
two-fermion wave function in many respects. This expresses the fact
that an ODLRO order parameter is not the thermal expectation value of
a physical observable but rather a complex pseudo-wave function describing
quantum phase correlations on the macroscopic scale of the SC
coherence length. Its phase is a direct signature of the broken gauge
invariance in the SC condensate. The Pauli principle then requires
$\Delta_{s_1s_2}$ to be antisymmetric under the interchange of particles  
\begin{equation}
{\Delta_{s_1,s_2}}({\hat {\bf k}})=                                            -{\Delta_{s_2,s_1}}{(-{\hat {\bf k}})} 
\label{eq:antisymm}
\end{equation}
In addition, it transforms like a two-fermion 
wave function under rotations in position and spin space and under gauge
transformations. The transformation properties yield a general
classification scheme for the superconducting order parameter which is
represented by a $2 \times
2$-matrix in (pseudo-) spin space. It can be decomposed into an 
antisymmetric (s) and a symmetric (t) contribution according to
${\bf \Delta}({\hat {\bf k}})={\bf \Delta}_s
({\hat {\bf k}})+{\bf \Delta}_t({\hat {\bf k}})$ with 
\begin{equation} 
{\bf \Delta}_s({\hat{\bf k}})  = 
\phi({\hat {\bf k}}) i\sigma_{2} 
\qquad \mbox{and} \qquad
{\bf \Delta}_t({\hat {\bf k}})  = 
\sum_{{\mu}=1}^3 d_\mu({\hat {\bf k}})\sigma_{\mu} i\sigma_{2}
\label{eq:Psi_decomp}
\end{equation}
where $\sigma_{\mu}$  denote the Pauli matrices. Antisymmetry 
${\bf \Delta}({\hat {\bf k}})\;=\;-\;{\bf \Delta}^T(-{\hat {\bf k}})$ 
requires
\begin{equation}
\phi({\hat {\bf k}})\;=\;\phi(-{\hat {\bf k}})  
\qquad \textrm{and} \qquad
d_\mu({\hat {\bf k}})\;=\;-d_\mu(-{\hat {\bf k}})
\end{equation} 
for the complex orbital functions $\phi({\hat {\bf k}})$ and 
$ d_\mu({\hat {\bf k}})$ ($\mu$ = 1-3). For brevity we will frequently
write $\Delta(\hat{\bf k})$ for $\phi({\hat {\bf k}})$ or $|\v d(\hat{\v k})|$.
The physical meaning of the order parameters $\phi$ and ${{\bf d}}$ for
singlet and triplet state respectively is evident from the identity
\index{singlet pairing} \index{triplet pairing}
\begin{equation} 
{1 \over 2} {\rm Tr} \Bigm|
{\bf \Delta}_s({\hat {\bf k}}) \Bigm|^2=
\Bigm| {\phi}({\hat {\bf k}}) \Bigm| ^2
\qquad \mbox{and} \qquad
{1 \over 2} {\rm Tr} \Bigm|
{\bf \Delta}_t ({\hat {\bf k}}) \Bigm|^2=
\Bigm| {{\bf d}} ({\hat {\bf k}}) \Bigm| ^2
\end{equation}                                                      
Therefore the modulus of $\phi$ and ${{\bf d}}$ is a measure for the total
gap amplitude of the Cooper pairs at a given point
$\hat{\v k}$ on the Fermi surface.
In addition, the direction of the vector ${{\bf d}}$ 
specifies the relative contributions of the three triplet pair states. 
The above relation holds for triplet states which satisfy ${\bf
d}^*\times{\bf d} =0$. They are called  `unitary states' because they
are invariant under time reversal. In this case the vector ${{\bf d}}$
defines a  unique direction in spin space for every point on the Fermi surface.

The order parameter can be chosen 
either as purely antisymmetric ${\bf \Delta}_s$ or purely 
symmetric ${\bf \Delta}_t$ when spin-orbit interaction can be neglected.
Then the total spin is a good quantum number and may be used to
classify the pair states. Accordingly the states ${\bf \Delta}_s$ and 
${\bf \Delta}_t$ are called spin singlet and spin triplet states, 
respectively. In crystals which have an inversion center, pair states
can also be classified with respect to their parity as even-parity
(${\bf \Delta}_s$) and odd-parity states (${\bf \Delta}_t$).

In the 4f- and 5f-based heavy fermion superconductors the spin-orbit
interaction is strong. As a consequence classification according to
physical pair spins cannot be used. If their high-temperature crystal
structures, however, have an inversion center then classification
according to parity is still possible. We note here that recently the
first example of a HF superconductor (CePt$_3$Si) which lacks
inversion symmetry was discovered \cite{Bauer03} and its theoretical
implications were discussed by \citeasnoun{Frigeri03}.

The simplest even-parity state is the isotropic             
state encountered in ordinary superconductors. This state is often   
referred to as "s-wave state". The isotropic order parameter    
does not depend on the direction ${\hat {\bf k}}$ and reduces to a
complex constant $\phi$=$|\phi|$e$^{i\varphi}$. Its only degree of
freedom is the Josephson phase $\varphi$.    
By far the most extensively studied examples of anisotropic pairing 
are the p-wave states realized in the superfluid phases of $^3$He, the
d-wave pair state in high-T$_c$ superconductors and the f-wave
states in UPt$_3$ and SrRu$_2$O$_4$. The odd parity (p,f) states among
these examples are characterised by more than one order parameter
component with internal phase degrees of freedom which appear in
addition to the overall Josephson phase.

The general classification scheme for superconducting order parameters
proceeds from the behavior under the transformations of the symmetry
group \index{symmetry group} ${\cal G}$ of the Hamiltonian. It
consists of the crystal point
group G, the spin rotation group SU(2), the \index{time reversal symmetry}
time-reversal symmetry group ${\cal K}$, and the gauge group U(1). The
latter two are respectively defined by
\begin{equation}
K{\bf \Delta}({\hat{\bf k}}) = 
\sigma_2{\bf \Delta}^*(-{\hat{\bf k}})\sigma_2 
\qquad \mbox{and} \qquad
\Phi{\bf \Delta}({\hat{\bf k}})=e^{i\varphi}{\bf \Delta}({\hat{\bf k}})
\end{equation}
Concerning rotations in \v k- and spin space we distinguish 
two different cases: (1) If spin-orbit coupling is negligible spatial
and spin rotations can be applied independently. For the elements g
$\in$ G of the point group which act on ${\hat {\bf k}}$ one has
\begin{equation}
g{\bf \Delta}({\hat {\bf k}})=\Delta(D_G^{(-)}(g){\hat {\bf k}})
\label{eq:space_rot}
\end{equation}
where $D_G^{(-)}(g)$ is the three-dimensional representation of G in 
${\hat {\bf k}}$-space. For the elements g $\in$ SU(2) of the spin
rotation group which act on the spin indices we obtain
\begin{equation}
g{\bf \Delta}({\hat {\bf k}})=D_{(s)}^T(g)\Delta({\hat {\bf k}})D_{(s)}(g)
\label{eq:spin_rot}
\end{equation}
where $D_{(s)}(g)$ denotes the representation of SU(2) for spin $\frac{1}{2}$.
This transformation leaves even-parity states invariant. For odd-parity
states where the order parameter can be represented by a vector ${\bf
d}$ in spin
space the transformation gives a conventional orthogonal rotation of the
${\bf d}$-vector according to
\begin{equation}
g{\bf d} ({\hat {\bf k}}) = D_G^{(+)}(g)  {\bf d} ({\hat {\bf k}}) 
\end{equation} 
(2) In the presence of spin-orbit interaction the transformations of 
${\hat {\bf k}}$ and the spin rotations are no longer
independent. The spins are frozen in the lattice and the operations of
the point group amount to simultaneous rotations in ${\hat {\bf k}}$
space and spin space:
\begin{equation}
g {\bf d} ({\hat {\bf k}}) = D_G^{(+)}(g) {\bf d} (D_G^{(-)}(g) 
{\hat {\bf k}}) 
\end{equation}
The appropriate choice of rotations corresponding to weak or strong
spin-orbit coupling case is determined by microscopic considerations.
Using the above transformation properties the singlet and triplet gap
functions $\phi(\hat{\v k})$ and $\v d(\hat{\v k})$ respectively may
further be decomposed into basis functions 
$\phi^n_\Gamma(\hat{\v k})$ or $\v d ^n_\Gamma(\hat{\v k})$ of the irreducible
representations $\Gamma$ (degeneracy index n) of G$\times$SU(2) (weak
s.o. coupling) or G (strong s.o. coupling).

The occurrence of long-range order at a phase transition          
described by an order parameter is most frequently associated with
spontaneous symmetry breaking. The simplest superconductors where only gauge 
symmetry is broken are called \index{conventional pairing} {\em
conventional}. In this case the SC order parameter has the same
spatial symmetry as the underlying  
crystal, i.e. it transforms as a fully symmetric even parity singlet
A$_{1g}$ representation of G. It should be noted, however, that
conventional is not a synonym for isotropic, for any G one can form
A$_{1g}$ representations from angular momentum orbitals of higher
order l, for example l $\geq$ 2 for tetragonal and hexagonal symmetry
and l $\geq$ 4 for cubic symmetry. On the other hand, a superconductor
with additional broken symmetries besides gauge symmetry is called
\index{unconventional pairing} {\em unconventional}. It can have
either parity. Any odd-parity SC state which has broken inversion symmetry is
unconventional in this sense. From the effective attractive
interactions as eq.~(\ref{HEFFSF}) obtained from model Hamiltonians
like  eq.~(\ref{HUBB}) containing `high energy' interaction
parameters the symmetry can be obtained by directly minimizing the free
energy and solving the resulting gap equations. These calculations
depend on the model parameters and approximation schemes and cannot
make predictions for real compounds.  

The usual procedure described in detail in the next section to
determine the symmetry of SC order parameters
is to select plausible candidate states corresponding to irreducible
representations (or mixtures) of the symmetry group, calculate the expected
(temperature, field) behavior of physical quantities and compare the
predictions with experiment. The selection of candidate states
exploits Michel's theorem \cite{Michel80} according to which stable points of
the free energy should correspond to states which are invariant under subgroups
of the full symmetry group. However the theorem was only proven for free
energy expressions which are polynomial in the order parameter. As long
as we work sufficiently close to the transition temperature, 
where a Ginzburg-Landau \index{Ginzburg-Landau theory} expansion is 
valid the systematic construction of gap functions in terms of basis
functions which are  invariant under subgroups is a useful guideline.
Exhaustive lists of the superconducting classes for the 
relevant crystal symmetries were given by \citeasnoun{Volovik85}, 
\citeasnoun{Ueda85} , \citeasnoun{Blount85}, 
\citeasnoun{Gorkov87}, \citeasnoun{Sigrist91}, \citeasnoun{Annett90} and 
\citeasnoun{Ozaki85}, \citeasnoun{Machida98}, \citeasnoun{Machida99} 
and references cited therein.
A recent summary is found in the textbook of \citeasnoun{MineevBook}. 

Quite generally the results as given above can be divided 
into two groups (1,2) according to their treatment of spin-orbit 
interaction. Group (1) treats the orbital and (pseudo-) spin degrees of
freedom separately, while group (2) assumes that the spin of the Cooper 
pair is frozen in the lattice. In mathematical terms: (1)
assumes the symmetry group to be the direct 
product G$\times$SU(2), while (2) considers the point 
group of the crystal to be the double group. The two schemes 
yield different predictions for the node structures. Treating the
Cooper pair spin or \v d as frozen in the lattice leads
to odd-parity \index{triplet pairing} order parameters which vanish only
in isolated points on the Fermi surface. This fact which was first pointed out
by \citeasnoun{Blount85} and is usually referred to as Blount's theorem. 
Even-parity \index{singlet pairing} states, on the other hand, have
gap functions which may vanish on lines and at isolated points on the
Fermi surface.

Blount's theorem \index{Blount's theorem} seems to rule out odd-parity
states in UPt\( _{3} \) at first
glance since there is strong evidence for node lines on the Fermi surface. As
a consequence, the majority of early order parameter models for UPt\(
_{3} \) adopted multicomponent even-parity states. However the anisotropy of
thermal conductivity, reversal of upper critical field
anisotropy and Knight shift results in UPt\( _{3} \) are better
accounted for by an odd-parity order parameter. For an extensive
discussion of this problem we refer to Sect.~\ref{Sect:UPt3}. Another
more recent case is UNi$_2$Al$_3$ where evidence for an odd parity
state exists. It seems that Blount's theorem is not respected in real HF
superconductors.

Concerning this ambiguity we add the following remark. Doubtlessly
spin-orbit \index{spin orbit coupling} interaction plays an 
important role in U compounds. However, the energy  $\xi_{so}\simeq$
2 eV associated with this relativistic effect is large compared to the
HF quasiparticle band width $k_BT^*\simeq$ 10 meV. In fact it is even
larger than the hybridization energies, leading to separate 5f LDA bands
belonging to different total angular momenta j = $\frac{5}{2}$ and
$\frac{7}{2}$ where the latter is almost empty \cite{Albers86}. With respect to
pair formation spin orbit coupling is therefore to be treated as a
high-energy effect. This suggests that spin orbit coupling
should be included already in the properties of the normal state
quasiparticle states. The latter can be classified according to the
pseudo spins connected with Kramers degeneracy of band states.

The transformation properties of the odd-parity order parameters under
spatial rotations is reduced to considering the behavior of the
quasiparticle states. To leading order in
the small ratio $kT^*/\xi_{so}$ we include the spin-orbit
interaction in the calculation of the local atomic basis states, 
in a second step they are coherently superposed to form extended
states. For bands derived from one doubly (Kramers-) degenerate orbital the
elements of the point group should act only on the propagation vector.
When the Cooper pairs, i. e., the two-particle states,
are formed the orbital and spin degrees of freedom can be treated 
independently. With the spin-orbit interaction already included in 
the normal state quasiparticles one can use Machida's states 
derived for vanishing spin-orbit interaction. 

The two different methods in treating the spin-orbit interaction have their
counterparts in atomic physics. The scheme put forward by Volovik and Gor'kov
parallels Russell-Saunders coupling where the orbital momenta and spins of the
individual electrons are coupled to the total orbital momentum \( {\bf L} \)
and total spin \( {\bf S} \), respectively. These two quantities are coupled
to form the total angular momentum \( {\bf J}={\bf L}+{\bf S} \). Including
the spin-orbit interaction in the quasiparticle states, on the 
other hand, closely parallels \( jj \)-coupling. \index{jj-coupling}

The symmetry classification of density wave order parameters may proceed in a
similar way. Charge and spin density are physical observables, hence
the CDW/SDW order parameter \index{spin density wave} \index{charge
density wave} describe diagonal long range order (DLRO)
meaning that they correspond to expectation values of diagonal
elements in the one-particle density matrix according to
\begin{eqnarray}
\Bigl\langle \v rs_1|\rho^{(1)}|\v r s_2\Bigr\rangle\rightarrow
A_{s_1s_2}\exp(i\v Q\v r)
\label{ONEDENS}
\end{eqnarray}
where \v Q is the wave vector of the density modulation and its 
amplitude is given by 
\begin{eqnarray}
\label{AMPLI}
A_{s_1s_2}=\sum_{\vk}F^{\v Q}_{s_1s_2}(\v k) 
\qquad \mbox{and} \qquad
F^{\v Q}_{s_1s_2}(\v k)=\la c^\dagger_{\v k s_1}c_{\vk +\v Q s_2}\ra
\end{eqnarray}
The electron-hole \index{electron-hole pairing} pair amplitude {\bf
F}$_{\v Q}$(\v k) may again be
decomposed into singlet (CDW) \index{singlet pairing} and triplet
(SDW) \index{triplet pairing} components $\rho_{\v
Q}(\v k)$ and \v s$_{\v Q}$(\v k) respectively in analogy to SC pair
amplitudes. Their corresponding gap functions \index{gap function} are
then given by the gap equations \index{gap equations}
(c.f. eq.~(\ref{DENSITY}))
\begin{eqnarray}
\label{CSGAP}
\Delta_C(\v k)= -I\la \rho_{\v Q}(\v k)\ra
\qquad \mbox{and} \qquad
\Delta_S(\v k)= -I\la s^z_{\v Q}(\v k)\ra 
\end{eqnarray}
for CDW and SDW (\v d vector along z) respectively, where I is the
on-site Hubbard interaction. Likewise the \v k-dependence of gap
functions may be classified according to the irreducible
representations of the point group G with even and odd parity type
when it contains the inversion symmetry. Several interesting aspects arise
here: (1) Since gauge symmetry is obviously not broken its phase cannot be
chosen arbitrarily, it rather is tied to the spatial transformation
properties of the representation considered. Commonly the gap
functions will either be real or imaginary for commensurate density
waves. (2) On the other hand for
electron-hole pairing there is no antisymmetry requirement hence the
gap function for singlet (CDW) or triplet (SDW) can {\em both} be odd
or even. (3) Conventional CDW and SDW states correspond to gap
functions $\Delta_C(\v k)$ and $\Delta_S(\v k)$ which transform as fully
symmetric (A$_{1g}$) representations, in the simplest case they are
constants, independent of \v k. In this case the amplitude of the
CDW/SDW density modulation in real space obtained from the summation
in Eq.~(\ref{AMPLI}) is nonzero because the gap function does not change
sign. If the gap functions belong to {\em nontrivial} representations,
e.g. d$_{x^2-y^2}$ in two dimensions they
do change sign and the sum in Eq.~(\ref{AMPLI}) is zero, i.e. there is
{\em no density modulation} although there is a `hidden' order
\index{hidden order}
parameter $\Delta_C(\v k)$ or $\Delta_S(\v k)$ which characterizes the
electron-hole condensate. Such states have been given the misleading
name \index{unconventional pairing} `unconventional density waves' in
analogy to the usage for SC states.
These hidden order parameters have recently been under intense
discussion for both underdoped cuprates, organic materials and U-HF
systems.

\subsubsection{Pairing model for coexistence \index{coexistence} of SC
and CDW/SDW}

In U-HF compounds the coexistence of SC and SDW order parameters is
frequently observed. In a purely itinerant picture this phenomenon is
well studied on the basis of a mean field (mf) pairing Hamiltonian
which allows for both type of instabilities. It has first been
proposed for SC/CDW coexistence \cite{Bilbro76} and then for
conventional SC/SDW \cite{Machida81a,Machida81b} \index{conventional pairing} 
coexistence and later generalized to unconventional anisotropic gap
functions \index{unconventional pairing} for both SC and CDW/SDW
\cite{Kato88,Thalmeier94}.

The effective mf Hamiltonian for pair formation in both electron-hole
(e-h or Peierls) and electron-electron (e-e or Cooper) channels is given by
\begin{eqnarray} 
H&=&\sum_{\vk,s}\epsilon_{\vk}
c^\dagger_{\vk s}c_{\vk s}+H_{e-h}+H_{e-e}
\end{eqnarray} 
Introducing the charge and spin-densities (polarised along z)
\begin{eqnarray}
\label{DENSOP} 
\rho_{\vk}(\vQ)&=&c^\dagger_{\vk\ua}c_{\vk +\vQ\ua}+
c^\dagger_{\vk\da}c_{\vk +\vQ\da}\nonumber\\
s_{\vk}(\vQ)&=&c^\dagger_{\vk\ua}c_{\vk +\vQ\ua}-
c^\dagger_{\vk\da}c_{\vk +\vQ\da}
\end{eqnarray} 
the electron-hole part can be separated in singlet (CDW) and triplet
(SDW) part
\begin{eqnarray}
\label{EHPAIR}
H_{e-h}&=&-\sum_{\vk}[\Delta_C(\vk)\rho_{\vk}(\vQ)+h.c.]
-\sum_{\vk}[\Delta_S(\vk)s_{\vk}(\vQ)+h.c.]
\end{eqnarray} 
Likewise the electron-electron part which, for example, is the mf
approximation of Eq.~\ref{HEFFSF} is given by 
\begin{eqnarray} 
H_{e-e}&=&-\sum_{\vk}\Delta^{ss'}_{SC}(\vk)
[c^\dagger_{\vk s}c^\dagger_{-\vk s'}+h.c.]
\end{eqnarray} 
Here $\Delta_C(\vk)$, $\Delta_S(\vk)$ and $\Delta^{ss'}_{SC}$ are the
CDW, SDW and SC gap functions which in general will depend on the
momentum \vk. The former two simply correspond to spin singlet (CDW) and
triplet pairing (SDW) in the e-h channel. Analogously the
superconducting gap matrix $\hat{\Delta}_{SC}$(\vk) may be decomposed into
singlet and triplet parts described previously.

If only a single nonzero OP is present diagonalisation of H leads to
quasiparticle excitations \index{quasiparticle excitations} (A $\equiv$ C,S)
\begin{eqnarray}
\label{QUASI}
E_{\v k}=[\epsilon_{\v k}^2 +\Delta^{\v k 2}_{SC}]^\frac{1}{2}; 
\qquad \mbox{or} \qquad
E_{\v k}=[\epsilon_{\v k}^2 +\Delta^{\v k 2}_{A}]^\frac{1}{2};
\end{eqnarray}
in the ordered phase. Here $\Delta^{\v k}_{SC}$ represents $\phi(\v
k)$ in the singlet case and $|\v d(\v k)|$ in the (unitary) triplet
case respectively. The general case of coexisting order parameters and
coupled gap equation is discussed below.

\begin{figure}
\includegraphics[width=6.5cm]{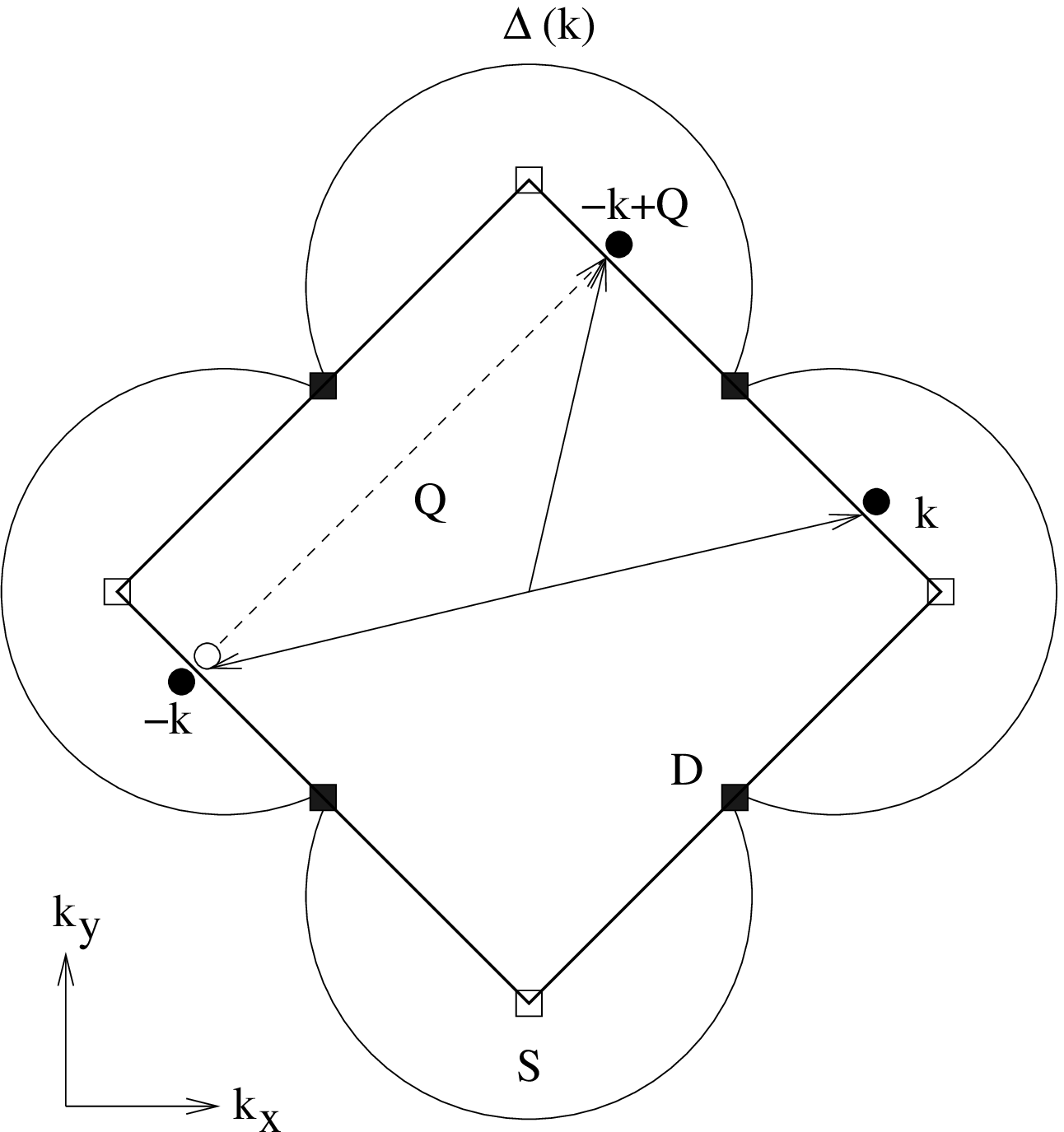}\hfill
\includegraphics[width=8.0cm]{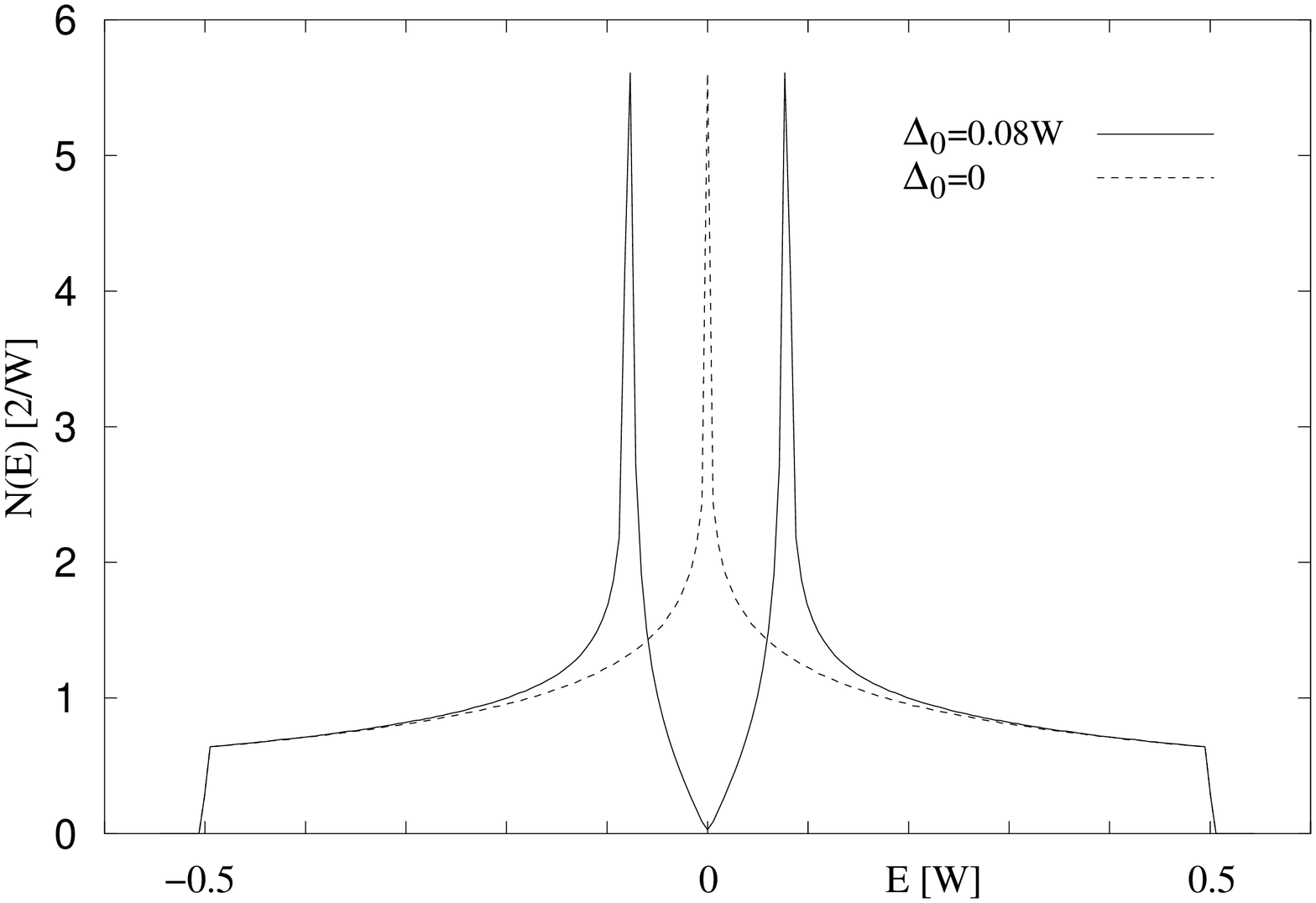}
\caption{Left panel: Cooper pairs (-\vk,\vk) and electron-hole (Peierls) pairs
(-\vk,-\vk+\vQ) for the \index{d-wave gap} n.n. tight binding
\index{Fermi surface} Fermi surface (thick line) with \index{nesting
property} perfect nesting vector \vQ. Saddle points (S)  of
$\epsilon$(\vk) at (0,$\pm\pi$) and
($\pm\pi$,0) lead to DOS peak at the Fermi energy. Therefore
unconventional pair states can only have nodes away from S, i.e. at
the `Dirac' points D ($\pm\frac{\pi}{2},\pm\frac{\pi}{2}$) where the
quasiparticle spectrum takes the form of eq.~(\ref{RDIRAC}). This is the
case for a d$_{x^2-y^2}$ - type gap function \De which is indicated
schematically. Right panel: corresponding \index{quasiparticle DOS}
quasiparticle DOS N(E) (W= tight binding band width) for normal state
(dotted) and with d$_{x^2-y^2}$- gap with amplitude $\Delta_0$. The
Fermi level is at E$_F$=0.}
\label{FIGOPDpair}
\end{figure}

In the spirit of the Landau theory of 2$^{nd}$ order
phase transitions one has to assume that each of the gap functions
transforms as an irreducible representation of the underlying
symmetry group which consists of gauge
transformations U(1), time reversal {\emph K}, spin rotations SU(2) and spatial
transformations G as introduced before. For e-h pairing the
former is preserved, for e-e pairing it is at least partly broken. 
If the gap functions belong to a fully symmetric or trivial
representation (A$_{1g}$ or $\Gamma^+_1$) i.e. if they are
invariant under all spatial symmetry operations (including parity) of
G then the condensed state describes a conventional CDW, SDW or
superconductor, for all other representations the pair state is called
unconventional as mentioned before. The most important difference in the
two cases is due to their generally different nodal structure. Whereas
for A$_{1g}$ in general $|\Delta_i(\vk)| > 0$ (i = A, SC) and hence
quasiparticle excitations are gapped, for the unconventional states
frequently zeroes of the gaps
$\Delta_i(\vk)$ = 0 or \vd(\vk) = 0 on points or lines on the Fermi
surface are {\em possible} but not necessary. 
In the SC case one basically has two types of nodes: (i) symmetry
enforced gap nodes which always appear when the gap function transforms as
an irreducible representation whose remaining symmetry group
in the SC phase contains elements composed of spatial
transformation and discrete gauge transformations and/or time reversal.
(ii) accidental nodes which depend on the specific choice of basis
functions within a single representation or which appear as a result
of superposition of different representations (`hybrid gap
functions'). Then gap nodes \index{gap nodes} exist only for
special superposition parameters (`fine tuning'). Such 
nodes (or at least very large gap anisotropies
$\Delta_{max}(\v k_{max})/\Delta_{min}(\v k_{min})$) may also be
present in conventional
superconductors where the hybrid gap function is constructed from
fully symmetric representations  A$_{1g}$ of various degree and only
U(1) symmetry is broken. Both types of nodes have
been found and examples will be discussed in later sections.
The precise conditions for the symmetry enforced gap nodes are discussed
in \citeasnoun{MineevBook}. On the
other hand in some compounds, for example \index{borocarbides}
borocarbides (sect.\ref{Sect:Boro}) and possibly \index{skutterudites}
skutterudites  (sect.\ref{Sect:PRSK}) the second type of SC gap nodes
is observed. The quasiparticle spectrum of eq.~(\ref{QUASI})
around the nodes defined by $\Delta^{\v k}_{SC}$ = 0 or $\Delta^{\v
k}_{A}$ = 0 determines completely
the low temperature physics of the ordered state. Knowledge about their
position, multiplicity and dimension (points or lines) is therefore essential.

\subsubsection{Coupled \index{gap equations} gap equations and results
for \index{coexistence} coexistence}

The interplay between superconductivity and antiferromagnetism
\index{antiferromagnetic order}
has been in the focus of interest for a long time. The coexistence of 
the two ordering phenomena is well understood in those materials 
where they occur in different electronic
subsystems which, in addition, are not coherently coupled. Wellknown 
examples are the Chevrel phases RMo$_6$S$_8$, the rhodium borides
RRh$_4$B$_4$ or the \index{RNi$_2$B$_2$C} borocarbides \RBC where the
influcence of
local moments of the lanthanide elements (R) on SC can be modelled by a
molecular magnetic field acting on the SC conduction electrons
(sect.~\ref{Sect:Boro}). This 
results in a weaker pair interaction as compared to the paramagnetic
phase and therefore leads to anomalies in the upper critical field
H$_{c_2}$ below the onset of magnetic order.

The interplay of superconductivity with itinerant-electron magnetism
where {\em both} SC and CDW/SDW are carried by conduction electrons
continues to be a theoretical challenge. Qualitative
phase diagrams reflecting the dominant order parameters and their
mutual coexistence or expulsion have been derived 
within a mf approximation in order to interpret experiments. 
The concepts developed for transition metal alloys or organic 
superconductors have been extended to the case of heavy-fermion
systems \cite{Kato88,Konno89}. The basic assumption is that the long-range
order, i.~e.~, superconductivity or spin-density wave results from an 
instability of the Fermi surface of the strongly renormalized heavy 
quasiparticles. This assumption implies that we have at least an 
approximate separation of energy scales with $T_c (T_{N}) \ll T^*$
where $T_c$, and $T_{N}$ denote the superconducting and magnetic
transition temperatures while $T^*$ is the HF quasiparticle band width. 

\citeasnoun{Kato88} adopt a two-dimensional (t,t')-tight binding
model \index{tight binding model} for the  quasiparticle bands
resulting in a Fermi surface that
exhibits nesting features with the commensurate wave vector \v Q =
($\frac{1}{2},\frac{1}{2}$,0). The effective interaction which consists of
an on-site repulsion and an attractive pairing leads to both SDW and
SC instabilities, the former supported by the nesting feature.  
The formation of a spin-density wave changes the symmetry of the
system. This requires a modified  classification scheme for the 
superconducting order parameters which can appear in a second order
phase transition within the SDW state. The presence of both types of order
parameters $\Delta_{A}$ (A=C,S) and $\Delta_{SC}$ may lead to induced SC
pair amplitudes with pair momentum \v Q which strongly influence the
coexistence behaviour. The competing order parameters for
various symmetry types are determined from coupled gap
equations for $\Delta_{A}$(\vk) and $\Delta_{SC}$(\vk). In
\citeasnoun{Kato88} the former was taken as constant (conventional
SDW) while the latter allowed to belong to any nonconventional
representation. In \citeasnoun{Thalmeier94} the coexistence study was
generalised to include also unconventional density waves
$\Delta_{A}$(\vk). In the general case coexistence, competition
as well as expulsion of the two order parameters may be observed.

For the tight binding model described above there are two types of
coupled gap equations which are due
to the different even/odd transformation properties of
$\Delta_{SC}$(\vk), $\Delta_{A}$(\vk) under the two discrete
transformations \v k $\rightarrow$ -\v k
(inversion) and \v k $\rightarrow$ \v k+\v Q (translation).\\ 
Case (I): there is no induced SC pairing with nonzero wave vector.
Case (II): a finite induced pairing amplitude $\la c_{\v k s_1}c_{\v
k+\v Q s_2}\ra$ for SC pairs with nonzero pair momentum exists.\\
Competition of SC and density wave pairs is much stronger in the
second case because the induced pairing inevitably leads to a large
loss in condensation energy.

In case I for the \index{nesting property} perfect nesting model (t'=
0) the gap equations are given simply by 
\begin{eqnarray}
\label{COEXI}
1=V\sum_{\v k}\frac{f_{SC}(\v k)^2}{2E_{\v k}}\tanh\frac{E_{\v k}}{2T}
\qquad \mbox{and} \qquad
1=I\sum_{\v k}\frac{f_{A}(\v k)^2}{2E_{\v k}}\tanh\frac{E_{\v k}}{2T}
\end{eqnarray}
where E$_{\v k}$ = $[\epsilon_{\v k}^2+
|\Delta_{SC}^{\v k}|^2+|\Delta_{A}^{\v k}|^2]^\frac{1}{2}$
is the quasiparticle energy \index{quasiparticle excitations} and the
form factors \index{form factor} f$_{SC}$(\v k),
f$_{A}$(\v k) which correspond to irreducible basis functions of D$_{4h}$ point
group are defined via $\Delta_{SC}$(\v k)=$\Delta^0_{SC}$f$_{SC}$(\v k) and
 $\Delta_{A}$(\v k)=$\Delta^0_{A}$f$_{A}$(\v k). Furthermore
interaction constants I and V determine whether
SDW or SC transition happens first. In HF compounds frequently we have
T$_N>$ T$_c$. In this weakly competitve case coupled gap equations are
formally the same as for
each of the gap functions individually. They influence each other
only through the quasiparticle energies where both gap functions
appear. Numerical solution shows that for this case coexistence is
always possible and (assuming T$_N>$T$_c$) the transition from SDW to
SDW+SC coexistence phase happens always in a second order phase
transition, irrespective how small T$_c$ is.
 
In case II the gap equations are formally different from individual
gap equations:
\begin{eqnarray}
\label{COEXII}
1&=&V\sum_{\v k}\frac{f_{SC}(\v k)^2}{4|\Delta^{\v k}_{SC}|}
\Bigl(\frac{|\Delta^{\v k}_{SC}|+|\Delta^{\v k}_{A}|}{E^+_{\v k}}
\tanh\frac{E^+_{\v k}}{2T}
+\frac{|\Delta^{\v k}_{SC}|-|\Delta^{\v k}_{A}|}{E^-_{\v k}}
\tanh\frac{E^-_{\v k}}{2T}\Bigr)\nonumber\\
1&=&I\sum_{\v k}\frac{f_{A}(\v k)^2}{4|\Delta^{\v k}_{A}|}
\Bigl(\frac{|\Delta^{\v k}_{A}|+|\Delta^{\v k}_{SC}|}{E^+_{\v k}}
\tanh\frac{E^+_{\v k}}{2T}
+\frac{|\Delta^{\v k}_{A}|-|\Delta^{\v k}_{SC}|}{E^-_{\v k}}
\tanh\frac{E^-_{\v k}}{2T}\Bigr)
\end{eqnarray}
where the two quasiparticle bands E$^{\v k}_\pm$ = 
$[\epsilon_{\v k}^2+(|\Delta_{SC}^{\v k}|\pm|\Delta_{A}^{\v
k}|)^2]^\frac{1}{2}$ are now split due to the presence of two order parameters.
Numerical solutions for case II show a much more competitve
behaviour of gap functions $\Delta_{A}$ and $\Delta_{SC}$. If T$_c$ is
comparable in size to  T$_N$ the SDW state may be destroyed at T$_c$
and replaced by the SC phase in a first order phase transition. For
smaller T$_c$ again a coexistence state SDW+SC appears in a second
order phase transition  but generally the order parameters in this
regime show strong competition resulting in nonmonontonic temperature
behaviour of gap functions.

\begin{figure}[h t b]
\begin{center}
\begin{minipage}{\columnwidth}
\includegraphics[width=.4\columnwidth,angle=0,clip]
{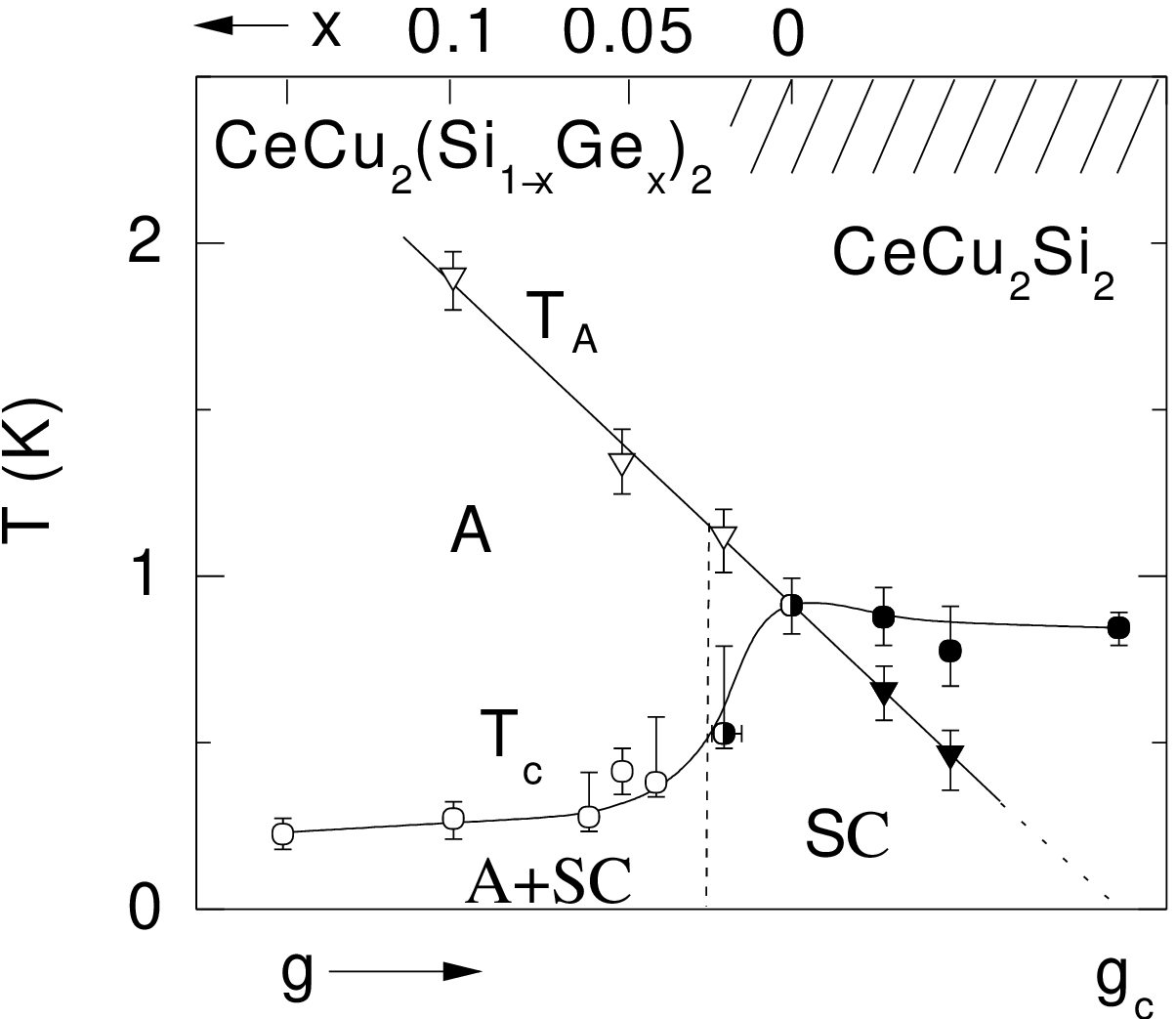}
\hfill
\includegraphics[width=.4\columnwidth,angle=0,clip]
{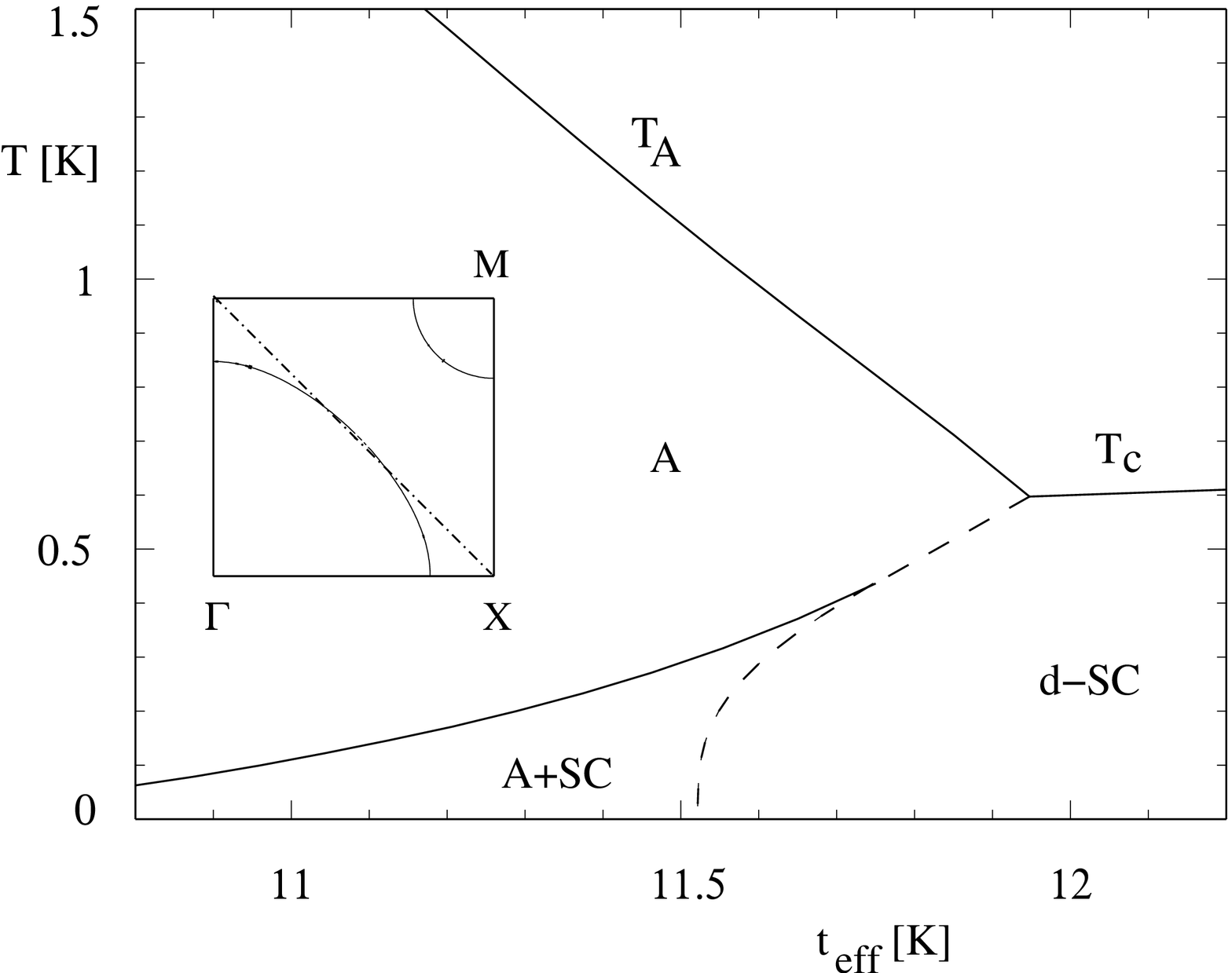}
\end{minipage}
\end{center}
\caption{Left panel: experimental results for \index{coexistence}
coexistence and competition between superconductivity and SDW in 
\index{CeCu$_2$(Si$_{1-x}$Ge$_x$)$_2$}
CeCu$_2$(Si$_{1-x}$Ge$_x$)$_2$ (Gegenwart et al. 1998). Ge
substitution acts like a (negative) chemical pressure. Here A = SDW
and SC is the superconducting state. T$_A$ depends linearly chemical
pressure, hence g $\sim$ 1-x. Right panel: two band model for A and
d-SC order in mf theory from Steglich et al, (2001b). \index{d-wave gap} 
Here A denotes a \index{spin density wave} conventional SDW state with
constant $\Delta_A$ and
d-SC has $\Delta_{SC}$(\v k) =  $\Delta^0_{SC}\sin k_x\sin k_y$; A+SC
denotes the coexistence region. Solid and dashed lines
correspond to second and first order transitions respectively. The
effective tight-binding hopping integral is t$_{eff}$, it should also scale
linearly with 1-x. Inset: FS topology with two bands. Dash-dotted line
indicates location of SDW gapping.}
\label{fig:CeCu2Si2PhaseDiagram}
\end{figure} 

The coexistence/competition behaviour of $\Delta_{A}$(\v k)  and
$\Delta_{S}$(\v k) is therefore quite different in case I and II and it also
depends strongly on the relative nodal structure of $\Delta_{A}$(\v k)
$\Delta_{S}$(\v k). Generally speaking if their nodes are on the same
positons on the FS (or if at least one of them is a conventional gap
function without nodes) this enhances
competition, nonmonotonic temperature behaviour and tendency to
expulsion of one OP. On the other hand an `orthogonal' nodal structure for both
(e.g. d$_{xy}$ symmetry for SDW and d$_{x^2-y^2}$ for SC) supports
coexistence because the two gaps become maximal at different parts of
the FS and this also favors monotonic temperature dependence of the gap
amplitudes.  

To illustrate this behaviour we briefly discuss an example which may
be relevant to \index{CeCu$_2$Si$_2$} CeCu$_{2}$Si$_{2}$.
Qualitatively the complex phase diagram of CeCu$_{2}$Si$ _{2}$ can be
described by a simple mean-field model for a conventional spin
density wave ground state (competing with d-wave \index{d-wave gap}
superconductivity within a two band model proposed by \citeasnoun{Dahm01}. The
calculated phase diagram displayed in 
fig.~\ref{fig:CeCu2Si2PhaseDiagram} shows that a state with
d$_{xy}$ symmetry can become stable within the SDW phase. This is in
contrast to the pairing states with d$_{x^2-y^2}$ symmetry considered by
\citeasnoun{Kato88} in a one band model. The experimentally observed
unusual stability of the (A+SC) phase might follow from the peculiar
Fermi surface of the heavy  quasiparticles displayed in fig.~
\ref{fig:CeCu2Si2fBandFermiSurface}. This is suggested by model
calculations which start from a more complex Fermi surface
model. Assuming two well-separated sheets centered at $\Gamma$ and M,
respectively, allows for a solution where the SDW gaps exist only on part
(centered around $\Gamma$) of the Fermi surface while the
quasiparticles on the remaining parts (centered around M) 
condense into a superconducting state. The calculated
temperature-'pressure' phase diagram (assuming that the electron band
width $\sim t_{eff}$ scales linearly with pressure) is qualitatively
in agreement with experimental results \cite{Gegenwart98} for
CeCu$_{2}$Si$_{2}$. Indeed the A-phase has now been
identified as conventional SDW state with an \index{incommensurate
order} incommensurate
propagation vector \v Q =(0.22,0.22,0.55) and a moment of 0.1$\mu_B$
per Ce-atom by neutron diffraction experiments \cite{Stockert03}. The
experimentally determined magnetic structure therefore requires a more
refined treatment of the SDW/SC coexistence behaviour.

\subsection{Methods to investigate the symmetry of order parameters}

\label{Sect:OrderParameters}

Phase transitions with the appeareance of a spontaneous symmetry
breaking are characterised by order parameters. In the spirit of
Landau theory they belong to representations of the high temperature
symmetry group and also characterise the class of remaining symmetries in
the low temperature phase. Furthermore they partly determine the
type of possible excitations in the ordered phase which in
turn influence the low temperature thermodynamic and transport properties.
The investigation of the order parameter symmetry is therefore of
singular importance. In the case of order parameters corresponding to
expectation values
of physical observables like spin- and charge densities this is in
principle straightforward as they reveal their existence and symmetry
in x-ray and neutron diffraction experiments. However for `hidden'
order e.g. quadrupole ordering or `unconventional' density waves
(orbital antiferromagnetism) there is no large obvious signature in diffraction
experiments and evidence for its presence has to be obtained
indirectly by other means. The superconducting order parameter does
not correspond to the expectation value of a  classical
observable but rather to the appearance of quantum mechanical phase
rigidity of Cooper pairs on a macroscopic scale (the coherence length). For
unconventional superconductors determination of the gap function
dependence on the momentum of paired
states is of fundamental importance. It is also extraordinarily
difficult. The classical experimental techniques which we discuss in
this section rely
on the interpretation of temperature dependences of physical
quantities obtained by {\em averaging} over the gap function momentum
dependence and, therefore, the conclusion about their proper symmetry is always
ambiguous. Recently a much more powerful method relying on the field
angle resolved measurement of \index{specific heat} specific heat and
\index{thermal conductivity} thermal conductivity
in the \index{vortex phase} vortex phase for T $\ll$ T$_c$ has been
developed. It leads
directly to the determination of the relative position of node lines
and/or points of the SC gap function with respect to crystal
axes. This leads to a strong reduction of the number of possible SC
order parameters and possibly allows a unique determination. Therefore we
devote an extra subsection to this new technique. On the other hand we
shall not discuss phase sensitive methods based on Josephson tunneling
which have been unsucessful in HF compounds and point contact
spectroscopy or $\mu$SR methods due to their difficult interpretation. 
For unconventional
density waves which also correspond to gap functions of nontrivial momentum
dependence in the particle-hole channel, the field angle
dependent investigation of magnetoresistance has similar potential but
is less developed.

\subsubsection{Detection of superconducting order parameter symmetry}
\label{SubSect:SCDetection}

In conventional electron-phonon superconductors with an almost isotropic
gap numerous thermodynamic, static and dynamic transport measurements
and also resonance methods can give information about the
superconducting gap. The SC transition affects these quantities in two ways:
i) the quasiparticle DOS exhibits a gap $\Delta$ with a square root
singularity of the DOS: N$_s$(E) = N$_0$E/(E$^2-\Delta^2$)
(E $\geq \Delta$) ii) coupling
of external probe fields to the quasiparticles involves a coherence factor
$\frac{1}{2}$(1$\mp\Delta^2$/EE') where $\mp$ corresponds to
perturbations even or odd under time reversal \index{time reversal
symmetry} respectively. In
the former case (I) the DOS singularity and the vanishing coherence
factor for E $\simeq\Delta$ compensate leading to a steep drop of the
corresponding physical quantity, e.g. ultrasonic attenuation below
T$_c$. In the latter case (II) the coherence factor is unity and
therefore a Hebel-Slichter type anomaly as e.g. in the NMR-relaxation rate
develops below T$_c$. At low temperatures (T $\ll$ T$_c$) invariably
experimental quantities determined by electron-hole excitations tend to zero
exponentially due to the finite superonducting gap.
Therefore the most obvious and seemingly easiest method to look for
evidence of unconventional superconductivity is the search for
deviations from the exponential low temperature dependence. This strategy
relies on the fact that nontrivial (unconventional) \vk-dependent
gap functions \De in many cases (though not always) exhibit node points or
lines on the Fermi
surface leading to a nonvanishing N$_s$(E) for all energies E $>$ 0 even in the
superconducting state. The asymptotic behaviour of N$_s$(E) for
E/$\Delta\ll$ 1 is $\sim$ E$^2$ for (first order) point nodes and
$\sim$ E for line nodes. Then this should ideally lead to \index{power
laws} `power law' behaviour $\sim$ T$^n$  for physical quantities like
specific heat, penetration depth, NMR relaxation rate and many others
for T $\ll$ T$_c$ . The experimentally observed exponent n
would then allow to determine whether \De has point nodes (n=3),
line nodes (n=2) or both if the exponent n is in between, or whether gapless
regions of the Fermi surface (n=1) exist. For nondirectional
quantities like C(T) this would however say nothing about the position
of nodes relative to the crystal axes. Unfortunately this picture is
so oversimplified as to make it useless. Firstly, experimental
determination of the exponent n very often includes temperatures where
T $\ll$ T$_c$ is not valid and more importantly impurity scattering in
anisotropic superconductors has dramatic effects \cite{Sigrist91,Fulde88a}. In
conventional superconductors Anderson's theorem ensures that the low
temperature thermodynamic properties are not affected by impurity
scattering. On the other hand normal impurities for unconventional
pairs act strongly pairbreaking. This is especially important in HF
compounds where the normal state electrons are already strongly
correlated and impurities act as unitary scattering centers. Their
effect for anisotropic superconductors has originally been studied by
 \citeasnoun{Buchholtz81} for p-wave states and later extended and applied 
 \cite{Hirschfeld86,Scharnberg86,Schmitt-Rink86,Hotta93,Sun95} for
other cases like d-wave. In the unitary scattering limit a resonant
residual quasiparticle DOS \index{quasiparticle DOS} at low energies
develops \cite{Hirschfeld86} invalidating the abovementioned power
laws. On the other hand hybrid nodal superconductors like s+g wave
type in borocarbides exhibit the opposite behaviour: Impurity
scattering immediately leads to the opening of a gap \cite{Yuan03a}
with resulting
low temperature exponential behaviour. It is therefore more reasonable 
to investigate the T-dependence of physical quantities
depending on the gap anisotropy for the whole temperature range below
T$_c$ and compare with
experiments, rather than looking  at the often ill-defined and
contradictory low-T power laws. This will be discussed now in a few cases.\\

\begin{figure}
\includegraphics[width=7.5cm]{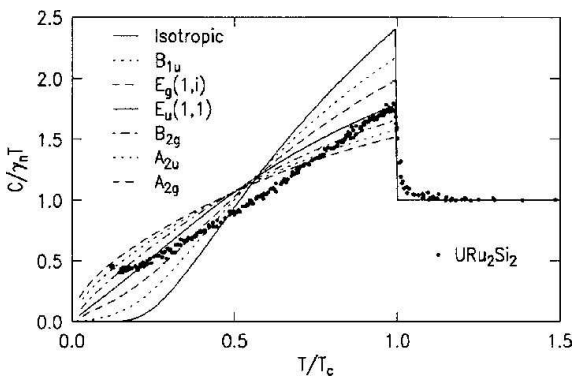}
\caption{Calculated specific heat for various superconducting order
\index{specific heat!URu$_2$Si$_2$}
parameters in comparison to experimental results for
URu$_2$Si$_2$ (Hasselbach et al. 1993). The specific heat jump $\Delta
C_s(T_c)$ agrees best for E$_u$(1,1) or B$_{1g}$ state.}
\label{FIGOPDspec}
\end{figure}

\index{specific heat}{\em Specific heat:}
The specific heat in the anisotropic SC state is given by an
expression formally identical to the s-wave case:
\begin{equation}
C_s=2k\beta\sum_{\vk}\Bigl(-\frac{\partial f_{\vk}}{\partial E_{\vk}}\Bigr)
\Bigl[E_{\vk}^2+\frac{1}{2}\beta\frac{\partial\Delta_{\vk}^2}
{\partial\beta}\Bigr] 
\qquad \mbox{and} \qquad
\frac{\Delta C_s}{C_n}=\frac{3}{2}\frac{8}{7\zeta(3)}
\frac{\la f^2_{\vk}\ra_{FS}}{\la f^4_{\vk}\ra_{FS}}
\end{equation}
where $\beta$ = 1/kT, 
E$_{\vk}$ = ($\epsilon_{\vk}^2+\Delta_{\vk}^2$)$^\frac{1}{2}$,
f$_{\vk}$ = f(E$_{\vk}$) is the Fermi function and C$_n$ = $\gamma$T.
The specific heat jump given by the second formula has the BCS value
$\frac{\Delta C_s}{C_n}$ = 1.43 for the isotropic s-wave gap. In
fig.~\ref{FIGOPDspec} model calculations for
C$_s/\gamma$T by \cite{Hasselbach93} for all possible gap symmetries
in tetragonal D$_{4h}$ group are 
shown and compared with experimental results for URu$_2$Si$_2$. The latter
show remarkably linear behaviour over the whole experimental
temperature range. It is obvious that a perfect fit is not possible
with any gap representation and at low T a tendency for saturation due
to an impurity induced residual DOS is visible. A comparison of
theoretical and experimental $\Delta$C(T$_c$) jumps however favors
either a E$_u$(1,1) state \De = $\Delta$(k$_x$+k$_y$)$^2$ or B$_{1g}$
state \De = $\Delta$(k$_x^2$-k$_y^2)^2$ for which one has
2$\Delta$(0)/kT$_c$ = 4.92 or 5.14 respectively. Since the presence of
nodes reduces T$_c$ compared to s-wave case for the same gap amplitude the BCS ratio (table~\ref{TABLEBCS}) is
always larger for anisotropic SC. However for real compounds this
increase may also be partly due to strong coupling effects. Likewise
the ratio $\Delta C_s/C_n$ is always maximal for isotropic SC.\\

\begin{table}
     \caption{\scriptsize BCS \index{BCS ratio} ratio
      2$\Delta(0)$/kT$_c$ and specific \index{form factor}
      heat ratio $\Delta C_s/C_n$ for some common order parameter
      models (Einzel 2002) characterised by the gap form factor
      f($\vartheta,\varphi$) as function of polar angles of \vk~ and 
      normalised on the unit sphere (a = 3$\sqrt{3}$/2).} 
     \vspace{0.5cm} 
     \begin{tabular}{cccccc}
     \hline
     & isotropic & axial  & polar & E$_{1g}$(D$_{6h})$ & E$_{2u}$(D$_{6h})$ \\
     \hline
     f($\vartheta,\varphi$) & 1 & $\sin\vartheta$ & $\cos\vartheta$ &
     2$\sin\vartheta\cos\vartheta$ & a$\sin^2\vartheta\cos\vartheta$\\
     2$\Delta(0)$/kT$_c$ & 3.52 & 4.06 & 4.92 & 4.22 & 4.26 \\
     $\Delta C_s/C_n$      & 1.43 & 1.19 & 0.79 & 1.00 & 0.97 \\
     \hline
     \end{tabular}
     \label{TABLEBCS}
\end{table}

\index{thermal conductivity}{\em Thermal conductivity:}
The thermal conductivity tensor is a directional quantity depending on
the orientation of temperature gradient and heat current. At low
temperatures its anisotropy may contain important information on the
relative position of nodal points or lines with respect to crystal
axes. For example this has proved decisive in the identification of
the E$_{2u}$ SC order parameter of UPt$_3$ (sect.~\ref{Sect:UPt3}). The
uniaxial thermal conductivity tensor in the unitary scattering limit
for a spherical Fermi surface is explicitly given 
\cite{Norman96,Kuebert98,Machida99,Graf00,Wu01} as 
\begin{eqnarray}
\label{THERMAL}
\frac{\kappa_i(T)}{\kappa_n}&=&\frac{9}{2\pi^2T_c}\int_0^\infty
d\omega(\frac{\omega}{T})^2 sech^2\frac{\omega}{2T}K_i(\omega,T)\nonumber\\
K_i(\omega,T)&=&
\frac{\la\hat{k}^2_i Re[\omega^2-|\Delta_{\vk}(T)|^2]^{\frac{1}{2}}\ra}
{\la Re[\omega^2-|\Delta_{\vk}(T)|^2]^{-\frac{1}{2}}\ra}
\end{eqnarray}
where $\hat{k}_i$ = ($\hat{\bf i\rm}\cdot\hat{\bf k\rm}$) =
$\cos\theta_i$ and $\hat{\bf i\rm}$ are the unit vectors along x,y,z
axes.

At low temperatures the \vk-averaging in $\kappa_i(T)$ is determined
only by the quasiparticles close to the nodal region. Because of the
direction cosines in the averages of eq.~(\ref{THERMAL}) coming from the
quasiparticle velocities, $\kappa_i$ depend
on the position of the nodal points or lines with respect to the axis
$\hat{\bf i\rm}$. When the nodal direction in \vk-space is parallel to the
direction $\hat{\bf i\rm}$ of the heat current a large contribution to
$\kappa_i(T)$ is obtained and a small one if it is perpendicular. This
may lead to a uniaxial thermal conductivity anisotropy characterised
by the ratio \cite{Norman96}
\begin{eqnarray}
\frac{\kappa_c(0)}{\kappa_a(0)}=\lim_{\omega\rightarrow 0}
\frac{Re\la\hat{k}^2_z[\omega^2-|\Delta_{\vk}(T)|^2]^{\frac{1}{2}}\ra}
{Re\la\hat{k}^2_x[\omega^2-|\Delta_{\vk}(T)|^2]^{\frac{1}{2}}\ra}
\end{eqnarray}
As discussed in sect.~\ref{Sect:UPt3} this anisotropy ratio of thermal
conduction has proved quite useful in the identification of the order
parameter symmetry in UPt$_3$ (see fig.~\ref{FIGacratio}).

\index{ultrasonic attenuation}{\em Ultrasonic attenuation:}
Sound attenuation has in addition to the propagation direction
($\hat{\vq}$) another polarization ($\hat{\ve}$) degree of freedom and
may therefore give even more
information on the gap nodes than the thermal conductivity.
In HF metals sound attenuation can generally be considered in the
hydrodynamic limit where $\omega\tau$, ql $\ll$ 1 with q, $\omega$
denoting wave number and frequency ($\sim$100 MHz) of the sound wave
respectively and l = v$_F\tau$ is the mean free path in the normal
state. In this case conduction electrons act like a viscous medium to
the sound waves and the attenuation coefficent may be expressed as a
correlation function of the electronic stress
tensor \cite{Tsuneto61,Kadanoff64}. Its evaluation for unconventional
SC states using a quasiclassical approximation leads
to \cite{Machida99,Graf00}
\begin{eqnarray}
\frac{\alpha_{ij}(T)}{\alpha_n}&=&\frac{1}{2T}\int_0^\infty
d\omega sech^2\frac{\omega}{2T}A_{ij}(\omega,T)\nonumber\\
A_{ij}(\omega,T)&=&\frac{1}{\la\Pi_{ij}^2\ra}
\la\Pi_{ij}^2\frac{1}{\omega} 
Re[\omega^2-|\Delta_{\vk}(T)|^2]^{\frac{1}{2}}\ra \\
\Pi_{ij}&=&\hat{k}_i\hat{k}_j-\frac{1}{3}\delta_{ij}\nonumber
\end{eqnarray}
%
\begin{figure}
\includegraphics[width=7.5cm]{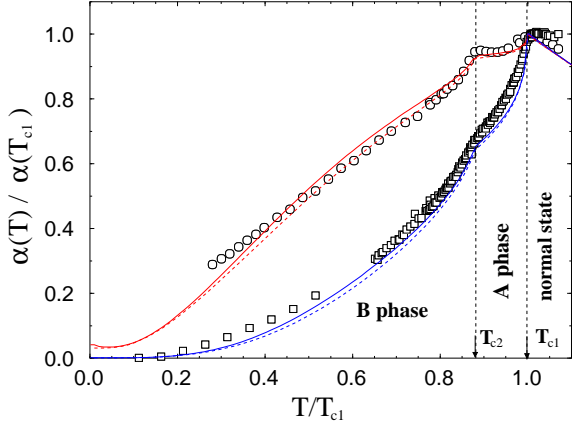}
\caption{Temperature dependence of ultrasonic attenuation in B-phase
of UPt$_3$ for transverse waves with propagation along a-axis and
polarisation along b-axis (ab: circles) or c-axis (ac: squares),
from Graf et al. (2000). Here T$_{c1,2}$ denotes the split SC
transition temperature. \index{ultrasonic attenuation!UPt$_3$}}
\label{FIGOPDatte}
\end{figure}
The projection factor $\Pi_{ij}$ (i = polarisation, j = propagation
direction) determines which of the quasiparticles having a
$\hat{\vk}$- vector 
described by polar and azimuthal angles $\vartheta,\varphi$
contribute strongly to the attenuation. Consider two cases with
ij = ab where $\Pi_{ij}^2\sim\sin^2(2\varphi)$ and ij = ac with 
$\Pi_{ij}^2\sim\sin^2(2\vartheta)$. The quasiparticle contribution to
$\alpha_{ij}$(T) therefore vanishes for
nodal directions $\varphi=n(\frac{\pi}{2})$ and $\vartheta=n(\frac{\pi}{2})$
respectively. For example if the gap has a node line in the
basal plane ($\vartheta=\frac{\pi}{2}$) it gives maximal contribution
to the attenuation for ij=ab but minimal contribution for
ij=ac. The same argument holds for point nodes on the c-axis where
$\vartheta$=0,$\pi$. This situation is indeed realized in the 
E$_{2u}$ type B phase of UPt$_3$ (sect.~\ref{Sect:UPt3}) and one can see in
fig.~\ref{FIGOPDatte} that the attenuation
for ac configuration is much smaller than for ab. The anomalies around
the split T$_c$ are complicated and analyzed in \cite{Graf00}. Therefore
polarization and propagation dependence of ultrasonic attenuation is a
powerful method to distinguish between gap representations with
different nodal structure. Finally it should be mentioned that
collective modes in unconventional superconductors hardly 
contribute to the attenuation since one has an off-resonance
situation due to v$_s\ll$ v$_F$ \cite{Kee00}.\\

\index{nuclear magnetic resonance}\index{Knight shift}
{\em NMR relaxation and Knight shift:}

In an external field nuclear spins exhibit a Larmor precession with a frequency
$\omega_0$. In metals this is modified by the hyperfine coupling to conduction
electrons which leads to spin-flip processes as witnessed by the NMR-relaxation rate T$_1^{-1}$ and the Knight shift of the resonance
frequency $\delta\omega_0$. A review of these important effects for HF
superconductors is given by \citeasnoun{Tou03}. The relaxation rate
T$_1^{-1}$ is determined by the availability of resonant electron-hole
excitations. In the normal state 
this leads to the Korringa law T$_{1n}^{-1}\sim$ T. In the
superconducting state the presence of \De should lead to a faster
decrease with temperature depending on the type of node structure,
i.e. the low energy behaviour of the quasiparticle DOS. According
to \citeasnoun{Sigrist91} for singlet pairs one has
\begin{eqnarray}
\label{NMR}
\frac{T_{1n}}{T_1}=\frac{2}{N_n^2}
\int_0^\infty dE\Bigl(-\frac{\partial f}{\partial E}\Bigr)
N_s(E)N_s(E+\omega_0)
\Bigl[1+\frac{|\la\Delta_{\vk}\ra_{FS}|^2}{E(E+\omega_0)}\Bigr]
\end{eqnarray}
For s-wave pairs the type II-coherence factor together with the
singular DOS leads to the appearance of the Hebel-Slichter
peak. Because for unconventional pair states the Fermi surface average
in eq.~(\ref{NMR}) vanishes, there is no difference between type I and
II coherence factors and the relaxation rate is determined by the DOS
alone. Because in the presence of nodes there is no divergence in
N$_s$(E) no Hebel-Slichter peak will appear below T$_c$.
Its absence may therefore be taken as a sign of an unconventional pair
state. Furthermore the low temperature behaviour of T$_1^{-1}$ should be
$\sim$ T$^3$ for line nodes and $\sim$ T$^5$ for point
nodes. Invariably the former is observed in HF systems, often in
conflict with `power laws' of other
quantities. \index{power laws} A puzzling feature is the absence of any
crossover for T $\ll$ T$_c$ to Korringa behaviour in the impurity
dominated gapless regime.\\

In addition to the relaxation the hyperfine interaction leads to a
Knight shift of the resonance frequency given by
\begin{equation}
\delta\omega_0(T)\sim|\psi_0|^2\chi_s(T)H 
\end{equation}
where $\psi_0$ is the conduction electron wave function at the nucleus
and $\chi_s(T)$ is the {\em spin} susceptibility of conduction
electrons. Without spin orbit coupling present in an s-wave singlet
state this drops to zero exponentially below T$_c$, leading to a
pronounced T-dependence of $\delta\omega_0$. For a triplet
SC state the condensate spin or the \vd-vector of the pair can be
freely rotated and $\chi_s(T)$ should not change leading to T-indepenent
$\delta\omega_0$. Therefore the existence or non-existence of
a T-dependent Knight shift below T$_c$ is usually interpreted as direct
evidence for singlet or triplet superconductivity
respectively. \index{singlet pairing}\index{triplet pairing} In the
former case the Knight shift should be proportional to the Yoshida
function given by the FS average
\begin{eqnarray}
Y(T)=\la\frac{1}{2T}\int_0^\infty 
d\epsilon\cosh^{-2}\frac{E_{\vk}}{2T}\ra_{FS}
\end{eqnarray}
with a low T behaviour Y(T) $\sim$ T$^2$ for point nodes and $\sim$ T for
line nodes. Unfortunately the situation may be completely changed
under the presence of strong spin orbit (s.o.) coupling, for both the
conduction electrons themselves and for the conduction electron-impurity
scattering. If the mean free
path due to conduction electron-impurity s.o. scattering is much smaller than
the coherence length, the Knight shift variation with temperature
approaches zero also for singlet pairing \cite{Abrikosov88}. On the other
hand in a triplet pair state a conduction
electron s.o. coupling may lead to the pinning of the triplet pair \vd-vector 
along one of the crystal axis. Then the T-dependence of 
$\delta\omega_0$ should be very
anisotropic, vanishing for \vH$\perp$\vd~ and large for
\vH$\parallel$\vd. A moderately anisotropic Knight shift is indeed
found in UPt$_3$ (sect.~\ref{Sect:UPt3}).
These complications make a quantitative analysis as for the previous
quantities difficult, in fact it is rarely performed for HF
compounds. The arguments for or against singlet or triplet pairing
using Knight shift results should therefore be taken with caution.\\ 

\index{upper critical field}{\em Upper critical field:}
The temperature dependence and anisotropy of the upper critical field
H$_{c2}$ may contain important information both about the question of singlet
vs. triplet pairing and the anisotropy of the gap function
\De. A general discussion of the problem within the semiclassical
approach has been given in \citeasnoun{Rieck91}.
Theoretical analysis in the context of a generalized Ginzburg
Landau theory depends very much on the crystal symmetry and
order parameter model. It will be briefly discussed in sect.~\ref{Sect:UPt3}
for UPt$_3$. In this case the importance of uniaxial anisotropy to decide
between singlet and triplet pairing has been realized
\cite{Choi91,Sauls94,Yang99}. In a pure isotropic superconductor the
upper critical field H$_{c2}$(0) is limited by orbital effects leading to 
\begin{equation}
\label{CRITORB}
\mu_BH_{c2}(0) = 0.58T_{c}|dH_{c2}/dT|_{T_c}
\end{equation}
The actual critical field may be much lower caused by the effect of
paramagnetic `Pauli-limiting' due to the Zeeman energy of the Cooper
pair. If the field is larger than $\mu_B$H$_{\mbox{P}}$ =
$\Delta_0/\sqrt{2}$ = 1.25T$_c$ the upper critical field is
limited by the breaking up of singlet Cooper pairs. According to
eq.~(\ref{CRITORB}) a large critical field slope should also lead to
a larger H$_{c2}$(0). In UPt$_3$ the opposite is true leading to a
'crossing' of H$_{c2}(T)$ curves for a- and c-directions. This led to
the proposal that  H$_{c2}$(0) is strongly reduced by Pauli-limiting
effects for field along c but not along a. This can only be explained
by assuming triplet pairing with strongly pinned \vd-vector parallel
to c. As will be discussed in sect.~\ref{Sect:UPt3} this interpretation is not
undisputed. In addition to the question of the spin state a nontrivial
\vk-dependence of the gap should also lead to anisotropies in the upper
critical field (slope) beyond those due to effective mass
anisotropies. This proved difficult or inconclusive for HF
superconductors investigated sofar \cite{Keller95}.

\subsubsection{Specific heat and \index{magnetotransport}
magnetotransport in the \index{vortex phase} vortex phase: a
genuine angular resolved method}

The classical methods to determine \De discussed above all
suffer under the same major deficiency: the angular dependence, or at
least nodal positions in \vk-space are never determined
directly. Instead the physical quantities investigated are obtained by
{\em averaging} over the quasiparticle and hence gap energies in
\vk-space. And then it is hoped that the averaged quantities still
show clear signatures depending on the gap symmetry. Naturally this leads to 
ambiguous results for any one method and only a comparison of all gap
models for a number of methods may lead to a conclusion on the gap symmetry.

However recently a new and powerful method has been established
that is able to locate directly the nodal positions of the SC gap in
\vk-space \citename{Izawa01} \citeyear{Izawa01,Izawa01a,Izawa02}. In
this method the {\em field-angle} dependence of
specific heat or thermal conductivity is investigated. Typical
oscillations are observed whose periodicity, phase and shape give
direct information on the type of \index{gap nodes} nodes (point- or
line-like) and their direction in \vk-space. Therefore possibilities
for allowed gap functions can be much more restricted. Notice however
that still the full \vk-dependence of \De cannot be determined.

\begin{figure}
\includegraphics[width=7.5cm]{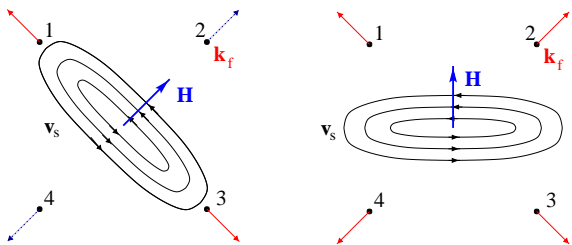}
\caption{Illustration of the Volovik effect \index{Volovik effect} 
(Vekhter et al., 2000). 
\vk$_i$ (i=1-4) denote the nodal directions of the gap. Left: field
\vH~ along nodal direction where \vv$_s$ is orthogonal to \vk$_2$, \vk$_4$;
these nodal directions do not contribute to the \index{Doppler shift}
Doppler shift, the \index{quasiparticle DOS}
residual DOS will be at minimum. Right: Field along antinodal
direction, all \vk$_i$ (i=1-4) directions contribute to the Doppler shift,
the residual DOS is maximal.}
\label{FIGOPDdopp}
\end{figure}

At the heart of this method is the `Volovik effect' in the vortex
state of the superconductor, it was first proposed by
\citeasnoun{Volovik93} for \index{d-wave gap} d-wave
superconductors. There it was shown that for \index{gap nodes} nodal
gap functions a continuum of quasiparticle states exists outside the
vortex cores
which dominate the specific heat and can easily carry a thermal
current even perpendicular to the vortex lines. These states are
due to quasiparticles channeling out of
the vortex core region through the nodal points or lines.
This is an important new aspect as compared to conventional
nodeless gap functions where one has only bound states in the vortex
core. For the quasiclassical limit $\xi_0k_F\gg$1 they also form a
quasi-continuum but they can carry a heat current only parallel to the
vortex direction. The theory of magnetothermal transport has
subsequently been developed by many authors
\cite{Barash97,Kuebert98,Vekhter00,Won00,Dahm00,Won01,Won01a}.
The Volovik effect can nicely be explained within the quasiclassical
picture where momentum (\vk) and position (\vcr) coordinates commute and the
quasiparticle energy E(\vk,\vcr) and occupation f(\vk,\vcr) depends on
both. The dependence on \vcr~ comes from the fact that the energy of
quasiparticles channeling into the inter-vortex region gets Doppler
shifted due to the \vcr-dependent superfluid velocity field
\vv$_s$(\vcr). This leads to E(\vk,\vcr)=
E(\vk)-\vv$_s$(\vcr)$\cdot$\vk. Here E(\vk) is the zero-field
quasiparticle energy and the second term is the \index{Doppler shift}
Doppler shift energy. This  position dependent shift leads to a
finite residual DOS \index{quasiparticle DOS} which will depend both
on field magnitude and direction:
\begin{equation}
\frac{N_s(E,\vH)}{N_n}\simeq\frac{1}{\Delta}
\la\la|E(\vk)-\vv_s(\vcr)\cdot\vk|\ra\ra
\end{equation}
The double average is performed both over the velocity field
coordinate \vcr~ and the quasiparticle momentum. It will depend on the
direction of the magnetic field \vH($\theta,\phi)$ with respect to the nodal
directions ($\theta$ and $\phi$ are polar and azimuthal angles of \vH~ with
respect to the c-axis).
An illustration of this direction dependence is given in fig.~\ref{FIGOPDdopp}.
The variation of the field angles ($\theta,\phi$) is therefore expected to
lead to 
a periodic variation in N(E,$\theta,\phi$) and hence in the specific
heat and thermal conductivity components. Oscillations will be strongest
i) in the low temperature limit T $\ll\tilde{v}\sqrt{eH}\ll\Delta(0)$
when only quasiparticle from the nodal regions contribute to the
residual DOS and ii) if the node structure is not smeared to much by
impurity scattering, i.e. one is in the \index{superclean limit}
`superclean limit' with
$(\Gamma\Delta)^\frac{1}{2}\ll\tilde{v}\sqrt{eH}$ where $\Gamma$
is the impurity
scattering rate and $\tilde{v}=\sqrt{v_av_c}$ with v$_a$ and v$_c$
giving the Fermi velocities along a-and c-axis in uniaxial symmetry.
This limit can also be expressed as $\Gamma/\Delta\ll H/H_{c2}\ll$ 1.
For hybrid nodal gaps such as s+g wave gap in the borocarbides
the superclean condition should be replaced by
$\Gamma\ll T\ll\tilde{v}\sqrt{eH}$ due to the completely different
effect of impurities in this case (sect.~\ref{Sect:Boro}).
Including the effect of \index{impurity scattering} impurity
scattering one obtains for the residual DOS at the Fermi level \cite{Won00}:
\begin{eqnarray}
\label{DENSVOR}
\frac{N_s(0,\vH)}{N_n}&=&\it Re[\rm g(0,\vH)]\nonumber\\
g(0,\vH)&=&\Biggl\la\frac{C_0-ix}{\sqrt{(C_0-ix)^2+f^2}}\Biggr\ra\\
C_0&=&\frac{\Gamma}{\Delta g(0,\vH)}\nonumber
\end{eqnarray}
Here f(\vk)=\De/$\Delta$ and x= $|\vv_s\cdot\vk|/\Delta$ are the
normalized gap function and Doppler shift energy respectively while
C$_0$ is a normalized scattering rate. The above condition for
the superclean limit can be expressed as C$_0\ll$ x.

The averaging is performed by restricting the \vk-integration to the
node region (since T $\ll\Delta$) and approximating the \vcr-integration
over the superfluid velocity field of the vortex lattice by that of a
single vortex (assuming H $\ll$ H$_{c2}$) with cutoff d=1/$\sqrt{eH}$
given by the half-distance between two vortices \cite{Won01}. It is
important to note that the averaging of the Doppler shift energy has
to be done simultaneously in \vcr~ and \vk. The field (angle-) 
dependent specific heat C$_s$, spin susceptibility $\chi_s$ and superfluid
density are then given by  

\begin{figure}
\includegraphics[width=7.5cm]{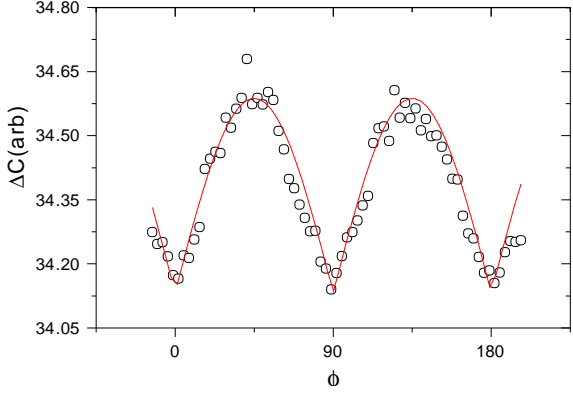}
\caption{Specific heat oscillation $\Delta$C($\phi$) in
\index{specific heat!YNi$_2$B$_2$C}
YNi$_2$B$_2$C for field \vH~ in the ab-plane as function of the azimuthal
field angle $\phi$. Full line is a fit using a phenomenological cusp
function (Park et al. 2003).}
\label{FIGOPDspecosc}
\end{figure}
%
\begin{eqnarray}
\frac{C_s(\vH)}{\gamma T}=\frac{\chi_s(\vH)}{\chi_n}=g(0,\vH)
\qquad \mbox{and} \qquad
\frac{\rho_s(\vH)}{\rho_s(0)}=1-g(0,\vH)
\end{eqnarray}
The explicit expression for C$_s$(H,$\theta,\phi$) depends on the node
structure of \De and is, up to numerical constants of order unity, obtained as
\begin{equation}
C_s(H,\theta,\phi)=\frac{\tilde{v}\sqrt{eH}}{\Delta}I(\theta,\phi)
\end{equation}
Where for example for the d-wave order parameter with
f(\vk)=$\cos(2\varphi)$ and two orthogonal
line nodes parallel to c one has for planar magnetic field \cite{Won00}
\begin{equation}
I(\frac{\pi}{2},\phi)\simeq 0.95 + 0.028\cos(4\phi) 
\end{equation}
Thus the specific heat should exhibit a fourfold oscillation with
field rotation in the ab-plane with a minimum in the nodal and the
maximum in the anti-nodal 
direction. The amplitude of the oscillation is not large because
of the \vk-averaging in eq.~(\ref{DENSVOR}) along the node line. It
may be larger for a gap function with point nodes as in the case of
borocarbides which is discussed in detail in sect.~\ref{Sect:Boro}. This has
been experimentally investigated by \citeasnoun{Park03} as shown in
fig.~\ref{FIGOPDspecosc}. It also shows the typical cusp-like minima
of point nodes as opposed to smooth minima observed for gap functions
with line nodes. The smoothing is also due to the additional
averaging along the line node. A most significant result is the
observed $\sqrt{H}$
dependence of the specific heat magnitude which is also a
fingerprint of nodal superconductivity (fig.~\ref{FIGPtsub}). However
one must keep in mind
that this should be observed for H $\ll$ H$_{c2}$, a global
approximate behaviour over the whole field range as sometimes found
in conventional superconductors has no relation to the above effect.

Similar field-angle dependence of \index{thermal conductivity} thermal
conductivity can be
calculated starting from the Ambegaokar-Griffin formula generalized to
anisotropic gap functions. In the d-wave case one obtains in the
superclean limit \cite{Won00}.
\begin{eqnarray}
\frac{\kappa_{xx}}{\kappa_n}&=&\frac{2}{\pi}\la x\ra^2
=\frac{2}{\pi}\frac{\tilde{v}^2}{\Delta^2}I(\frac{\pi}{2},\phi)^2
\end{eqnarray}
%
\begin{figure}
\includegraphics[width=7.5cm]{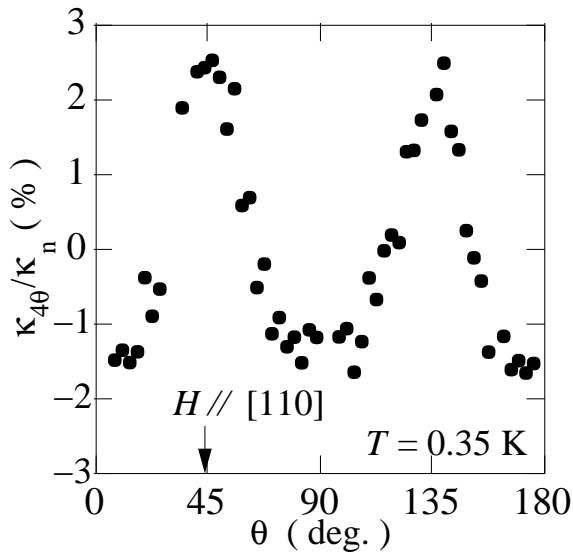}
\caption{Thermal conductivity \index{thermal conductivity!CeCoIn$_5$}
$\kappa_{xx}(\phi)$ for field in the
ab-plane as function of the azimuthal angle $\phi$ for
CeCoIn$_5$ (Izawa et al 2001a). A fourfold oscillation compatible with
d$_{x^2-y^2}$ SC state \index{d-wave gap} is clearly seen.}
\label{FIGOPDosc}
\end{figure}
Due to the magnetic field there is also an off-diagonal `thermal Hall
conductivtity' $\kappa_{xy}$ which has been identified in high-T$_c$
compounds \cite{Ocana02}. The fourfold oscillations in
$\kappa_{xx}(\phi)$ have been found in the HF superconductor
CeCoIn$_5$ with node directions $\phi=0,\frac{\pi}{2}$ corresponding
to maxima in $\kappa_{xx}(\phi)$ in the anti-nodal directions
$\phi=\pm\frac{\pi}{4}$ as shown in fig.~\ref{FIGOPDosc}. This lead to
the proposal of a d$_{x^2-y^2}$ gap functions as in high-T$_c$
compounds. For this geometry, where both field and heat current are in
the ab-plane, there is also a term that shows a twofold rotation
$\sim\cos(2\phi)$ which is possibly due to the heat current
contribution of vortex core states. This contribution is zero or
finite for heat current perpendicular or parallel to vortex
lines respectively. Therefore a more favorable geometry is to rotate the field
around the heat current direction, e.g. measuring $\kappa_{zz}$ as
function of the azimuthal angle for a given polar angle $\theta$. In
this case only oscillations due to the nodes should appear. Various 
examples will be discussed later (sect.~\ref{Sect:UPt3},
sect.~\ref{Sect:Boro} and sect.~\ref{Sect:PRSK})

Finally we note that the field angle-oscillations in specific heat
and thermal conductivity cannot be explained by normal state Fermi surface
effects. Calculations based on the Kubo formula indicate that the
angle variation of thermal conductivity is determined by the
anisotropy of the gap function and not by that of the Fermi velocity
which enters only as secondary effect. This can be checked
experimentally by measuring the angle dependence in the {\em normal}
state for fields above H$_{c2}$. Also the (planar) angle dependence of
H$_{c2}$ is usually of a few percent and for fields H $\ll$ H$_{c2}$
this does not contribute significantly to the angle dependence of the
thermal conductivity.

The field angle dependence of specific heat and thermal transport in the
vortex phase have revolutionized the investigation of superconducting
gap structure in nodal superconductors. Many ambiguites from the
previous methods which have persisted over years have been resolved in
a short time. This method certainly has enormous potential, it remains
to be seen whether it can also be tried for other quantities,
e.g. ultrasonic attenuation in the vortex phase as function of
magnetic field direction.

\subsubsection{Detection of density wave type order parameters}

\label{SubSect:DensityWaves}

Density wave order parameters are of `diagonal long range order'
(DLRO) type and correspond to the expectation \index{charge density wave} 
value of a physical observable, i.e. the charge or spin density
operator in eq.~(\ref{DENSOP}). \index{spin density wave}
Their gap functions for e-h pair states are determined by
selfconsistent equations formally identical to the BCS gap equations
as obvious from eq.~(\ref{COEXI}). The physical meaning of
the gap function in the e-h pairing \index{electron-electron pairing}
case becomes obvious by
calculating the expectation values of charge (C) and spin (S) densities in
eq.~(\ref{DENSOP}):
\begin{eqnarray}
\label{DENSITY}
\la\rho_{\vQ}\ra&=&\sum_k\Delta_C(\vk)F(\vk)
\qquad \mbox{and} \qquad
\la M^z_{\vQ}\ra=\sum_k\Delta_S(\vk)F(\vk)
\end{eqnarray}
with F(\vk) = $\tanh(E_{\vk}$/2T)/2E$_{\vk}$ and
E$_{\vk}=\sqrt{\epsilon(\vk)^2+\Delta_i(\vk)^2}$ (i = C,S) denoting
the quasiparticle energy and the gap functions $\Delta_i(\vk)$ are
given by eq.~(\ref{CSGAP}).

\index{conventional pairing}{\em Conventional density waves :}\\
In the conventional `s-wave' CDW and SDW case with \vk-independent
gap functions $\Delta_{C,S}$(\vk) = $\Delta$ the sum in eq.~(\ref{DENSITY})
leads to a finite charge or spin density Fourier component for the
modulation vector \vQ. This also holds for a fully symmetric
(A$_{1g}$) gap function which is \vk-dependent but satisfies
$\Delta(g\vk)=\Delta(\vk)$ for any group element g $\in$ G. Due to
nonvanishing $\la\rho_{\vQ}\ra$ and $\la M^z_{\vQ}\ra$
conventional CDW and SDW states are straightforward to identify because
macroscopic densities either couple to the electric and magnetic
fields of external probes or, as in the case of CDW, lead to observable
distortions of the underlying crystal lattice. 
In neutron and (magnetic) x-ray diffraction experiments one can therefore
observe additional lattice or magnetic superstructure Bragg
reflections that originate in the scattering from density modulations
$\la\rho_{\vQ}\ra\cos(\vQ\cdot\vcr)$ and $\la
M^z_{\vQ}\ra\cos(\vQ\cdot\vcr)$. This
is the standard method to observe broken spatial
and time reversal symmetries, i.e. structural and magnetic phase
transitions. There are also numerous other physical quantities,
both static and dynamic which are affected by the appearance of
conventional CDW and SDW modulations for which we refer to the book by
\citeasnoun{Gruener94} and references cited therein.

\begin{figure}
\includegraphics[width=6cm]{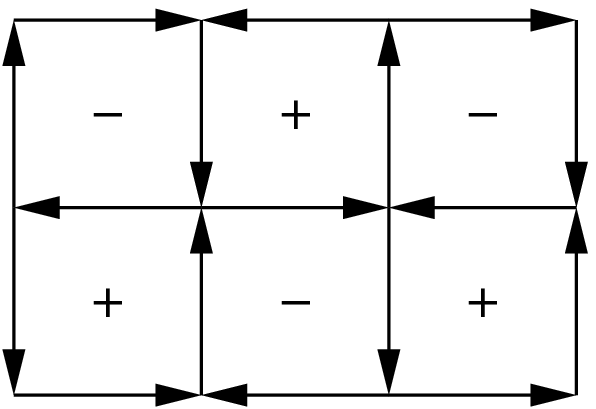}\hfill
\includegraphics[width=7.5cm]{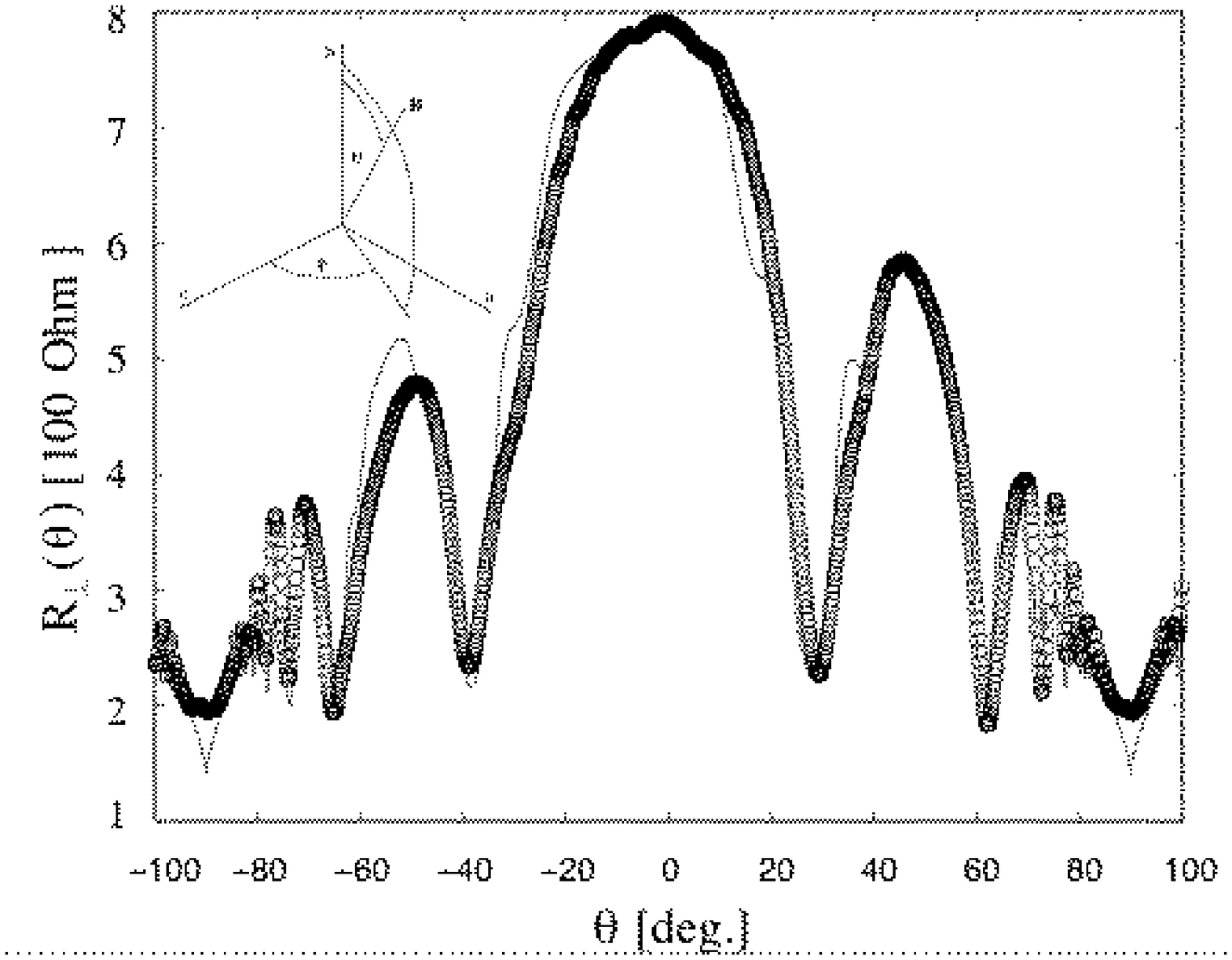}
\caption{Left: illustration of staggered charge or spin current pattern in
the square lattice for d-CDW or d-SDW. Right: magnetoresistance (\v
j $\perp$ ac-plane) \index{magnetotransport}
oscillations in the low temperature d-CDW state of the organic
quasi-1D conductor $\alpha$-(BEDT-TTF)$_2$KHg(SCN)$_4$. The
oscillations result from the \index{Landau levels} Landau quantisation
of energy levels
eq.~(\ref{RLANDAU}) around the nodal Dirac points in
fig.~\ref{FIGOPDpair}. The circles are from experiment (T = 1.4 K, B = 15
T, $\phi$ = 45$^\circ$), and the full line from d-CDW calculations (Maki et
al. 2003).}
\label{FIGOPDcurr}
\end{figure}

\index{unconventional pairing}{\em Unconventional density waves:}\\
These order parameters also correspond to DLRO of density operators
but the gap functions $\Delta_{\vk}$ now belong to a nontrivial
representation of G sometimes called `unconventional density
waves'. In this case there will
be elements g$\in$ G with $\Delta(g\vk)=-\Delta(\vk)$ signifying a
sign change of $\Delta_i(\vk)$ between different sectors of \vk-space. Under
these circumstances cancellation in eq.~(\ref{DENSITY}) occurs and
macroscopic charge and spin-densities
$\la\rho_{\vQ}\ra$ and $\la M^z_{\vQ}\ra$ vanish even though the order
parameter corresponds to a physical observable (therefore the name convention
for these states is misleading). A more formal proof of this
cancellation effect was given in \citeasnoun{Thalmeier94}. Now one has a
perplexing situation of \index{hidden order} `hidden order'. Although
there is a condensation of e-h pairs associated with quasiparticle energies
and thermodynamic signatures
similar to conventional (s-wave) CDW and SDW, the order parameter cannot be
identified by the usual diffraction type experiments. This makes the
search for such `hidden order' a difficult task, and in fact it has
not been unambiguously successful in any specific material. Some of the
most prominent candidates for unconventional density waves are; the HF compound
\index{URu$_2$Si$_2$} URu$_2$Si$_2$ \cite{Ikeda98} to be discussed in
sect.~\ref{Sect:URu2Si2}; 
the underdoped pseudo-gap phase of high-T$_c$ materials
\cite{Benfatto00,Chakravarty01}; and the organic conductors
\cite{Dora01,Basletic02,Korin-Hamzic02}, especially Bechgaard-salts
(\citename{Maki03} \citeyear{Maki03,Maki03a}). To discuss the properties
of unconventional CDW and SDW e-h pair states and their physical
signature we restrict ourselves in the following to the  
2D tight binding model with nearest neighbor (n.n.) hopping which in
this respect is the only model sofar investigated in any detail. In
this case the \index{tight binding model}
tight-binding band $\epsilon_{\vk}$= -2t($\cos k_x +\cos k_y$) has the
important \index{nesting property} perfect nesting property
$\epsilon_{\vk\pm\vQ}$= - $\epsilon_{\vk}$
leading to parallel Fermi surface sections at half filling (E$_F$=0)
as shown in fig.~\ref{FIGOPDpair}. This also leads to a singular
behaviour of the DOS according to N(0)$\sim\ln(t/|\epsilon|)$ due to
the saddle points (S) of $\epsilon_{\vk}$ at \vk=(0,$\pm\pi$) and
($\pm\pi$,0). The various 
possibilities of density wave and superconducting pair states assuming
three types of interactions (U = on-site Coulomb, V = n.n. Coulomb, J =
n.n. exchange) were studied and classified 
 \cite{Gulacsi87,Schulz89,Ozaki92}. From the many possible phases the most
important ones are conventional s-SC, s-CDW,SDW, and unconventional
d-SC,d-CDW,SDW states. Historically the latter two states have also
been called orbital antiferromagnet (OAF) and spin-nematic (SN) states
respectively \cite{Nersesyan89,Nersesyan91,Gorkov92} which points to their
physical origin in staggered charge or spin currents on the 2D
square lattice respectively. Their pairing amplitudes or gap functions
have a d-wave \index{d-wave gap} like \vk-dependence given by 
\begin{eqnarray}
\Delta_{C,S}^d(\vk)= i\Delta_0(\cos k_x-\cos k_y) 
\end{eqnarray}
and illustrated in fig.~\ref{FIGOPDpair}. The d-wave gaps have nodes
which are located at the four `Dirac' points (D) \vk=
($\pm\frac{\pi}{2},\pm\frac{\pi}{2}$). The node points are in
'orthogonal' (rotated by $\frac{\pi}{4}$) configuration to the saddle
points (S) because at the latter with their singular DOS the gap value of
a density wave state (and also SC state) has to achieve maximum value
to be stable. In other words the ordered state consists mostly of e-h
pairs (\vk,\vk+\vQ) with \vk~ close to the saddle point regions. This
allows one to study the problem in a simplified continuum
theory \cite{Schulz89,Thalmeier96}. 
As indicated the phase of gap function for d-density waves with
commensurate nesting \v Q is not arbitrary. Because of the symmetry
$\Delta_{C,S}^d(\vk)^*$= -$\Delta_{C,S}^d(\vk)$ the
gap function has to be purely imaginary. Due to this property the
ordered states are connected with staggered persistent charge (C) or spin
(S) currents around the lattice plaquettes. In real space the pattern
of currents at the lattice links is indicated in
fig.~\ref{FIGOPDcurr}.  

As a means of identification of d-density wave states one then has to
look for signatures of persistent currents in the ground state and for
evidence of nodes in the quasiparticle excitation \index{magnetotransport}
spectrum in thermodynamic and (magneto-) transport properties.\\ 

\index{neutron diffraction}{\em Neutron diffraction:}
For d-CDW the charge currents break \index{time reversal symmetry} time
reversal symmetry while the spin-currents in d-SDW do not (under time
reversal both current and spin direction are reversed leaving the
spin current invariant), they do, however, break spin rotational
symmetry. Therefore d-CDW state may lead to the
appearance of staggered {\em orbital} moments. Because the currents
are spread out over a whole plaquette, the generated magnetic fields
are of the order $\sim$ 10 G associated with orbital moments $\sim$
10$^{-2}\mu_B$ per plaquette. The contribution to
elastic and inelastic neutron cross section has been discussed
\cite{Chakravarty01} in
context with the d-CDW scenario for the spin gap phase of underdoped
cuprates \cite{Chakravarty01a}, but it is generally valid for the
tight binding model used here. While the
scattering from spin moments is determined by the form
factor g(\vq) of atomic electron densities which falls off moderately
fast with momentum transfer \vq, the scattering cross section from orbital
moments is determined by the orbital current distribution \v j(\vq)
namely proportional to $|\la\bf j\rm(\vq)\ra|^2/\vq^2\sim$ 1/q$^4$
which falls off much more quickly with momentum transfer due to the
spread-out orbital moment density. Although this method would be rather
straightforward it is quantitatively difficult due to the small
orbital moment size and only applicable for d-CDW.\\ 

A more indirect but more versatile method consists in looking for
signatures of the gap nodes in the d-density wave states. For low
temperatures T$\ll\Delta$ the nodal regions dominate thermodynamics and
transport properties. Specific heat and thermal conductivity is equivalent to
that of d-SC which have been discussed before. Susceptibility and frequency
dependent electrical conductivity however are characteristic  for the
d-density wave states. Of course there is no vortex phase and the
Doppler shift method cannot be applied.\\ 

\index{giant diamagnetism}{\em Giant diamagnetism:} 
The susceptibility has been
analysed in detail \cite{Nersesyan89,Nersesyan91}. Strong anomalies
in the diamagnetic susceptibility for both d-CDW and -SDW are
predicted at low fields. This is due to the peculiar conical or `relativistic'
quasiparticle spectrum around the nodal Dirac points (D) in
fig.~\ref{FIGOPDpair}. For T $\ll\Delta_0$ the spectrum can be linearized and
consists of two bands 
\begin{eqnarray}
\label{RDIRAC}
E_\pm(\vk)=\pm\sqrt{v_F^2 k_x^2 +v_{\Delta_0}^2 k_y^2}
\end{eqnarray}
with v$_F$ = 2$\sqrt{2}$ta and v$_\Delta$ = $\sqrt{2}\Delta_0$a giving the
group velocity perpendicular and parallel to the original Fermi
surface in fig.~\ref{FIGOPDpair} respectively which are related to the
effective mass via m$^*$ = $\Delta_0$/2v$_F$v$_\Delta$. Here the lower band (-)
is completely filled and the upper band (+) empty, thus the
\index{Fermi surface} Fermi
surface in the ordered state ($\Delta_0>0$) consists of two inequivalent points
\vk$_{1,2}$=($\frac{\pi}{2},\pm\frac{\pi}{2}$), the other Dirac points
are equivalent since \vQ~is now a reciprocal lattice vector. The
linearization corresponds to restricting to a (2+1) dimensional
continuum field theory of chiral massless relativistic fermions with
'anisotropic' velocities v$_F$ and v$_\Delta$ \cite{Nersesyan89}.
The \index{quasiparticle DOS} quasiparticle DOS close to the Fermi
level E$_F$ = 0 is given by
N(E) = $|E|/(\pi$v$_F$v$_\Delta$) therefore one has low-T power laws 
C $\sim$ T$^2$/(t$\Delta_0$) and $\chi_P^\parallel\sim\mu_B^2T/\Delta_0$
for the specific heat and the in-plane spin susceptibility. The most striking
effect however should appear in the diamagnetic susceptibility anomaly
caused by the persistent plaquette currents. This is obtained
through the relativistic Landau quantisation of the spectrum which
leads to Landau levels \index{Landau levels}
\begin{eqnarray}
\label{RLANDAU}
E^\pm_n=\pm\sqrt{|\Omega_\perp|n}
\qquad \mbox{and} \qquad
\Omega_\perp=2eH_\perp v_Fv_\Delta/c
\end{eqnarray}
which have a degeneracy $\nu$(H) = $|eH|$/2$\pi c$ (c = speed
of light) per unit area. Furthermore, H$_\perp$ = H$\cos\theta$ where
$\theta$ = polar field angle with respect to the plane normal vector
$\hat{\v n}$. This equation holds for fields with
$\sqrt{|\Omega_\perp|}\ll\Delta_0$. From eq.~(\ref{RLANDAU}) the
H-dependent total energy is obtained leading to a (2D) diamagnetic susceptibility
\begin{equation}
\label{2DSUS}
\chi_{2D}\sim \frac{\sqrt{\Delta_0}}{|H|^\frac{1}{2}}
\cos^{\frac{3}{2}}\theta
\end{equation}
which predicts an enhancement factor of $(\Delta/\Omega_c)\gg$ 1
compared to the normal state diamagnetic susceptibility. The
divergence in eq.~(\ref{2DSUS}) will be arrested below a tiny
crossover field where 3D behaviour due to interlayer
hopping sets in and perfect diamagnetism with $\chi_{3D}(H\rightarrow 0)=-1/4\pi$
will be approached. Thus for temperatures T $\ll\Delta_0$ and almost zero
field the staggered currents of d-CDW or SDW should
achieve a similar perfect screening effect as supercurrents in the d-SC state.\\

The conical quasiparticle spectrum and associated `relativistic'
Landau quantisation should also leave its mark on the (magneto-)
resistance which we only briefly
mention since sofar it has only been applied to d-CDW candidates
among quasi-1D organic conductors. In
this case a finite carrier concentration ($|E_F|>0$) should be present. This is
achieved by including a finite next-nearest-neighbor (n.n.n.) hopping element
t' in the model. It does
not appear in the linearized spectrum eq.~(\ref{RLANDAU}) however. Analogous
to the susceptibility, the magnetoresistance \index{magnetotransport}
is determined by the Landau
levels of the d-CDW state leading to typical oscillations
(fig.~\ref{FIGOPDcurr}) in its angular dependence which have been
identified in two examples \cite{Dora02,Basletic02,Maki03}.\\ 

{\em Finite frequency probes:} 
Finally we discuss finite frequency probes for d-DW states like
optical conductivity \cite{Yang01a}. It exhibits non-Drude like
behaviour at low frequencies because of arbitary low excitation
energies for \vq = 0 interband (E$_-\leftrightarrow$ E$_+$) transitions
at the nodal (Dirac) points. For perfect nesting (E$_F$=0) at low
temperatures one obtains (for $\omega\tau\gg$ 1, $\tau$ = quasiparticle
lifetime)  
\begin{eqnarray}
\sigma(\omega,T)=(\ln 2)e^2\alpha T\delta(\omega)
+\frac{1}{8}e^2\alpha|f(-\frac{\omega}{2})-f(+\frac{\omega}{2})|
\end{eqnarray}
where $\alpha\simeq(v_F/v_\Delta)$ and f denotes the Fermi
function. The first part is the Drude term and the second is the
interband term which should extend to arbitrary low frequencies and
contribute to the d.c. conductivity because of the vanishing gap at
the node points. The weight of the Drude peak vanishes for
T$\rightarrow$ 0 
while the interband contribution stays finite at zero frequency. Since
the thermal conductivity has only a Drude contribution, the
\index{Wiedemann-Franz law}
Wiedemann-Franz law $\kappa\sim\sigma$T should be strongly violated in
the d-DW state
and one would rather expect $\kappa\sim\sigma T^2$. This result holds
as long as E$_F<$ kT and also for finite quasiparticle scattering rate
 $\Gamma$ = 1/$\tau$ if $\Gamma <$ E$_F$. In addition the d-DW state
may lead to anomalous field dependence of electrical and thermal Hall
conductivity.

The peculiar excitation spectrum eq.~(\ref{RLANDAU}) of a d-DW should
also leave its imprint in the inelastic neutron scattering cross
section \cite{Chakravarty01} which is proportional to the imaginary
part of the dynamical spin susceptiblilty.

\section{Ce-based heavy-fermion superconductors}

The discovery of superconductivity in CeCu\( _{2} \)Si\( _{2} \) by 
\citeasnoun{Steglich79} initiated the rapid development of 
heavy fermion physics. For nearly two decades, this material was the
only Ce-based heavy fermion superconductor at ambient pressure.
Only recently, superconductivity at ambient pressure was found in the
new class of heavy fermion materials Ce\( _{n} \)M\( _{m} \)In\(
_{3n+2m} \) where M stands for the transition metal ion Ir or Co 
and \( m=1 \) while \( n=1 \) or \( n=2 \) \cite{Thompson01}.
Typical examples are CeCoIn$_{5}$ \cite{Petrovic01a} and CeIrIn$_{5}$
\cite{Petrovic01}. 

The SC phases are characterized by BCS-type pair
correlations among the heavy quasiparticles. In addition the strong
on-site correlations introduced by the partially filled Ce-4f shells
are reflected in
several aspects. Firstly, the SC gap functions are anisotropic due to
the local quasiparticle repulsion. Secondly,
the competition between the non-magnetic Fermi liquid state and
magnetically ordered phases leads to pronounced strong-coupling
effects. They highlight the complex low-frequency dynamics in these
systems which result from the competition between Kondo effect
and long-range magnetic order.

The main emphasis of the current experimental and theoretical studies
is on the fundamental question which factors actually determine the
low-temperature phases, i.~e., when does a Ce compound become a
heavy fermion superconductor or why does it order magnetically. The
subtle interplay between Kondo effect and long-range magnetic order
can be monitored experimentally in pressure studies where isostructural
relatives of the heavy-fermion superconductors are tuned from magnetic
phases at ambient pressure to superconducting states. Similar behavior
is found in doping experiments where constituents of the metallic host
are successively replaced. Examples for pressure-induced
superconductors are   CeCu\( _{2} \)Ge\( _{2} \)
\cite{Jaccard92}, CePd\( _{2} \)Si\( _{2} \) \cite{Grosche96,Mathur98},
CeNi\( _{2} \)Ge\( _{2} \) \cite{Grosche00} and CeRh\( _{2} \)Si\( _{2} \) 
\cite{Movshovich96} CeSn\( _{3} \) \cite{Walker97} and CeIn\( _{3} \)
\cite{Grosche96}.

Finally, a different type of competition between superconductivity and
magnetic order may be realized in CeCu$_2$Si$_2$. In this
compound, the strongly renormalized Fermi liquid seems to be unstable
with respect to the formation of both superconducting pairs and an
(incommensurate) spin-density wave. The actual ground state depends
sensitively on the composition of the sample. 

In the present section, we review the properties of the stoichiometric
Ce compounds exhibiting competition between heavy fermion
superconductivity and long-range magnetic order. We shall not discuss
the highly complex phase diagrams obtained in alloy systems. In
addition, we summarize recent attempts at a theoretical description of
the complex low-frequency dynamics resulting from the competition
between Kondo effect and RKKY interaction.

\subsection{CeM$_2$X$_2$}
\index{CeM$_2$X$_2$}\index{CeCu$_2$Ge$_2$}\index{CePd$_2$Si$_2$}
\index{CeNi$_2$Ge$_2$}\index{CeRh$_2$Si$_2$}

\begin{figure}[tbh]
\includegraphics[angle=0,width=7cm]{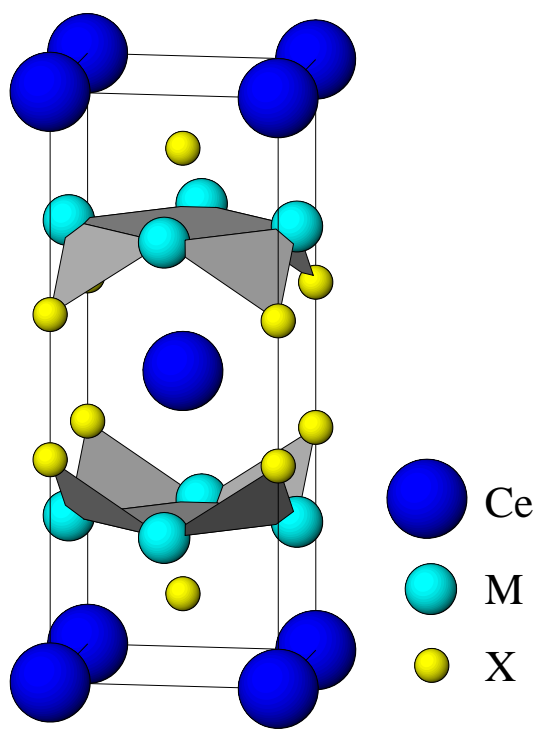}
\caption{Conventional unit cell of the ThCr$ _{2}$Si$ _{2} $ and
CeM$_2$X$_2$ structure where M = Cu, Ni, Ru, Rh, Pd, Au, .. and X =
Si, Ge. This is also the unit cell of URu$_2$Si$_2$ (sect.~\ref{Sect:URu2Si2})}
\label{fig:ThCr2Si2Structure}
\end{figure}

The archetype heavy fermion superconductor CeCu$ _{2} $Si$ _{2}$ as well
as the pressure-induced superconductors CeCu$ _{2}$Ge$ _{2}$, 
CePd$ _{2}$Si$_{2}$, CeNi$_{2}$Ge$ _{2}$ and CeRh$_{2}$Si$_{2}$
crystallize in the tetra-gonal ThCr$_{2}$Si$_{2}$ structure. The
unit cell is shown in fig.~\ref{fig:ThCr2Si2Structure}. The isostructural
compounds CeM$_{2}$X$_{2}$ with M=Ru, Ni, Pd, Cu, Ag, Au and X=Si,
Ge exhibit a great variety of possible ground
states and have been extensively studied to clarify the interplay between
the formation of magnetic order and heavy fermion behavior. 

\subsubsection{Electronic properties, Fermi surfaces and heavy quasiparticles}
To study the electronic structure, we compare the results for two
different models treating the Ce 4f degrees of freedom as localized 
(atomic like) states and as delocalized yet strongly renormalized
electrons. The first procedure provides a
good quantitative description of the properties at elevated
temperatures, high excitation energies, and above the metamagnetic
transition. The latter ansatz yields a model for the Fermi liquid
state.

With respect to the electronic properties, the compounds 
CeM$_{2}$X$ _{2}$ fall into two distinct categories, the key feature
being the presence or absence of transition metal d-states at the
Fermi level. We apply this criterion to classify the CeM$_{2}$X$ _{2}$
compounds exhibiting superconductivity. The first group is formed by 
CePd$_{2}$Si$_{2}$, CeNi$_{2}$Ge$_{2}$ and CeRh$_{2} $Si$_{2}$ which 
have a large conduction electron density of states at the Fermi
level. A well-studied model for the normal state of these systems is 
provided by CeRu$_{2}$Si$_{2}$ which we shall consider in detail below. 
The macroscopic properties of this compound closely resemble
those of  CeNi$_{2}$Ge$_{2}$ and CePd$_{2}$Si$_{2}$. The general
conclusions derived for CeRu$_{2}$Si$_{2}$  should therefore apply
also to the pressure-induced superconductors. The second group is
given by CeCu$_{2} $Si$_{2}$ and CeCu$_{2}$Ge$_{2}$ where the metal d-bands 
are filled.

The low-temperature behavior of CeRu$_{2}$Si$_{2}$ is well described
by a paramagnetic Fermi liquid with weak residual interactions. The relevant
low-energy excitations are heavy quasiparticles as inferred from the 
linear specific heat coefficient $\gamma\simeq$ 350 mJ/molK$^2$ 
\cite{Steglich85b}. The electronic structure was calculated both for
the local moment regime at high temperatures and for the Fermi liquid
state. Both types of calculations start from ab-initio crystal
potentials which are determined selfconsistently within the
LDA. Details of the calculation can be found in \cite{Runge95}. In the
local moment regime, the Fermi surface is determined exclusively by
the conduction states. The strongly renormalized Fermi liquid state,
on the other hand, is described by the renormalized band method using 
$\tilde{\Delta}_f\simeq$ 10 K in eq.~(\ref{eq:efren}) for the
instrinsic width of the quasiparticle
band. The value is consistent with inelastic neutron data
\cite{Regnault88} as well as thermopower and specific heat data
\cite{Steglich85}. CEF effects are accounted for by adopting a
$\Gamma_7$ ground state. The details of the calculation are described
in \cite{Zwicknagl92}.

The renormalized band scheme gives the correct Fermi surface topology
for CeRu$ _{2} $Si$ _{2} $ and thus consistently explains the measured 
dHvA data \cite{Zwicknagl90,Zwicknagl93a}.
The Fermi surface consists of five separate sheets among which are four closed
hole surfaces centered around the Z point. 
The remaining one is a complex multiply-connected
surface where extremal orbits of rather different character exist. 
In particular,
one can find typical particle-like orbits as well as others for which a hole
picture would be more appropriate. The agreement between calculated
and experimental results with respect to general topology is rather
good (for further details see \cite{Zwicknagl92,Zwicknagl93a}).

\begin{figure}[tbh]
\begin{minipage}[htb]{150mm}
\begin{center}
\includegraphics[angle=0,width=6cm]{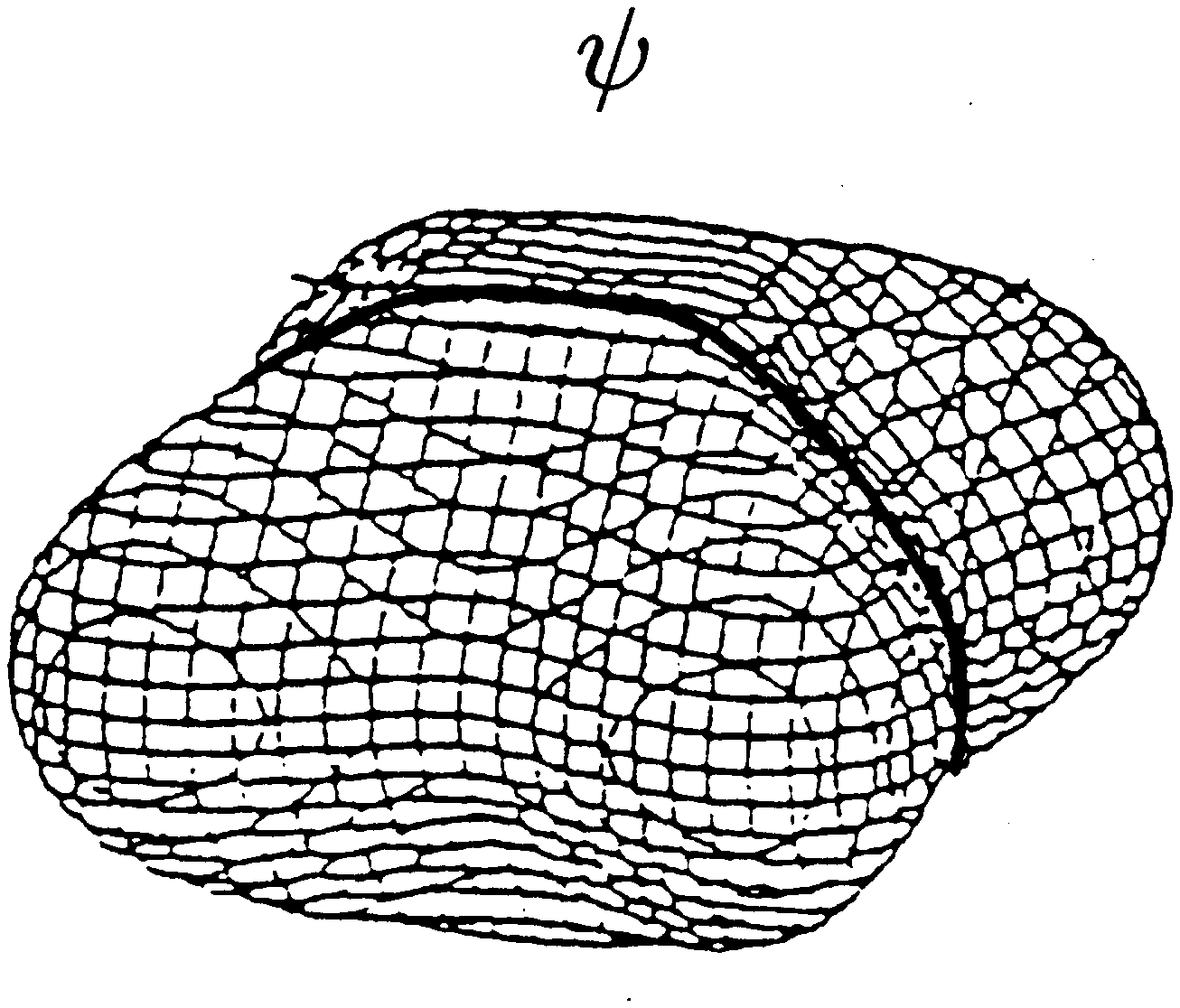}
\hfill
\includegraphics[angle=0,width=6cm]{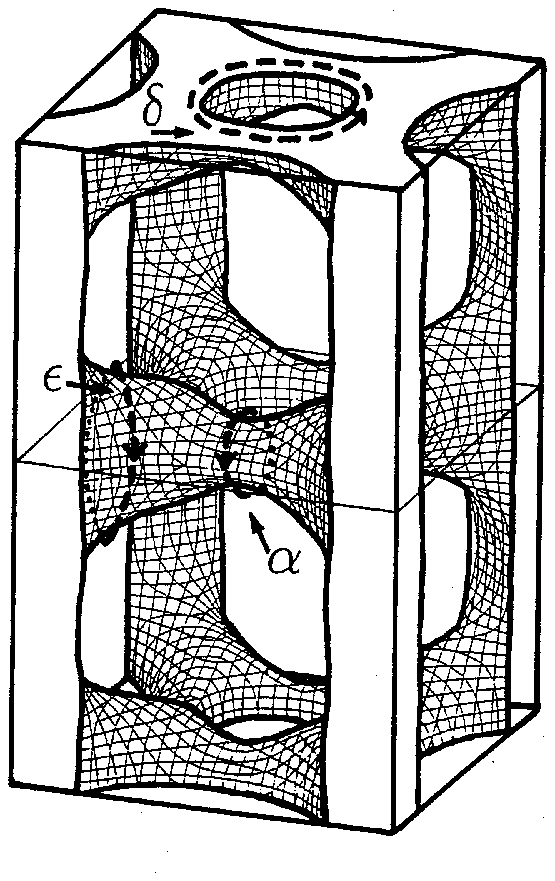}
\end{center}
\end{minipage}
\caption{
CeRu$_2$Si$_2$: Fermi surface \index{Fermi surface!CeRu$_2$Si$_2$}
sheets for quasiparticles with
f-character. The labels $\psi$ and $\alpha$, $\delta$, $\epsilon$
refer to the branches observed in dHvA experiments (Lonzarich 1988,
Albessard et al. 1993). \index{de Haas-van Alphen effect} 
Left: hole surface centered around the Z-point
of the Brillouin zone with effective mass m$^*\sim$ 100 m which dominates the
specific heat $\gamma$-value. For localized f-electrons at elevated
temperatures, the hole surface expands extending to the boundaries of
the Brillouin zone while the multiply connected electron-like surface 
shrinks. The expansion of $\psi$ is confirmed by photoemission
experiments. Right: multiply-connected electron-like sheet}
\label{fig:CeRu2Si2fBandFermiSurface}
\end{figure}

The character of quasiparticles in CeRu$ _{2} $Si$ _{2} $ 
varies quite strongly over the Fermi surface. There are three Z-centered
hole ellipsoids with rather light quasiparticles.
The states on the pillow-shaped Z-centered hole surface displayed in
fig.~\ref{fig:CeRu2Si2fBandFermiSurface}, however, have predominantly 
f-character and therefore yield the dominant contribution to the specific
heat. Experimentally, one finds heavy and light quasiparticles
coexisting on the multiply-connected sheet.

The validity of the Fermi liquid picture is concluded from a
comparison of the effective masses on the fourth pillow-shaped sheet
as given in table \ref{tab:CeRu2Si2HeavyMasses}.
From the large linear specific heat the renormalized
band scheme deduces a characteristic energy kT$^*\simeq$ 10 K which
in turn implies heavy masses of the order of m$^{*}$/m $\simeq$ 100.
This value was confirmed by experiments
\cite{Albessard93,Aoki93,Tautz95} where the $\psi$ orbit
with m$^{*}$/m $\simeq$ 120 was observed. The corresponding Fermi surface
cross section is in agreement with estimates from the renormalized band theory.
\index{band structure} This proves that the heavy quasiparticles exhaust
the low-energy excitations associated with the f-states in
HF systems. The shape of the Fermi surface of heavy
quasiparticles implies the existence of nesting vectors. This fact
suggests that the Fermi liquid state might become unstable with
respect to particle-hole pairing, leading for example to the
formation of a SDW state.

\begin{table}
\caption{CeRu$_2$Si$_2$: Effective masses of the heavy
quasiparticles. Comparison of experiment 
(Albessard et al. 1993) and theory (Zwicknagl et al. 1990). 
The heavy quasiparticles explain the specific heat at low
temperatures. There are no further low-energy excitations involved.}
\label{tab:CeRu2Si2HeavyMasses}
\vspace*{0.3cm}
\begin{tabular}{l r r}
\hline 
$\psi$ & experiment & theory (approx.)\\[0.5mm]
\hline
area [MG] & 53.6 & 62 \\[0.5mm]
m$^*$/m & 120 & 100 \\
\hline
\end{tabular}
\end{table}

\begin{figure}[h t b]
\includegraphics[angle=0,width=0.4\columnwidth]
{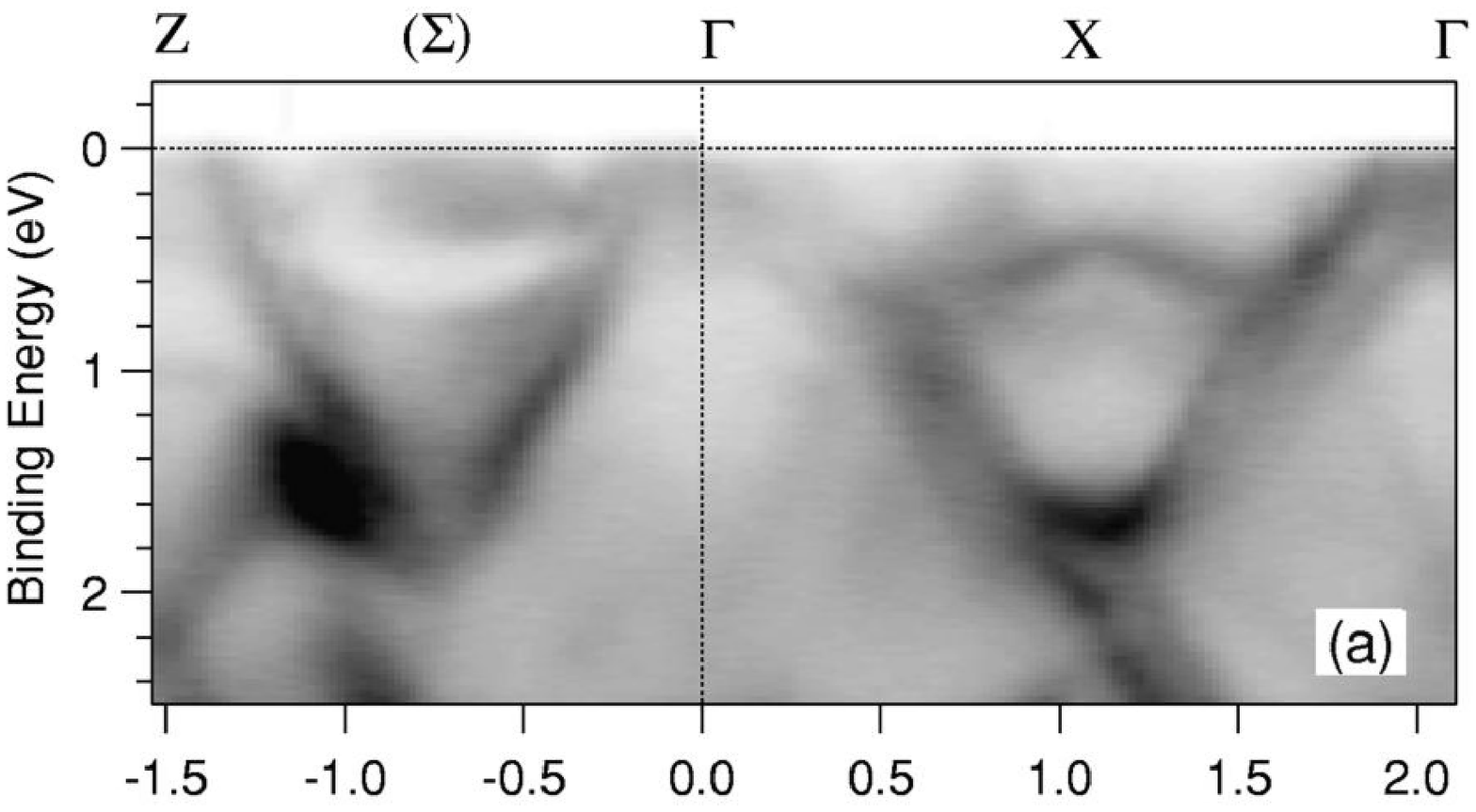}\hfill
\includegraphics[angle=0,width=0.4\columnwidth]
{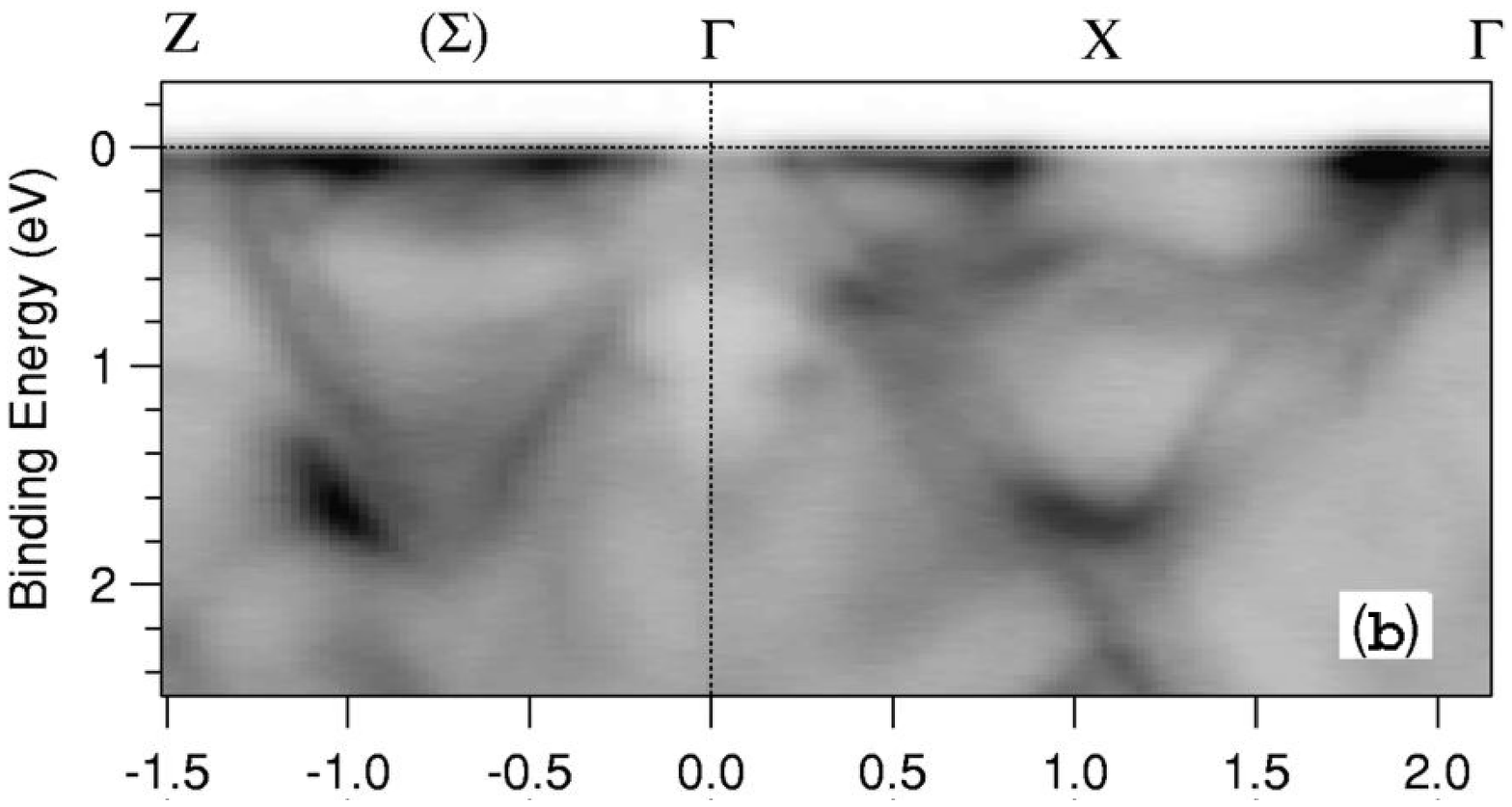}
\caption{Photoemission results \index{LaRu$_2$Si$_2$} for
LaRu$_2$Si$_2$ (a) in comparison to \index{CeRu$_2$Si$_2$}
CeRu$_2$Si$_2$ (b) at T= 25 K above the Kondo temperature T$^*$ = 15
K of CeRu$_2$Si$_2$. Band structures are similar for both
compounds (Denlinger et al. 2001).}
\label{fig:CeRu2Si2flocalizedPES}
\end{figure}

The change in the volume of the Fermi surface
when going from $ T\ll T^*$ to $ T\gg T^{*} $ is
observed by comparing the Fermi surface of CeRu$_{2}$Si$_{2}$ to that
of its ferromagnetic isostructural \index{CeRu$_2$Ge$_2$} counterpart 
CeRu$_{2}$Ge$_{2}$
where the f-states are clearly localized. In a series of beautiful experiments
\cite{King91} it was demonstrated that the \index{Fermi surface} Fermi
surfaces of these two 
compounds are rather similar. However, the enclosed Fermi volume
is smaller in the case of CeRu$_{2}$Ge$_{2}$, the difference being
roughly one electron per unit cell. 
More direct evidence is provided by recent photoemission experiments 
(see fig.~\ref{fig:CeRu2Si2flocalizedPES}).
\citeasnoun{Denlinger01} have shown  that at temperatures around 25 K,
the Fermi surface of CeRu$_{2}$Si$_{2}$,
is that of its counterpart LaRu$_{2}$Si$_{2}$ which has no f-electrons.

The similarity of the Fermi surface topologies for localized and
itinerant Ce 4f states is a characteristic feature of the 
 CeM$ _{2} $X$ _{2} $ systems with partially filled M d-bands. In
addition, we mention that the topology of the Fermi surface 
for the heavy quasiparticles is qualitatively reproduced by standard
band structure calculations based on the LDA \cite{Zwicknagl88a}.

This, however, is not the case for CeCu$_{2}$Si$_{2}$. As the 
Cu d bands are filled there are only two major twofold degenerate 
bands crossing the Fermi energy, and the effective hybridization becomes 
rather anisotropic. To calculate the quasiparticle bands in
CeCu$_{2}$Si$ _{2}$ by means of the renormalized band
method, we adopt the doublet-quartet CEF scheme suggested by
\citeasnoun{Goremychkin93}. The ground state is separated from the excited
quartet by $\delta\simeq$ 340 K.

The results for the Fermi surface \cite{Zwicknagl93,Pulst93} can be summarized
as follows: two separate sheets of the Fermi surface for heavy
and light quasiparticles are found. The light quasiparticles have effective 
masses of the order of $m^{*}/m\simeq$ 5. They can be considered as
weakly renormalized 
conduction electrons and the corresponding Fermi surface is rather similar to
the LDA prediction \cite{Harima91}. The observed Fermi surface cross 
sections \cite{Hunt90} can be explained by both the renormalized band 
structure as well as by the LDA bands.
There is, however, a characteristic difference between the Fermi
surfaces for the light quasiparticles as derived from the two schemes:
The LDA calculation predicts a small closed surface centered around 
the $\Gamma$-point which is absent in the dHvA-experiment although the
corresponding cross sections are rather small and the masses are
expected  to be light.

\begin{figure}[tbh]
\begin{minipage}[htb]{150mm}
\includegraphics[angle=0,width=7cm,clip]{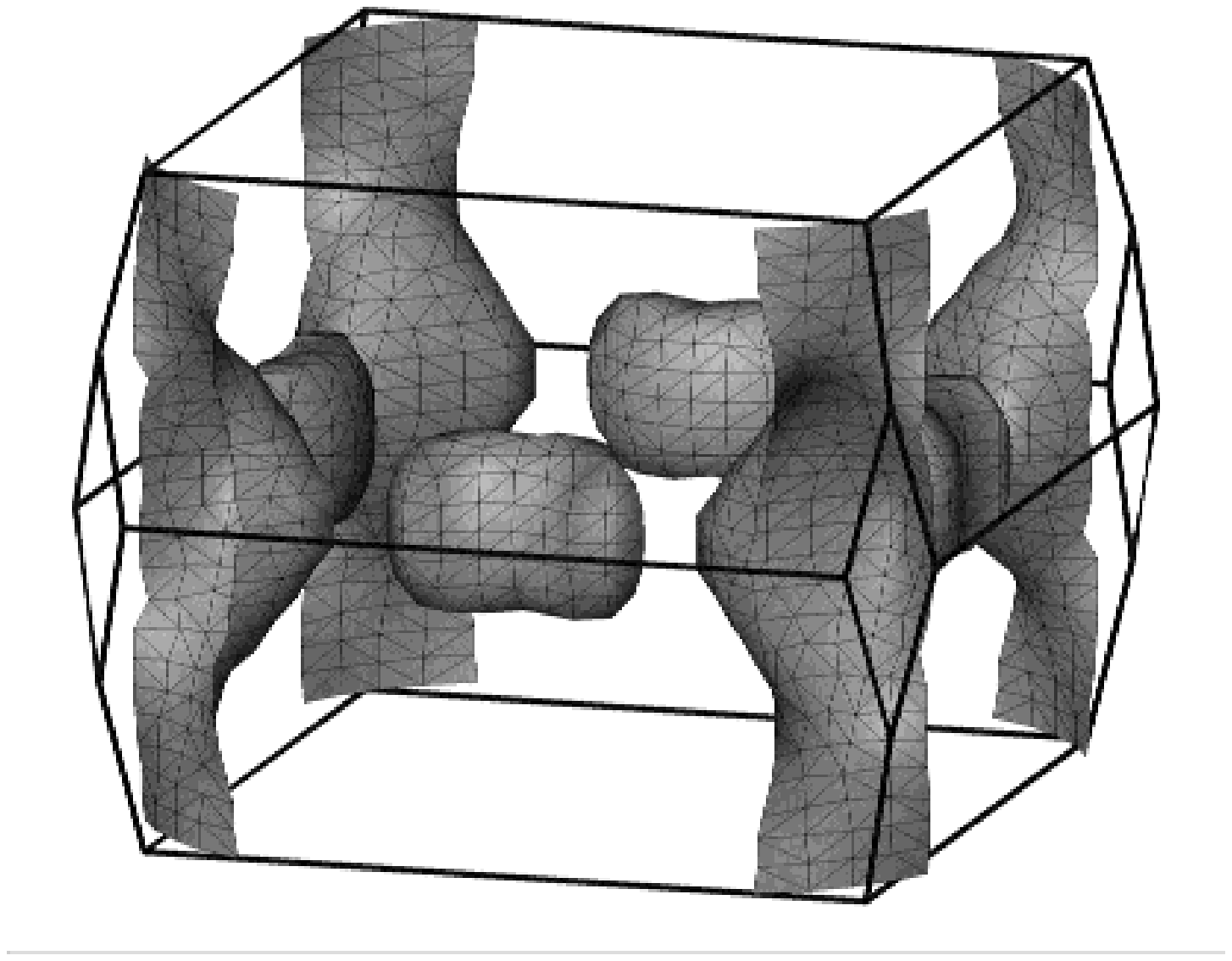}
\includegraphics[angle=0,width=4cm,clip]{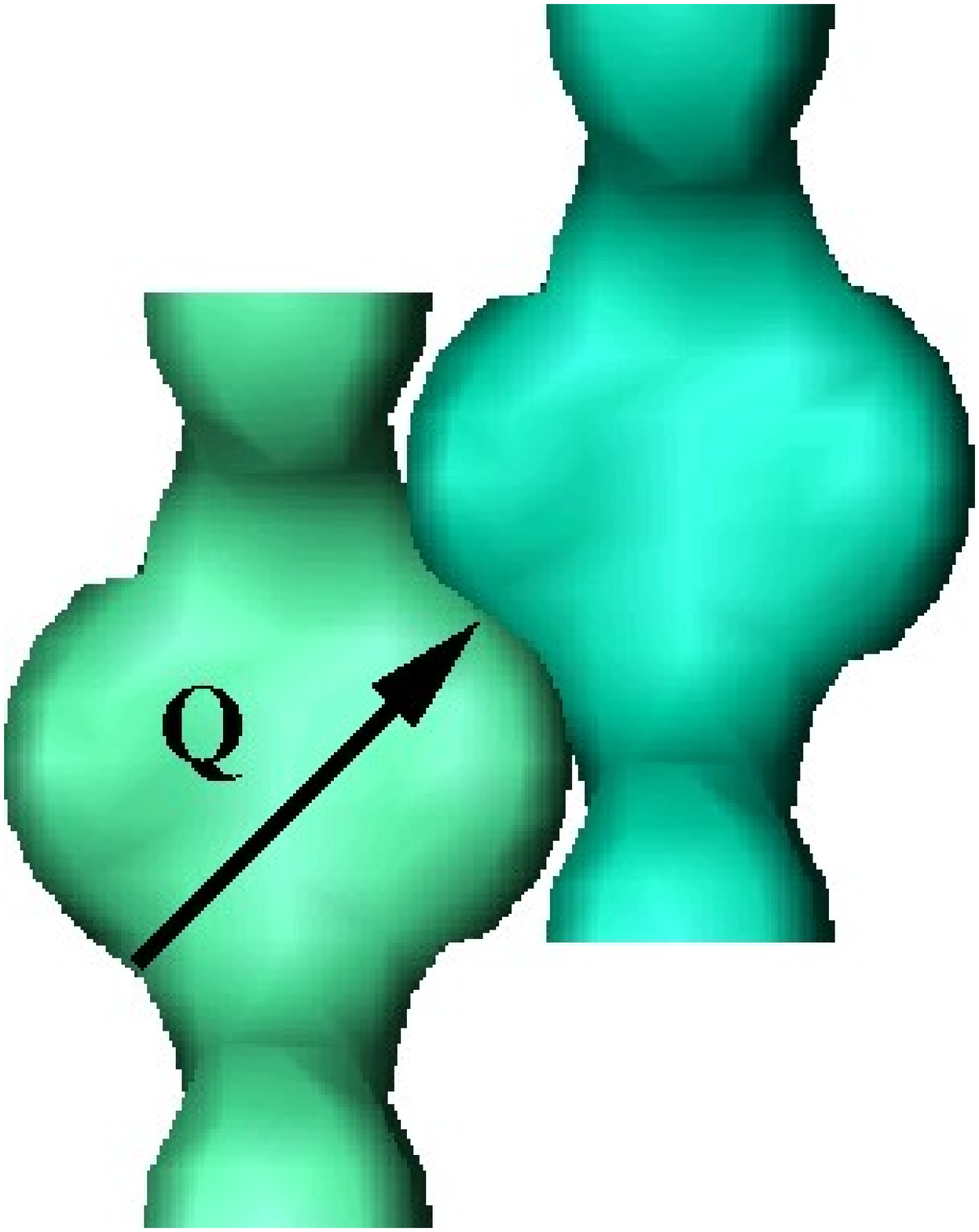}
\end{minipage}
\caption{
CeCu$_2$Si$_2$: Left: FS \index{Fermi surface!CeCu$_2$Si$_2$} 
of \index{heavy fermions} heavy quasiparticles (m$^*$/m $\simeq$ 500)
calculated by using the renormalized band method. It consists of
ellipsoids and modulated columns which are oriented parallel to the
tetragonal axis. The calculations adopt the CEF scheme suggested by
Goremychkin and Osborn (1993) consisting of a singlet \index{singlet CEF state}
ground state separated from an excited quartet by a CEF splitting
$\delta\sim$ 340 K. \index{crystalline electric field excitations}
Therefore $\delta\gg$ T$^*\simeq$ 10 K (obtained
from the $\gamma$- value). Consequently quasiparticle properties are strongly
anisotropic. The topology of the renormalized band FS differs
qualitatively from LDA results. Right: enlarged FS columns exhibiting
nesting between a column and its counterpart shifted by the nesting vector
${\bf Q}=(0.23,0.23,0.52)$ (Zwicknagl and Pulst 1993).}  
\label{fig:CeCu2Si2fBandFermiSurface}
\end{figure}

Of particular interest are heavy quasiparticles of effective masses 
$ m^{*}/m\simeq$ 500
which are found on a separate sheet. The topology of
this surface is rather different from the corresponding LDA result as
can be seen from fig.~\ref{fig:CeCu2Si2fBandFermiSurface}. It mainly
consists of columns parallel to the tetragonal axis and small pockets.
The topology of the Fermi surface suggests that the strongly
correlated Fermi liquid state should become unstable at sufficiently
low temperatures. Firstly, it exhibits pronounced nesting features which
may eventually lead to the formation of a ground
state with a spin density modulation. This will be discussed in detail
below. Secondly, the topology of this surface depends
rather sensitively on the position of the Fermi energy. The band
filling and hence the f-valence are critical quantities. Reducing the 
f-occupancy from the initial value of $n_f\simeq 0.95$ by $\simeq
2\%$ leads to changes in the topology as shown by
\citeasnoun{Pulst93} and \citeasnoun{Zwicknagl93}. As a result, the
quasiparticle density of states (DOS)  should exhibit rather
pronounced structures  in the immediate vicinity of the Fermi energy
which, in turn, can induce instabilities \cite{Kaganov79}.

An external magnetic field couples to the f-degrees of freedom.
It leads to a Zeeman splitting of order $\mu_{eff}B$ of the heavy
quasiparticle bands and hence can move the structures in the DOS
relative to the Fermi energy. The effective moment $\mu_{eff}$ 
of the f-states which contains the CEF effects depends
on the direction of the external magnetic field.
From the renormalized band calculation we
determine the values of the critical magnetic fields when the
pronounced structures are moved to the Fermi energy. Due to CEF
effects on the f-states this critical value of the external magnetic
field will depend upon its orientation. The explicit values of the critical
fields $B_{crit}^c$ and
$B_{crit}^a$ for field directions parallel to the c-and a-axes as 
listed in table \ref{tab:CeCuSiBcrit} agree rather well with the
observed critical magnetic fields for the transition into the
`B-phase' (see fig.~\ref{fig:CeCu2Si2BTPhaseDiagram}).

\begin{table}[h t b]
\caption{Low-temperature values of the critical magnetic fields for 
the transition into the high-field `B-phase'.
Comparison with experiment of the  renormalized band
prediction for the approximate critical magnetic fields for field
directions along the tetragonal axis (B$_{crit}^c$) and in the basal
plane  (B$_{crit}^a$) (Zwicknagl and Pulst 1993)}
\label{tab:CeCuSiBcrit}
\vspace*{0.3cm}
\begin{tabular}{l c r}
\hline 
critical & renormalized & experiment\\
field & bands &  \\[0.5mm]
\hline
B$_{crit}^{a}$ [T] & 8.0 &   7.0 \\
B$_{crit}^{c}$ [T] & 6.5 &   4.0 \\
\hline
\end{tabular}
\end{table}

Having discussed the basic electronic structure of the 
CeM$ _{2}$X$ _{2}$ heavy fermion compounds we now turn to their
low-temperature properties. 

\subsubsection{Superconductivity and itinerant
antiferromagnetism in CeCu$ _{2}$Si$_2$}

CeCu$_2$Si$_2$ exhibits highly complex phase diagrams at low
temperatures. This fact partially results from an extreme sensitivity
of the physical properties with respect to variations of the
stoichiometry. Thorough investigations of the ternary chemical phase
diagram (for a review see \cite{Steglich01b}) showed that the actual
ground state is mainly determined by the average occupation of the
non-f sites surrounding the Ce-ion.

\begin{figure}[t b h]
\includegraphics[angle=-90,width=7cm]{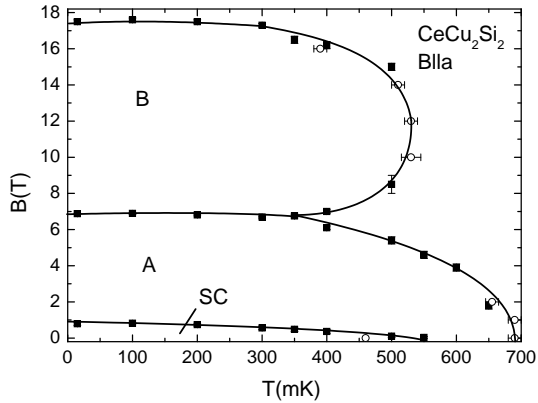}
\caption{B-T phase diagram \index{B-T phase diagram!CeCu$_2$Si$_2$}
of CeCu$_2$Si$_2$ for B $\parallel$ a. The original version of this
phase diagram is from Bruls et al. (1990)
while the completed version is from
Weickert et al. (2003). For this sample the A-phase is expelled from
the SC region (no coexistence).}
\label{fig:CeCu2Si2BTPhaseDiagram}
\end{figure}

Apart from the composition, the phases are strongly affected by
magnetic fields. The B-T-phase diagram \cite{Bruls94} for a high-quality
single-crystal is displayed in fig.~\ref{fig:CeCu2Si2BTPhaseDiagram}.
The phase transitions were deduced from anomalies in thermodynamic
quantities including the elastic constants, thermal
expansion and magnetostriction coefficients. In the low temperature
regime, there is a SC phase surrounded by a phase
`A' which becomes
unstable in high magnetic fields. Similar phase diagrams are obtained
from resistivity measurements, e.~g.~\cite{Steglich00a}. In
addition to these `A-SC-type' crystals, there exist also `A-type'
systems which do not superconduct as well as `SC-type' compounds
where the A phase is suppressed
(fig.~\ref{fig:CeCu2Si2PhaseDiagram}). The high-field phase B, on
the other hand, is present in all three cases. It is interesting to
note that the transition  into the B-phase occurs close to the calculated
critical fields B$_{crit}^a$ and B$_{crit}^c$ listed in table
\ref{tab:CeCuSiBcrit}. At these values, the Fermi surface of the heavy
quasiparticles is expected to change drastically. 

Much effort has been devoted to the characterization of the A
phase which orignally had the appearance of another `hidden order'
phase. However later the spin density wave character first inferred
from resistivity results \cite{Gegenwart98} was supported by
specific heat and high-resolution magnetization measurements
\cite{Steglich00}.
The transition temperature T$_A$ is suppressed by increasing the
4f-conduction electron hybridization and
eventually vanishes. This can be achieved  by applying hydrostatic
pressure or choosing a few percent excess of Cu. 
The ordered moments are expected to be rather small. 
The propagation vector ${\bf Q}$ can be estimated by extrapolation
from neutron diffraction studies in
\index{CeCu$_2$(Si$_{1-x}$Ge$_x$)$_2$} single-crystalline
CeCu$_2$(Si$_{1-x}$Ge$_x$)$_2$. The central result is that the wave
vector of the magnetic order changes only slightly with the Ge
concentration while the ordered moment per Ce site, however, decreases
strongly from $\mu \simeq 0.5 \mu_B$ for x = 0.5 to 
$\mu\simeq 0.1\mu_B$  for x = 0. 

\begin{figure}[tbh]
\begin{minipage}[htb]{150mm}
\includegraphics[clip,angle=0,width=8cm]{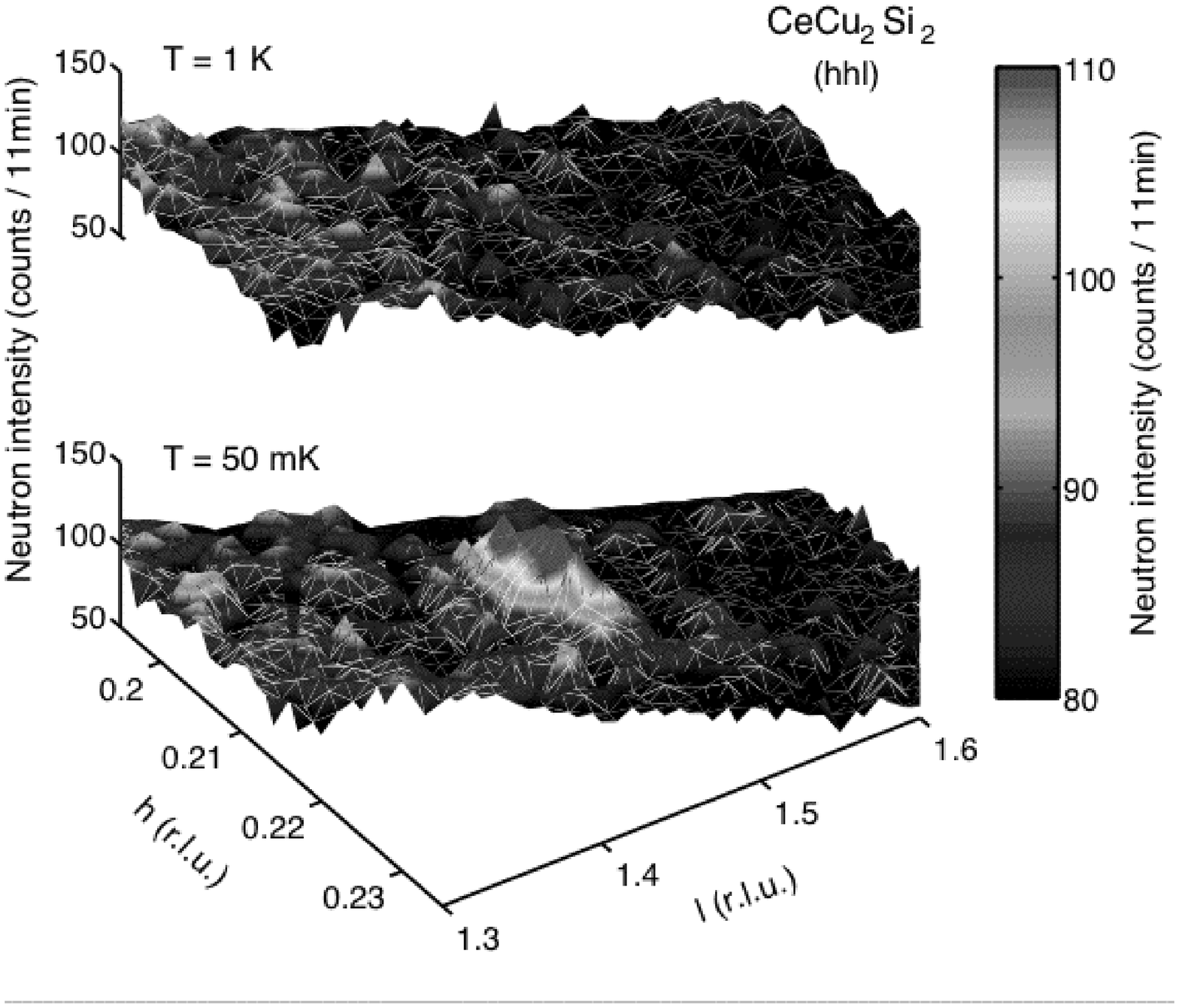}
\hfill
\includegraphics[clip,angle=0,width=6.5cm]{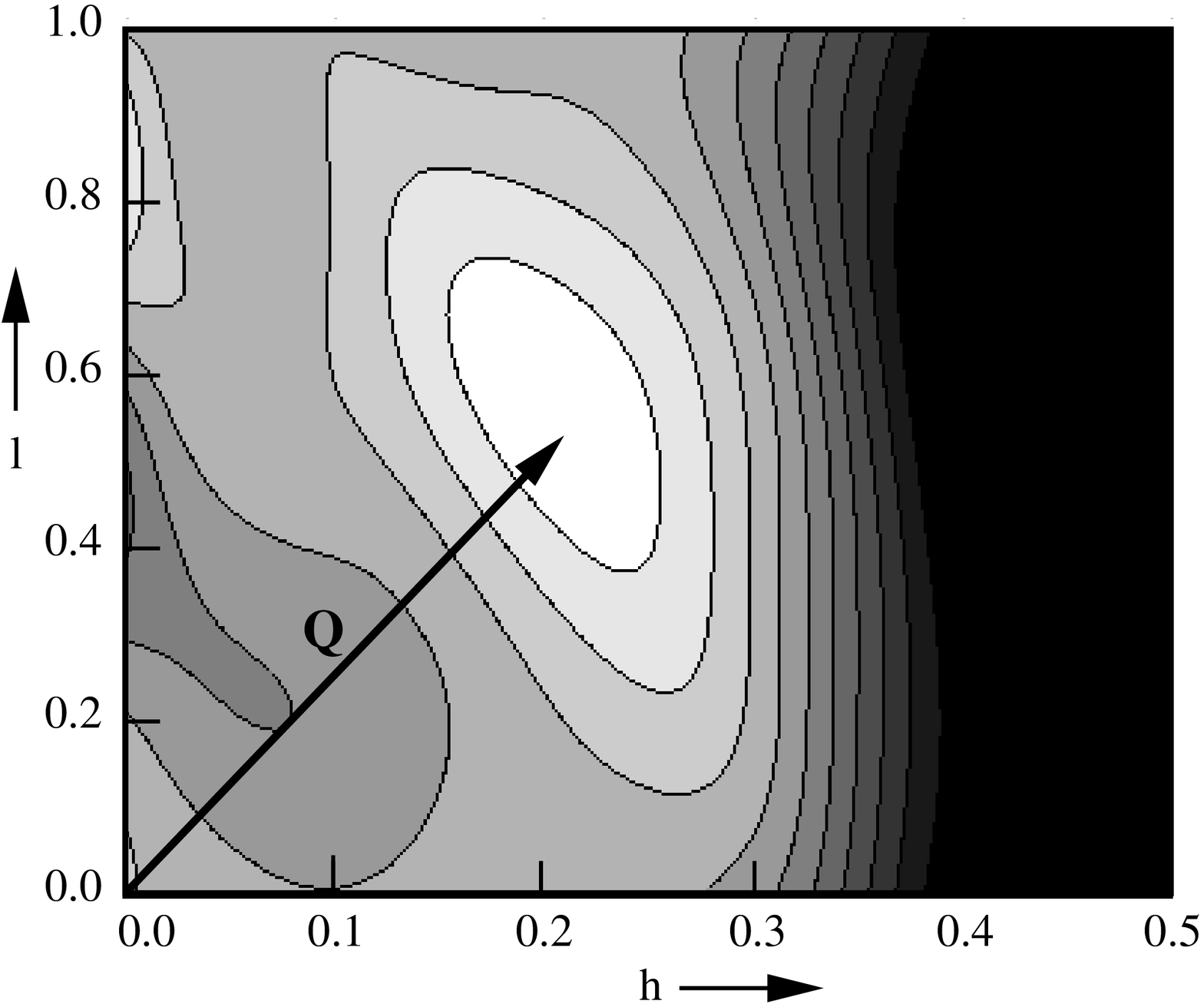}
\end{minipage}
\caption{Left panel: Neutron diffraction intensity in 
CeCu$_2$Si$_2$ at temperatures above 
\index{neutron diffraction!CeCu$_2$Si$_2$} and below the A-phase transition
temperature T$_A$. Incommensurate peak is at \v Q=
(0.22,0.22,0.55) (Stockert et al. 2003). Right panel:
\index{incommensurate order}\index{nesting property} 
Nesting of heavy FS columns
(fig.~\ref{fig:CeCu2Si2fBandFermiSurface}) leads to a peak
in the static susceptibility $\chi({\bf q})$ at {\bf q}={\bf
Q}. The intensity map of $\chi({\bf q})$ (value increasing from dark to
bright) in the reciprocal (h,h,l)-plane as
calculated for the renormalized bands at T = 100 mK. The
{\em experimental} \v Q at 50 mK from the left panel shows perfect
agreement with the calculated maximum position of $\chi({\bf q})$.} 
\label{fig:CeCu2Si2fBandFermiSurfaceNesting}
\end{figure}

Recent neutron scattering experiments \cite{Stockert03} 
(fig.~\ref{fig:CeCu2Si2fBandFermiSurfaceNesting}) for the 
stoichiometric compound (x=0) show a spin density wave which forms
below T$_N\simeq$ 0.7 K. The experimental propagation vector ${\bf Q}$
is close to (0.22,0.22,0.55) and the ordered moment amounts to $\mu \simeq
0.1 \mu_B$. The instability 
of the Fermi liquid state with respect to the SDW state follows from
the nesting properties of the heavy quasiparticle surface shown in
fig.~\ref{fig:CeCu2Si2fBandFermiSurface}
which leads to peaks at \v q = \vQ~ in the static susceptibility
$\chi$(\v q) with \v Q close to the above experimental value
(fig~\ref{fig:CeCu2Si2fBandFermiSurfaceNesting}).

\subsubsection{Pressure-induced superconductivity in 
CePd$_{2}$Si$_{2}$ and CeNi$_2$Ge$_2$}
\index{CePd$_2$Si$_2$}\index{CeNi$_2$Ge$_2$}

The compounds  CeNi$_2$Ge$_2$ and CePd$_2$Si$_2$ offer
the possibility for studies of the antiferromagnetic instability in a
heavy-fermion system. In the vicinity of the quantum critical point,
non-Fermi liquid behaviour and superconductivity are observed. 
\index{quantum critical point}
At ambient pressure, CePd$_2$Si$_2$ orders in an antiferromagnetic
structure below a N{\'e}el temperature T$_N\simeq$ 8.5 K. The magnetic
structure is characterized by a propagation vector 
${\bf Q}=\left(1/2,1/2,0\right)$. The ordered moment lying in the basal 
plane is reduced to $\sim$ 0.66 $\mu_B$ by CEF effects and possibly
Kondo screening \cite{vanDijk00}. The coefficient 
of the linear specific heat $\gamma\simeq$ 250 mJ/mol K$^2$ and the
quasielastic line width suggest a Kondo temperature T$^*\simeq$ 10 K
which is close to the  N{\'e}el temperature T$_N$. The interplay
between the Kondo effect and the RKKY interaction is reflected in the
depression of T$_N$ with increasing pressure. The antiferromagnetic
order is suppressed by a critical pressure p$_c$ between 28 and 29.5
kbar. There is a superconducting phase between 23 and 33 kbar with a
maximum T$_c\simeq$ 290 K close to p$_c$. This fact indicates that
the appearance of superconductivity should be related to the magnetic
instability \cite{Demuer01}.

CeNi$_2$Ge$_2$ has a slightly smaller lattice constant than
CePd$_2$Si$_2$ and  exhibits non-Fermi liquid behavior at ambient pressure
\cite{Gegenwart99}. No phase transition is found down to lowest
temperatures. However magnetic correlations appear gradually with
decreasing temperature \cite{Fak00}. They have a characteristic energy 
of 4 meV and an incommensurate wave vector 
${\bf Q}=\left(0.23,0.23,0.5\right)$ which is rather similar to the ordering 
vector in CeCu$_2$Ge$_2$. They also exhibit
quasi-two dimensional character with strong correlations between moments in 
the $\left[110\right]$ planes. The electronic structure was recently
studied by angle-resolved photoemission \cite{Ehm01}. The
low-temperature data (T $\simeq$ 20 K) exhibit strongly dispersing
bands in agreement with LDA band structure calculations. Close to the
Fermi level two features with high spectral intensity can be
distinguished one of which has predominantly Ni-3d character. The
other one can be assigned to a non-dispersive Kondo resonance. These
findings indicate that CeNi$_2$Ge$_2$ is in the intermediate temperature
regime below the single-site Kondo temperature T$^*$ and above 
the coherence temperature where extended heavy band states form. The
magnetic moments of the Ce ions are already screened by the Kondo
effect as indicated by the
f-spectral weight at the Fermi level. The coherence between the sites
in the periodic lattice, on the other hand, is not yet fully developed.

Traces of possible superconductivity at ambient pressure have been
found in resistivity measurements \cite{Steglich97,Gegenwart99,Grosche00}.
The preliminary phase diagram \cite{Braithwaite00} shows that the transition
temperature T$_c$ goes through a maximum around 30 kbar and
vanishes above 65 kbar. The upper critical field data H$_{c2}$(T) of the
system under pressure are analyzed by comparing them with calculated 
curves assuming weak coupling in the clean limit. The analysis reveals 
two interesting points. First, the values of H$_{c2}$ at low
temperatures considerably exceed the Pauli limit which indicates the
possibility of triplet pairing. Second, the initial slope
$\left(-\frac{dH_{c2}}{dT}\right)_{T=T_c}$ exhibits only a weak
decrease in the range between 0 kbar and 23 kbar which is much weaker
than the change expected from the variation of the specific heat.

\subsection{CeMIn$_5$}
\index{CeMIn$_5$}

The compounds CeMIn$ _{5}$ (M=Co,Ir,Rh) are a newly reported family
of heavy fermion systems
\cite{Hegger00,Petrovic01} which exhibit a superconducting 
transition at comparatively high temperatures. The values of the critical 
temperatures reach T$_c$ = 2.3 K in CeCoIn$_5$ \cite{Petrovic01a}. We
mention that the isostructural compound PuCoGa$_5$ is
the first Pu-based superconductor with T$_c$ = 18.5 K \cite{Sarrao03}. The 
discovery of this new class of heavy-fermion compounds has opened a
way to systematically investigate the evolution from the
antiferromagnetic to superconducting state as a function of pressure.

\begin{figure}[h t b]
\includegraphics[angle=90,width=12cm,clip]{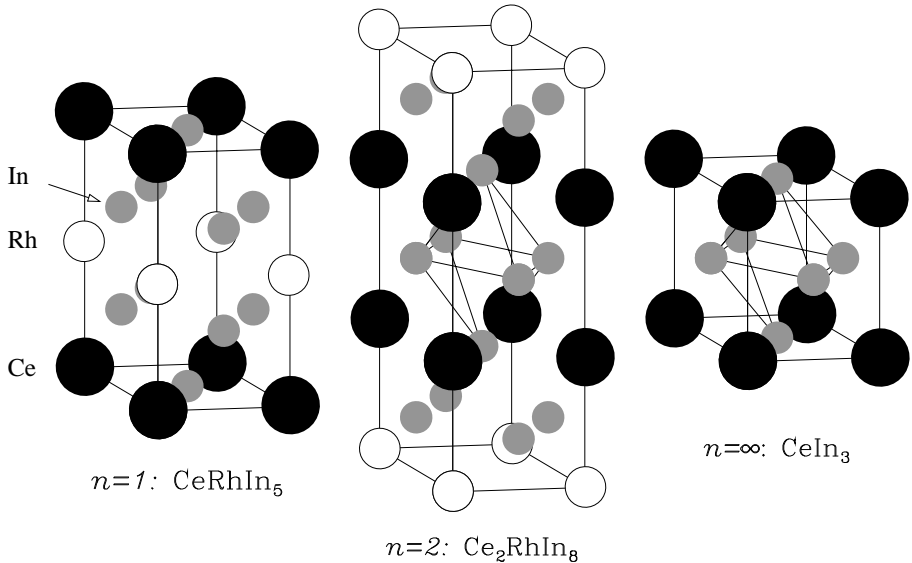}
\caption{Unit cell of typical Ce$_n$M$_m$In$_{3n+2m}$ (here for M = Rh)
compounds which are built from n `CeIn$_3$' and m `MIn$_2$' subunits
stacked along the tetragonal axis. CeCoIn$_5$ has n = 1.\index{CeCoIn$_5$}}
\label{fig:CenMmIn3n+2m}
\end{figure}

The structurally layered compounds CeMIn$_5$ crystallize in the
tetragonal HoCoGa$_5$-structure \cite{Grin79} which is
displayed in fig. \ref{fig:CenMmIn3n+2m}. They belong to the 
broader familiy of materials Ce$_n$M$_m$In$_{3n+2m}$ which can be
considered as built
from `CeIn$_3$' and `MIn$_2$' layers stacked along the tetragonal
c-axis. The infinite-layer parent CeIn$_3$ crystallizes in the cubic
AuCu$_3$ structure. The layering introduces pronounced anisotropies in the
electronic and structural properties which, in turn, influence the
magnetic and superconducting properties as well as the behavior in the
quantum critical regime. Varying
the stacking sequence, i.~e.~, the parameters n and m, allows for a
study of the general role on dimensionality in the competition between the
various ground states. 

The subsequent section is mainly devoted to the
properties of the n = m = 1 materials commonly referred to as
single-layer systems. We begin, however, by briefly summarizing the 
properties of the infinite- layer parent CeIn$_3$.
At ambient pressure it  orders antiferromagnetically at T$_N\simeq$ 10 K with 
an ordering vector ${\bf Q}=\left(0.5,0.5,0.5\right)$ \cite{Morin88}
and an ordered moment of 0.4$\mu_B$.
The magnetic ordering temperature decreases monotonically with
pressure. In a narrow range around a critical pressure p$_c$ = 2.6 GPa
resistivity vanishes below the onset temperature at 0.2 K. Recent
measurements of T$^{-1}_1$ and the ac-susceptibilities
\cite{Kawasaki02a} confirm the 
bulk character of superconductivity. The results, however, yield a
much lower value T$_c$ = 95 mK than the onset temperature.
The absence of the coherence peak is taken as evidence of an
unconventional pair state.

The single-layer (n=1) members with M = Rh, Ir, Co and the bilayer
(n=2) member with M = Rh can be considered as pressure variants of the parent 
CeIn$_3$. This is suggested by the lattice constants of CeIn$_3$ (a =
4.689 \AA), CeRhIn$_5$ (a = 4.652 \AA, c =7.542 \AA), CeIrIn$_5$ (a =
4.668 \AA, c = 7.515 \AA), CeCoIn$_5$ (a = 4.62 \AA, c = 7.56 \AA)
and Ce$_2$RhIn$_8$ (a = 4.665\AA, c = 12.244\AA). Like the parent, the
compounds CeRhIn$_5$ and Ce$_2$RhIn$_8$ order \index{CeRhIn$_5$}
\index{Ce$_2$RhIn$_8$} antiferromagnetically at ambient
pressure and become pressure-induced superconductors. 
The magnetic ordering temperatures T$_N$, however, are reduced while the
superconducting transition temperatures T$_c$ are an order of
magnitude higher than the maximum value found in CeIn$_3$. 
CeIrIn$ _{5}$ (T$_c$ = 0.4 K) and CeCoIn$ _{5}$ (T$_c$ = 2.3 K) become
superconducting at ambient pressure. \index{CeIrIn$_5$}\index{CeCoIn$_5$}

Structure data as well as pressure studies indicate that superconductivity in 
CeMIn$ _{5}$ occurs close to a quantum critical point (QCP). The
lattice parameters suggest that the antiferromagnetic compound
CeRhIn$_{5}$ and the superconductor CeCoIn$ _{5}$ fall on different
sides of the QCP in the Doniach phase diagram.

Finally we want to add two comments. The layered crystal structure
of the materials seems to suggest that the compounds may be
approximately described as 2- dimensional systems. However, this is not
justified in view of magnetic properties. Neutron scattering data on
CeRhIn$_5$ and Ce$_2$RhIn$_8$ have shown that these materials are magnetically
3- dimensional despite their layered structure.
The Fermi surface topologies of the alloys Ce$_x$La$_{1-x}$ RhIn$_5$
\cite{Alver01} are found to be nearly independent of the composition
parameter x.

\subsubsection{Electronic properties and Fermi surfaces}

The electronic properties of the CeMIn$_5$ compounds have been studied 
theoretically by band structure calculations based on the LDA. Of
particular interest are the Fermi surfaces which can be directly
compared with experimental dHvA data. The central result is the
confirmation of the qualitative picture that Ce 4f-degrees of
freedom should be considered as localized moments in CeRhIn$_5$ (at
ambient pressure) while a description in terms of itinerant, strongly
renormalized fermions is more appropriate in  CeIrIn$_5$ and 
CeCoIn$_5$. The conclusions are based on the following findings
by \citeasnoun{Shishido02} and references therein: For the 
reference compound LaRhIn$_5$ without f electrons the Fermi surface as
well as the
effective masses of the quasiparticles are well described by the LDA
band structure. The quasi-two-dimensional character reflects the
tetragonal lattice structure. The topology of the main Fermi surfaces
is also in good agreement with that of the Ce-counterpart CeRhIn$_5$
indicating the localized nature of the Ce-4f electron in this
system. A detailed quantitative comparison between calculated and measured
cross-sections, however, is difficult due to the complicated
antiferromagnetic structure. Angle-resolved
photoemission experiments \cite{Fujimori02} performed at 15 K showed
rather small 4f weight close to E$_F$. The presence of the localized 4f-states
in CeRhIn$_5$
is reflected in the rather large quasiparticle masses m$^*$ which are
enhanced by a factor 7-9 over those of LaRhIn$_5$. The Fermi surfaces
of LaRhIn$_5$ and the localised 4f-compound CeRhIn$_5$ are highly different
from those of the HF systems CeIrIn$_5$ and CeCoIn$_5$
which are reasonably well explained by the itinerant 4f-band model.

Let us now turn to the properties of the 4f-electrons in the local
moment regime. In the tetragonal CEF the degeneracy of the 4f j=5/2
state is lifted. The level scheme has been determined using
susceptibility and thermal expansion coefficient data for CeRhIn$_5$,
CeIrIn$_5$ and CeCoIn$_5$ \cite{Takeuchi01,Shishido02}. In all three
compounds, the CEF ground state has predominantly 
$j_z = \pm 5/2$- character while the first and second excited states are 
mainly derived from $j_z = \pm 3/2$- and $j_z = \pm 1/2$- states. However
the CEF level schemes differ with respect to their excitation energies. In
the Rh and Ir systems, the $j = 5/2$- manifold is split into three Kramers 
doublets with the excitation energies E($\pm\frac{3}{2})\sim$ 60 K and
E($\pm\frac{1}{2})\sim$ 300 K whereas these excited states are (almost)
degenerate in the Co-compound separated by $\sim$ 150 K from the ground state.

Finally, we mention angle-resolved photoemission studies of CeRhIn$_5$
and CeIrIn$_5$ \cite{Fujimori02,Fujimori02a}. The spectral
contributions from the ligand
states, i.e., from the Rh, Ir and In states, agree rather well with
the results of LDA band structure calculations. However, no feature
originating from the Ce 4f-states is observed near the Fermi
level. This fact may have an explanation within  the Kondo scenario:
The experiments were carried out at T = 14 K which is already equal to
the estimated Kondo temperature of CeIrIn$_5$ given by T$^*\simeq$ 15
K.

\subsubsection{Unconventional superconductivity in CeCoIn$_{5}$}
\index{CeCoIn$_5$}

\begin{figure}[t b h]
\includegraphics[angle=0,width=0.4\columnwidth]{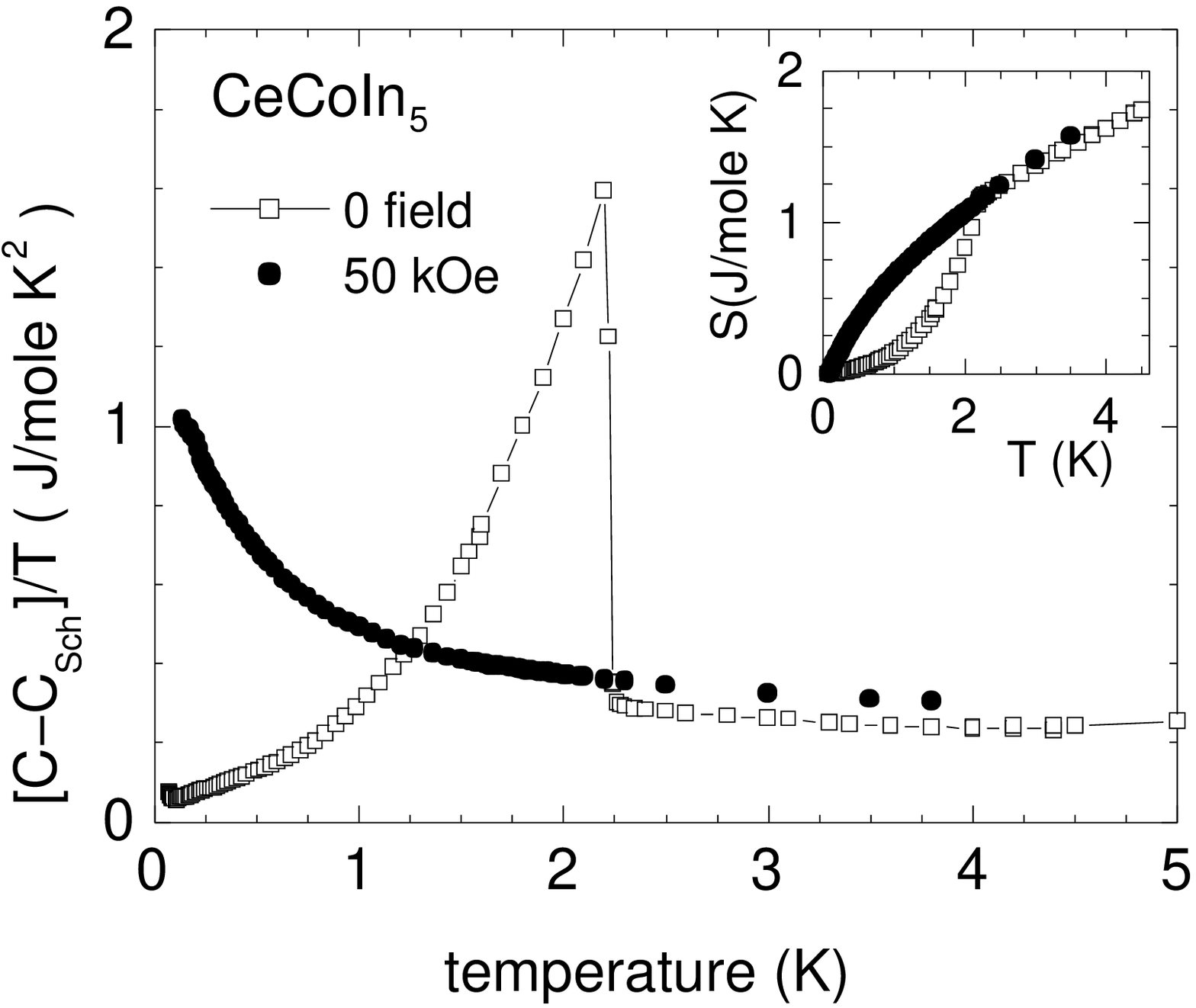}
\hfill
\includegraphics[angle=0,width=0.4\columnwidth,clip]{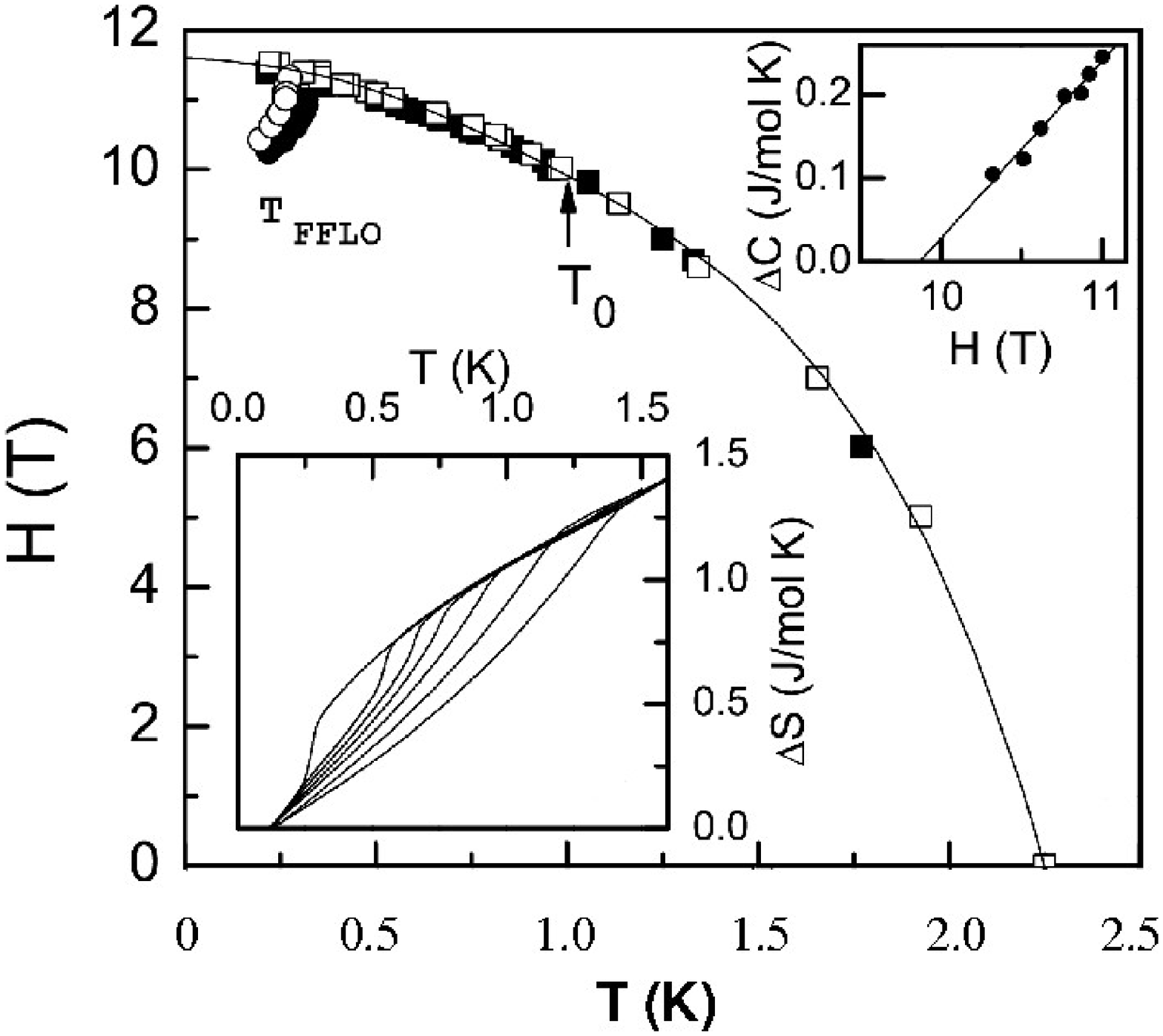}
\caption{Left panel: Linear specific \index{specific heat!CeCoIn$_5$}
heat coefficient of CeCoIn$_5$ vs 
temperature, for zero field anomalous nFl behaviour is observed. 
$\Delta C(T_c)/\gamma T_c\simeq 5$ is strongly enhanced over the
isotropic BCS value 1.43. Inset shows corresponding entropies.
Right panel: H$_{c2}$ curve from specific heat data. 
\index{upper critical field!CeCoIn$_5$} At T$_0\simeq$ 1.1 K the
transition changes from second to first
order. lower inset shows entropy gain as function of T starting from 0.13 K,
for increasing field (right to left: 8.6 T - 11.4 T) a step is evolving.
Upper inset shows specific heat jump at the transition line T$_{FFLO}$
to the Fulde-Ferrell-Larkin-Ovchinnikov state (Bianchi et al. 2003). 
\index{FFLO phase}}
\label{fig:CeCoIn5CJump}
\end{figure}

CeCoIn$_{5}$ appears to be a typical heavy-fermion material. The 
effective magnetic moment of $\sim 2.59 \mu_B$ as determined at high 
temperatures (T $>$ 200K) from a polycrystalline average is consistent 
with the free-ion value for Ce$^3+$ ($2.54 \mu_B$). The $\rho$(T)
dependence is rather weak above a 
characteristic temperature $T^*\simeq$ 30 K. The crossover to a 
coherent state is reflected in a rapid decrease at lower
 temperatures. The electronic contribution to the low-temperature 
specific heat in the normal state (T $>$ 2.3 K) varies linearly 
C $\sim \gamma$ T with a large coefficient 
$\gamma \sim$ 350 mJ/molK$^2$ indicating substantial
mass renormalization \cite{Petrovic01}.  The normal Fermi liquid
becomes unstable at T$_c$ = 2.3 K where the heavy quasiparticles form
Cooper pairs. This can be seen  from the discontinuity in the specific
heat with $\Delta C(T_c)/\gamma T_c\simeq 5$ as shown in
fig.~\ref{fig:CeCoIn5CJump}. The enhancement
over the universal BCS values for isotropic SC points to pronounced
strong coupling effects. \index{strong coupling effects} The BCS  value
of $1.43$ for isotropic superconductors should be considered as an
upper bound on weak-coupling theory which is reduced for anisotropic
pair states as obvious from table~\ref{TABLEBCS}. \index{BCS ratio}

The superconducting state seems to be of unconventional d$_{x^2-y^2}$
singlet type. \index{d-wave gap}\index{singlet pairing} The presence of
line nodes is 
inferred from the power-law behavior exhibited by various
thermodynamic and transport properties at low temperatures. The
latter include the specific heat \cite{Petrovic02,Hegger00} and NMR 
relaxation rate \cite{Kohori01,Zheng01}. The most direct 
evidence, however, is provided by thermal conductivity measurements 
in a magnetic field rotating within the a-b plane
\cite{Izawa01,Matsuda03} (fig.~\ref{FIGOPDosc}). In addition, the
observation of a zero-bias 
conductance anomaly in point-contact spectroscopy supports
the  hypothesis of an unconventional pair state \cite{Goll03}. This is
also suggested by the Knight shift measurements of
\citeasnoun{Kohori01} which give support for an even singlet SC order
parameter with line nodes such as the d$_{x^2-y^2}$ gap function
deduced from the above thermal conductivity results.
 
The properties of this unconventional superconductor are of great
interest. A first issue is the sensitivity to nonmagnetic
impurities which are expected to act as strong pairbreakers. 
Substitution of Ce-ions with La-ions results in a depression of the 
transition temperature T$_c$ \cite{Petrovic02}. However the measured 
T$_c$-reduction is weaker than the value expected for 
isotropic (s-wave) scatterers in a d$_{x^2-y^2}$-state. This 
suggests that La-substitution introduces anisotropic scattering. 
Anisotropies in the scattering matrix element may strongly reduce 
pairbreaking as compared with pure s-wave scattering \cite{Fulde90,Fulde88a}.
The upper critical field $H_{c_2}$ is simply reduced by the impurities
while its anisotropy remains unaltered.

A magnetic field can suppress superconductivity by acting on the
orbits of the charged quasiparticles and by 
acting on their spins. Both mechanisms determine the upper critical
magnetic field $H_{c2}(T)$ which separates the normal state for  
 $H>H_{c2}(T)$ from the superconducting \index{vortex phase} vortex
phase at $H<H_{c2}(T)$.
 The relative importance of the two depairing effects determines the 
order of the phase transition in a type-II superconductor. While the
omnipresent and usually dominating orbital pairbreaking is associated 
with a second order phase transition, a first order transition is 
anticipated for dominant spin depairing. In superconductors with sufficiently
weak orbital pairbreaking a change from second order at 
$T_0<T\leq T_c$ to first order at $T<T_0$ should be observed 
(for a review see \citeasnoun{SaintJamesBook}).  The crossover, 
however, was not observed in conventional superconductors due to 
the presence of strong spin-orbit scattering. CeCoIn$_5$ appears 
to be an ideal candidate for the observation of the first order 
phase transition in a magnetic field. Firstly, it is a very clean 
type-II superconductor where the orbital pairbreaking is rather 
weak due to the high effective mass of the quasiparticles. Secondly, the 
magnetic susceptibility is enhanced
which further lowers the Chandrasekhar-Clogston field. The first 
order phase transition at T$_0$ shown in the phase diagram of
fig.~\ref{fig:CeCoIn5CJump} was originally inferred from a step
in thermal  conductivity by \citeasnoun{Izawa01}. It is reflected in a
pronounced sharpening of both the specific heat and thermal expansion, a large
magnetoelastic effect  and a steplike change in the sample volume
\cite{Bianchi02}.

The change from the second-order nature of the transition observed at zero and 
low magnetic field to first order at high fields occurs at T$_0 \simeq
$ 1.1 K = 0.48 T$_c$. This value should be compared with the estimate
of 0.33 T$_c$ deduced from the  Chandrasekhar-Clogston field of a
d-wave order parameter and the orbital critical field obtained from
extrapolating the behavior close to the zero-field transition to low
temperatures.

The structure of the flux line lattice which forms at low temperatures 
in the presence of an external magnetic field is of particular
interest. A comprehensive review of this topic for s-wave
superconductors has been given by 
\citeasnoun{Brandt95}. An ideal isotropic type-II superconductor 
favors the hexagonal flux lattice. However, as
the energy gain with respect to the square lattice is only 
2\%, relatively weak perturbations like FS or gap anisotropies may lead to 
distorted hexagonal or even square vortex lattices. CeCoIn$_5$ 
is a clean type-II superconductor for which one anticipates a rather 
complex variation with magnetic field of the flux line lattice. 
Recent small-angle neutron scattering experiments \cite{Eskildsen02} have
imaged the vortex lattice of CeCoIn$_5$. At low fields, a hexagonal 
lattice is observed. The latter undergoes a ``phase transition'' to a 
square lattice when the field is increased. The orientation of the 
square flux line lattice relative to the crystal lattice is consistent 
with the expectations for the suggested d-wave order
parameter. The square lattice for \index{d-wave gap} d-wave
superconductors has been
predicted theoretically by \citeasnoun{Won95} and 
\citeasnoun{Berlinsky95}. Finally at low temperatures and close to
H$_{c2}$ specific heat measurements \cite{Bianchi03} seem to indicate
the presence of another inhomogeneous phase for both a-and c-field
directions which was interpreted as the long-sought
Fulde-Ferrell-Larkin-Ovchinnikov (FFLO) phase \index{FFLO phase} which
is characterised
by Cooper pairs with finite net momentum. The FFLO phase for d-wave
SC order parameter of CeCoIn$_5$ was studied theoretically in
\citeasnoun{Won03a}.


\subsection{Superconductivity close to a quantum critical point}
\index{quantum critical point}

The symmetry and mechanism of SC in the CeMIn$_5$ compounds have not
yet been identified in any comprehensive way. However there are strong
indications of unconventional order parameters. The interest in the SC of 
these compounds results from the observation that under pressure and
chemical substitution the SC pair correlations develop in the vicinity
of a QCP. Recently a few U-compounds have been found which fall also
in this class. Theories to describe SC at the QCP are at present highly
controversial and will be discussed there to some extent
(sect.~\ref{Sect:UGe2}).
 
Here we shall briefly focus on the genuine QCP phenomenon in
Ce-systems and its theoretical description. The compounds under
consideration
have a tendency to AF order. Two different scenarios are discussed for
the transition from the HF state to a phase with long-range magnetic order.
 
The first approach views the magnetic order as itinerant magnetism of the 
heavy quasiparticle system. The theory starts 
on the Fermi liquid side which is characterized by a large Fermi
surface including the f-degrees of freedom. The low-energy excitations 
are fermionic quasiparticles and their collective excitations. 
Close to the QCP, the static susceptibility is 
assumed to diverge at a specific wave vector ${\bf Q}$ which signifies 
the transition to a spin density wave state. The quasiparticles
are strongly scattered by spin fluctuations along ''hot lines'' which
are connected by ${\bf Q}$. This scattering modifies the 
low-temperature thermodynamic and transport 
properties which exhibit anomalous scaling relations close to
QCP. They differ from those of the familiar Fermi liquid and therefore
are at the focus of experimental interest. 
The theoretical derivation of the QCP scaling relations
procedes in close analogy to the spin-fluctuation 
\index{spin fluctuation} theory of
ferromagnetism \cite{MoriyaBook} starting from the self consistent 
RPA expression for the magnetic susceptibility of the heavy Fermi liquid 
 \cite{Moriya95}. The theory was recast
in the renormalization group language \cite{Hertz76,Millis93}. 
The ansatz yields scaling relations for the specific heat, 
the susceptibility, and the Wilson ratio which differ from the
well-known Fermi liquid results
\cite{ContinentinoBook,SachdevBook}. The scaling
behavior obtained for various different models and many non-Fermi
liquid compounds are summarized in \citeasnoun{Stewart01}. 
The anomalous transport properties exhibited by metals near
antiferromagnetic quantum phase transitions are of particular interest 
in the experimental study of Ce-based heavy fermion systems. The
theory \cite{Rosch99} exploits the fact that in the vicinity of the magnetic
instability the magnetic susceptibility reduces to the simple form
\begin{equation}
\chi^{-1}({\bf q},\omega) \simeq
\frac{1}{\left(q_0 \xi\right)^2}+\omega_{\bf q}-\frac{i \omega}{\Gamma} 
\label{eq:ChiQCP}
\end{equation}
with
\begin{equation}
\frac{1}{\left(q_0 \xi\right)^2} \simeq 
r+c \left(\frac{T}{\Gamma}\right)^{3/2} 
\qquad \mbox{and} \qquad
\omega_{\bf q} \simeq \frac{1}{q_0^2} \left({\bf q}-{\bf Q}\right)^2
\quad 
\end{equation}
Here r measures the (almost T-independent) distance from the quantum 
critical point, $\Gamma$ denotes the characteristic energy scale of
the fluctuations and c is a constant. The frequency
$\omega_{\bf q}$
vanishes at the ordering vector ${\bf Q}$ of the antiferromagnetic
structure. In the dirty limit the variation with temperature of the
resistivity 
\begin{equation}
\rho(T)=\rho(0)+\Delta \rho (T) \qquad \mbox{with} \qquad \Delta
\rho (T) \sim T^{3/2} 
\label{eq:DirtyRhoAtQCP}
\end{equation}
In the clean limit, on the other hand, the transport is dominated by
the carriers with the longest life times, i.~e.~, by the
quasiparticles far away from the `hot'  regions on the Fermi surface 
yielding $\rho(T) \sim T^2$.
Eq. (\ref{eq:DirtyRhoAtQCP}) is valid only at very low temperatures
$T/\Gamma < 1/\hat{\rho}_0$ where $\hat{\rho}_0$ denotes the residual
resistivity ratio $\hat{\rho}_0$ = $\rho(0)/\rho(273K)$.
At intermediate temperatures $1/\hat{\rho}_0 < T/\Gamma <
1/\sqrt{\hat{\rho}_0}$ the resistivity rises linearly with temperature
\begin{equation}
\Delta\rho (T) \sim T \sqrt{\rho} \quad   
\label{eq:DirtyRhoLinear}
\end{equation} 
The important feature of the  weak-coupling picture as described above 
is that $\Gamma$, which is related to the characteristic energy of the 
Fermi liquid phase still plays the role of the scaling energy. This 
fact is not surprising since only parts of the Fermi surface contribute to the
anomalous properties. One should note however that this theory is
not fully self consistent because the AF fluctuation wave vector is
not connected with any FS feature since the latter is assumed as spherical. 

The loss of the energy scale $\Gamma$ is a characteristic feature of
an alternative scenario which emphasizes the role of local
correlations in quantum critical behavior
\cite{Si99,Si01a,Coleman99}. The dynamical competition between
the Kondo and RKKY interactions \index{Kondo lattice}\index{RKKY
interaction} is analyzed by an extension
\cite{Smith00,Chitra00} of the dynamical mean-field theory (see
\cite{Georges96} and references therein). The resulting
momentum-dependent susceptibility has the general form
\begin{equation}
\chi^{-1}({\bf q},\omega)=M(\omega)+J({\bf q})
\label{eq:ChiEDMFT}
\end{equation}
where the spin self energy $M(\omega)$ is calculated from the effective 
impurity problem. A zero-temperature phase transition will occur when 
$\chi({\bf q},\omega=0)$ diverges. Anomalous behavior is expected
whenever the (effective) local dynamics has an anomalous frequency
dependence. Different scenarios are discussed in \citeasnoun{Si01}. We
finally mention that the \index{YbRh$_2$Si$_2$} compound
YbRh$_2$Si$_2$ (where Yb has one
f-hole instead of one f-electron as in Ce) has been identified as
one of the most promising and cleanest QCP systems where these theoretical
models can be tested \cite{Custers03}
\section{U-based heavy-fermion superconductors}

\label{Sect:UHFS}

Heavy Fermion superconductivity is found more frequently in 
intermetallic U-compounds than in Ce-compounds. This may be related to
the different nature of heavy
quasiparticles in U-compounds where the 5f-electrons have a
considerable, though orbitally dependent, degree of delocalisation. 
The genuine Kondo mechanism is not appropriate for heavy
quasiparticle formation as in Ce-compounds. This may lead to more
pronounced delocalised spin fluctuations in U-compounds which mediate
unconventional Cooper pair formation as discussed in
sect.~\ref{Sect:Theory}. The AF quantum critical point scenario invoked
for Ce compounds previously also does not seem to be so important for
U-compounds with the possible exception of \UGE. On the other hand AF
order, mostly with small moments of the order 10$^{-2}\mu_B$ is
frequently found to envelop and coexist with the SC phase in the B-T plane . 

A common trait of U-compounds is the varying degree of localisation of
5f-electrons. It has recently become clear that Fermi surface
properties of e.g. UPt$_3$ \cite{Zwicknagl02} and
UPd$_2$Al$_3$ \cite{Zwicknagl03} can be well described by treating two of
the three 5f-electrons as localised in orbitals of specific
symmetry and small CEF splitting of the order of meV. This approach
also leads to a natural mechanism for the formation of heavy
quasiparticles via the scattering of delocalised 5f-electrons by the
internal CEF excitations of their localised 5f-electrons. This type of
`dual model' seems much closer to the truth than
either a purely delocalised LDA type descripition which fails to
explain the large masses or Kondo lattice models with fully localised
5f-states whose associated Kondo type anomalies above the Kondo
temperature in transport and thermodynamics are not present in U-compounds.

In the following we will discuss to some length a
number of prominent examples of U-compounds which have extensively been
investigated experimentally and partly theoretically, mostly in the context of
phenomenological Ginzburg-Landau (GL) theories or BCS type theories
with magnetic effective pairing interactions. One is however far from
being able to predict
the spin state (parity) or even symmetry type of the gap function. Simple
models of AF spin fluctuations would predict singlet even parity gap
functions but well characterised examples of triplet odd parity
U-superconductors exist.

Indeed the hexagonal compound UPt$_3$ 
\cite{Stewart84} exhibits triplet pairing and it sticks out as the
most interesting case of unconventional SC with a multicomponent
order parameter whose degeneracy is lifted by a symmetry breaking
field due to the small moment AF order. On the other hand
in UPd$_2$Al$_3$ \cite{Geibel91a} SC coexists with large moment AF and
probably spin singlet pairing is realised, it also exhibits a new kind
of magnetic pairing mechanism mediated by propagating magnetic exciton
modes. The sister compound
UNi$_2$Al$_3$ \cite{Geibel91b} is an example of coexistence of large
moment AF with a SC triplet order parameter. In
URu$_2$Si$_2$ \cite{Palstra85} the
SC order parameter symmetry is still undetermined. The interest in
this compound is focused more on the enveloping phase with a `hidden'
order parameter presumably of quadrupolar type or an `unconventional'
SDW. The oldest cubic U-HF superconductor UBe$_{13}$ \cite{Ott83} and
its thorium alloy
U$_{1-x}$Th$_x$Be$_{13}$ is also the most mysterious one. While for
the pure system there is a single SC phase of yet unknown symmetry,
in the small Th concentration range two distinct phases exist which may 
either correspond to two different SC order parameters or may be related to a
coexistence of SC with a SDW phase. In addition in UBe$_{13}$ SC
order appears in a state with clear non-Fermi liquid type
anomalies. 
More recently the coexistence of ferromagnetism and SC in
UGe$_2$ \cite{Saxena00} has been found. This is the only case of
U-compounds where quantum critical fluctuations might be involved in
the SC pair formation. Due to the FM polarisation the triplet gap function
contains only equal spin pairing.

\subsection{Heavy fermion multicomponent superconductor UPt$_3$}
\index{multiphase superconductivity}\index{UPt$_3$}
\label{Sect:UPt3}

The intermetallic compound UPt\( _{3} \) which crystallizes in the
hexagonal  Ni\( _{3} \)Sn structure is regarded as the archetype of a
strongly correlated Fermi liquid. The existence of heavy
quasiparticles is inferred from enhanced thermodynamic and transport
coefficients, for example $\gamma \sim 420 mJ/(mol U
\ K^2$) \cite{Frings83} and anisotropic Pauli-like susceptibility
$\chi_c \sim 50 \cdot 10^{-9} m^3/(mol U)$ and 
$\chi_{a,b} \sim 100 \cdot 10^{-9} m^3/(mol U)$ 
\cite{Frings83,Stewart84}. In addition, the resistivity rises quadratically
with temperature with a $T^2$- coefficient $A \simeq 0.49 \mu \Omega /(cm 
\ K^2)$ \cite{Taillefer88}. The formation of the coherent heavy Fermi
liquid state is clearly seen in the optical
conductivity \cite{Marabelli87,Degiorgi99,Dressel02} and in the
dynamical magnetic susceptibility as measured by neutron scattering
\cite{Bernhoeft95}. Direct evidence for the existence of heavy
quasiparticles comes from the observation of quantum oscillations
\cite{Taillefer87,Taillefer88,McMullen01,Kimura98} where Fermi
surfaces and effective masses up to m$^*$ = 135 m were
determined. The average dHvA mass enhancement m$^*$/m$_b\simeq$ 20
compared to the band mass m$_b$ is found to be consistent with
the specific heat enhancement $\gamma_{exp}/\gamma_{calc} = 17$
\cite{Norman88}. These results prove convincingly that UPt$_3$ in its
normal state is a heavy Fermi liquid.

The normal Fermi liquid state becomes  unstable at low temperatures. At T$_N
\simeq$ 5.8 K an antiferromagnetic phase with extremely small ordered
moments $\mu \simeq$ 0.035 $\mu_B$/U develops 
\cite{Heffner89,Hayden92,Lussier96}. The most exciting phenomena 
are associated with the superconducting state. The existence of 
two clearly distinct transition
temperatures $T_{c1}$ = 530 mK and  $T_{c2}$ =480 mK \cite{Fisher89,Brison94}
implies that the superconducting phases are characterized by a 
multicomponent order parameter. The discontinuities in the specific heat at the
superconducting transitions show that the Cooper pairs are formed by
the heavy quasiparticles of the normal heavy Fermi liquid state.

From this we conclude that a quantitative theory for the origin of the 
heavy quasiparticles in UPt$_3$ is a prerequisite for a 
detailled understanding of the fascinating low-temperature properties
of this compound.

\subsubsection{Dual model and heavy quasiparticles}
\index{dual model}\index{heavy fermions}

Although a comprehensive picture of the low temperature ordered phases of
UPt\( _{3} \) has emerged, a complete theoretical understanding
of the origin of 5f-derived heavy quasiparticles is
still missing. The number of itinerant U 5f electrons as well as the
microscopic mechanism yielding the high effective masses are still
controversial. It has
been considered a success of the LDA that de Haas van Alphen (dHvA) 
frequencies can be related to the areas of extremal orbits on the
Fermi surface obtained by band-structure calculation
which treats the U 5f states as itinerant \cite{Albers86,Norman88}. 
From these findings, however, one should not conclude that the 
U 5f states are ordinary weakly correlated band states. The 
calculated energy bands are much too broad for explaining the effective
masses while on the other hand, they are too narrow to fit the 
photoemission data \cite{Allen92}. 

\begin{figure}[htb]
\begin{minipage}[t]{70mm}
\includegraphics[width=64mm]{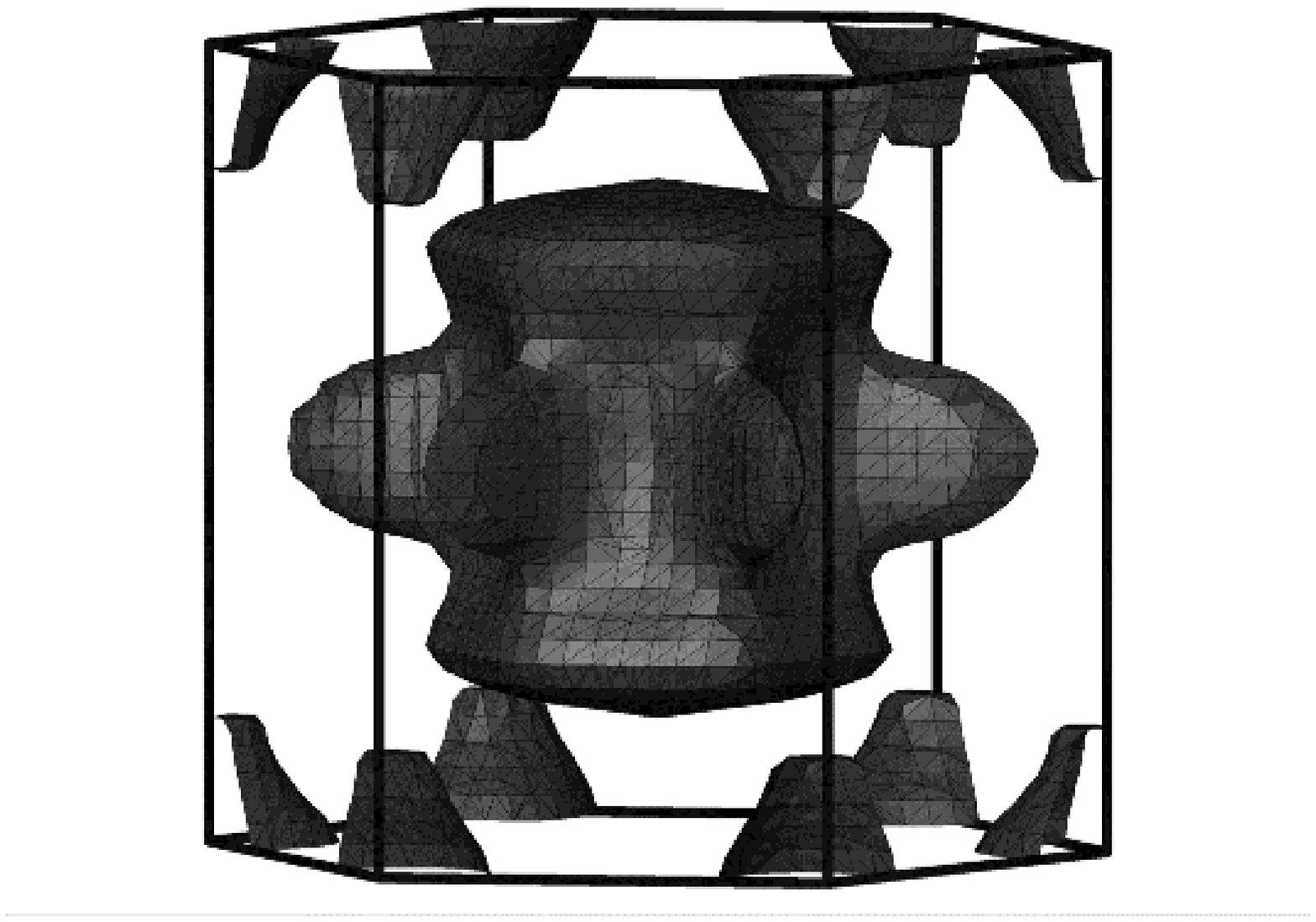}
\end{minipage}
\hspace{\fill}
\begin{minipage}[t]{70mm}
\includegraphics[width=64mm]{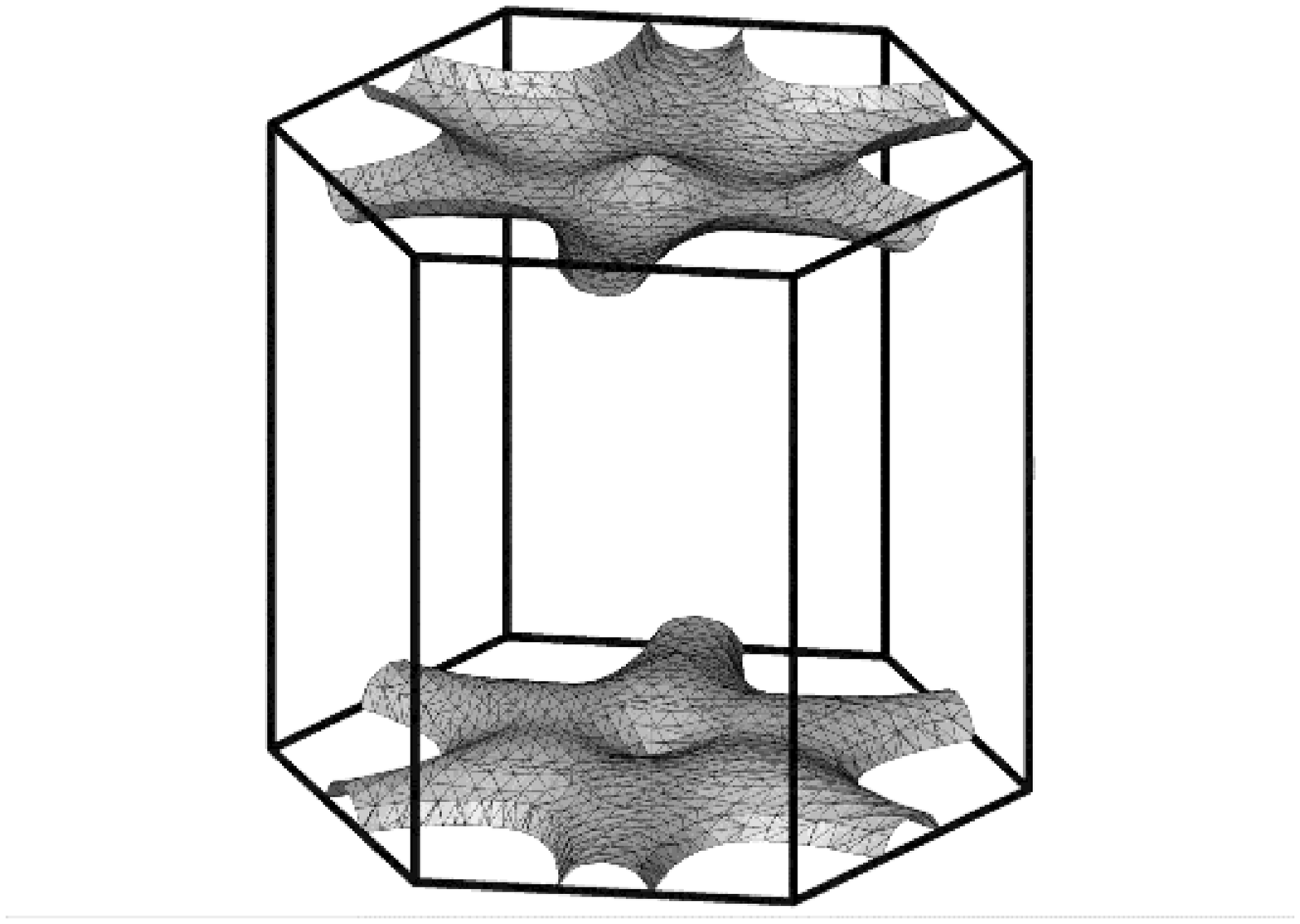}
\end{minipage}
\caption{
UPt$_3$: FS \index{Fermi surface!UPt$_3$} of heavy quasiparticles 
\index{heavy fermions} as calculated
within the dual model (Zwicknagl et al. 2002). One of the U
5f-electrons (j = 5/2, j$_z$ = 3/2) is treated as itinerant and is
included in the FS while the remaining two are localized. Then two bands
contribute to the FS. Left: $\Gamma$-centered heavy electron sheet which
dominates $\gamma$. Right: A-centered open orbit sheet consistent
with magnetoresistance results (Mc Mullen et al. 2001).}
\label{fig:DualModelFermiSurfaceUPt3}
\end{figure}

The key feature is likely the dual nature of the U $5f$ states,
i. e., the presence of both localized and delocalized U $5f$
electrons. The theoretical investigation proceeds in three
steps. Firstly, band-structure calculations have been performed
starting from the self-consistent LDA potentials but excluding the 
U 5\( f \) j=\( \frac{5}{2} \), j\( _{z} \)=\( \pm \frac{5}{2} \)
and j\( _{z} \)=\( \pm \frac{1}{2} \) states from forming bands. The localized
5\( f \) orbitals are accounted for in the self-consistent density
and likewise in the potential seen by the conduction
electrons. The 5\( f \) bands are calculated
by solving the Dirac equation \cite{Albers86}. The intrinsic bandwidth of the
itinerant U 5\( f \) j=\( \frac{5}{2} \), j\( _{z} \)=\( \pm \frac{3}{2} \)
is taken from the LDA calculation while the position of the
corresponding band center is chosen such that the density
distribution of the conduction states as obtained
within LDA remains unchanged. The resulting position of the \( f \)
band relative to the calculated Pt d states is consistent with
photoemission data. It was found that the U 5\( f \) bands with 
j\( _{z} =\pm \frac{3}{2} \)
hybridize strongly near the Fermi level so that the \( f \) occupancy per U
atom for the delocalized 5\( f \) electrons amounts to n\( _{\textrm{f}} \)
= 0.65 indicating that we are dealing with a mixed valent situation.
The theoretical Fermi surface shown in
fig.~\ref{fig:DualModelFermiSurfaceUPt3} is formed by two bands which
are doubly degenerate and which are derived from the 5\( f \)
$|j=\frac{5}{2},j_z=\pm\frac{3}{2}\rangle$ states. The
thermodynamically most important orbit is assigned to the 
\( \Gamma  \)-centered strongly anisotropic electron surface shown in
the left part of 
fig.~\ref{fig:DualModelFermiSurfaceUPt3}. The A-centered part of the Fermi
surface shown in the right part of
fig.~\ref{fig:DualModelFermiSurfaceUPt3} has open orbits spanning the entire
Brillouin zone for magnetic fields oriented along the a-direction in
the basal plane. This feature is consistent with magnetoresistance
measurements \cite{McMullen01}.
%
%
Now we turn to the discussion of the localized U 5\( f \) states. 
The multiplet structure of the localized \( f^{2} \) states is
calculated by diagonalizing the Coulomb matrix. The spin-orbit
splitting is rather large and therefore \index{jj-coupling}
a jj-coupling scheme is used. This simplification gives six 2-particle states
built from \( |j=\frac{5}{2},j_{z}=\pm \frac{5}{2}\rangle  \) and 
\( |j=\frac{5}{2},j_{z}=\pm \frac{1}{2}\rangle  \).
The resulting eigenstates are generally no longer eigenstates of the 
total angular momentum \v J\( ^{2} \), but remain eigenstates of 
J\( _{\textrm{z}} \). The Coulomb matrix elements are calculated
following \citeasnoun{CondonShortley}.
Inputs are the Slater-Condon parameters F\( ^{K} \) (Coulomb integrals) and
G\( ^{K} \) (exchange integrals) which are evaluated with the radial function
R\( ^{U}_{f,\frac{5}{2}}(\textrm{r}) \) for U which is determined from 
a self-consistent band structure potential.  Diagonalization of the
matrix yields a doubly degenerate ground state J\( _{z}=\pm 3 \) which
must be an eigenstate of J = 4. Note that the Pauli principle permits
even values of J only, i.e., J = 0, 2, 4 in our case. The states \(
|j=\frac{5}{2},j_z=\frac{5}{2};J=4,J_{z}=\pm 3\rangle \)
have an overlap of 0.865 with the Hund's rule ground state \( ^{3}H_{4} \)
derived from the LS-coupling scheme. Therefore the choice of jj vs. LS coupling
should only weakly affect the results obtained for the ground-state multiplet.
The two-fold degeneracy of the ground-state is lifted by a CEF yielding the
two singlet states
\begin{eqnarray}
|\Gamma _{3}\rangle  & = & \frac{1}{\sqrt{2}}(|J=4;J_{z}=3\rangle
 +|J=4;J_{z}=-3\rangle )\nonumber \label{2} \\
|\Gamma _{4}\rangle  & = & \frac{1}{\sqrt{2}}(|J=4;J_{z}=3\rangle
 -|J=4;J_{z}=-3\rangle ).
\end{eqnarray}
Note that \( |\Gamma _{4}\rangle  \) has also been suggested as ground state of
UPd\( _{2} \)Al\( _{3} \) \cite{Grauel92,Boehm92}. The 
\index{Coulomb interaction} excited eigenstates of the Coulomb matrix
are neglected, they are 
assumed to be separated by a rather high excitation energy from the
ground state. The coupling between the localized and delocalized \( f
\) electrons is directly obtained from the expectation values of the Coulomb 
interaction H$_C$ (eq.~\ref{eqCoulomb}) in the 5\( f^{3} \)
states yielding the transition matrix element 
\begin{equation}
\label{3}
\left| M \right| = \left|\langle f^{1};\frac{5}{2},\frac{3}{2}|\otimes \langle 
\Gamma _{4}|H_{\textrm{C}}|\Gamma _{3}\rangle 
\otimes |f^{1};\frac{5}{2},\frac{3}{2}\rangle\right| = 0.19~eV.
\end{equation}
%
\begin{figure}[tb]
\begin{minipage}[t]{150mm}
\includegraphics[width=75mm]{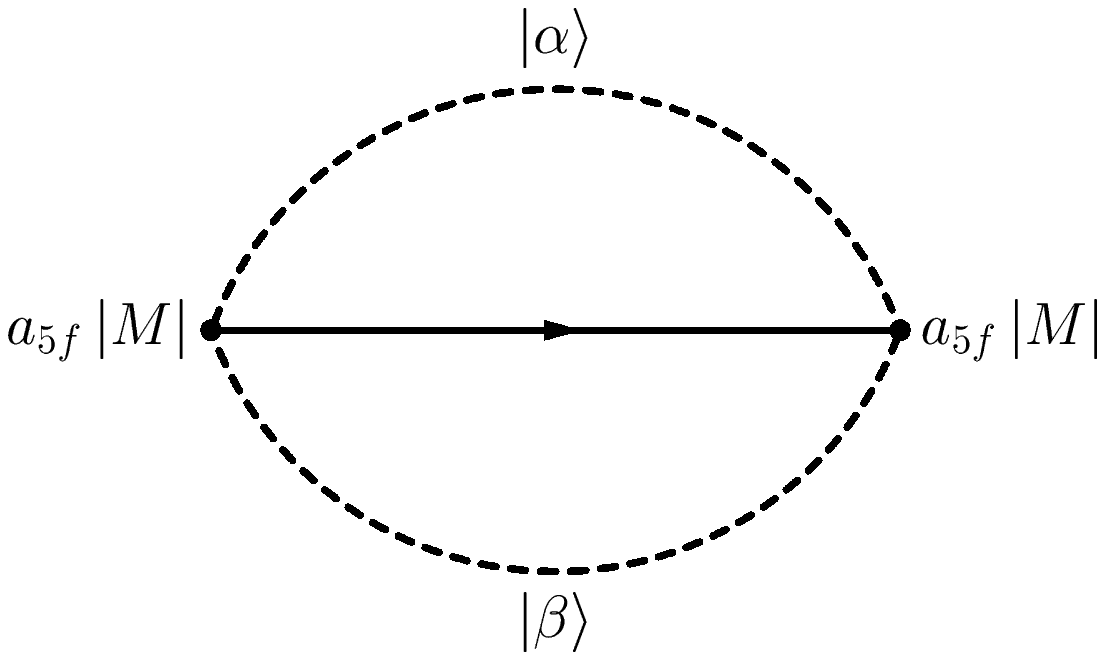}
\end{minipage}
\caption{
Self energy due to virtual 
\index{crystalline electric field excitations}
CEF excitations $|\alpha\ra\leftrightarrow|\beta\ra$ leading to the
enhanced m$^*$. Full line and dashed bubble represent conduction electron
propagator and local 5f susceptibility of eq.~(\ref{EFFMASS2}),
respectively. The self consistent solution of eq.~(\ref{EFFMASS1}) yields
the leading order in an expansion in terms of the small ratio
$\delta$/W$^*$ ($\delta$ = CEF-splitting, W$^*$ = effective
conduction band width). Input
parameters: matrix element $\alpha_{5f}|M|$, DOS N(E) and (effective)
CEF splitting $\delta$. The former two are determined from first
principles calculations while $\delta$ is taken from experiment.}
\label{fig:SelfEnergyCEF}
\end{figure}

Finally, we determine the mass enhancement \index{mass enhancement}
which arises from the
scattering of the itinerant U $5f$ electrons due to virtual excitations of
the localized $f^2$ subsystem. The latter are characterized by a CEF splitting
energy $\delta$ while the relevant energy scale for the `itinerant'
$5f$ states is set by the effective band width W$^*$. 
To leading order in the ratio $\delta/W^*$ the mass enhancement
m$^*$/m is obtained  from the fluctuation exchange (FLEX)
contribution to the band self energy as shown in
fig.~\ref{fig:SelfEnergyCEF}  in close analogy to the case
of Pr metals \cite{White81}. This leads to an expression 
\begin{eqnarray}
\label{EFFMASS1}
\frac{m^*}{m_b}&=&
1-\bigl(\frac{\partial\Sigma}{\partial\omega}\bigr)_{\omega =0}\nonumber\\
\Sigma(i\omega_n)&=&a^2_{5f}|M|^2T\sum_{n'}\chi(i\omega_n-i\omega_{n'})
g(i\omega_{n'})
\end{eqnarray}
where $\omega_n$ are Matsubara frequencies, m$_b$ is the LDA band mass
and a$_{5f}$ denotes the 5f weight of conduction electron states close
to E$_F$. Furthermore the local 5f susceptibility and conduction
electron propagator are given, respectively, by 
\begin{eqnarray}
\label{EFFMASS2}
\chi(i\omega_n -i\omega_{n'})&=&\tanh\frac{\delta}{2T}
\frac{2\delta}{\delta^2-(i\omega_n-i\omega_{n'})^2}\nonumber\\
g(i\omega_n)&=&\int dE\frac{N(E)}{i\omega_n-E-\Sigma(i\omega_n)}
\end{eqnarray}
Here 2N(E) is the total electronic DOS obtained from the LDA
bandstructure with two of the 5f electrons kept localised. 

\begin{figure}[htb]
\begin{minipage}[t]{70mm}
\includegraphics[width=75mm]{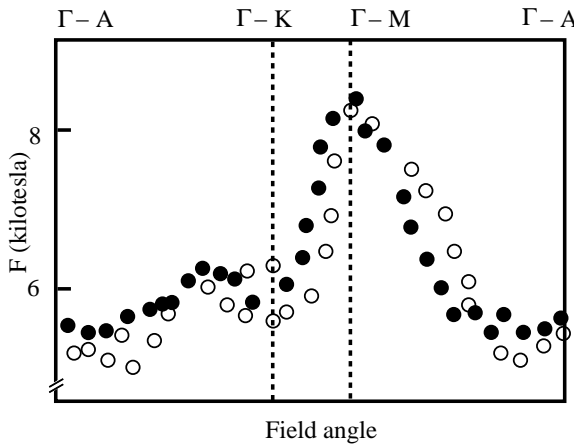}
\end{minipage}
\caption{
UPt$_3$: dHvA cross sections of  heavy quasiparticle FS ($\Gamma$-
centered closed sheet in fig.~(\ref{fig:DualModelFermiSurfaceUPt3}) from
experiment (filled symbols) (Kimura et al. 1998) and from dual model
calculation (Zwicknagl et al. 2002) (empty symbols). Field direction
is turned from c-direction ($\Gamma$-A) to a-direction ($\Gamma$-K)
and then rotated in the basal plane from a-direction to b-direction
($\Gamma$-M) and finally turned back to c-direction. The
calculated splitting for magnetic field orientations in the a-c-plane
($\Gamma$-A to $\Gamma$-K sector) is indeed observed in recent dHvA
experiments (McMullen et al. 2001). 
\index{de Haas-van Alphen effect!UPt$_3$}}  
\label{fig:HeavyFermiSurfaceCrossSectionsUPt3}
\end{figure}

This calculation procedure leads to a good agreement of theoretical
and observed quasiparticle masses \cite{Zwicknagl02} for the main
Fermi surface sheets. The results are summarized in
fig.~\ref{fig:HeavyFermiSurfaceCrossSectionsUPt3} as well as in
Table~\ref{tab:EffMassesUPt3}.
%
%
\begin{table}[htb]
\caption{Mass renormalisation in \UPT from dHvA experiments and dual model
calculations for field along hexagonal symmetry directions.}
\label{tab:EffMassesUPt3}
\begin{tabular}{lccc}
\hline
$m^*/m$          & c & a & b  \\
\hline
 Experiment   & 110 & 82 & 94  \\
Theory        & 128 & 79 &104  \\
\hline
\end{tabular}\\[2pt]
\end{table}
%
If the dependence upon the wave vector ${\bf k}$ is neglected one
obtains an isotropic enhancement of the effective mass $m^*$ 
over the band mass $m_b$. The explicit evaluation requires as input
parameters the density of states and the f-weight per spin and U-atom
at the Fermi energy, $N(0)$ and $4a_{5f}^2$, respectively, as well as the
average energy ${\delta}$ of CEF excitations. Solving for the
selfconsistent conduction electron propagator
with the input determined from 
the ab-initio calculations of the electronic structure 
(N(0)= \( \simeq  \) 15.5 states/ (eV cell),
\( | \)M\( |^{2} \) = 0.036 eV\( ^{2} \), 4a$_{5f}^2$ = 0.13) and the
estimate \( \tilde{\delta }\simeq  \) 20 meV) we finally obtain the
effective masses listed above.

The dual model and its partial 5f localisation scenario therefore
provides for a satisfactory explanation for both the FS
structure and mass renormalisation of the heavy quasiparticles
observed in UPt\( _{3} \). The physical reason for the strongly
orbital dependent effect of local electron correlations on the different
LDA hybridization matrix elements has been investigated separately 
(sect.~\ref{Sect:Theory}). Finally we want to point out that the prescribed
way of treating the \( 5f \) electrons in UPt\( _{3} \) is also applicable
to other uranium compounds, notably UPd$_2$Al$_3$ (sect.~\ref{Sect:UPd2Al3}).

\subsubsection{Pairing and the spin-orbit coupling problem}
\index{spin orbit coupling}

There is little doubt that UPt\( _{3} \) is an unconventional superconductor
with a multicomponent superconducting order parameter. 
Theoretical models which have been proposed are
reviewed by \citeasnoun{Sauls94} and \citeasnoun{Joynt02} and will be
discussed in the next section. Here we
shall briefly comment on the issue of the
importance of spin orbit coupling in the pairing scheme.
In all theoretical treatments the superconducting phases of
UPt\( _{3} \) are characterized by BCS-type gap functions $\Delta
_{s1,s2}(\hat{\v k})$ which satisfies the antisymmetry relation
eq.(\ref{eq:antisymm}). The wave vector \v k = k$_F(\hat{\v k})\hat{\v k}$
is restricted to the Fermi surface since the coherence length $\xi_0$ is much
larger than the interatomic distance or, equivalently, the Fermi wave
length $\lambda_F = 2\pi/k_F$ satisfies 
\begin{equation}
\xi _{0}=\frac{\hbar v_{F}}{k_{B}T_{c}}\gg \lambda _{F}\quad .
\end{equation}
As a consequence, the gap function ${\bf \Delta}(\hat{\v k})$ is a
matrix function of $\hat{\v k}$ only or its corresponding polar and
azimuthal angles $\vartheta$ and $\varphi$. We have
neglected indices for interband pairing because the available phase
space is much smaller than in the case of intraband pairs. 

The indices s$_1$, s$_2$  refer
to the (pseudo-) spin labels of the quasiparticles. Due to the large spin-orbit
interaction the quasiparticle states are no longer eigenstates of the spin.
In the absence of an external magnetic field the states \( \left| {\bf
k}s \right\rangle  \) and\( \left| -{\bf k}-s \right\rangle  \) show
Kramers degeneracy. In
UPt\( _{3} \), the crystal structure has inversion
symmetry. This fact implies that all four quasiparticle states \(
\left| {\bf k}s \right\rangle  \),
\( \left| -{\bf k}s \right\rangle  \), \( \left| {\bf k}-s \right\rangle  \)
and \( \left| -{\bf k}-s \right\rangle  \) are degenerate. As a result,
the superconducting order parameter can be decomposed into the
contributions \index{singlet pairing}\index{triplet pairing}
\begin{eqnarray}
\mbox{even-partiy (pseudo-spin singlet, S = 0)}:
&&\Delta _{s_1s_2}({\bf k}_{F})=
\Delta ({\bf k}_{F})\left( i\sigma _{y}\right) _{s_1s_2} \nonumber\\
\mbox{odd-parity (pseudo-spin triplet, S = 1)}:
&&\Delta _{s_1s_2}({\bf k}_{F})=
\Delta ({\bf k}_{F})\cdot 
\left( i{\bf \sigma }\sigma _{y}\right) _{s_1s_2}
\end{eqnarray}
Because the energy splitting of j=$\frac{5}{2}$ and j=$\frac{7}{2}$ states
due to spin orbit coupling is larger than the bandwidth due to
hybridisation and overlap \cite{Albers86} their mixing can be
neglected. Consequently the pseudospin indices s$_1$, s$_2$ are good
quantum numbers and effectively the classification of the gap function
can proceed as for the weak spin orbit coupling case (sect.~\ref{Sect:Theory}).

\subsubsection{Multicomponent superconducting order parameter}

The hexagonal heavy Fermion compound \UPT can justly be called the flagship
of unconventional superconductivity, despite having a critical
temperature less
than one Kelvin. It is set aside from all other unconventional
superconductors sofar because it exhibits {\em two} superconducting
phase transitions which have to be interpreted as a direct signature
of the fact that the SC order parameter is a complex vector with more than one
component . In all other unconventional superconductors, e.g. in
high-T$_c$ compounds the order parameter is still a complex scalar as
in ordinary superconductors, albeit belonging to a nontrivial
(but one-dimensional) representation of the symmetry group as
witnessed by nodes in the anisotropic gap function. Only recently
another example of a multicomponent  superconductor, the Pr-
skutterudite compound \PRS (sect.~\ref{Sect:PRSK}),
may have been found. The exciting
discovery of the split SC transitions in \UPT at T$_{c1}$ = 530 mK and
T$_{c1}$ = 480 mK in specific heat measurements by \cite{Fisher89} has
lead to an enormous amount of experimental and theoretical work on
\UPT whose historical evolution has been described in detail in a recent
review by \citeasnoun{Joynt02}. Therefore in the following
discussion we shall restrict to the essential
aspects of SC in \UPT. The additional small moment AF observed in \UPT
(T$_N$ = 5.8 K, $\mu$ = 0.035$\mu_B$) plays a key role in the
identification of the SC
order parameter since the in-plane staggered magnetisation acts as a
symmetry breaking field (SBF) to the SC multicomponent order
parameter. The SBF is believed to be responsible for the appearance of
two SC transitions which otherwise would merge into one, in fact this
has later been proven by specific heat pressure
experiments \cite{Trappmann91} on the SC T$_{c1,2}$(p) and by complementary
neutron diffraction under pressure \cite{Hayden92}. 
Naturally there are also two critical-field curves. They intersect at a
tetracritical point in the B-T plane that is present for all field
directions. Therefore one can identify three distinct SC phases A,B,C in
the B-T plane (fig.~\ref{FIGhcrit}) corresponding to different choices of the
orientation of the complex vector SC-order parameter. These phase
boundaries have first been identified by ultrasound velocity measurements
\cite{Bruls90} and have been confirmed using the same
method \cite{Adenwalla90}. Subsequently many different methods like
thermal expansivity \cite{vanDijk93,vanDijk94}, field dependent
specific heat \cite{Loehneysen94} etc. have yielded equivalent B-T phase
diagrams. An important observation is a crossover of the anisotropy ratio of
upper critical fields as function of temperature indicating that
H$_{c2}$ shows Pauli limiting for \v H$\parallel$c but not for \v
H$\perp$c \cite{Shivaram87,Choi91}. Again the asymptotic (T $\ll$ T$_c$)
low temperature behaviour of thermodynamic and transport coefficients,
most notably thermal conductivity \cite{Lussier96,Graf99} with
power law behaviour points to a gap function with
nodes. The precise node structure is different for the three phases
due to the internal degrees of freedom for a multicomponent order parameter.
Despite the wealth of experimental results on \UPT there is no
unequivocal consensus on the symmetry and node structure of the
superconducting gap. The many different proposals that have been put
forward at various times are summarized in \cite{Joynt02}. Here we
restrict to the most commonly accepted E$_{2u}$ (f-wave) model for the order
parameter which was originally proposed by \cite{Choi91,Sauls94} and
in the form used here by \cite{Norman96}. This odd parity spin triplet
order
parameter is consistent with $^{195}$Pt Knight shift measurements and
the observed linear in T dependence of the thermal conductivity for
both a and c directions. The corresponding B-phase (e.g. H = 0,
T $<$ T$_{c2}$) has a gap function with an equatorial node line and polar
second order node points. The anisotropic paramagnetic limiting of the
upper critical field suggests that its \v d-vector is pinned
along the c axis by an anisotropy potential due to spin orbit
coupling. This interpretation is however in conflict with a nearly
isotropic constant Knight shift for
larger fields \cite{Kitaoka00}. Further investigations, notably
field-angle dependent thermal conductivity experiments are necessary 
to confirm the node structure of the E$_{2u}$ model.

The underlying normal heavy fermion state of \UPT, which was the first
system where heavy quasiparticles have directly been seen in dHvA
oscillations \cite{Taillefer88}, has been described above.
It was argued that the picture of heavy quasiparticle mass generation in 5f-metals has to be revised. This is due to a a considerably different
degree of localisation of 5f electrons in different orbital states as
opposed to the simple LDA picture which assumes complete
delocalisation for all 5f orbitals.

\subsubsection{Small moment AF order}
\index{antiferromagnetic order!UPt$_3$} 

\begin{figure}
\includegraphics[width=75mm]{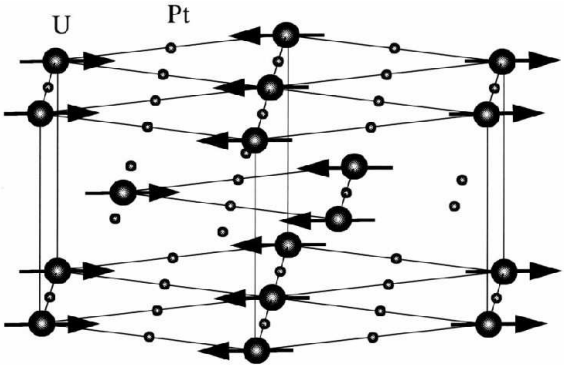}
\vspace{1cm}
\caption{Crystal structure and AF magnetic structure of \UPT below
T$_N$ = 5.8 K. The hexagonal axis defines the \v c direction and the
moments are oriented along \v b $\perp$ \v a (using orthogonal \v a, \v b \v c
lattice vectors where a = b = 5.764 \AA~ and c = 4.884 \AA). Hexagonal
planes are separated by c/2.
The AF order leads to a doubling of the unit cell along \v b (moment
direction) resulting in orthorhombic (D$_{2h}$) symmetry of the AF
crystal.}
\label{FIGmagstruc}
\end{figure}

\UPT crystallizes in the hexagonal MgCd$_3$ structure (D$_{6h}$ point
group) and has two inequivalent U-sites with C$_{3v}$ site symmetry.
As in most other U-based superconductors the SC phase is embedded in
an antiferromagnetic phase with very small moments ($\mu$ =
0.035$\mu_B$). The commensurate
AF order of hcp \UPT with moments parallel to the hexagonal b-axis is
indicated in fig.~\ref{FIGmagstruc}. The D$_{6h}$ sixfold symmetry of the
paramagnetic phase is reduced to an orthorhombic D$_{2h}$
symmetry by the AF order with modulation vector \v Q
=($\frac{1}{2}$,0,0). Although the AF intensity shows mean field
behaviour $\sim$(T$_N$-T) \cite{Aeppli88} AF Bragg peaks retain
a finite width pointing to imperfect AF ordering with a correlation
length of $\xi_{AF}\sim$ 150 \AA. Only below 50 mK $\xi_{AF}$ starts to
diverge \cite{Koike98} and true long
range magnetic order develops at 28 mK according to specific heat
measurements \cite{Schuberth92}. Still in the imperfect ordered regime
the Bragg intensity starts to decrease below the superconducting transition
\cite{Aeppli88} which signifies a coupling of superconducting and AF
order parameters, this has also been observed in \URU,\UPD and \UND
\cite{Metoki01}. Thus the small moment AF itself is a
rather unconventional magnetic state but in the following discussions
its influence on the superconducting state will be treated as 
ordinary AF order though with small moment.

\subsubsection{The superconducting state, coupled with AF order}

\begin{figure}
\includegraphics[width=75mm]{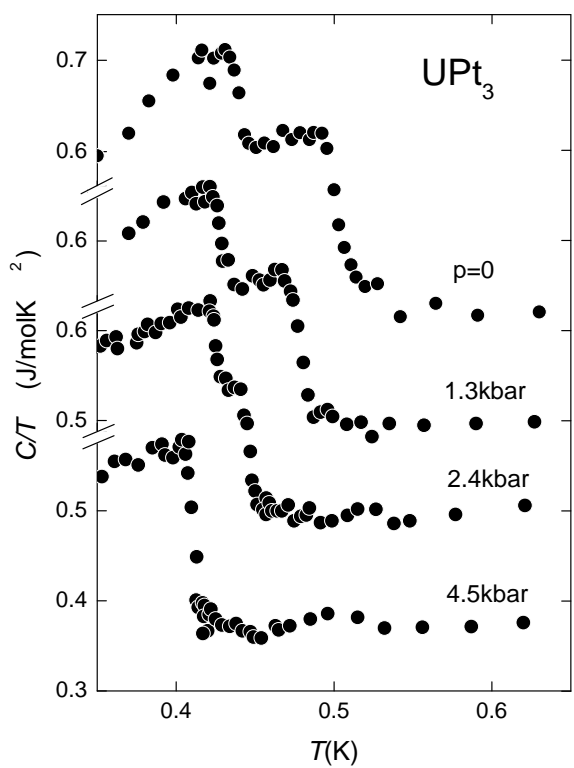}
\caption{Disappearance of the double-peak structure of specific heat
under hydrostatic pressure (Trappmann et al. 1991) from which the pressure
dependence of T$_{c1,2}$(p) shown in fig.~\ref{FIGafsplit} is
obtained. \index{specific heat!UPt$_3$}\index{hydrostatic pressure}
\index{T$_c$-splitting}}
\label{FIGspec}
\end{figure}

\begin{figure}
\includegraphics[width=75mm]{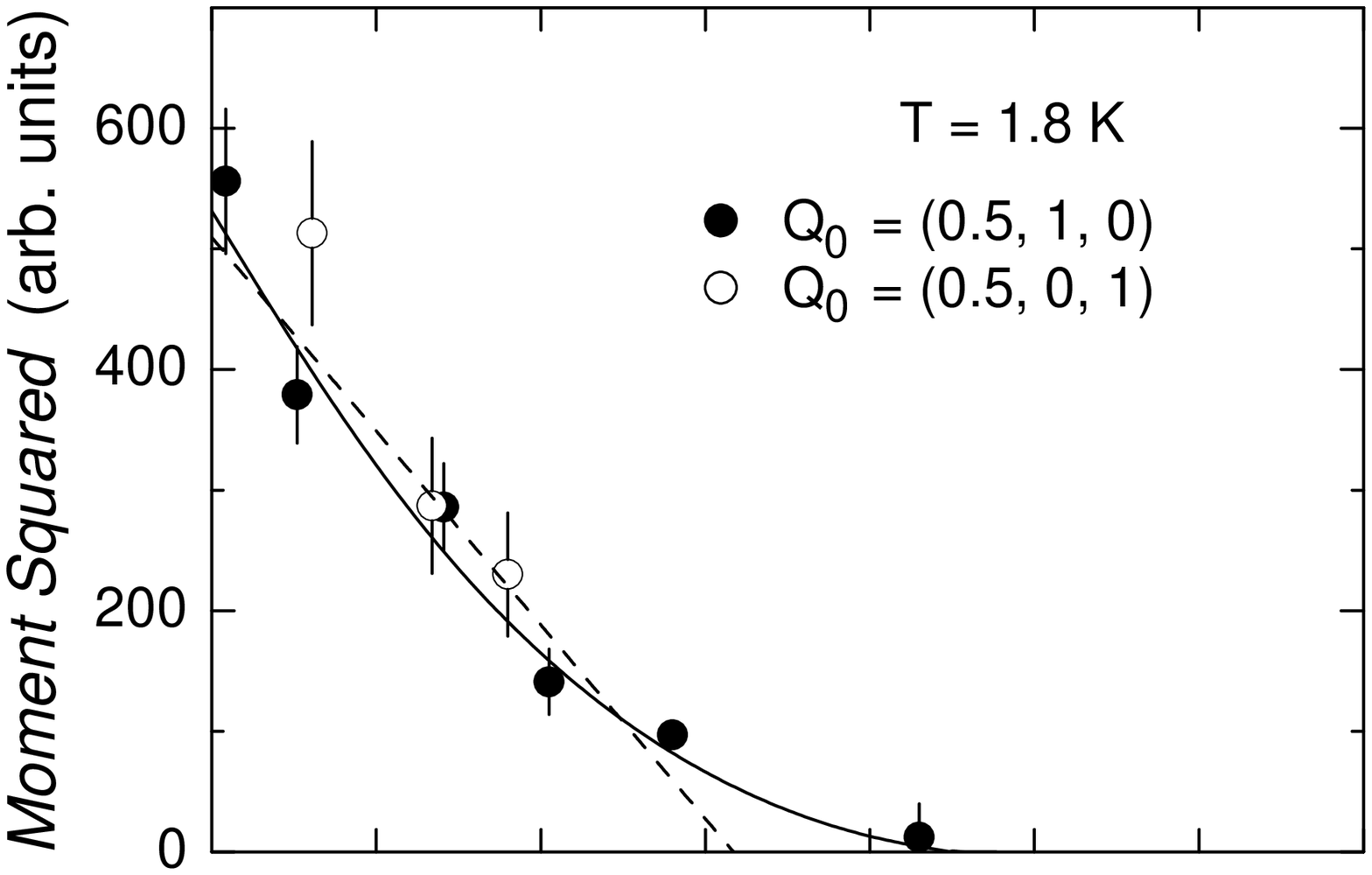}\\
\includegraphics[width=75mm]{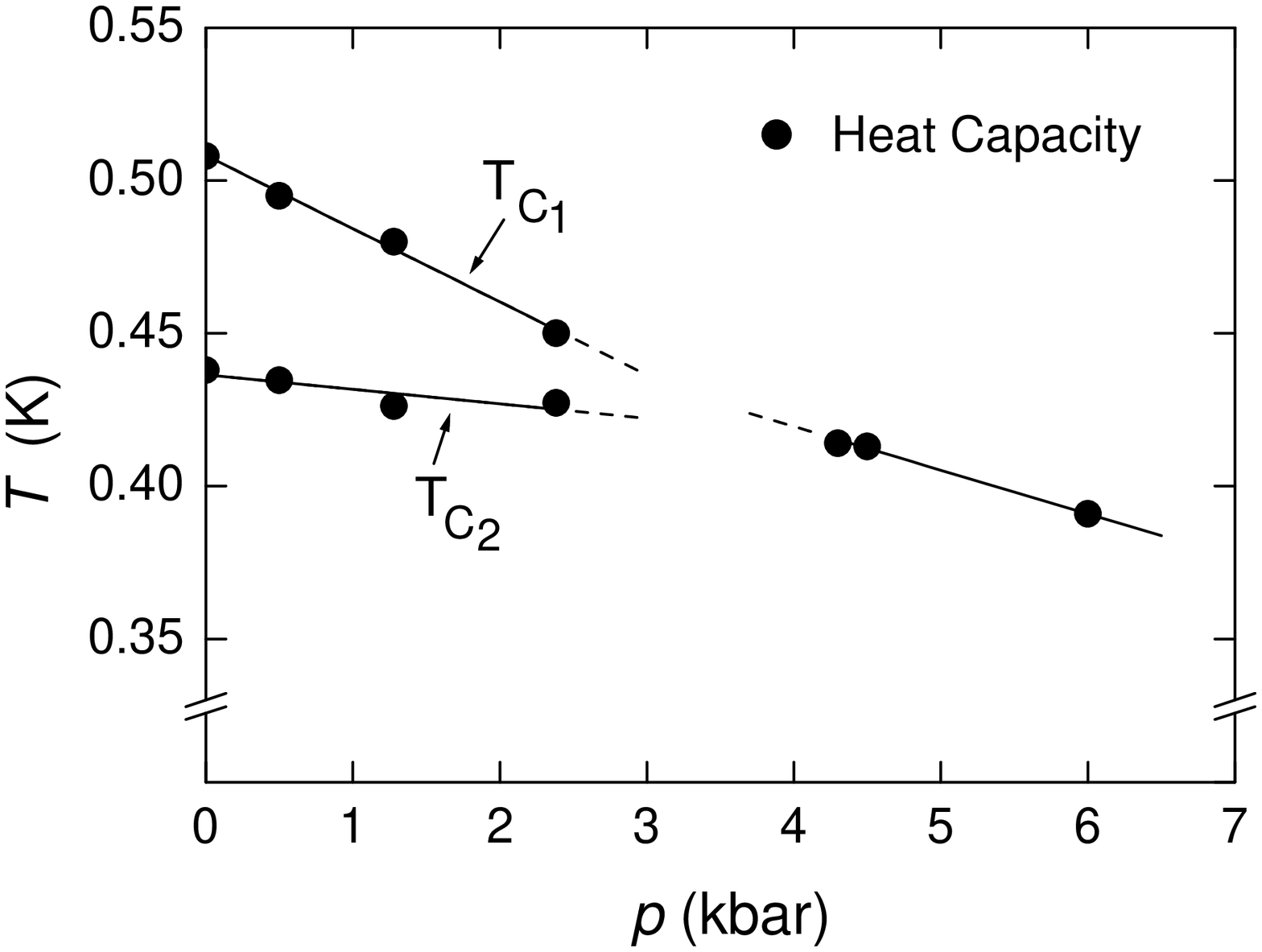}
\caption{Reduction of the squared AF moment in \UPT with hydrostatic
pressure p (above), it vanishes at a critical pressure of
p$_c\simeq$ 3.8 kbar. Simultaneously the SC T$_{c}$ splitting is
reduced and also vanishes at p$_c$ staying zero for larger
pressures (Hayden et al. 2002). This supports the idea that a twofold
degenerate SC order parameter (e.g. E$_{2u}$) is split by an AF
symmetry breaking field according to eqs.~(\ref{LAN}),(\ref{SPLIT})
\index{hydrostatic pressure}\index{T$_c$-splitting}
\index{antiferromagnetic order!UPt$_3$}}
\label{FIGafsplit}
\end{figure}

The exceptional nature of the superconducting state was already
obvious from the two specific heat jumps first observed by \cite{Fisher89}
of comparable size indicating two SC transitions at T$_{c1,2}$. Later
it was shown \cite{Trappmann91} that under hydrostatic pressure the two
transitions merge at a critical pressure of p$_c\simeq$ 3.8
kbar (fig.~\ref{FIGspec}). It is important to note that above p$_c$ the two
transitions do not  separate again. Together with the pressure
dependence of the AF moments observed in neutron diffraction
\cite{Hayden92} this gives a clue about the nature of the two
superconducting transitions as well as the role of AF order. If
the two transitions would correspond to two SC phases whose order
parameters belong to one-dimensional representations with accidentally
close transition temperatures then one would expect simply a crossing
of T$_{c1,2}(p)$ at p$_c$, there would be no symmetry reason why the
two transitions should merge above p$_c$ as is actually seen in
(fig.~\ref{FIGafsplit}). The observation \cite{Hayden92} that the magnetic
moment vanishes at about the same critical pressure (fig.~\ref{FIGafsplit})
then suggests a natural scenario: at large hydrostatic
pressure p$>$p$_c$ which preserves hexagonal symmetry
the SC order parameter belongs to a two-dimensional hexagonal
representation,
E$_1$ or E$_2$, i.e. it is a complex 2-component vector
$\boldeta$ = ($\eta_1$,$\eta_2$) = ($|\eta_1|,|\eta_2|e^{i\Phi})e^{i\varphi}$
rather than a complex scalar $\eta e^{i\varphi}$ as for
one-dimensional representations associated with a single T$_c$. Here
$\varphi$ is the overall Josephson-phase of the order parameter and
$\Phi$ is an intrinsic phase which is a
novel feature of the 2-component order parameter.
At pressures p $<$ p$_c$ the AF moments appear, since they are aligned
in the b-direction they reduce hexagonal symmetry to orthorhombic
symmetry. The AF order parameter acts as a symmetry
breaking field (SBF), any small coupling between the AF and SC order parameters
will then lead to a splitting of T$_c$ into T$_{c1,2}$ which are
associated with two SC order parameters of different symmetry with
respect to the
orthorhombic group. Therefore the small moment magnetism with its
easy-plane anisotropy is a lucky coincidence which helps to unveil the
complex vector nature of the SC order parameter. This scenario may
generically be described within the Landau theory characterised by a
\index{free energy} free energy density \cite{Machida89,Hess89,Joynt90} 
\begin{eqnarray}
f_L(\boldeta,\v M_{\v Q})&=&\alpha_0(T-T_c)\boldeta\cdot\boldeta^*
+\beta_1(\boldeta\cdot\boldeta^*)^2
+\beta_2|\boldeta\cdot\boldeta|^2
-\gamma_0 \v M_{\v Q}^2(\eta_x^2-\eta_y^2)
\label{LAN}
\end{eqnarray} 
Here \v M$_{\v Q}$= M$_{\v Q}\hat{\v y}$ is the AF order parameter which
has already
reached saturation in the SC temperature range, furthermore a shift of
the critical temperature T$_c\sim$ M$_{\v Q}^2$ has already been
included in T$_c$. There are two fourth order terms
characterized by the \index{Landau parameters} Landau parameters
$\beta_1$, $\beta_2$ as compared
to only one in the case of a scalar SC order parameter. The last term
is a phenomenological coupling of the SBF to the E-type
SC order parameter. This type of theory is the same for each of the
even or odd two-dimensional E-representations. The different \v
k-dependence of the respective SC-gap function enters only via the
parameters of the Landau free energy in eq.~(\ref{LAN}) where
$\alpha_0, \beta_1 >$ 0. The ratio $\beta_2/\beta_1$ selects the SC
ground state in the two dimensional manifold of possible order
parameters $\boldeta$. First we discuss the decoupled case $\gamma_0$ = 0
where a single transition at T$_c$ into a state belonging
to that manifold takes place. Then all possible SC ground states have
$|\eta_1|$ = $|\eta_2|\equiv |\eta|$. For -1$<\beta_2/\beta_1<0$ the
'unitary' state \index{unitary state} with intrinsic phase $\Phi$=0 is
stable whereas for  
$\beta_2/\beta_1>0$ the `non-unitary' state with $\Phi=\pm\frac{\pi}{2}$
and $\boldeta\times\boldeta^*\neq$ 0 is the stable one which is the case
in \UPT. The SBF splits the degenerate E-representations through the
last term in eq.~(\ref{LAN}). Whereas at low temperature the stable
state is determined by the fourth order terms, close to T$_{c1}$ the
second order terms $\sim\gamma_0>$ 0 dominate and favor a single
component state, only below T$_{c2}$ will the second component be
nonzero. One has two phases (A) $\boldeta$ = $\eta$(1,0) below T$_{c1}$
and (B) $\boldeta$ = $\eta$(1,ai) below T$_{c2}$. The T$_{c1,2}$-splitting
\index{T$_c$-splitting} and the difference in amplitudes (a $<$ 1) in
the B-phase are determined by the Landau parameters according to
\begin{eqnarray}
\label{SPLIT}
\Delta T_c&=&T_{c1}-T_{c2}=(1+\frac{\beta_1}{\beta_2})\lambda_0 T_c\nonumber\\
a&=&(1-\frac{\beta_1}{\beta_2}\lambda_0)/(1+\frac{\beta_1}{\beta_2}\lambda_0)\\
\lambda_0&=&\frac{\gamma_0 M^2_{\v Q}}{\alpha_0T_c}\nonumber
\end{eqnarray} 
Using the somewhat smaller transition temperatures T$_{c1,2}$ of
\citeasnoun{vanDijk94} with $\Delta$T$_c$ = 0.054 K (fig.~\ref{FIGhcrit})
one obtains for the dimensionless coupling parameter of SC and AF
order parameters
$\lambda_)$ = 0.038. The smallness of this parameter is primarily due to
the small AF moment in
\UPT. If it would be only slightly larger the lower B-phase of \UPT
would be completely suppressed and only one SC transition would be
observable. Furthermore one has
$\frac{\beta_1}{\beta_2}$=$\frac{T_c-T_{c2}}{\lambda_0 T_c}\simeq$ 2 and
the amplitude ratio a = 0.86 which will be set to 1 in the following. The
ratio of the SC specific heat jumps at T$_{c1,2}$ is given by
\begin{eqnarray} 
\frac{\Delta_2C}{\Delta_1C}&=&\frac{T_{c2}}{T_{c1}}
(1+\frac{\beta_2}{\beta_1})
\end{eqnarray} 
Here $\Delta_iC$ (i=1,2) are both counted from the normal state. Using
the experimental values this also leads to a ratio
$\frac{\beta_1}{\beta_2}$=2. This type of phenomenological theory can
be extended to include also the pressure dependence
\cite{Thalmeier91}. From the two different slopes of T$_{c1,2}$
vs. pressure one may infer a magnetic Gr\"uneisen parameter
$\Omega_M$= -($\partial\ln M^2_{\v Q}/\partial\epsilon_v$)= -385 which is
in reasonable agreement with $\Omega_M$= -260 as determined directly from
the pressure dependence of Bragg intensities which are $\sim M^2_{\v
Q}$. This gives further support that the AF order provides the SBF
that leads to the observed splitting into (1,0) and (1,i)
superconducting phases at zero field.

\subsubsection{The critical field curves and Ginzburg-Landau theory}
\index{Ginzburg-Landau theory}\index{upper critical field}

\begin{figure}
\includegraphics[width=90mm]{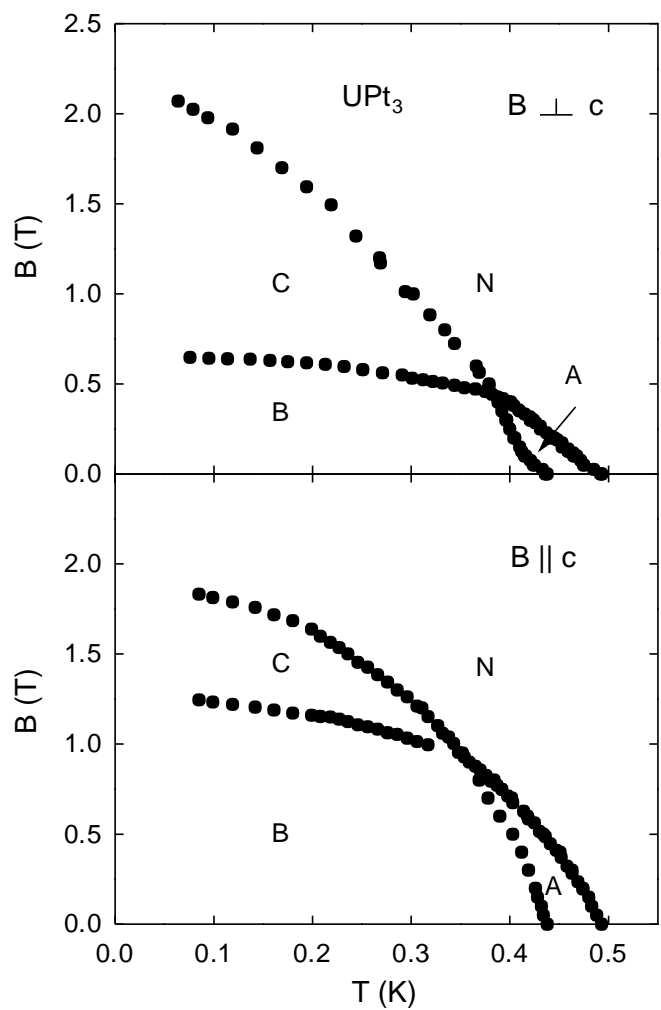}
\caption{Superconducting B-T phase diagram as obtained from thermal
expansion and magnetostriction (van Dijk et al. 1993a and
1993b, van Dijk 1994). Critical field curves
cross at a tetracritical point which is present for all field
directions, therefore three distinct SC phases A(1,0), B(1,i)
and C(0,1) with different E$_{2u}$ order parameters $\boldeta$ exist.
\index{B-T phase diagram!UPt$_3$}\index{thermal expansion!UPt$_3$}
\index{magnetostriction!UPt$_3$}\index{E$_{2u}$-model}}
\label{FIGhcrit}
\end{figure}

Naturally the two superconducting transtitions at T$_{c1,2}$ will lead to two
different critical field curves H$_{c2}^{1,2}$ which have been
investigated by experimental methods mentioned before. The most
obvious feature is a crossing of the critical field curves at a
tetracritical point for all field directions. As a consequence there
are three distinct SC regions in the H-T plane with phases A (high T,
low H), B (low T, low H) and C (low T, high H). This important
result has first been obtained by \citeasnoun{Bruls90} with ultrasonic
experiments. In fig.~\ref{FIGhcrit} we show the phase diagram for
H$\perp$c and H$\parallel$c obtained by \citeasnoun{vanDijk94}
using thermal expansion measurements. Its most important properties
may be summarized as follows: (i) \index{tetracritical point} The
existence of a tetracritical
point (T$_t$,H$_t$). Purely thermodynamical analysis of its vicinity
\cite{Yip91} leads to the conclusion that if all four phase boundaries
meeting at (T$_t$,H$_t$) are of second order, relations between their
slopes and the specific heat jumps across them exist. They were
confirmed by \citeasnoun{vanDijk94} which proved the second order
nature of transitions. The change of slopes leads to pronounced kinks
in the phase boundaries for H$\parallel$a. (ii) Defining the upper
critical field H$_{c2}$(T) as the upper phase boundary in
fig.~\ref{FIGhcrit} for a,c
directions one notices a reversal of the a-c anisotropy ratio
H$^a_{c2}$(T)/H$^c_{c2}$(T)as function of temperature. This ratio is $<1$ for
T/T$_c>$ 0.4 and $>1$ for T/T$_c<$ 0.4. This reversal has been advanced as a
major argument for spin triplet pairing by \citeasnoun{Sauls94}.
For the explanation of the structure of the A,B,C phase
diagram an appropriate Ginzburg-Landau theory for a SC 2-component
vector order parameter has to be constructed
\cite{Machida89,Hess89}. It is obtained by adding all
appropriate hexagonal invariants formed from $\eta_j$ and its covariant
gradient operators $D_j=\partial_j$-iA$_j$ with \v A denoting the vector
potential corresponding to the field \v B given in units of
$\phi_0/2\pi$ where $\phi_0=\frac{hc}{2e}$ is the flux quantum. The homogeneous
terms of the GL free energy density are again given by eq.~(\ref{LAN})
to which we have to add the gradient terms
\begin{eqnarray}
\label{GL}
f_G(\boldeta,\v M_{\v Q})&=&
K_1(|D_x\eta_1|^2+|D_y\eta_2|^2)+K_2(|D_x\eta_2|^2+|D_y\eta_1|^2)\nonumber\\
&&+K_3(D_x\eta_1D_y^*\eta_2^*+D_x\eta_2D_y^*\eta_1^* +c.c)
+K_4(|D_z\eta_1|^2+|D_z\eta_2|^2)\\
&&-\kappa M_{\v Q}^2(|D_x\eta_1|^2-|D_y\eta_1|^2+|D_x\eta_2|^2-|D_y\eta_2|^2)
\nonumber
\end{eqnarray}
The expression for the total GL free energy density f$_{GL}$ = f$_L$ +
f$_G$ is valid for both E$_1$ and E$_2$ type order parameters of odd and
even parity but in the whole section we restrict discussion to the most likely
E$_{2u}$ case. The last term in eq.~(\ref{GL}) with K$_{M}$= $\kappa
M_{\v Q}^2$ describes the coupling of the SBF to the gradients
of the SC order parameter. In a conventional GL theory for a scalar
order parameter one has only one GL parameter K=$\frac{\hbar c}{2m^*}$
determined by the effective electron mass. For the vector order
parameter $\boldeta$ one has altogether 5 parameters. It is therefore
obvious that a GL theory for a multicomponent superconductor has 
little real predictive power, but it may be used as a convenient frame
for qualitative discussion. In weak coupling BCS theory the coefficients K$_i$
can be expressed as angular averages of Fermi
velocities \cite{Sauls94} which leads to the conclusion that in the
E$_{2u}$- case one has K$_2$, K$_3\ll$ K$_1$. In this case the terms
mixing $\eta_x$ and $\eta_y$ can be neglected and the solution
of the linearized GL-equations corresponding to eqs.~(\ref{LAN}),(\ref{GL})
lead to two parallel critical field curves without the term $\sim$K$_{M}$
for all directions of the field. The tetracritical point in this model
is therefore entirely due to the coupling of gradient terms D$_{x,y}\eta_i$ to
the SBF where the sign of $\kappa$ is negative, i.e. it has to be
opposite to that of $\gamma$ in the homogeneous coupling term of
eq.~(\ref{LAN}) in order to  obtain a crossing at T$_t$. Using eq.(\ref{GL})
and the quadratic part of eq.(\ref{LAN}) one obtains the critical field curves
\begin{eqnarray}
H^1_{c2}&=&\frac{\phi_0}{2\pi}\frac{\alpha_0(T_{c1}-T)}
{\sqrt{(K_1+K_{M})K_4}}\nonumber\\
H^2_{c2}&=&\frac{\phi_0}{2\pi}\frac{\alpha_0(T^0_{c2}-T)}
{\sqrt{(K_1-K_{M})K_4}}
\end{eqnarray} 
Here T$^0_{c2}>$T$_{c2}$ is the second transition temperature without
inclusion of quartic terms of eq.(\ref{LAN}) \cite{Machida89}. 
For fields parallel to the c-axis similar expressions hold with
$\sqrt{(K_1\pm K_{M})K_4}$ replaced by ($K_1\pm K_{M}$). Ignoring
K$_{M}$ for the moment the a-c anisotropy of H$^1_{c2}$ is given
by a slope ratio 
H'$^{1c}_{c2}$/H'$^{1a}_{c2}$= $\sqrt{K_4/K_1}$= $\sqrt{\la
v_c^2\ra/\la v_a^2\ra}$= $\sqrt{m_a/m_c}$. The experimental slope ratio
1.64 is indeed exactly equal to the root of the mass anisotropy
m$_a$/m$_c$= 2.7 as obtained from transport measurements. The
linearized version of the E$_{2u}$ model has however no natural
explanation for the slope changes (B$\perp$c) at the tetracritical
point. On the other hand it explains the basic observation of the three (A,B,C)
phases in the B-T plane for all field directions. In addition to the
low- (or zero-) field phases (A,B) discussed before a high
field phase (C) with an order parameter $\boldeta$= $|\eta|$(0,1)
appears which is stabilized by the gradient coupling term $\sim K_{M}$. 

\subsubsection{The superconducting gap function}
\index{gap function}

A detailed understanding of thermodynamics and transport properties
requires the \v k-dependent gap functions of the three
phases. Their node structure determines the low temperature behaviour of
thermodynamic and transport coefficients. The most commonly proposed
triplet f-wave gap function with E$_{2u}$ symmetry is given by 
\index{E$_{2u}$-model}
\begin{eqnarray}
\v d(\v k)=\eta_1[k_z\hat{\v z}(k_x^2-k_y^2)+\eta_2(2k_z\hat{\v z}k_xk_y)]
\end{eqnarray}
The orientation of the \v d-vector is assumed to be \v k-independent and
pinned along the hexagonal c-axis ($\hat{\v z}$) by an anisotropy
potential acting
on the (Kramers degeneracy) pseudospin of the heavy quasiparticles
which has its origin in the spin-orbit coupling of 5f-electrons. This
pinning effect leads to a large paramagnetic
limiting effect for \v B$\parallel$\v d, $\hat{\v z}$ and to no effect for \v
B$\perp$\v d, $\hat{\v z}$ which was suggested \cite{Graf00} as origin of the
crossover in the a-c upper critical field anisotropy mentioned
before. This is claimed as strong evidence for triplet \index{triplet pairing}
pairing \cite{Graf00} and as argument against the 
\index{singlet pairing} singlet d-wave (E$_{1g}$) gap
function. Defining \v d(\v k)=$\Delta(\v k)\hat{\v z}=
\Delta(\vartheta,\varphi)\hat{\v z}$ where $\vartheta,\varphi$ are the
polar and azimuthal angles of \v k one has explicitly the following
E$_{2u}$ gap functions (without normalization) for the A,B(a=1) and C
phases ($\Delta_0\equiv|\eta_{1,2}|$)
\begin{eqnarray}
\label{GAPFUN}
A(1, 0):~\v \Delta(\vartheta,\varphi)&=&\Delta_0k_z(k_x^2-k_y^2)
=\Delta_0\cos\vartheta\sin^2\vartheta\cos(2\varphi)\nonumber\\
B(1,i):~\v \Delta(\vartheta,\varphi)&=&\Delta_0k_z(k_x+ik_y)^2
=\Delta_0\cos\vartheta\sin^2\vartheta\exp(2i\varphi)\\
C(0, 1):~\v \Delta(\vartheta,\varphi)&=&2\Delta_0k_zk_xk_y
=\Delta_0\cos\vartheta\sin^2\vartheta\sin(2\varphi)\nonumber
\end{eqnarray}
%
\begin{figure}
\begin{minipage}[t]{\columnwidth}
\includegraphics[width=.35\columnwidth]{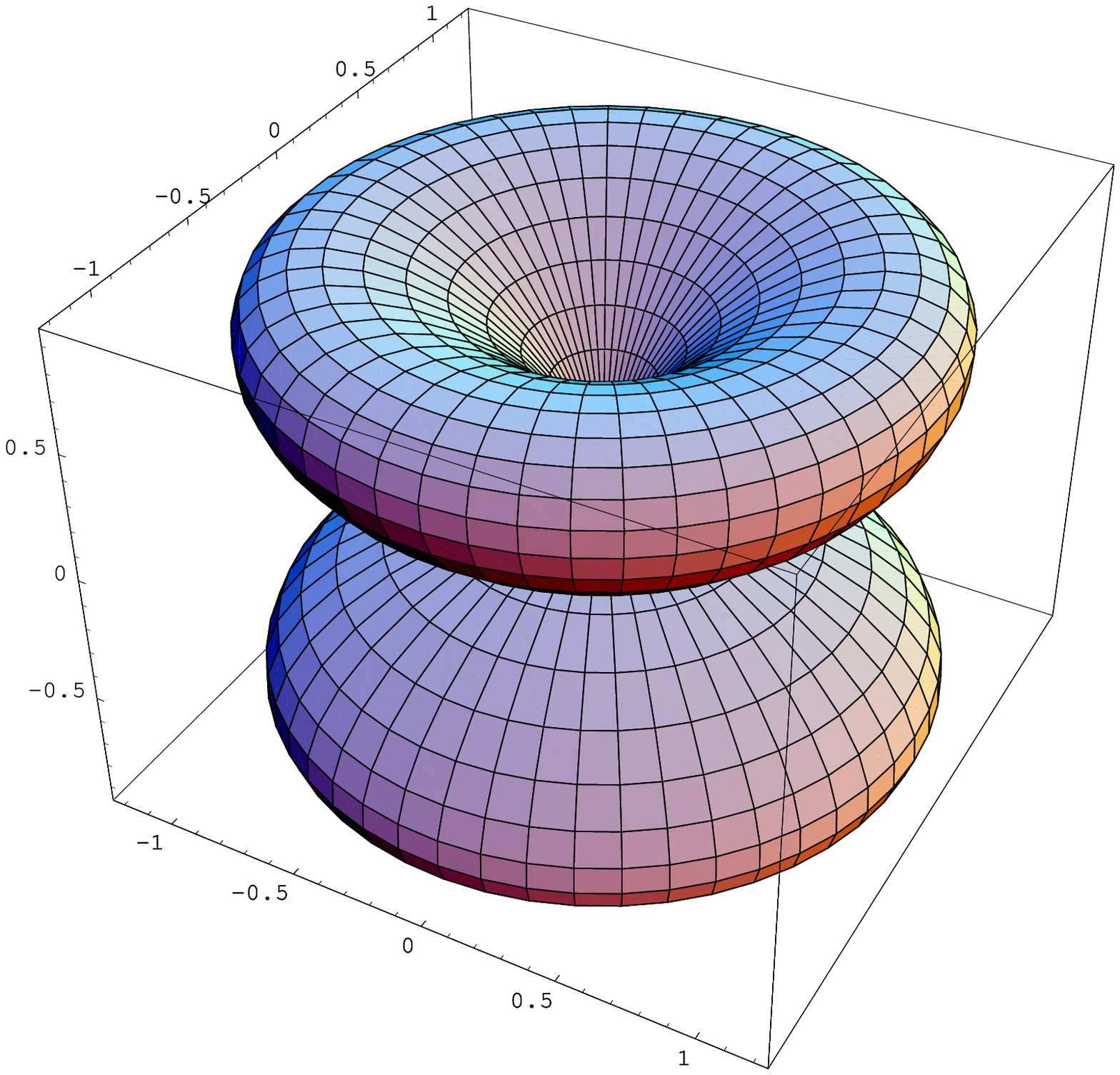}\hspace{1.5cm}
\includegraphics[width=.35\columnwidth]{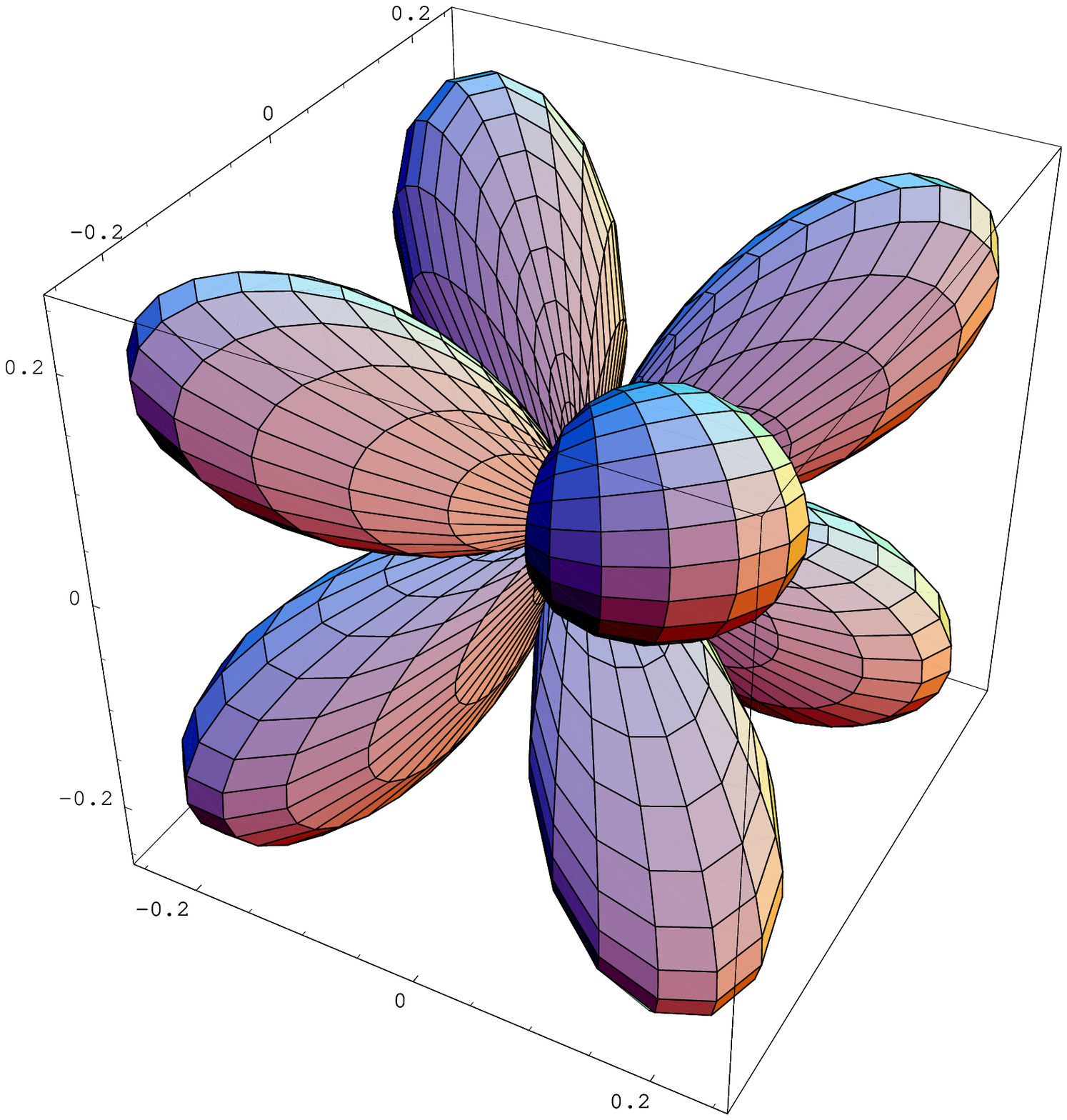}
\end{minipage}
\caption{Spherical plots of $|\Delta(\vartheta,\varphi)|$ for B-phase (left)
and C-phase (right). B-phase has node line $\vartheta =\frac{\pi}{2}$
(k$_z$=0) and 2$^{nd}$ order node points along c-axis ($\vartheta =
0,\pi$). C-phase has additional node lines at k$_x$=0, k$_y$=0
($\varphi$ = 0,$\frac{\pi}{2}$) (Yang and Maki 1999).}
\label{FIGgap}
\end{figure}
%
An angular plot of these gap functions is shown in fig.~\ref{FIGgap} which
also exhibits their node structure described in the caption. In the
above equation the equal amplitude approximation a = 1 was made for the
B-phase which leads to a gap with full rotational symmetry around
c. As an effect of the SBF which leads to a $<$ 1 the rotational
symmetry is slighly broken, one has to multiply the (B-phase) gap
function by a factor $(1-\epsilon^2\sin^2(2\varphi))^\frac{1}{2}$ with
fourfold symmetry around c and $\epsilon^2$=1-a$^2$ = 0.26. The node
structure is unchanged, it \index{point nodes}
consists of an equatorial line node and two second order node points
at the poles. Both contribute to the linear behaviour of the
quasiparticle DOS N(E) for E $\ll\Delta_0$. It was calculated 
\cite{Yang01} including the effect of impurity scattering which
quickly leads to a
residual DOS N(0). Indeed in UPt$_3$ large residual $\gamma$
values are usually obtained by extrapolation from temperatures above
the magnetic ordering at 28 mK. Finally one should note that the
presence of a node line is not required by symmetry if one assumes the
strong spin-orbit coupling case \cite{Blount85}, 
\index{Blount's theorem} it is rather a result of a
special choice of the E$_{2u}$ order parameter in the allowed
subspace. Forming a linear combination of all possible E$_{2u}$
representations there are in general only point nodes. However, as
argued above and in sect.~\ref{Sect:Theory}, the weak spin orbit
coupling case is more realistic in the effective quasiparticle pseudo spin
picture for \UPT and then the appearance of node lines is natural.
It is obvious from fig.~\ref{FIGgap} and eq.(\ref{GAPFUN}) that the
non-unitary B-phase has fewer nodes than the unitary A- or C-phases. 
This general feature of non-unitary phases means that they are prefered as
stable low temperature states in unconventional SC because fewer nodes
lead to a larger condensation energy.

The ratio of fourth order Landau coefficients $\beta_1/\beta_2\simeq$
2 was taken as evidence for the weak
coupling nature of superconductivity. Therefore the question arises
whether it is possible to calculate the symmetry of the order
parameter within a microscopic weak coupling Hamiltonian starting from
an on-site effective quasiparticle interaction which is
repulsive. Such attempts have originally been made without the
inclusion of the orbital degeneracy of U-5f states and within one
band models with the static susceptibility tensor used as an
input. However, not surprisingly the gap functions obtained are 
sensitive to the input function and also the E$_{2u}$ gap function is
not the favored one \cite{Norman91,Norman94} in such models. Attempts
to include orbital degrees of freedom to remedy this situation have
been made \cite{Norman94}. As mentioned before it has now become
clear that 5f-states in \UPT have also a dual nature, partly localised
in 5f$^2$ configurations and partly itinerant states which have heavy
masses due to renormalisation by low lying CEF excitations of the
5f$^2$ states. It is therefore possible that, as in the case of
UPd$_2$Al$_3$, exchange of localised 5f$^2$ excitation modes play
an essential role for the formation of unconventional Cooper pairs in \UPT.

\subsubsection{Low temperature transport properties}

\begin{figure}
\includegraphics[width=75mm]{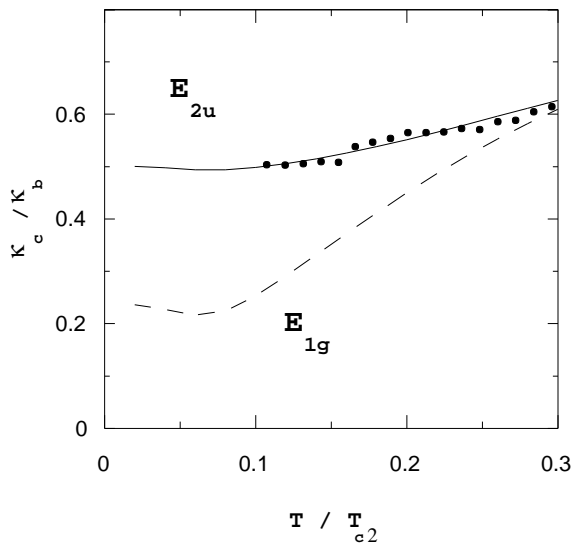}
\caption{Anisotropy ratio $\kappa_c$/$\kappa_b$ of thermal
conductivities as function of the reduced temperature. Data from
(Lussier et al. 1996).\index{thermal conductivity!UPt$_3$}}
\label{FIGacratio}
\end{figure}

The quasiparticle energies vanish at the nodes of the superconducting
gap. Therefore the low temperature (T $\ll$ T$_c$) behaviour of
thermodynamic and transport coefficients may give information on the
node structure. Specific heat analysis has proved of little use due to
the presence of the huge 28 mK peak caused by the onset of true long
range magnetic order of the small moments. Thermal conductivity
measurements for 0.1 $<T/T_{c2}<$ 0.3 \cite{Lussier96} have been more
succesful. The temperature dependence in this asymptotic regime 
requires the presence of node lines and/or second order node points,
furthermore comparison of the anisotropy ratio
$\kappa_c(T)/\kappa_a(T)$ shown in fig.~\ref{FIGacratio} with the calculated
ones \cite{Norman96} gives good agreement for the E$_{2u}$ gap
function of the B-phase in eq.~(\ref{GAPFUN}). Extrapolation for zero
temperature leads to a ratio $\sim$ 0.5 contrary to the older E$_{1g}$
model which extrapolates to
zero. This difference is due to the presence of second order node
points with $\Delta(\vartheta)\sim\vartheta^2$ or $(\pi-\vartheta)^2$ at the
poles in the E$_{2u}$ case. Finally for T/T$_{c2}<$ 0.1 one reaches the gapless
regime where the pair breaking energy scale
$sqrt{\hbar\Gamma kT}>$ kT with $\Gamma$ denoting the
normal state impurity scattering rate. In this case a behaviour
$\kappa/T\sim\alpha +\beta T^2$ is predicted \cite{Graf96} in good
agreement with observations \cite{Suderow97}.

\subsubsection{NMR Knight shift results}
\index{nuclear magnetic resonance}\index{Knight shift!UPt$_3$}

The Knight shift (KS) in the superconducting regime is a direct
measure of the change $\delta\chi_s$ in the spin susceptibility of
the SC condensate. It can be measured by $\mu$SR or NMR experiments. In
the latter the NMR resonance frequency is shifted due to the contact
interaction of nuclear and electronic spin momemt
(sect.~\ref{Sect:Theory}). Ideally when spin orbit coupling can be
neglected, the KS should drop
to zero for a singlet SC state and should be unchanged 
for a triplet state. Spin orbit coupling complicates the interpretation and may
lead to only a partial reduction for a singlet state and an observable
anisotropic reduction in the triplet state \cite{MineevBook}. In the
latter case for a rigid \v d(\v k) with respect to \v H one has for
the susceptibility tensor (i,j=x,y,z):
\begin{equation}
\chi_{sij}=\chi_n\int\frac{d\Omega}{4\pi}
[(P^\perp_{ij}(\vk)+Y(\hat{\v k},T)P^\parallel_{ij}(\v k)]         
\end{equation}
Where $\chi_n$ is the normal state spin susceptibility (assumed isotropic
for simplicity). Furthermore 
P$^\perp_{ij}$(\v k)= d$^*_\alpha$(\v k)d$^*_\beta$(\v k)/
$|\v d(\v k)|^2$ and  
P$^\perp_{ij}$(\v k)= [$\delta_{\alpha\beta}$ - P$^\perp_{ij}$(\v k)]
are the projectors to the direction perpendicular and parallel to \v
d(\v k) respectively and Y($\hat{\v k}$,T) is related to the Yoshida
function of sect.~\ref{Sect:Theory}.

\begin{figure}
\includegraphics[width=75mm,angle=0]{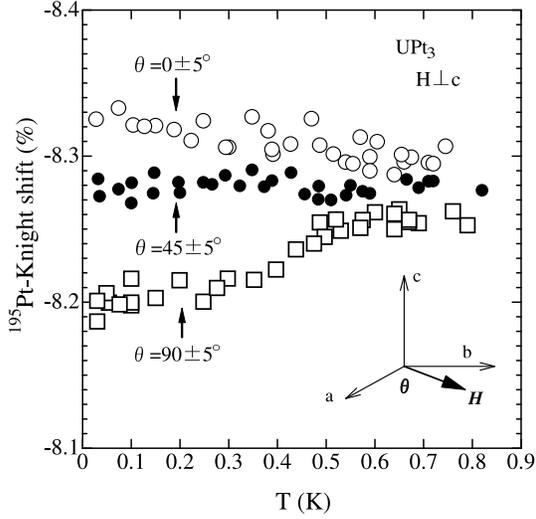}
\caption{Ansiotropic Knight shift for field in the hexagonal plane
(H = 0.19 T) (Kitaoka et al. (2000)). The field angle dependence of $\sim$
1\% of the Knight shift is possibly a sign of weak \v d-vector
pinning by the SBF.}
\label{FIGupt3knight}
\end{figure}

In the strong spin-orbit coupling case \index{spin orbit coupling}
with the \v d-vector pinned along c one would expect a suppresion of
the KS for \v H applied parallel to the c direction due to the effect of
Y($\hat{\v k}$,T) but not for \v H $\perp$ c, i.e. for \v H in the 
hexagonal plane. The Knight shift should therefore be quite
anisotropic for pinned \v d-vector. However it was found
\cite{Tou96,Tou98,Kitaoka00} that for large fields (in the C-phase) there is
almost no change for both field
directions. This was interpreted as evidence that the weak
spin orbit coupling case is realized in \UPT. In this case the \v d-vector
may be rotated by the field. For large fields it is always
perpendicular to \v H and therefore no KS reduction will take
place. On the other hand for low fields (in the B-phase) it may be
pinned by the SBF (parallel to a) and therefore rotation of \v H in the
hexagonal ab-plane will produce an anisotropic Knight shift as shown
in fig.~\ref{FIGupt3knight}. The weak spin orbit coupling scenario has
also been
theoretically investigated \cite{Ohmi96a,Ohmi96b}. On the other hand
this interpretation seems to be in conflict with the anisotropic paramagnetic
limiting which leads to the reversal for the upper critical field
ansisotropy as explained before. This discrepancy has sofar not been resolved.

\subsubsection{Magnetothermal properties in the vortex phase}
\index{vortex phase}\index{magnetotransport}

The a-c anisotropy of the zero-field thermal conductivity in the
B-phase has provided a major argument for the E$_{2u}$ SC order parameter
with its second order nodal points at the poles. These node points (and also
the equatorial node line) have, however, not been seen directly until
now. As explained in sect.~\ref{Sect:Theory} magnetothermal
conductivity or
specific heat measurements in the vortex phase as function of field
angles can provide such direct evidence. Due to the comparatively low
T$_c$ of \UPT experiments have to be done at 50 mK or below, sofar
they have not been performed. Calculations of the field angle
dependence for the B-phase \cite{Maki02a} and the
C-phase \cite{Thalmeier03a} have been carried out however. For simplicity
we discuss only the former since the modulus of the gap function in
the B-phase is isotropic in the hexagonal plane according to
eq.~(\ref{GAPFUN}). This means that specific heat and c-axis thermal
conductivity $\kappa_{zz}$ depend only on the polar field angle
$\theta$. They are given in the low temperature and superclean limit
of sect.~\ref{Sect:Theory} by
\begin{eqnarray}
\frac{C_s(\theta)}{\gamma T}&=&
\frac{1}{\sqrt{3}}\tilde{v}\sqrt{eH}I_B(\theta)\nonumber\\
\frac{\kappa_{zz}(\theta)}{\kappa_n}&=&\frac{2}{3}
\frac{v_av_c}{\Delta^2}(eH)I_B(\theta) F_B^{zz}(\theta)\\
I_B(\theta)&=&\alpha\sin\theta +\frac{2}{\pi}E(\sin\theta)\nonumber
\end{eqnarray}
%
\begin{figure}
\includegraphics[width=75mm]{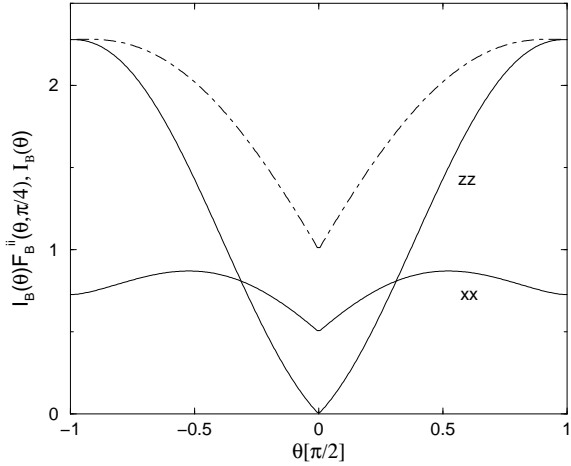}
\caption{Angular dependence of specific heat (dot-dashed) and thermal
conductivity $\kappa_{ii}$ (i=x,z) polar field angle $\theta$. Cusps
at $\theta$=0 are due to polar node points of the gap in
eq.~(\ref{GAPFUN}). While $\kappa_{zz}$ is independent of $\phi$,
$\kappa_{xx}$ is shown for $\phi=\frac{\pi}{4}$ (Thalmeier and Maki 2003a).}
\label{FIGangular}
\end{figure}
%
where F$_B^{zz}(\theta)=\sin\theta$ and E($\sin\theta$) is the complete
elliptic integral. The $\theta$- dependent factors in the above equation
are shown in fig.~\ref{FIGangular}. Sharp cusps for $\theta$ = 0,$\pi$
are indeed seen in these functions which if observed in the
corresponding specific heat C$_s(\theta)$ and thermal conductivity
$\kappa_{zz}(\theta)$ would constitute a direct and unambiguous proof
for the polar node points \index{point nodes} in the B-phase of
\UPT. Likewise the
equatorial line node of the B-phase will lead to a flat minimum
(without a cusp) for $\theta=\frac{\pi}{2}$ in
$\kappa_{xx}(\theta)$. Observing these features in \UPT as ultimate
proof for the node structure is an experimental challenge.

\subsection{Magnetic exciton mediated superconductivity in \UPD}
\label{Sect:UPd2Al3}\index{magnetic exciton}\index{UPd$_2$Al$_3$}

Among the U-based HF superconductors \UPD \cite{Geibel91a}
is a rather special case. In this HF compound with a moderate $\gamma$
= 120 mJ/mole K$^2$ there is also AF order below T$_N$ = 14.3 K 
with almost atomic size local moments ($\mu = 0.85\mu_B$) in contrast
to the small moments in other U-compounds. The
entropy release is $\Delta S(T_N)$ = 0.67Rln2 per mole, much larger
than in the itinerant SDW sister compound \UND.
The AF order coexists with superconductivity
below T$_c$ = 1.8 K. This suggests that in addition to the heavy
itinerant quasiparticles nearly localised 5f-electrons should
be present. They result from the dominating 5f$^2$ configuration of the
U$^{4+}$ ion \cite{Grauel92}. This dual nature of 5f-electrons is even
more obvious than in \UPT as is seen from various
experimental investigations like susceptibility \cite{Grauel92},
Knight shift \cite{Feyerherm94} and optical measurements \cite{Dressel02}. 
The former exhibit a pronounced a-c axis
anisotropy \cite{Grauel92} shown in fig.~\ref{FIGsuscep} with a much larger
$\chi_a(T)$ whose T-dependence is very reminiscent of CEF-effects. A
CEF scheme with two low lying singlets split by $\delta$ was proposed
for the 5f$^2$ localized states. Later
investigations \cite{Shiina01} indicate that the first excited state
is rather a doublet with $\delta$ = 6 meV. Since the ground state is a
singlet the local 5f-moment magnetism must be of the induced type
caused by the mixing with the excited doublet via inter-site
exchange. A direct confirmation of this dual nature of 5f-electrons
in \UPD was obtained from inelastic neutron scattering (INS) \cite{Mason97}
which found excitations that originate
in local CEF transitions of energy $\delta$ and disperse into bands of
`magnetic excitons' due to intersite exchange. This band extends up to 8
meV and along the c direction the modes are propagating with little
damping. Later
high resolution INS experiments \cite{Sato97,Sato97a,Bernhoeft98,Bernhoeft00} 
have shown that below T$_c$ a resonance like structure in the
dynamical structure function of localised moments
appears which is linked to the superconducting quasiparticles. 
Complementary tunneling experiments probe the response
of the itinerant quasiparticles and their superconducting gap. In a
breakthrough experiment this has been achieved the first time for a HF
superconductor using an epitaxially grown UPd$_2$Al$_3$-AlO$_x$-Pb tunneling
device \cite{Jourdan99}. Typical strong coupling features in the
tunneling DOS have been observed which appear at an
energy related to the excitations of local moments seen in
INS. Together both experiments strongly suggest that the magnetic
excitons identified in INS are the bosonic `glue' which binds the
electrons together to Cooper pairs \cite{Sato01}. 
This is a new mechanism for superconductivity distinctly different
from both the electron-phonon and spin fluctuation mechanism known
sofar. The pairing potential is mediated by a propagating boson
(the magnetic exciton) as in the former but depends on the spin state
of conduction electrons as in the latter. It is the main purpose of
this section to present the evidence for this important new mechanism
for unconventional superconductivity. Before considering this in
detail we summarize some essential physical facts known about \UPD.

\subsubsection{AF structure and superconducting properties}

\begin{figure}
\includegraphics[width=90mm]{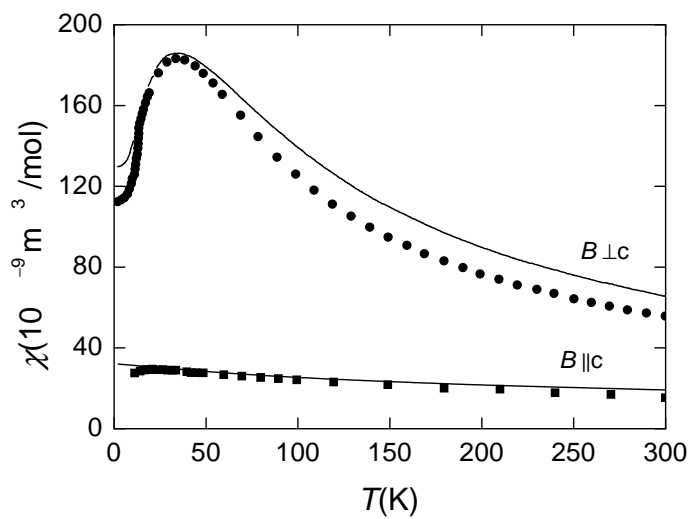}
\caption{Magnetic susceptibility of \UPD for field parallel to a- and
c-axis. The pronounced maximum for $\chi_\perp$(T) is a
typical signature of CEF split localised 5f-states (Grauel et
al. 1992). The full line is a fit using a U$^{3+}$ CEF level scheme
containing singlet ground state and first excited state (33 K).
\index{susceptibility!UPd$_2$Al$_3$}}
\label{FIGsuscep}
\end{figure}

\begin{figure}
\includegraphics[width=75mm]{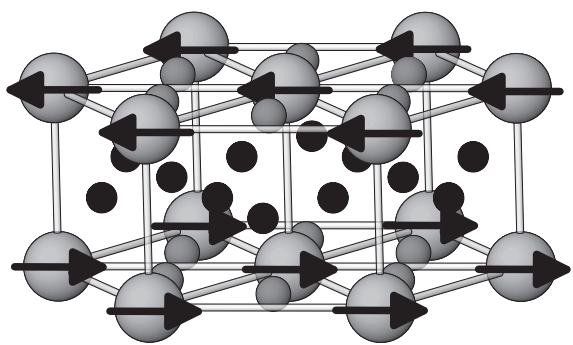}
\caption{Conventional unit cell of \UPD (a = 5.350 \AA, c = 4.185 \AA)
and simple AF magnetic structure with propagation vector \v Q =
(0,0,$\frac{1}{2}$).
The large and small grey spheres in hexagonal planes correspond to the
U and Pt atoms respectively, intercalated by Al atoms (small black
spheres).\index{antiferromagnetic order}}
\label{FIGstructure}
\end{figure}

The AF magnetic structure of \UPD consists of FM ordered hexagonal
planes with moments ($\mu$ = 0.83$\mu_B$) in [100] direction and stacked
antiferromagnetically along the c-axis \cite{Krimmel92,Kita93}. It
corresponds to an AF wave vector \v Q=(0,0,$\frac{1}{2}$) (in
r.l.u.) and is shown in fig.~\ref{FIGstructure}. As expected from the large 
a-c anisotropy of the susceptibility in fig.~\ref{FIGsuscep} the AF structure
does not change in applied magnetic fields \v H along c. For fields in the easy
ab-plane the hexagonal in-plane anisotropy is much smaller and moment
reorientation can be seen. The magnetic B-T phase diagram has been
determined by \citeasnoun{Kita93} and no effects of the superconducting
transition at T$_c$ = 1.8 K on the magnetic structure has been
found supporting the idea of two separate superconducting (itinerant)
and magnetic (localised) 5f-subsystems.

The superconducting state of \UPD has been investigated by
thermodynamic and transport measurements \cite{Caspary93,Hessert97,Hiroi97},
NMR \cite{Tou95,Matsuda97}, $\mu$sR experiments \cite{Feyerherm94}
and tunneling studies \cite{Jourdan99}. Despite this great effort the
symmetry of the order parameter is not reliably known sofar.

The upper critical field exhibits flattening  for low
temperatures which has been interpreted as a Pauli limiting effect and
hence evidence for singlet pairing \cite{Hessert97}. The angular dependence
H$_{c2}(\theta)$, where $\theta$ is the polar angle, changes
dramatically from T = 1.5 K to T = 0.5 K which was interpreted as a
gradual change from d-wave to s-wave character of the order parameter
when temperature is lowered. Within this picture the effect of
background AF order was ignored which may not be permissible because
the critical fields for localized 5f-spin reorientation are of the
same order as H$_{c2}$ \cite{Grauel92,Kita93}. 

\begin{figure}
\includegraphics[width=75mm,clip]{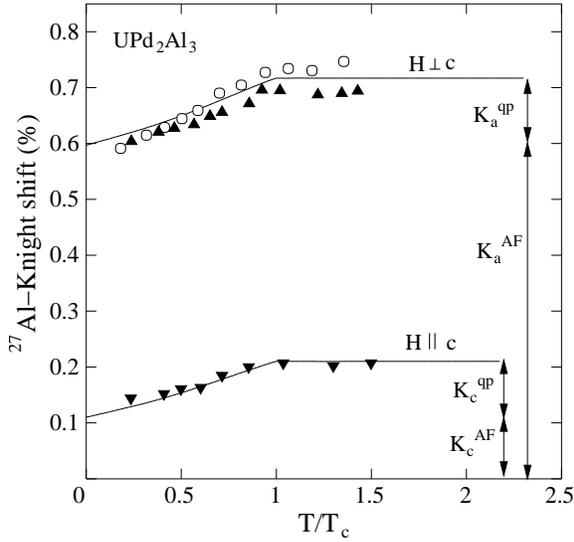}
\caption{$^{27}$Al Knight shift from NMR below T$_c$ (Kitaoka et
al. (2000)). The arrows
indicate the suggested separation in localised (AF) and itinerant
quasiparticle (qp) contributions. The AF and qp parts cannot be
independently determined experimentally. Therefore the complete
suppression of K$^{qp}$ is a conjecture. \index{Knight shift}}
\label{FIGknight}
\end{figure}

The Knight shift in the
SC phase is a direct probe for the reduction of the spin
susceptibility $\chi_s$(T) in the SC state \cite{Tou03}. Ideally 
there should be a complete suppression of $\chi_s$(T) for singlet
pairing and no effect for triplet pairing
(sections~\ref{Sect:Theory},\ref{Sect:UPt3}). This picture is however
complicated by the effect of spin-orbit coupling which may lead to
intermediate cases. The result for $\Delta K_s$(T) as
obtained from $^{27}Al$ NMR experiments \cite{Tou95} is shown in
fig.~\ref{FIGknight}. Although a reduction is clearly visible, large
(anisotropic) residual Knight shift values remain. This is attributed to the
localized 5f-susceptibility which cannot be independently
determined. Therefore these results and similar ones from $\mu$SR
experiments \cite{Feyerherm94} are difficult to interpret. It is not
possible to say with certainty how strong the Knight shift caused by the
itinerant quasiparticles is actually reduced in the SC phase. But the
conclusion concerning singlet pairing seems unaffected by this amibguity.   

The $^{27}$Al NMR relaxation rate T$_1^{-1}$ was found to exhibit a T$^3$
behaviour in the SC state over four orders of magnitude down to 0.1
T$_c$ \cite{Tou95} where the gap amplitude will be constant. This
means that \De should have a node line implying a low quasiparticle
energy DOS N$_s$(E) $\sim |E|$. This conclusion was confirmed by
$^{105}$Pd NMR/NQR experiments \cite{Matsuda97} which also observed a
T$^3$ behaviour of the relaxation rate and the absence of a coherence
peak immediately below T$_c$ was noticed, both facts are naturally
explained by the existence of a node line in \De.

Commonly in HF compounds the low temperature behaviour of the specific
heat is not easily explained by simple models that involve only the
topology of gap nodes. Very often residual $\gamma$- values exist
where they should not due to node points or lines in \De. This is
ascribed to the presence of a residual density of states induced by
impurity scattering. Extracting
the true low temperature behaviour of the electronic specific heat
C$_s$ is also complicated by the presence of nuclear terms due to the
hyperfine splitting which lead
to a low temperature upturn $\sim\alpha T^{-2}$. This situation
has also been encountered in \UPD. Originally the existence of a
residual $\gamma$- value was concluded in \cite{Caspary93} but later
measurements  extending to lower temperatures \cite{Sato93} have shown that
the data can be well fitted without the residual $\gamma$ but assuming
instead a dominant T$^2$ term in C$_S$(T) which is compatible with the
node line hypothesis for \UPD. It is not known yet whether the field
dependence of C$_S$(T,H) is dominated by field induced
$\gamma$(H)$\sim\sqrt{H}$ expected in the presence of nodes.

Thermal conductivity measurements for parallel heat current and field
lying in the hexagonal plane \cite{Hiroi97} have shown a
dominating temperature dependence $\kappa$(T) $\sim$ T$^2$ and a linear
behaviour $\kappa$(T,H) $\sim$ H at low temperatures. This
is again in agreement with the existence of node lines in \De. However
their exact position on the FS is not yet known. This has
to be investigated by field-angle dependent thermal conductivity
measurements  which has been successful in this task in other
unconventional superconductors as described in sect.~\ref{Sect:Theory}.

\subsubsection{Electronic structure, Fermi surface and effective mass}
\index{band structure} 

\begin{figure}
\includegraphics[width=75mm]{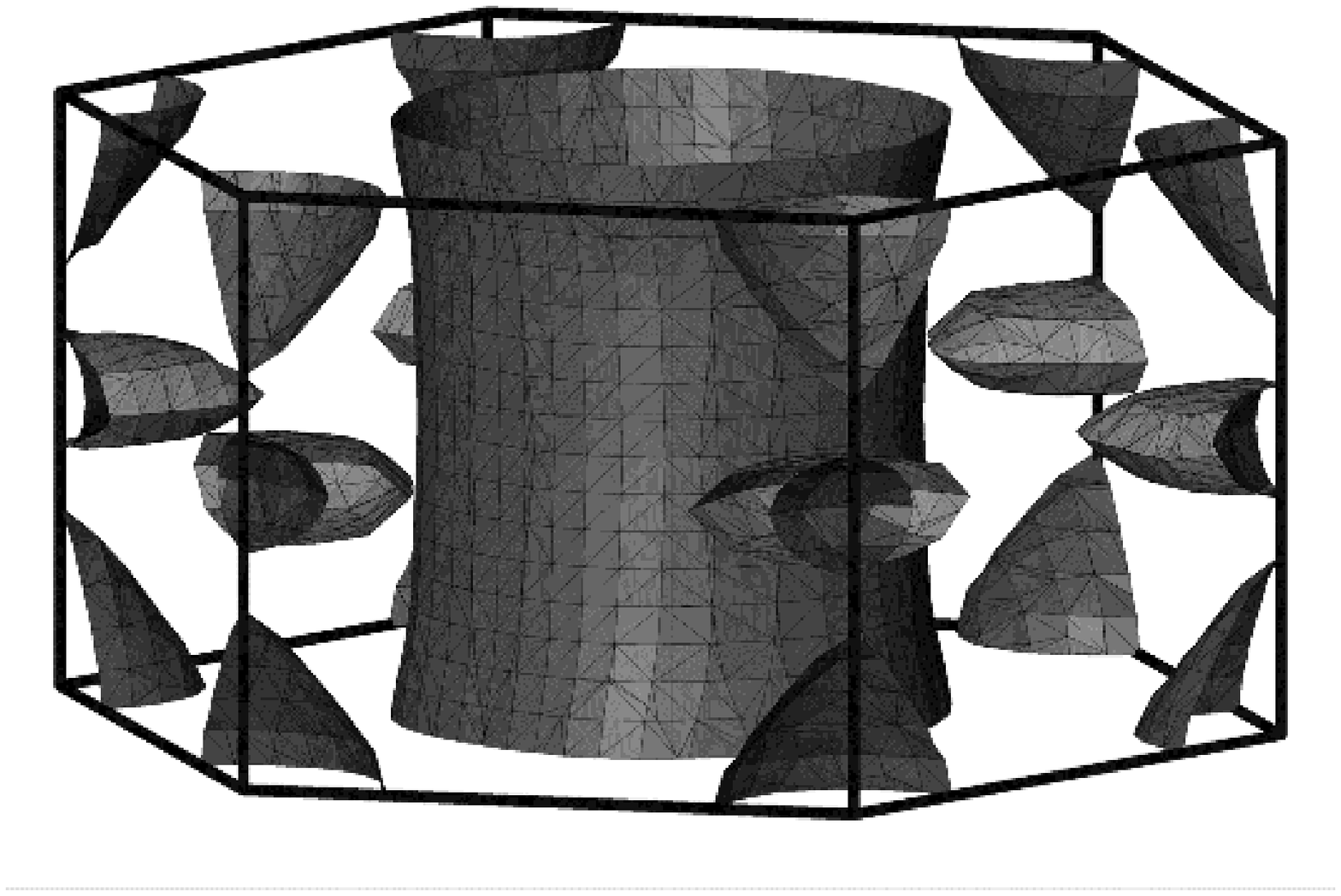}
\includegraphics[width=75mm]{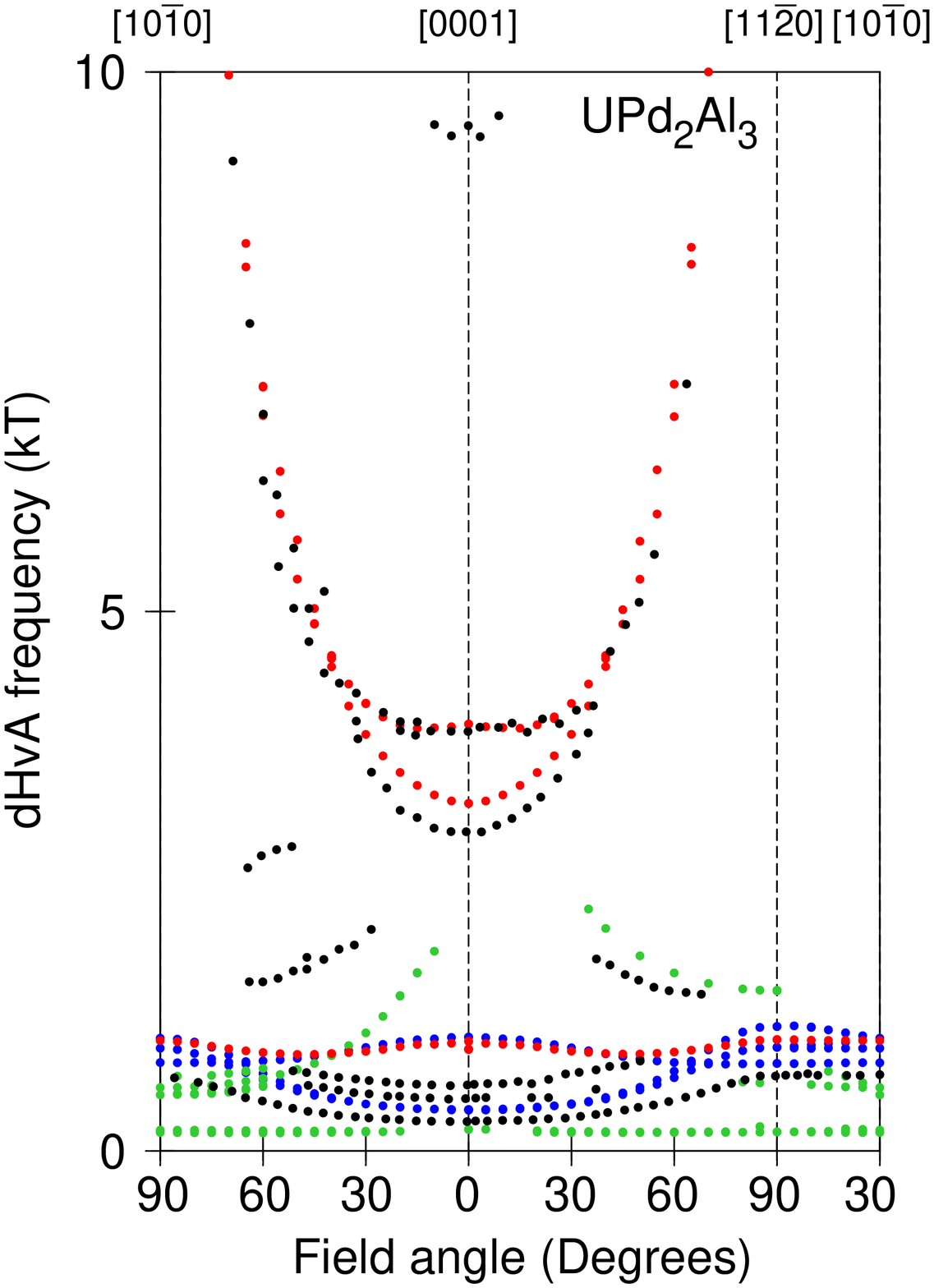}
\caption{Left panel: Fermi surface of \UPD calculated within the dual
approach method (Zwicknagl et al 2003). The main cylinder part has
also a heavy mass with m$^*$ = 19 - 33 m. Right panel: comparison of
experimental dHvA frequencies (black symbols)
from (Inada et al. 1999) and calculated frequencies (color symbols
compatible with FS sheets on left panel) from the dual approach model
(Zwicknagl et al 2003). Large parabola corresponds to the main FS
cylinder. \index{Fermi surface!UPd$_2$Al$_3$}\index{dual model}
\index{de Haas-van Alphen effect!UPd$_2$Al$_3$}}
\label{FIGfermi}
\end{figure}

In LDA type electronic band structure calculations for intermetallic
U-compounds all 5f-electrons are treated in the same way and their 
degree of (de-)localisation is not much different. This is in
contradiction to the behaviour observed in \UPD just as described
above. Nevertheless
LDA calculations give usually reasonable results for the Fermi surface
topology. This can be understood in the context of model calculations
as a result of many body correlations. For example within the Anderson
lattice model it may be shown that the Fermi volume comprises both the
f-electrons and conduction electrons (`Luttinger's theorem') , even in
the Kondo limit when the f-electron is
localised \cite{Fazekas99}. Therefore it looks as if the Fermi wave
vector, and more generally the Fermi surface is the same as if both
type of electrons were delocalised. This is indeed assumed in LDA
calculations for \UPD which we discuss now. FS
sheets and associated angle dependent de Haas-van Alphen (dHvA)
frequencies have first been
calculated by \citeasnoun{Sandratskii94}, \citeasnoun{Knoepfle96}
and compared to the experimental dHvA results \cite{Inada95}. Similar
calculations \cite{Inada99} obtain the same main FS sheets but differ
in the positions of smaller sheets. The larger sheets are also those
with experimentally found
heavy masses, which the LDA calculation naturally cannot explain. The
largest sheet which also has the strongest 5f-admixture is shown in
fig.~\ref{FIGfermi} in the AF Brillouin
zone. It has the shape of a corrugated cylinder if one ignores the
hexagonal in-plane anisotropy. In the subsequent model calculations we
assume this simplified FS for \UPD and neglect all other sheets.
The LDA approach to the 5f-electronic structure does not include 
the different degree of localisation evident from experiments
discussed before. Progress in the treatment of partial
localisation of 5f-electrons has recently been made for a number of
U-compounds including UPt$_3$ \cite{Zwicknagl02}
(sect.~\ref{Sect:UPt3}) and \UPD \cite{Zwicknagl03}. Due to the strong
spin orbit coupling
5f electrons occupy total angular momentum orbitals
$|j=\frac{5}{2},j_z\ra$. The two localised 5f electrons are put into 
j$_z=\pm\frac{5}{2}$ and j$_z=\pm\frac{1}{2}$ localized states and the
remaining j$_z=\pm\frac{3}{2}$ states is included in the LDA basis of
band electrons. This treatment is justified by the larger hybridsation
of the j$_z=\pm\frac{3}{2}$ orbitals as compared to j$_z=\pm\frac{1}{2},
\pm\frac{5}{2}$ already on the LDA level. It has
been shown within model calculations \cite{Efremov03} that intraatomic
correlations strongly increase the orbital dependence of the effective
hybridisation, leading to the dual character of the 5f-electrons
(sect.~\ref{Sect:Theory}). Although two
of the 5f-electrons are localised, the Fermi
surface obtained from this calculation shows good agreement with the
experimental results of the dHvA experiments. They are presented in
fig.~\ref{FIGfermi} together with the theoretical calculations.

This is reassuring for the dual 5f model approach, however the 
FS is also in good agreement with standard LDA calculations as discussed
before. The decisive advantage of the former is that it also provides
a basically parameter free explanation for the mass enhancement which
cannot be obtained within LDA. The enhancement is due to the coupling
of the delocalised with the localised 5f-electrons
characterised by a matrix element $\alpha$ = 2a$_{5f}$M.
As explained below the latter are split into two low lying CEF singlet
states with an excitation energy $\delta$ which is known from
INS. Then the global mass enhancement
factor ($m^*$/m$_b$) (independent of the FS sheet) with respect to the
band mass m$_b$ may be calculated according to
eqs.~(\ref{EFFMASS1}),(\ref{EFFMASS2}) as in the case of \UPT. The
comparison of experimental and theoretical total mass enhancement
m$^*$/m is presented \index{mass enhancement} in table~\ref{tab:UPDmass}.
%
\begin{center}
\begin{table}[htb]
\caption{Effective masses for \v H $\parallel$ c. Notation for FS
sheets and experimental values from (Inada et al. 1999). Theoretical
values from (Zwicknagl et al. 2003)}
\vspace{0.5cm}
\label{tab:UPDmass}
\begin{tabular}{ccc}
\hline
FS sheet     & m$^*$/m (exp.)  & m$^*$/m (theory) \\
\hline
$\zeta$      &   65      & 59.6 \\
$\gamma$     &   33      & 31.9 \\
$\beta $      &  19       & 25.1\\
$\epsilon_2$  &  18       & 17.4\\
$\epsilon_3$  &  12       & 13.4\\
$\beta $      &  5.7       & 9.6\\
\hline
\end{tabular}\\[2pt]
\end{table}
\end{center}
%
\subsubsection{The dual model for \UPD and induced moment AF}

The dual model for \UPD which comprises both localised 5f$^2$
electrons with total angular momentum \v J and itinerant heavy 5f-electrons
created by c$^\dagger_{\v k\sigma} $ is described by the model Hamiltonian
\begin{eqnarray}
\label{HAM}
H&=&H_c+H_{CEF}+H_{ff}+H_{cf}\nonumber\\
H&=&\sum _{\v k\sigma}\epsilon_{\v k\sigma}c^\dagger_{\v k\sigma}c_{\v k\sigma}
+\delta\sum _i\mid e\ra\la e\mid _i \\
&&-\sum _{\ll ij\gg}J_{ff}(ij)\v J_i\v J_j\nonumber
-2I_0(g-1)\sum _i\v s_i\v J_i
\end{eqnarray}
where
\begin{equation}
\label{DIS}
\epsilon_{\v k\sigma}=\epsilon_{\perp}(\v k_{\perp}\sigma)
-2t_\parallel\cos k_z 
\end{equation}
is a model for the heavy conduction band energies whose FS is a
corrugated cylinder along the hexagonal c-axis. Here $t_\parallel$ is
the effective hopping along c which determines the amount of
corrugation. The form of the dispersion $\perp$ c is not important.
The localised 5f$^2$ electrons show a CEF splitting $\delta$ = 6 meV
into a \index{crystalline electric field excitations}
singlet ground state $|g\ra$ and an excited singlet $|e\ra$ at $\delta$. The
remaining terms describe a superexchange J$_{ff}$ between localised
and an on-site exchange I = I$_0$(g-1) between itinerant and localised
5f-electrons. The total effective inter-site exchange has therefore
an additional RKKY contribution:
\begin{eqnarray}
\label{EFF}
J(\v q)&=&J_{ff}(\v q)+I_0^2(g-1)^2\chi_e(\v q)
\end{eqnarray}
Here $\tensor{\chi}_0(\v q)$ is the conduction electron susceptibility. 
$\tensor{J}(\v q)$ may be fitted to the experimentally observed magnetic
excitations.

\subsubsection{Induced moments and magnetic exciton dispersion in
\UPD} \index{induced moment magnetism}\index{magnetic exciton}

First we consider the magnetism of the localised 5f moments without 
dynamic effects of coupling to itinerant 5f electrons.
In a nonmagnetic singlet ground state system with $\la g|\v J|g\ra
\equiv$ 0 the moments have to be induced via nondiagonal matrix
elements $\la e|J_x|g\ra$ = -i$\la e|J_y|g\ra$ = $\frac{1}{2}\alpha$
between ground state and excited state singlets. In \UPD the maximum of  
J(\v q) is at the AF wave vector \v Q = (0,0,$\frac{1}{2}$) with
J$_e\equiv$ J$_e$(\v Q). In the resulting AF state the CEF ground
state will then be a superposition of $|g\ra$ and $|e\ra$. This type of
magnetism is well known for Pr-metal and its compounds where one has two 
CEF split singlets from the Pr$^{3+}$(4f$^2$) configuration \cite{Jensen91}.
This induced moment AF is only possible if the control parameter
\begin{equation}
\label{XSI}
\xi=\frac{\alpha^2J_e}{2\delta}
\end{equation}
exceeds a critical value, i.e. $\xi>\xi_c$ = 1. Then the N\'eel
temperature T$_N$ and saturation moment $\la J\ra_0$ oriented along the
a-axis (x-direction) are given by 
\begin{eqnarray}
\label{CRI}
T_N&=&\frac{\delta}{2\tanh^{-1}(\frac{1}{\xi})}
\qquad \mbox{and} \qquad
\la J\ra_0=\frac{1}{2}\alpha\frac{1}{\xi}(\xi^2-1)^{\frac{1}{2}}
\end{eqnarray}
When $\xi$ is only slightly larger than $\xi_c$ = 1 (T$_N/\delta\ll$1) 
the saturation moment will be $\sim\alpha\exp(-\delta/2T_N)$ and hence
decreases exponentially with T$_N$. This is the major difference
from the usual local moment AF where the saturation moment is
independent of T$_N$.

\begin{figure}
\includegraphics[width=75mm]{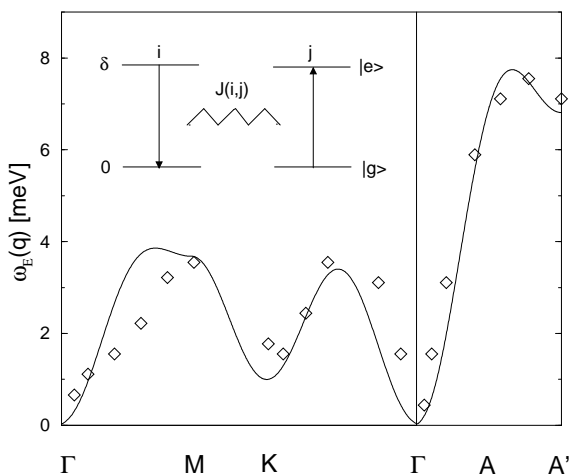}
\caption{Magnetic exciton dispersion of \UPD in the AF BZ. The diamonds
are data from Mason and Aeppli (1997). The solid line is a calculation using a
generalized version of eq.~(\ref{DI1}) with an appropriate model for the
exchange function J(\v q) (Thalmeier (2002)). The inset illustrates the
principle of magnetic (Frenkel-) exciton propagation: A CEF singlet-singlet
excitation ($\delta$) propagates between lattice sites i, j through
the action of intersite exchange J(i,j) and thereby acquires a
dispersion $\omega_E(\v q)$. \index{magnetic exciton}
\index{RKKY interaction}}
\label{FIGexciton}
\end{figure}

The signature of singlet-singlet induced moment magnetism is the existence of a
paramagnetic excitation that grows soft on approaching T$_N$. The
excitation spectrum is obtained from the dynamical local moment susceptibility
\begin{equation}
\label{CHI}
\chi_{ij;\alpha\beta}^{\lambda\mu}(\tau)=
-\la T\{J_{i\alpha}^\lambda(\tau)J_{j\beta}^\mu(0)\}\ra
\end{equation} 
where i,j= lattice site, $\lambda,\mu$=  AF sublattices (A,B) and
$\alpha,\beta$= x,y,z denotes cartesian components. Within RPA its
Fourier transform is obtained as
\begin{equation}
\label{RPA}
\tensor{\chi}(\v q,\omega)=[1-\tensor{u}(\omega)\tensor{J}(\v q)]^{-1}
\tensor{u}(\omega)
\end{equation}
Here $\tensor{\chi}(\v q,\omega)$ and the single ion dynamical
susceptibility $\tensor{u}(\omega)$ are tensors in both sublattice (A,B) and
transverse (xy) cartesian coordinates. The poles of eq.~(\ref{RPA})
determine the collective excitations of 5f-local moments. In the
paramagnetic phase they are given by the {\em magnetic exciton} dispersion
\begin{eqnarray}
\label{DI1}
\omega_E(\v q)&=&\delta[1-\frac{\alpha^2J(\v q)}{2\delta}
\tanh\frac{\beta}{2}\delta]
\end{eqnarray}
The physical origin and nature of magnetic excitons is illustrated in the inset
of fig.~\ref{FIGexciton} and described in the caption. When T$_N$ is approached
the exciton energy at \v Q (PM zone boundary or AF zone center)
becomes soft as a precursor to the appearance of the AF moment
according to 
\begin{equation}
\label{DI3}
\omega_E(\v Q,T)=\frac{1}{2}(\frac{\delta}{T_N})^2(\xi-\frac{1}{\xi})
(T-T_N)
\end{equation}
%
%
Below T$_N$ an induced staggered moment leading to two magnetic
sublattices appears. Its saturation value is given by eq.~(\ref{CRI}). 
Then $\v Q=(0,0,\frac{1}{2})$ becomes the center of the new AF Brillouin zone
$[\frac{\pi}{2c},-\frac{\pi}{2c}]$. In reality the softening of
$\omega_E(\v Q,T)$ will be arrested at a finite value at T$_N$ by Curie
type contributions to the static susceptibility which contribute to
the AF instability but not to the exciton dispersion. Therefore a
magnetic excitation gap at \v Q will appear. Below T$_N$ the
dispersion is somewhat modified due to the effect of the molecular
field, however for $\xi$ only marginally above the critical value this
modification is not important. This is indeed the case for the dual
model of \UPD where T$_N$/$\delta$= 0.22 and hence $\xi$= 1.015. The magnetic
exciton dispersion including the molecular field and exchange
anisotropy has been derived \cite{Thalmeier02} and compared to the
experimental results obtained in INS experiments \cite{Mason97}. In
this work the excitations were measured up to 10 meV in the whole BZ
and it was found that well defined propagating modes exist along the
hexagonal c$^*$-axis. For wave vector in the hexagonal plane
$\perp$c$^*$ the line
width of excitations was found to be much bigger. No excitation gap
at the AF zone center \v Q could be identified in these early
experiments. This conclusion had to be revised in later high
resolution experiments around \v Q as discussed below. The comparison
to the theoretical calculation using a parametrized exchange function
J(\v q) was given in \citeasnoun{Thalmeier02} and is shown in
fig. \ref{FIGexciton}. There the extended BZ was used which allows one
to plot only
the acoustic mode. It should be mentioned that it is not clear whether the
dip at the K-point is realistic because the line width of magnetic
excitons becomes rather large.
 
The damping of magnetic exciton modes has two sources: (i) intrinsic dynamical
effects in the localised moment system beyond RPA, e.g. damping by
thermal fluctuations in the singlet occupation. (ii) extrinsic damping
due to the coupling to conduction electrons described by the last term
in eq.~(\ref{HAM}). For low temperatures the latter is the dominating
part. In addition it leads to a renormalized exciton mode
frequency. Both effects can be described in extending the previous
RPA approach including the last coupling term in eq.~(\ref{HAM}). This
leads to coupled RPA equations \cite{Buyers85}, the solution for the
localised dynamical susceptibility \index{susceptibility} is given by
\begin{eqnarray}
\label{RPASUS}
\chi(\v q,\omega)=
\frac{u(\omega)}{1-J_{ff}(\v q)u(\omega)-I^2u(\omega)\chi_e(\omega)}
\end{eqnarray}
with 
\begin{eqnarray}
\label{DUALSUS}
u(\omega)&=&\frac{\alpha\la S\ra}{\delta-\omega}
\qquad \mbox{and} \qquad
\chi_e(\omega)=\frac{1}{i\omega-\Gamma}
\end{eqnarray}
where u($\omega$) is the single ion local moment susceptibility and
$\chi_e(\omega)\equiv\chi_e(\v Q,\omega)$ is the low frequency conduction
electron susceptibility at \v q = \v Q and $\Gamma$ is a phenomenological
damping rate. If one neglects its imaginary part and uses $\chi_e$(\v
q) $\sim\Gamma^{-1}$ for \v q $\simeq$ \v Q then, with
eqs. (\ref{DUALSUS}), (\ref{EFF})
the poles of eq.~(\ref{RPASUS}) again lead to the undamped magnetic exciton
dispersion $\omega_E$(\v q) of eq.~(\ref{DI1}). Inclusion of the
imaginary part of $\chi_e(\omega)$ leads to a shift of the mode frequency and a
damping. For a large coupling constant I, a part of the spectral weight of the
magnetic exciton is shifted to low energies leading to an additional
quasielastic peak. 

\begin{figure}
\includegraphics[width=75mm]{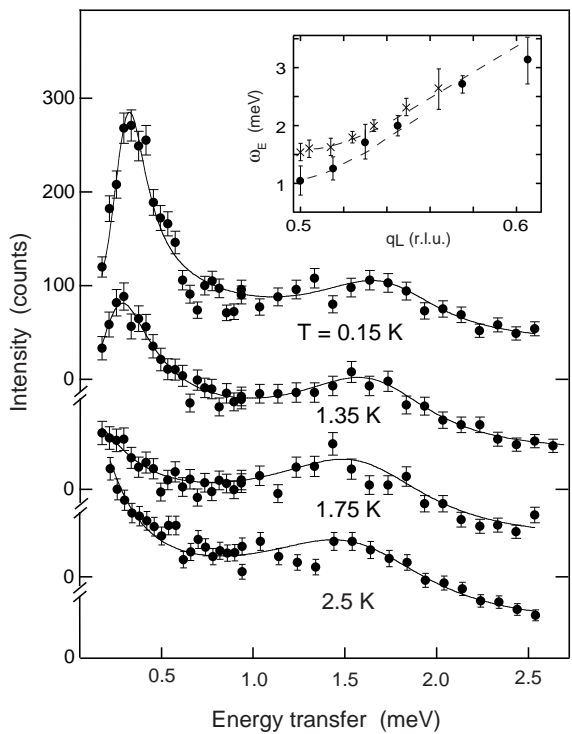}
\caption{INS spectrum for various temperatures above and below
T$_c$ = 1.8 K (Sato et al. 2001). The momentum transfer was varied
around the AF point \v q = (0,0,q$_L$). The solid line is the result of
a calculation using eq.~(\ref{RPASUS}) from which the magnetic exciton
energy $\omega_E(\v q)$ is
determined for every \v q. The result is shown in the inset for T = 2.5
K (crosses) and T = 0.15 K (circles). 
\index{inelastic neutron scattering!UPd$_2$Al$_3$}}
\label{FIGneutron}
\end{figure}

The INS intensity is then proportional to the {\em localised}
5f-dynamical structure factor 
S(\v q,$\omega$) = (1+n($\omega$)){\em Im}$\chi(\v q,\omega$) where
n($\omega$) = $(\exp(\beta\omega)-1)^{-1}$ is the Bose factor. The
experimental high resolution magnetic INS intensity for wave vectors
close to \v Q as obtained by \cite{Sato97,Sato97a,Bernhoeft98}  is
shown in fig.~\ref{FIGneutron}. At the AF vector \v Q above T$_c$ (2.5 K) 
one can observe a strongly broadened inelastic peak at around 1.5 meV which is
interpreted as the (upward) shifted magnetic exciton energy. In
addition, as a signature of the strong coupling to conduction
electrons one observes a quasielastic peak around
$\hbar\omega\simeq$ 0. The full line is a calculation of S$(\v
Q,\omega$) using eq. (\ref{RPASUS}) for $\chi(\v q,\omega)$ and taking
I and $\omega_E$(\v Q) as adjustable parameters. The INS intensity was
also measured for other wave vectors close to \v Q along
c$^*$. Repeating the same
fitting procedure one may determine the {\em unrenormalized} magnetic
exciton dispersion $\omega_E$(\v q) which is shown in the inset of
fig.~\ref{FIGneutron}. Obviously the magnetic excitons have a gap at
the AF zone
center of about 1 meV. As mentioned before in the original experiments
\cite{Mason97} this gap was not identified due to insufficient
resolution. In the singlet-singlet CEF model, even including a
uniaxial anisotropy of the exchange, this relatively large gap cannot
be explained. Its most probable origin is an arrested
softening of $\omega_E$(\v Q,T) due to higher lying CEF states.

In fig.~\ref{FIGneutron} another important feature is obvious: As temperature
is lowered below T$_c$ = 1.8 K the quasielastic peak evolves into a
low energy inelastic peak. This is due to the appearance of the SC
quasiparticle gap function \De which possibly may have
node lines or points. Its influence on the conduction electron
susceptibility may be described in a simple manner by shifting the
diffusive pole at -i$\Gamma$ in $\chi_e(\omega)$ to
-i$\Gamma$+$\Delta_{av}$ where the real part corresponds to the \v
k-averaged SC gap. The calculated intensity in the SC phase (e.g. at T
= 0.15 K) is then again shown as a full line from which the size of
$\Delta_{av}$ may be exctracted. The result
$\Delta_{av}\simeq \omega_E(\v Q)$ shows
that magnetic exciton energy and SC gap are almost degenerate, which
means that there will be a strong mixing of magnetic excitons with SC
quasiparticle excitations. This explains why the lower peak appears
with a large intensity.

\subsubsection{Magnetic exciton anomalies in quasiparticle tunneling
spectra} \index{quasiparticle tunneling!UPd$_2$Al$_3$}

\begin{figure}
\includegraphics[clip,width=85mm]{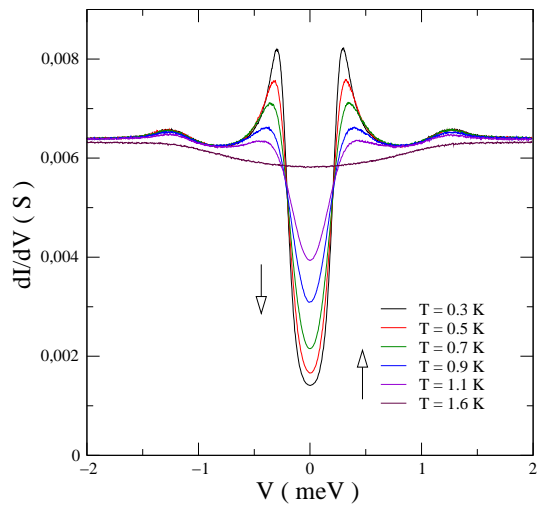}
\caption{Differential conductivity dI/dV of a UPd$_2$Al$_3$-AlO$_x$-Pb
tunneling contact as function of the voltage across the tunneling
contact for different temperatures (Jourdan et al. 1999). At low
temperatures dI/dV is proportional to the quasiparticle DOS.
\index{quasiparticle tunneling!UPd$_2$Al$_3$}}
\label{FIGdiffcond}
\end{figure}

The near degeneracy of SC gap and magnetic exciton energy suggests
that the latter might play a role in the formation of Cooper pairs in
\UPD. Knowing only the INS results this would only be a
hypothesis. However in a breakthrough experiment \cite{Jourdan99} the
first SC quasiparticle tunneling in a HF system by using epitaxially grown
\UPD was achieved. This experiment is complementary to INS
measurements as it probes the
itinerant 5f-electrons. The resulting tunneling spectra which ideally are
proportional to the SC quasiparticle DOS are shown in
fig.~\ref{FIGdiffcond}. The tunneling current is parallel to the c-axis and a
real gap is seen in dI/dV in this direction. Therefore, if node lines
of \De exist they should be perpendicular to the c-axis at k$_z$
positions where the velocity $v_c(k_z)$ vanishes. The most striking
result is the presence of typical `strong coupling anomalies' around 1
meV which are well known from ordinary electron-phonon superconductors
like Pb. These anomalies are connected to the frequency spectrum of
the exchanged boson which is responsible for the formation of
Cooper pairs in \UPD. The Debye energy of \UPD k$\theta_D$ = 13 meV is
much too large to be connected with the observed modulation
around 1 meV, however it agrees perfectly with the magnetic exciton
energy $\omega_E(\v Q)$ found by INS (inset of
fig.~\ref{FIGneutron}). Furthermore the average gap energy and mode energy
determined in INS are of the same order. This leads to the conclusion
that \UPD is a magnetic exciton mediated strong coupling
superconductor. This is the first time that a non-phononic mechanism for
superconductivity has been proven in a direct way by identifying the
non-phononic boson (magnetic exciton) that provides the `glue' for the
formation of Cooper pairs in \UPD. These arguments can be made more
quantitative by using Eliashberg theory \index{Eliashberg equations}
to calculate the SC quasiparticle
DOS. The retarded effective potential due to magnetic exciton exchange
can be described by a factorized ansatz \index{effective interaction}
\begin{eqnarray}
\label{POT}
V_E(\v q,\omega)=
\frac{V_0\omega_0}{(\omega-\omega_E(\v Q))^2+\omega_0^2}
f(\v k)f(\v k')
\end{eqnarray}
%
\begin{figure}
\includegraphics[width=100mm]{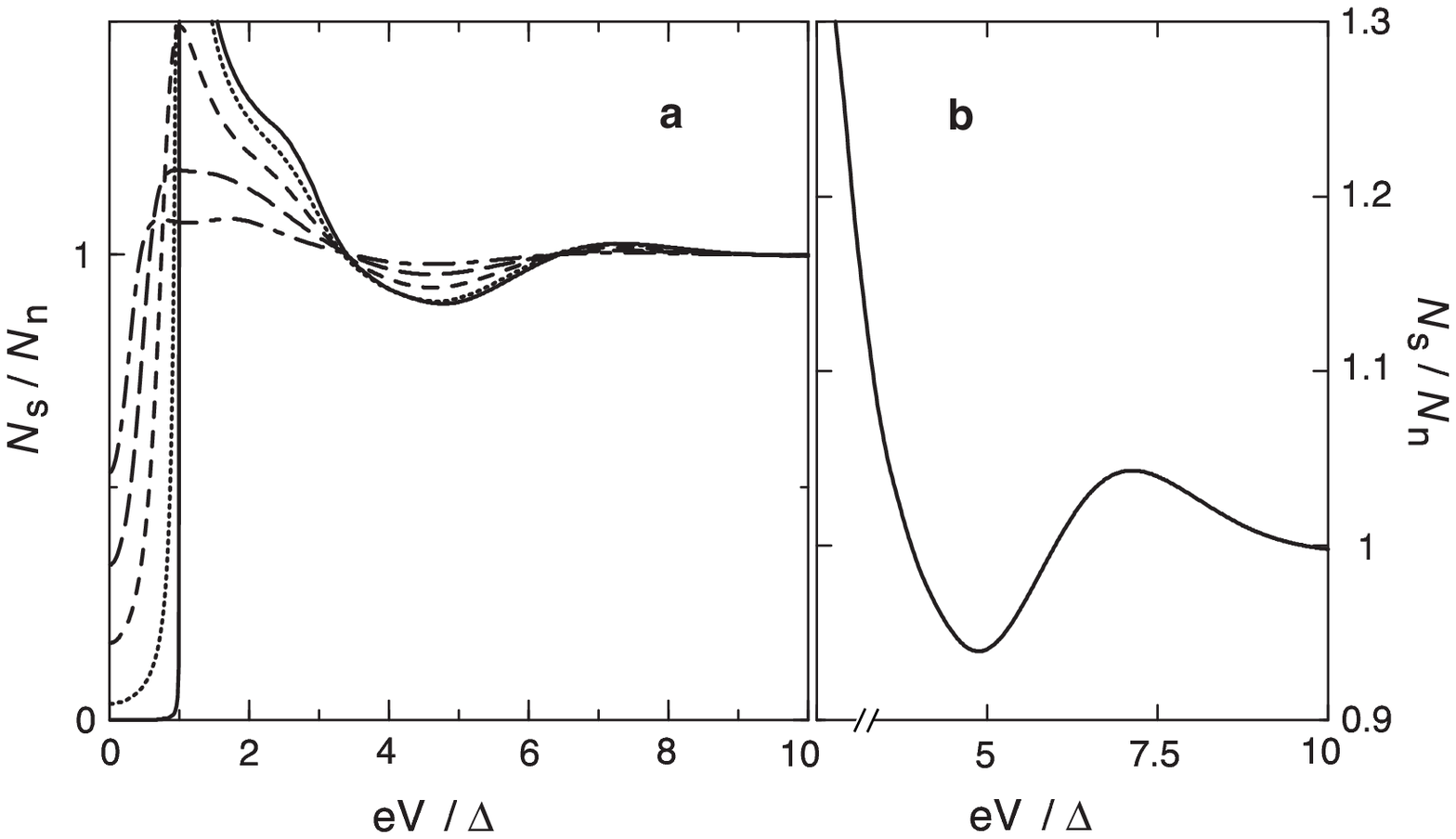}
\caption{Calculated tunneling DOS N$_s$ using isotropic Eliashberg equations
and the retarded potential of eq.(\ref{POT}). N$_n$ is the DOS in the normal
state. (a) The various lines correspond to temperatures T/T$_c$ = 0.25,
0.6, 0.9, 0.95 from top to bottom at eV/$\Delta$ = 2. (b) Enlarged
strong coupling anomaly due to magnetic exciton exchange for
T = 0.25T$_c$ (Sato et al. 2001). \index{Eliashberg equations}
\index{quasiparticle DOS}\index{strong coupling effects}}
\label{FIGtunneldos}
\end{figure}
%
which exhibits a maximum at the observed exciton frequency $\omega_E(\v
Q)$ with a width $\omega_0$ and strength V$_0$. Here \v q =\v k'-\v k and  
f(\v k) is the form factor which describes the \v k-dependence of the
superconducting singlet gap function \De = $\Delta$ f(\v
k). For an isotropic conventional e-p superconductor f(\v
k) $\equiv$ 1. As explained in the next section in \UPD the gap function
should have a node line perpendicular to
the c-axis. For tunneling current along c the tunneling DOS can be
determined from the {\em isotropic} Eliashberg equations using eq.~(\ref{POT})
if one replaces f(\v k) by the averaged constant $\la f(\v
k)^2\ra_{FS}^\frac{1}{2}$ which may then be absorbed into the coupling strength
V$_0$ that is adjusted to obtain the experimental T$_c$. The result of
this analysis is shown in fig.~\ref{FIGtunneldos} which nicely explains
the experimental results in fig.~\ref{FIGdiffcond}, especially: (i) The
width of the gap does not depend much on temperature but is rather
filled up with increasing
temperature. This is characteristic for a strong coupling
superconductor, the ratio $\Delta$/2T$_c$ = 5.6 obtained here is rather
high and agrees well with the ratio $\Delta$/2T$_c$ = 6 obtained
previously from the analysis of INS spectra. (ii) In the calculated
tunneling DOS of fig.~\ref{FIGtunneldos} typical modulations above the
gap due to the retardation of the potential in eq.~(\ref{POT}) appear
at a voltage that corresponds to the magnetic exciton energy at \v Q
which is about 1 meV.    

\subsubsection{Possible symmetries of the superconducting order parameter}

The existence of node lines of \De in \UPD was suggested by various
thermodynamic properties discussed previously. Other evidence was
proposed by \citeasnoun{Bernhoeft00} where it is claimed that the symmetry
property $\Delta(\v k\pm\v Q)=\pm$\De has to be fulfilled to explain
the large intensity of the low energy quasiparticle-like peak in
fig.~\ref{FIGneutron}. This implies a node line orthogonal to the c-axis which
is compatible the appearance of a gap in the c-axis tunneling geometry as
used for fig.~\ref{FIGdiffcond}. In the ansatz of eq.~(\ref{POT}) the
energy and momentum
dependences are factorized and therefore the form factor f(\v k),
i.e. gap function symmetry is put in by hand. Considering the presence
of a strong AF ordered moment $\parallel$ x-axis the appropriate
symmetry group is orthorhombic D$_{2h}$. The simplest spin singlet
even parity irreducible 
representation appropriate for the FS with cylindrical symmetry
(fig.~\ref{FIGfermi}) has the form 
\begin{eqnarray}
\label{DSING}
\Delta(\v k)=\Delta\cos k_z \qquad \mbox{with} \qquad
A_{1g}(\Gamma_1^+) - \mbox{symmetry}
\end{eqnarray}
This gap function is independent of k$_x$ and k$_y$. A more general
\De with 
the same symmetry can be obtained by multiplying eq.(\ref{DSING}) with
any fully symmetric function f$_{\Gamma^+_1}$(k$_x$, k$_y$). \De has the
required behaviour under the transformation \v k $\rightarrow$ \v
k$\pm$\v Q and it has a node line at the AF zone boundary
k$_z= \pm\frac{1}{2}Q_z= \pm\frac{\pi}{2c}$. A theoretical
investigation of possible gap functions based on a nonretarded weak coupling
theory of the magnetic exciton mediated pair potential was
undertaken in \citeasnoun{Thalmeier02}. It was found that due to the
CEF-anisotropy and the action of the AF local moment order the
degeneracy
of odd parity triplet states is lifted and one of them  would be
favored against the above even parity singlet state. However
this may be due to the neglect of retardation which is not a good
approximation for \UPD as evident from the previous discussion. As
mentioned before Pauli limiting of H$_{c2}$ and Knight shift reduction
below T$_c$ favor an even parity A$_{1g}$ (singlet) state.

As in the borocarbides the
gap function will be modified by the background magnetic order due to
the reconstruction of Bloch states close to the magnetic Bragg planes
k$_z$= $\pm\frac{1}{2}Q_z$. The AF modified gap function for the even
parity state is given by eq.~(\ref{MODGAP}) which in the
present case reduces to
\begin{eqnarray}
\Delta(\v k)=\Delta\cos k_z\bigl(\frac{\cos^2k_z}
{\lambda_{AF}^2+\cos^2k_z}\bigr)^\frac{1}{2} 
\end{eqnarray}
Here the dimensionless interaction parameter
$\lambda_{AF}=(I\la J\ra _0)/(\alpha W_\parallel)$ describes the
strength of the AF influence on \De. It is interesting to note that
despite the presence of a node line the above order parameter
is fully symmetric (A$_{1g}$) because it is invariant under all
symmetry transformations of the {\em magnetic} unit cell. Thus in a
strict sense \UPD cannot be called a superconductor with
unconventional gap symmetry, although the novel magnetic exciton mechanism
certainly is unconventional. Various possible nodal structures in
addition to eq.~(\ref{DSING}) have been discussed by
\citeasnoun{Thalmeier02a} in the context of magnetothermal properties.

\subsubsection{\UND: a possible triplet superconductor}
\index{UNi$_2$Al$_3$}\index{triplet pairing}

The HF superconductor \UND \cite{Geibel91b} ($\gamma$ = 120 mJ/K$^2$
mole) is isostructural to its hexagonal sister compound \UPD
(fig.~\ref{FIGstructure}) but it has rather different
physical properties. It exhibits an incommensurate SDW below T$_m$ =
4.6 K ($\mu$ = 0.2 $\mu_B$) with a modulation wave vector \v Q$_m$
= (0.39,0,0.5), like in \UPD the moments are pointing to n.n. along the a
direction of the hexagonal plane. Superconductivity sets in below
T$_c$ = 1.2 K and coexists with magnetic order. The T-dependence of
1/T$_1$ in NMR \cite{Kyogaku93} and the 
small entropy release of $\Delta_S$ =0.12Rln2 at T$_m$
\cite{Tateiwa98} points to an itinerant character of magnetic order
in contrast to \UPD ($\Delta_S$ = 0.67Rln2) which exhibits local
moment magnetism of the induced type with large ordered
moment. This view is supported by INS experiments \cite{Gaulin02}
which do not show any evidence for propagating collective modes in the
SDW state unlike the magnetic exciton mode discussed previously for
\UPD. The magnetic excitation spectrum in \UND consists of quasielastic
spin fluctuations for all wave vectors and
its energy width of about 6 meV corresponds to the coherence
temperature $\sim$ 80 K associated with the $\gamma$ value. For
temperatures not too close to T$_m$ the spin fluctuations are centered
around odd multiples of the commensurate AF wave vector \v Q
= (0,0,0.5) and also extending along ridges in q$_x$- direction.

From the spin fluctuation spectrum of \UND observed in INS which is
located around an AF Bragg point one might expect it should be a
textbook example of spin fluctuation mediated
superconductivity. However as in the case of UPt$_3$ the spin singlet
pairing predicted by AF spin fluctuation type theories \cite{Miyake86}
is probably not realised. According to the $^{27}Al$-Knight shift
measurements of \cite{Ishida02} which did not observe any drop in the spin
susceptibility below T$_c$ \UND should have a spin triplet SC order
parameter. This is supported by the observed lack of Pauli
paramagnetic limiting effect on H$_{c2}$(T) \cite{Sato96}. As in the
case of \UPD the position of gap nodes is not known with
certainty, but NMR results suggest the presence of a node line in
\UND as in \UPD. In the simplest scenario then the latter might correspond
to the singlet gap function of eq.~(\ref{DSING}) while the former has one of
the triplet gap functions .
\begin{eqnarray}
\label{DTRIP}
\v d(\v k)=\Delta\hat{d}\sin k_z
\end{eqnarray}
With the possibilities $\hat{d}=\hat{x},\hat{y},\hat{z}$ due to the AF
orthorhombic symmetry.
These gap functions have all an equatorial node line at
k$_z$ = 0. Confirmation of the node structure has to await the results
of angle resolved magnetotransport or specific heat measurements in
the vortex phase and for a confirmation of the spin state the Knight
shift measurements on high quality single crystals are necessary.

\subsection{Ferromagnetism and Superconductivity in \mbox{\UGE}}
\index{ferromagnetic order}\index{UGe$_2$}

\label{Sect:UGe2}

The possibility of coexisting ferromagnetism (FM) and
superconductivity was first considered by Ginzburg \cite{Ginzburg57}
who noted that this is only possible when the internal FM field is
smaller than the thermodynamic critical field of the
superconductor. Such a condition is hardly ever fulfilled except
immediately below the Curie temperature T$_C$ where coexistence has been
found in a few superconductors with local moment FM and T$_C<$ T$_c$ such as
ErRh$_4$B$_4$ and HoMo$_6$S$_8$. If the temperature drops further below T$_C$
the internal FM molecular field rapidly becomes larger than H$_{c2}$ and SC is
destroyed. The reentrance of the normal state  below T$_C$ has
indeed been observed in the above compounds. The only compound known
sofar where local moment FM coexists homogeneuously with SC for all
temperatures below T$_c$ is the borocarbide compound ErNi$_2$B$_2$C
\cite{Canfield96} (sect.\ref{Sect:Boro}).
The competition between FM and superconductivity becomes
more interesting if FM order is due to itinerant electrons which also
form the SC state. In Hartree-Fock approximation the Stoner condition for
the reduced interaction constant $\lambda$ = IN(0) = 1-S$^{-1}\geq$ 1
(S = Stoner parameter, I = exchange constant) of conduction electrons
determines
the existence of itinerant FM order. If the interaction is slightly
above the critical value $\lambda_c$ = 1 one has weak FM order such as
in ZrZn$_2$ with large longitudinal \index{spin fluctuation}
ferromagnetic spin fluctuations. \citeasnoun{Fay80} have shown that in
this case p-wave superconductivity may actually be mediated by the exchange of
FM spin fluctuations and
coexist with the small FM moments. According to this theory p- wave
superconductivty should exist on both (FM and PM) sides of the
critical value $\lambda_c$ = 1 for some range of the interaction
parameter $\lambda$. Until recently this remained only a theoretical
scenario. The discovery of unconventional superconductivity under
pressure in the itinerant FM UGe$_2$ \cite{Saxena00} and later for FM
URhGe \cite{Aoki01} and the 3d FM ZrZn$_2$ \cite{Pfleiderer01} under
ambient pressure has opened this field to experimental investigation. 

\begin{figure}
\includegraphics[width=3.5cm]{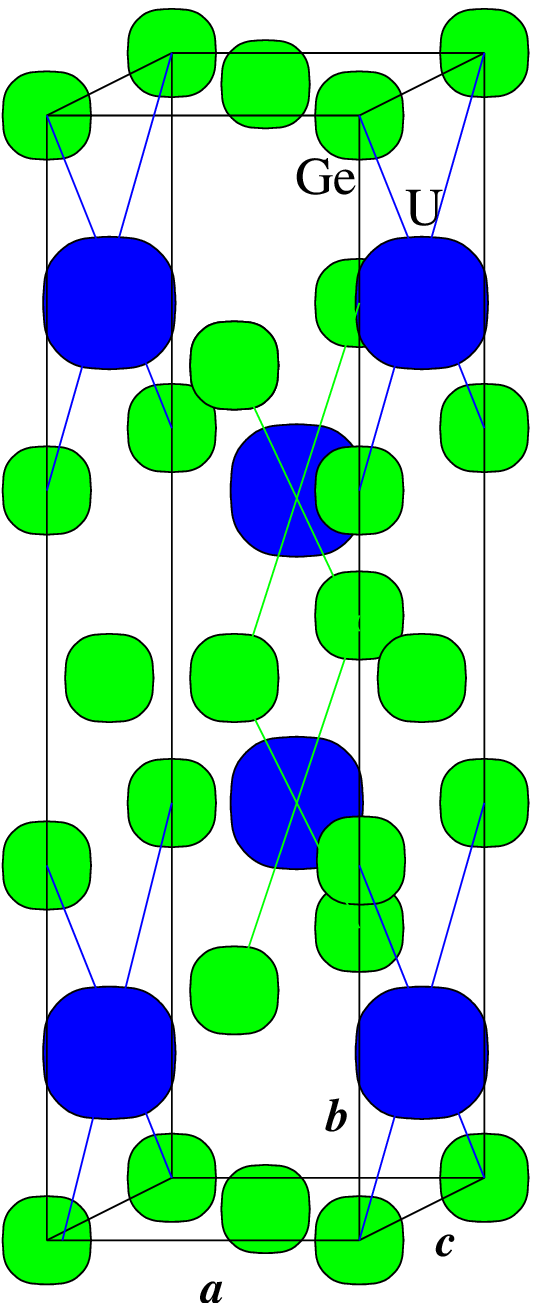}
\caption{Conventional orthorhombic unit cell of \UGE with lattice
parameters given by a = 3.997\AA, b = 15.039\AA~ and c = 4.087\AA. U
atoms (large spheres) form zig-zag lines along a. 
\index{hydrostatic pressure!UGe$_2$}\index{ferromagnetic order}}
\label{FIGuge2struc}
\end{figure}

\begin{figure}
\includegraphics[width=9cm]{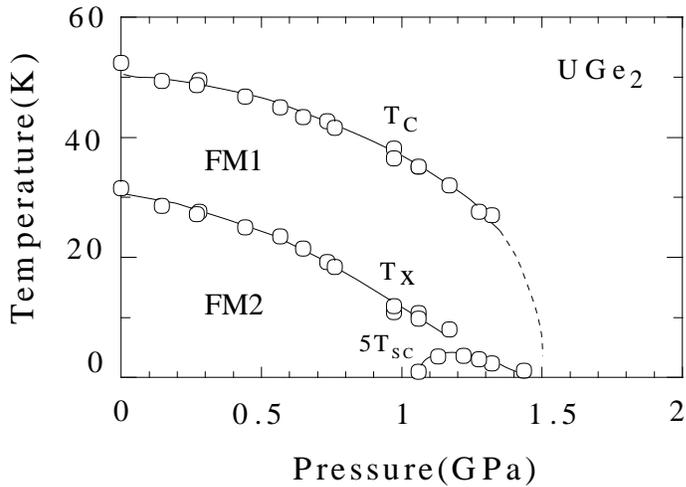}
\caption{pressure - temperature phase diagram of \UGE from resistivity
measurements. T$_C$ = FM Curie temperature, T$_x$ = boundary of new
x-phase. T$_{SC}$ = superconducting transition temperature (note scale
enhancement) with optimum pressure at p$_m$ = 1.25 GPa. Ferromagnetic
FM1 and FM2 regions have different moments ($\simeq$ 1 $\mu_B$ and
1.45 $\mu_B$ at low T respectively) (Kobayashi et al. 2001).}
\label{FIGuge2ptdiagr}
\end{figure}

It became quickly clear however that the physics of UGe$_2$ is
not as simple as suggested by the FM spin fluctuation model. Firstly
the maximum of SC T$_c$ occurs at a pressure where the FM moment
per U is still 1$\mu_B$ as compared to 1.5$\mu_B$ at ambient
pressure thus $\lambda_c$ should still be far from the critical value
where spin fluctuations are important. Secondly the SC phase
diagram is not approximately symmetric around the critical value as
expected but the region of SC lies completely inside the FM
phase. This problem has been further investigated by resistivity
measurements \cite{Huxley01} where an additional phase within the FM
region appeared at lower temperature witnessed by a resistivity
anomaly. Its critical temperature T$_x$(p) precisely hits the maximum of the SC
T$_c(p)$ curve as seen in fig.~\ref{FIGuge2ptdiagr}. 
The x-phase has been associated with a combined CDW/SDW transition. This is
also suggested by electronic structure
calculations \cite{Shick01} which show nesting features of the Fermi
states. In addition the main FS sheet consists mostly of majority spin
(m$_s=\uparrow$, m$_l$=0) electrons. Therefore the SC pairs must
have triplet character. It is suggestive that the unconventional
Cooper pairing then appears close to the quantum critical point of the
alledged CDW/SDW transition and therefore T$_c$ should rather be
associated with the T$_x$ transition instead of the FM transition at
T$_C$. However sofar no indication of a CDW/SDW order parameter below
T$_x$ has been identified in neutron diffraction \cite{Kernavanois01} and
the subject remains open.
 
Various theoretical scenarios for UGe$_2$ have been formulated. In the
context of Ginzburg-Landau theories the symmetries of possible SC
order parameters and their coupling to the FM have been
discussed, e.g. \cite{Machida01,Mineev02}. However, as already mentioned in
the context of UPt$_3$ the predictive power of such theories is
limited. The traditional approach of microscopic theories invoking FM
spin fluctuations has been extended beyond Hartree Fock with the
inclusion of mode coupling terms in order to explain the asymmetric
phase diagram around the critical coupling
strength \cite{Kirkpatrick01}. The new aspect of the
T$_x$ transition line was introduced by \citeasnoun{Watanabe02} who, in order
to comply with the experimental evidence from
\citeasnoun{Kernavanois01}, interpret
it as a crossover line into a region of enhanced SDW/CDW fluctuations
rather than an ordered phase. When T$_x$(p) approaches zero these
fluctuations then would mediate formation of a triplet p-wave SC state.

\subsubsection{Electronic structure and band magnetism}
\index{band structure}\index{ferromagnetic order}

The itinerant FM \UGE crystallizes in the base-centered orthorhombic
ZrGa$_2$ type structure (Cmmm) which is shown in
fig.~\ref{FIGuge2struc}. It may be viewed consisting of antiphase
zig-zag U chains running along the a-axis. This is also the easy axis
for the FM moments ($\mu$ = 1.43$\mu_B$ per formula unit) below the Curie
temperature of T$_C$ = 52 K at ambient pressure. Magnetisation and
susceptibility are extremely anisotropic \cite{Onuki92a} corresponding to
almost Ising like behaviour of U-moments. The specific heat
coefficient is $\gamma$ = 32 mJ/molK$^2$ which corresponds to an enhancement
factor of 2.7 as compared to the band mass \cite{Shick01}. This points
to some degree of mass renormalization by dynamic correlation
effects. Individual masses observed in dHvA experiments exhibit
an enhancement factor of m$^*$/m $\simeq$ 15 - 25 compared to the free electron
mass. Therefore \UGE is an itinerant 5f metal with sizable
correlation effects but has an order of magnitude smaller effective
masses than real U-heavy fermion metals.

The electronic band structure of \UGE has first been calculated within
the LDA+U approach \cite{Shick01} where U $\simeq$ 0.7 eV is a strongly
screened on-site Coulomb interaction that was fitted to reproduce the
proper FM ground state moment of 1.43$\mu_B$. On the other hand this value is
too small to reproduce the observed photoemmission spectrum. 
The calculated Fermi surface has a main sheet which mostly consists of
(m$_l$=0, m$_s=\uparrow$)-states. Bands crossing the Fermi energy have
mostly U-5f majority spin character, indicating almost complete
polarisation, i.e. strong FM. The LDA+U Fermi surface also displays
a nesting feature for \v Q = (0.45,0,0) (in r.l.u.) which is
due to bands that have little dispersion along b, i.e. perpendicular
to the zig-zag U chains. The proper easy a axis for FM moment and a
strong magnetocrystalline anisotropy of the moment due to spin orbit
coupling was also correctly
found. These calculations describe the ambient pressure strong band
ferromagnetism in \UGE which is still far from the superconducting
pressure region with reduced moments.
 
\begin{figure}
\includegraphics[width=0.40\columnwidth]{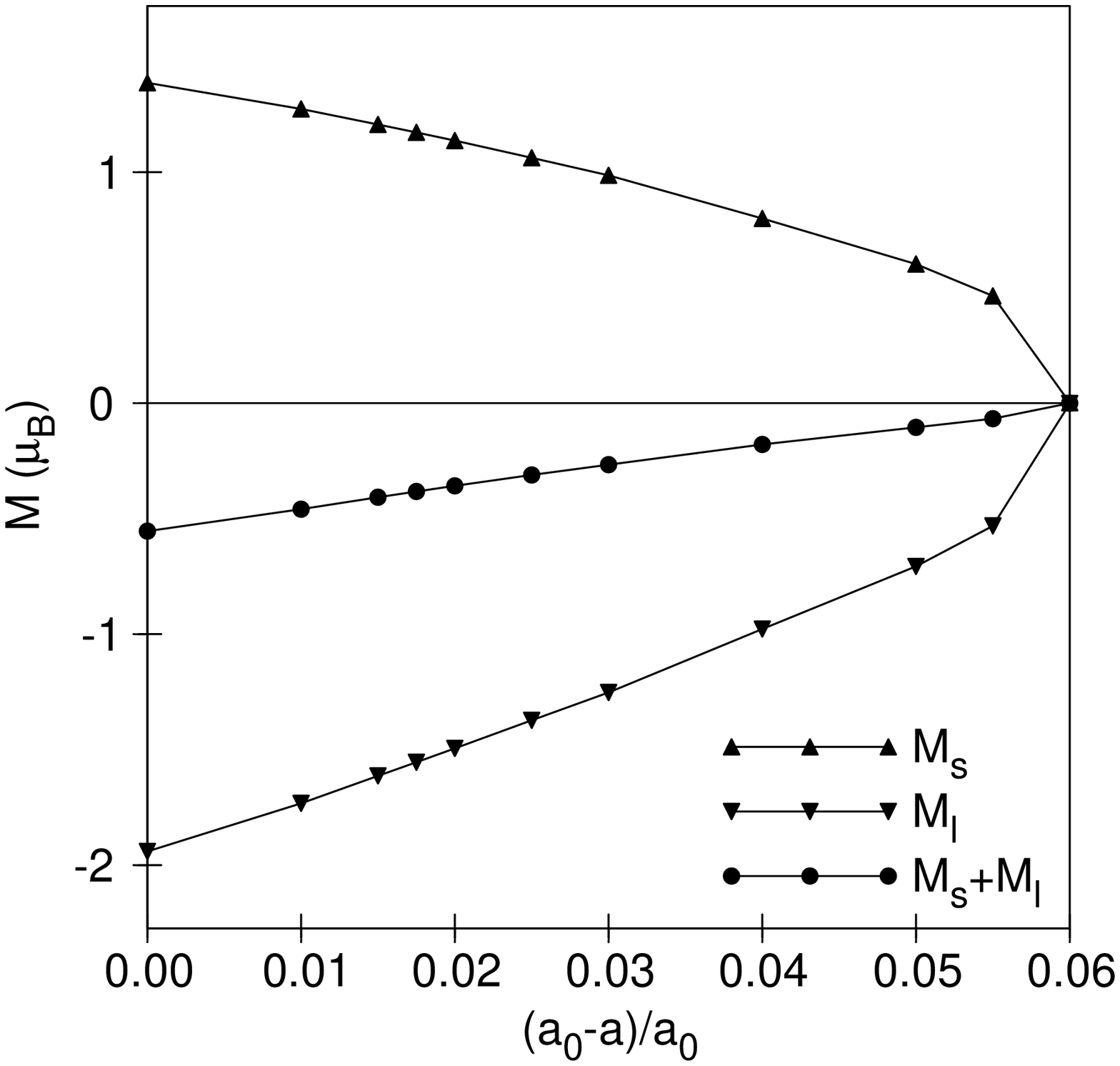}\hfill
\includegraphics[width=0.40\columnwidth]{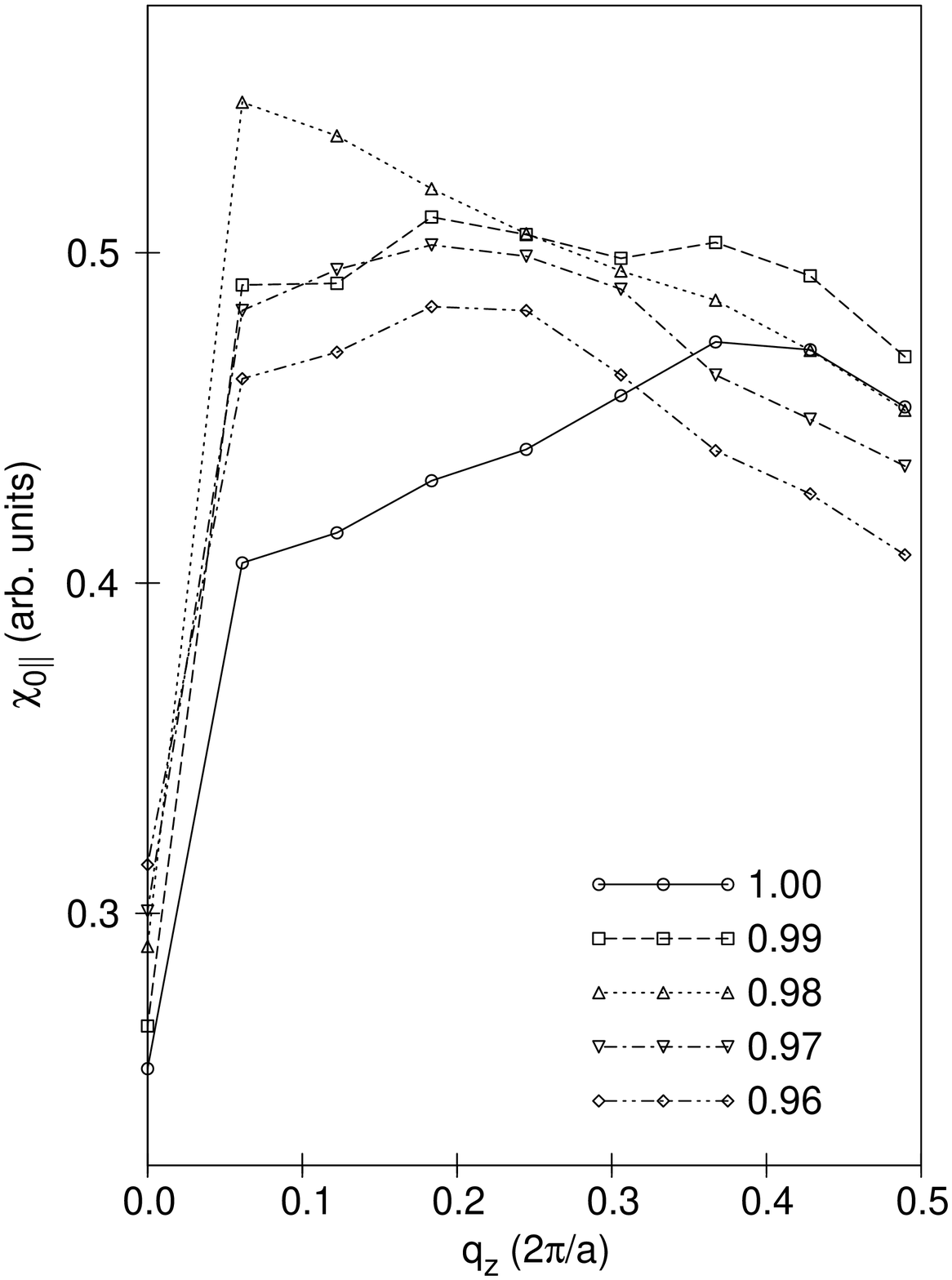}
\caption{Left panel: calculated pressure dependence of spin (M$_S$),
orbital (M$_L$) and total moment (M$_S$+M$_L$). Right panel: pressure
dependence of calculated longitudinal susceptibility. The steep drop
at q$_z$=0 is caused by the loss of intra-band transitions due to
finite q$_z$ resolution. The enhancement of $\chi_{0\parallel}$ 
under pressure for small q$_z$ suggests the evolution of a SDW
instability within the FM phase (Yaresko and Thalmeier 2003).
\index{susceptibility}}
\label{FIGuge2psusc}
\end{figure}

A more recent theoretical investigation of the \index{hydrostatic pressure}
collapse of the FM state under hydrostatic pressure and the
simulataneous appearance of another instability has been given
in \citeasnoun{Yaresko03}. In this study a relativstic LSDA calculation was
performed for isotropically compressed lattice constants simulating the
application of hydrostatic pressure p. The experimental lattice
constants corresponding to p = 0 are given in
fig.~\ref{FIGuge2struc}. In agreement
with the experimental results a FM ground state with the a-axis
as easy axis is obtained. However the U spin (-1.39
$\mu_B$) and orbital (1.94 $\mu_B$) magnetic compensate partly leading
to a total moment of only 0.55 $\mu_B$ significantly smaller than the
experimental value (1.43$\mu_B$). This was the main reason for
performing the above mentioned LDA+U calculation. However the 
LSDA calculation reproduces quite well the angular dependence of the
cross section of a major FS sheet with an area of F $\simeq$ 9 kT. The
calculated cyclotron masses are about 8 m and smaller than the
experimental values of 15.5 m. This 
confirms that UGe$_2$ has itinerant 5f-electrons with only moderately
strong correlations. 

The hydrostatic pressure simulation shows that FM U moments decrease
continuously with pressure until at compression factor x = 0.94 only
a nonmagnetic solution is obtained (fig.~\ref{FIGuge2psusc}). The
magnetocrystalline anisotropy energy E$_{\v M\parallel a}$- E$_{\v
M\parallel b}$ also decreases upon compression. On the other hand the
DOS N(0) first increases to a maximum at x $\sim$ 0.98 and then
decreases again. In order to check whether a tendency to a CDW/SDW
instability exists which might be related to the empirical observation
of the T$_x$ phase boundary inside FM the staggered static
susceptibility \index{susceptibility} within LSDA has been calculated
\cite{Yaresko03}. It is given by 
\begin{eqnarray}
\chi_{0\alpha}(\v q,0)=\sum_{ij,\v k}
[f(\epsilon_{i\v k})-f(\epsilon_{j\v k +\v q})]
\frac{|M^\alpha_{i\v k,j\v k+\v q}|^2}
{\epsilon_{i\v k}-\epsilon_{j\v k +\vq}}
\end{eqnarray}
where $\epsilon_{i\v k}$ are the band energies, M$^\alpha_{i\v k,j\v k+\v q}$
are the transition matrix elements ($\alpha =\parallel,\perp$ \v M)
and f($\epsilon$) is the Fermi function. The q$_z$ wave-number
dependence of $\chi_{0\perp}$ and $\chi_{0\parallel}$ has been
calculated as function of the lattice compression. Whereas
$\chi_{0\perp}$(q$_z$)
shows monotonic decrease with compression $\chi_{0\parallel}$(q$_z$) is very
sensitive to lattice constant change, it develops a strong enhancement
at small wave vector at the same intermediate
compression x = 0.98 where N(0) also is at its maximum. This is an
indication that for an intermediate compression (smaller than the
critical compression x = 0.94 where FM vanishes) \UGE has a tendency to
develop a long wavelength longidudinal SDW-like instability. It is possible
that the tendency towards instability for the susceptibility associated with
an unconventional SDW is even stronger. However it should be noted
that the susceptibility enhancement does not correspond to any
obvious nesting in the FS sheets obtained in LSDA calculations. This raises the
question whether the observed nesting feature in the LDA+U FS of
\citeasnoun{Shick01} has a real physical significance to the alleged hidden
order phase.

\subsubsection{Coexistence of FM order and superconductivity under
pressure}

\begin{figure}
\begin{minipage}[t b]{\columnwidth}
\includegraphics[width=.48\columnwidth]{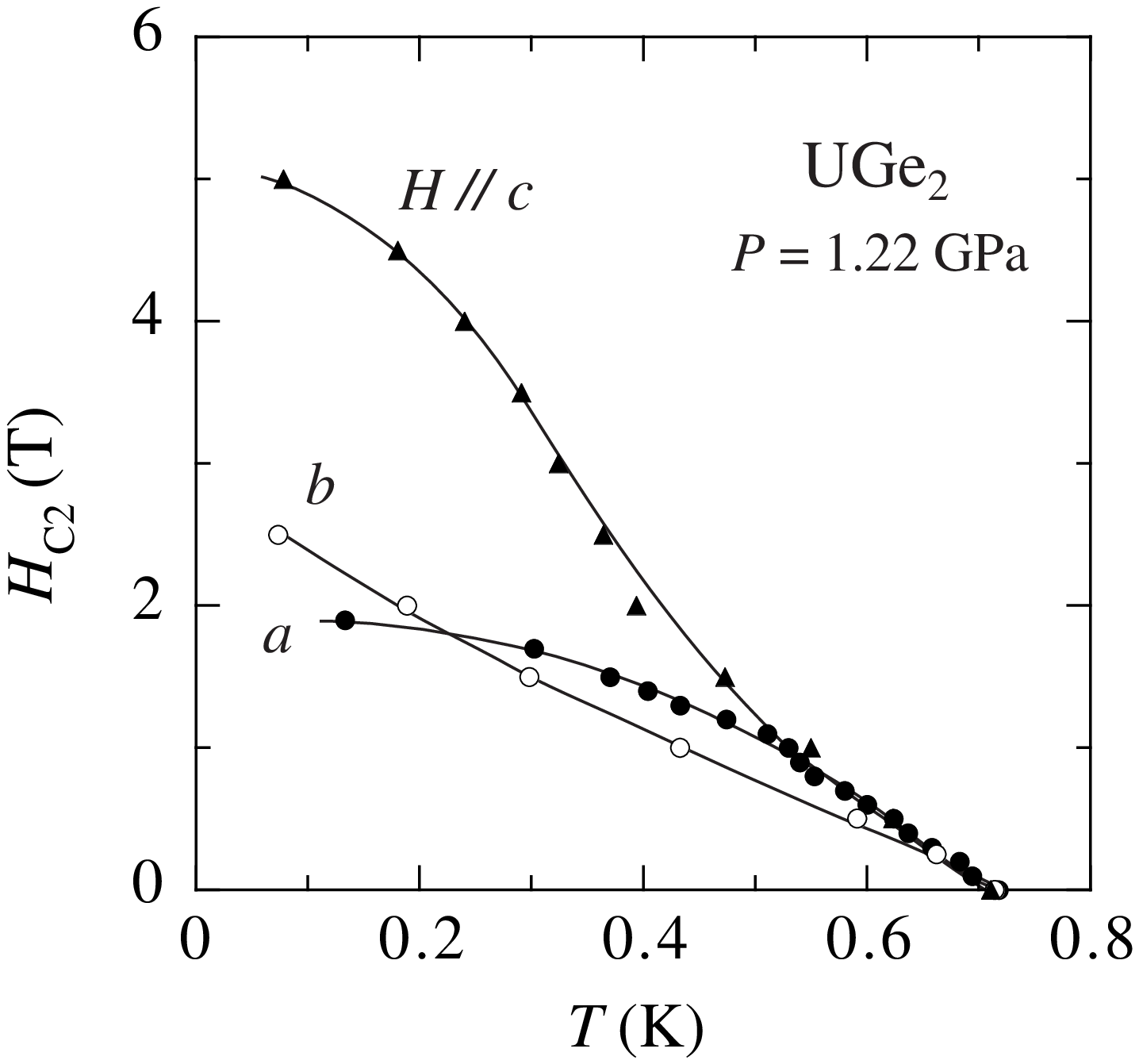}\hfill
\includegraphics[width=.48\columnwidth]{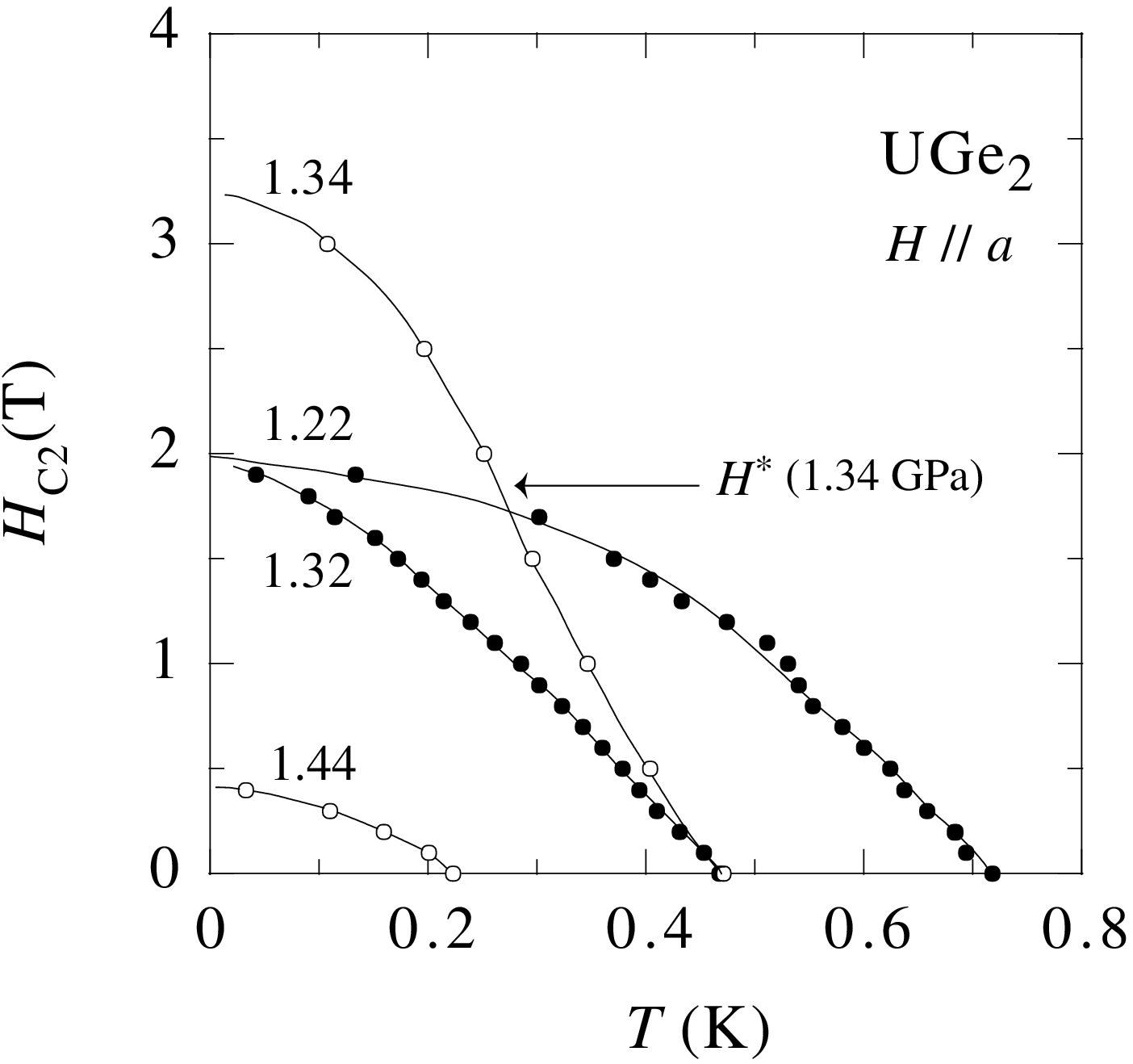}
\end{minipage}
\caption{Left panel: H$_{c2}$ - curves for pressure close to p$_m$ and
field along the crystal axes.
Right panel: H$_{c2}$ for field parallel to easy a-axis at four
different pressures below and above p$_m$ = 1.25 GPa. Unusual H$_{c2}$
at 1.34 GPa leads to SC reentrance as function of pressure
for low temperature (Kobayashi et al. 2001). 
\index{upper critical field!UGe$_2$}\index{reentrance behaviour}}
\label{FIGuge2hc2tp}
\end{figure}

The experimental phase diagram of FM collapse under pressure and
simultaneous appearance of superconductivity is shown in
fig.~\ref{FIGuge2ptdiagr}. The critical pressure for disappearance of
FM order is p$_{c2}$ = 16 - 17 kbar. The SC phase appears between
p$_{c1}$ = 10 kbar and p$_{c2}$ = 16 kbar which is also the critical
pressure for the FM-PM transition. The
critical temperature T$_x$(p) of the x-phase hits the maximum of T$_c$(p) at
the optimum pressure p$_m$ = 12.5 kbar. As mentioned before the nature of
the order parameter in the x-phase remains elusive. The coincidence
of maximum T$_c$ with
vanishing x-phase order parameter suggests that the collective
bosonic excitations of the x-phase which supposedly become soft at
p$_m$ mediated superconductivity and not quantum critical FM spin
fluctuations which are absent due to the persisting large FM molecular
field. Associated with T$_x$(p) phase boundary is an abrupt first
order change of the magnetic moment $\mu(p)$ around p$_m$ whose different size
distinguishes the FM2 region ($\mu(0)\simeq$ 1.45 $\mu_B$) from the FM1
region ($\mu(p_m)\simeq$ 1 $\mu_B$) \cite{Pfleiderer02}.
The connection between SC and the x-phase is also
witnessed by the anomalous upper critical field behaviour around
p$_m$ which displays a reentrance behaviour as function of pressure
seen in fig.~\ref{FIGuge2hc2tp}. Apparently close to the T$_x$ line the
superconducting state is strongly stabilized as shown by the
dramatically increased value of H$_{c2}$(0). The existence of phases
with `hidden order parameters' which show up in thermodynamic and
transport coefficients but do not appear as spin- or charge-density
modulations in neutron diffraction is not uncommon for U-compounds. As
discussed in the next section URu$_2$Si$_2$ has a \index{hidden order}
similar phase. Two kinds of proposals for such hidden order may be
considered: (i) if the U 5f electrons have a partly localised
character as in the latter case the hidden order may be of quadrupolar
or more generally multipolar character. (ii) If 5f-electrons are
strongly itinerant, as apparently in \UGE, one may have an unconventional
density wave order parameter which has been introduced in
sect.~\ref{Sect:Theory}. These phases have \vk-dependent gap
functions which do not belong to identity representations as in the
CDW/SDW case. This type of order has also been proposed for the yet
unidentified x-phase in \UGE and its coexistence behaviour with FM
and unconventional superconductivity was studied \cite{Varelogiannis03}.

\subsubsection{Theoretical scenarios for superconductivity in \UGE}

From the previous discussion of experimental evidence the 
superconducting mechanism via ferromagnetic fluctuations can be ruled
out since the ferromagetic polarization at p$_m$ where T$_c$ is
largest is still 65\% of the maximum value at ambient
pressure \cite{Pfleiderer02}. Therefore models based on the FM quantum
critical point scenario \index{quantum critical point} that follow the
original work of \cite{Fay80}
are not relevant for UGe$_2$. On the other hand the phenomenological
classification of possible SC order parameters based on Landau
expansion of the free energy is simplified by the remaining large spin
polarisation of conduction electrons, only spin triplet pairing is
possible in order to avoid the effect of the large exchange field.

\subsubsection{Symmetry properties of gap states and Ginzburg-Landau
theory} \index{Ginzburg-Landau theory}

Triplet pair states are characterised by the vector gap function
\vd(\vk) defined in sect.~\ref{Sect:Theory}. Due to the constraint of equal
(pseudo-) spin pairing (i.e. $\Delta_{\ua\da}\equiv$ 0) the \v d-vector is
confined to the bc-plane perpendicular to the FM moment \v M$_0$ which is
parallel to the easy a-axis. Neglecting the small orthorhombicity of
the ab-plane the gap function can be written \index{triplet pairing}
as \cite{Machida01,Fomin01} 
\begin{eqnarray}
\v d(\v k) &=&f(\v k)\boldeta =f(\v k)(\hat{\v a}+i\hat{\v b})
\end{eqnarray} 
Here $\boldeta$=($\eta_x,\eta_y,\eta_z$)= (1,i,0) is the vector order
parameter and f(\vk) the orbital part which transforms as a
representation of the (approximate) tetragonal group. As in the case
of UPt$_3$ it is assumed that the pseudo spins describing the Kramers
degeneracy of quasiparticle states are only weakly coupled to the
orbital momentum,
then f(\v k) may belong to any tetragonal representation which have
typical point or line node structures. Presently there is no
experimental information on the nodes. The equal spin pairing state
is nonunitary, i.e. it breaks time reversal symmetry. \index{time
reversal symmetry} This property is
directly enforced by the nearly complete FM polarisation. Therefore
the Cooper pairs have a net spin moment \v S = i\v d(\v k)$\times
\v d(\v k)^*$ which aligns with the ordered FM moment. In a Ginzburg
Landau expansion of
the free energy it is sufficient to consider only the aligned component 
$\eta_+ =\eta_x+i\eta_y$ and neglect $\eta_-=\eta_+^*$ because of the
large \v M$_0$ even around the critical pressure. The total GL
functional  should then only be expanded in terms of $\eta_+$ and
should also not contain \v M$_0$ explicitly except in the vector potential. One
obtains \cite{Machida01} for \v H along the c - axis ($\perp$ \v M$_0$), 
\begin{eqnarray}
\label{UGEGL}
f_{GL}&=&
\alpha_0(T-T_c)|\eta_+|^2+\frac{1}{2}\beta|\eta_+|^4\nonumber\\
&&+K_1\Bigl(\frac{d\eta_+}{dx}\Bigr)^2+\Bigl(\frac{2\pi}{\phi_0}\Bigr)^2
K_2(M_0+\mu_0H))^2x^2|\eta_+|^2
\end{eqnarray} 
because in this case the magnetic induction in the FM state is given by \v B
=(0,0,$\mu_0$H+M$_0$). Similar expressions hold for \v H along a-and
b-axis. Minimization of eq.~(\ref{UGEGL}) leads to the critical field
curves close to T$_c$. The term $\sim$ M$_0^2|\eta_+|^2$ only renormalizes
T$_c$. Then one obtains \index{upper critical field} the upper
critical fields H$^a_{c2}\sim$(T$_c$-T)
and H$^{b,c}_{c2}\sim$(T$_c$-T)$^\frac{1}{2}$. The remarkable root
singularity for H$^{b,c}_{c2}$ at T$_c$ is due to the presence of a finite FM
moment. Experimentally the critical field curves are quite anomalous
close to the optimal pressure p$_m$ although different
exponents for different field directions are not observed, the
H$^{c}_{c2}$-curve, however, does show an anomalous strong upturn around
0.5T$_c$ \cite{Kobayashi01}, see fig.~\ref{FIGuge2hc2tp}. In any case a
proper theory of H$_{c2}$
has to include the influence of the hidden order parameter close to
maximum T$_c$ (fig.~\ref{FIGuge2ptdiagr}). The symmetry considerations of this
section gave some constraints one should expect to hold for the SC
order parameter, however the simple Landau theory approach only
indicates that H$_{c2}$ should behave anomalous, without being able to
give a quantitative account close to the optimum T$_c$.

\subsubsection{Microscopic approaches}

Aside from the FM QCP scenarios now known to be irrelevant for
\UGE\cite{Pfleiderer02} little theoretical investigation has been
undertaken sofar with the exception of \citeasnoun{Watanabe02}. This theory
starts from the assumption that T$_x$(p) represents the phase boundary
of a coupled (conventional) CDW-SDW transition. Because of the strong
background FM polarisation these two order parameters have to appear
simultaneously. Inversely, if one is close to the critical pressure
p$_m$ their critical fluctuations will be coupled to fluctuations in
the FM polarisation by a mode-mode coupling term in the free energy,
which couples the amplitudes of FM, CDW and SDW fluctuations 
at the commensurate nesting vector \v Q of a  n.n.n tight binding
model used for the majority bands of \UGE. 
The strong coupling theory for this mechanism has been used to
calculate H$_{c2}$ and apparently qualitative anomalies like those
close to p$_m$ (fig.~\ref{FIGuge2hc2tp}) are obtained. 

However, no evidence for CDW or SDW formation central to this
theory has been seen below T$_x$ sofar. On the other hand in specific
heat measurements a pronounced anomaly $\Delta$C(T$_x$)
at 1.15 GPa suggests that T$_x$(p) is a real phase line \cite{Tateiwa03}. This
leads one to the natural suggestions that one should look for more
general particle-hole pairing ('unconventional density wave')
discussed in sections~\ref{Sect:Theory} and \ref{Sect:URu2Si2} as an
alternative for the phase below T$_x$ since they do not lead to charge
or magnetic superstructures \cite{Varelogiannis03}.

\subsection{A case of `Hidden Order' in \URU}
\label{Sect:URu2Si2} \index{hidden order}\index{URu$_2$Si$_2$}

This moderate HF compound ($\gamma$ = 110 mJ/moleK$^2$) has mystified both
experimentalists and theorists alike since the discovery of
AF order at T$_m$ and another still unidentified ('hidden order')
phase at T$_0$ which both seem to appear at the same temperature T$_0$
= T$_m$ = 17.5K, at least for annealed samples. In addition the
compound becomes a nodal superconductor below T$_c\simeq$ 1.2K
\cite{Palstra85,Schlabitz86,deVisser86}.
The simple tetragonal AF order with wave vector \v Q = (0,0,$\frac{\pi}{c}$)   
has tiny moments $\mu\simeq$ 0.02$\mu_B$ along c-axis
\cite{Broholm87,Walker93} which are of the same
order as in UPt$_3$. However, unlike in UPt$_3$, very large
thermodynamic effects,
e.g. in specific heat ($\Delta$ C/T$_0\simeq$ 300 mJ/molK$^2$),
thermal expansion etc., occur which are hard to reconcile with the
small ordered moments. The pronounced anomalies at T$_0$ were
interpreted as evidence for the presence of a second `hidden order'
parameter which cannot be seen in neutron or x-ray
diffraction. Many different types of models for hidden order have been
proposed, where the
5f-electrons of \URU are considered as essentially localized
\cite{Santini94}, or itinerant \cite{Ikeda98} or, of dual
nature \cite{Okuno98,Sikkema96}. In the former case local quadrupoles of
the CEF states are supposed to show staggered order below T$_0$ akin
to the many examples of such order in 4f-compounds.
Quadrupolar order does not break time reversal symmetry and cannot be directly
seen by neutron diffraction. The small dipolar moments are considered
as an unrelated secondary order parameter with accidentally the same
transition temperature. 
In the itinerant models the order parameter is due to an unconventional
pairing in the particle-hole channel leading to an unconventional SDW
which has vanishing moment in the clean limit and also does not break
time reversal invariance. The small staggered moments may then 
be induced in the condensate due to impurity scattering. Finally in
the dual models one assumes a localised singlet-singlet system in
interaction with the itinerant electrons to cause induced moment
magnetism with small moments but large anomalies. 

In all models it was previously taken for granted that both the primary
`hidden' order parameter and AF order coexist homogeneously within the
sample. However, hydrostatic and uniaxial pressure experiments
\cite{Amitsuka99,Amitsuka02} have radically changed this view,
showing that the order parameters exist in different parts of the sample
volume; the tiny AF moment is not intrinsic but due to the small
AF volume fraction under ambient pressure. Applying hydrostatic pressure
or lowering the temperature increases the AF volume fraction and hence
the ordered moment until it saturates at an atomic size moment of
0.4$\mu_B$/U. This means that the evolution of AF arises from the
increase of AF volume with pressure rather than the increase of the
ordered moment $\mu$ per U-atom. This interpretation is supported by the
observation of a comparatively weak increase of T$_0$ with
pressure \cite{Amitsuka99}.

\subsubsection{Electronic structure and 5f-states}
\label{sub1URU}

Inelastic neutron scattering \cite{Park02} using the time-of-flight method has
shown that the valence state in \URU is U$^{3+}$ corresponding to
5f$^2$, judging from the observation of the  $^3$H$_4\rightarrow^3$F$_2$
transition of this configuration. In tetragonal (D$_{4h}$) symmetry
the ninefold degenerate $^3H_4$ multiplet should be split by the CEF
potential into five singlets and two doublets.
There is indeed an indication of four strongly broadened CEF transitions
at energies ranging from $\delta$ = 5 meV to 159 meV. However, the data 
are not sufficient to determine the CEF potential and states. There is a
transition at 49 meV which might be associated with the $\Gamma_3$-
$\Gamma_5^1$ transition from the $\Gamma_3$ ground state
\index{crystalline electric field excitations}
which has been invoked in the model of \citeasnoun{Santini94}. Earlier,
low energy triple axis INS \cite{Broholm91} has shown that the
transitions have a considerable dispersion. 
In addition the c-axis susceptibility exhibits clear CEF anomalies from
which a CEF level scheme has been derived \cite{Santini94}. However
the result is not unique and the overall splitting obtained is by a factor two
smaller as compared to the above INS results. In this analysis the
low energy group of CEF states consists of three singlets shown in
table~\ref{tab:URUCEF}.

\begin{center}
\begin{table}[htb]
\caption{Model for low lying CEF states in \URU (tetragonal structure
as in fig.~\ref{fig:ThCr2Si2Structure}). 
The complete CEF level system for concentrated \URU is comprised
of five singlets (the low lying states are listed with $\epsilon$ and
$\gamma$ being adjustable parameters) and two doublets. The dilute
(x$\ll$1) Th$_{1-x}$U$_x$Ru$_2$Si$_2$ system may have a different
single-ion CEF level scheme with a doublet ground state 
(Amitsuka et al. 1994). \index{crystalline electric field excitations}
\index{Th$_{1-x}$U$_x$Ru$_2$Si$_2$}}
\vspace{0.5cm}
\label{tab:URUCEF}
\begin{tabular}{lcr}
\hline
symmetry     & CEF-state & energy (meV) \\
\hline
$\Gamma_3$   & $|0\ra =\frac{1}{\sqrt{2}}(|2\ra +|-2\ra)$  & 0 \\
$\Gamma^1_1$ & $|1\ra =\epsilon|4\ra +|-4\ra +\gamma|0\ra$ & 3.8\\
$\Gamma_2$   & $|2\ra =\frac{1}{\sqrt{2}}(|4\ra -|-4\ra)$  & 9.6\\
\hline
\end{tabular}\\[2pt]
\end{table}
\end{center}

The clear evidence for CEF states supports the view of mostly localised
5f-states in \URU. This raises the question of the origin of HF
behaviour in this compound. The specific heat coefficient in the low
temperature ordered (magnetic+hidden) phase is reduced to only a
moderate $\gamma$ = 64 mJ/molK$^2$, 40\% of the
paramagnetic value and only one tenth that of typical HF
values. Magnetoresistance measurements have shown that \URU is a
compensated metal without open orbits \cite{Ohkuni99}. 
Three of the closed Fermi surface sheets have been 
\index{de Haas-van Alphen effect} determined by dHvA experiments
\cite{Ohkuni99}. Their 
moderately heavy masses are in the range m$^*$= 8-25 m which are,
however, still larger by a factor ten as compared to the LDA band
masses. For a nonmagnetic 5f singlet ground state
system as given in table~\ref{tab:URUCEF} there is no Kondo mechanism to
generate heavy quasiparticles. It is tempting to assume that rather
the same mechanism as recently proposed for UPt$_3$ \cite{Zwicknagl02}
and \UPD \cite{Zwicknagl03} \index{mass enhancement}
is important: mass renormalization through virtual excitations within
a pair of low lying 5f singlet states which leads to \cite{White81} 
\begin{eqnarray}
\label{EFFMASS}
\frac{m^*}{m}= 1+N(0)\frac{2\alpha^2}{\delta}
\end{eqnarray}
where $\alpha$=$\la 0|J_z|1\ra$ is a dipolar matrix element and
$\delta$ the singlet-singlet splitting. This is a simplified version
of the mass renormalisation in eqs.~(\ref{EFFMASS1}),(\ref{EFFMASS2})
with $\alpha\equiv$ 2$a_{5f}|M|$.
For several singlet pairs as in the present case one has a sum of such
contributions. As discussed for UPt$_3$ for realistic parameters of
U-compounds this mechanism may easily lead to a mass enhancement
factor $\frac{m^*}{m}\simeq$ 10 which would be an appropriate value for \URU. 
 
\subsubsection{Phase transitions, field and pressure dependence}
\label{sub2URU}

The discrepancy between small AF moment in \URU and large
thermodynamic anomalies has led to the postulation of a hidden
(non-dipolar) order parameter. The crucial questions about its nature
are: i) is the order
primarily involving the localised 5f-CEF split states or the
heavy itinerant conduction electrons. ii) Does the hidden order parameter
break time reversal invariance or not. In the former case it may
induce AF as secondary order parameter, in the latter the two order
parameters are unrelated and their appearance at the same temperature
T$_m$= T$_0$ has to be considered as accidental.

\begin{figure}
\includegraphics[clip,width=7.5cm]{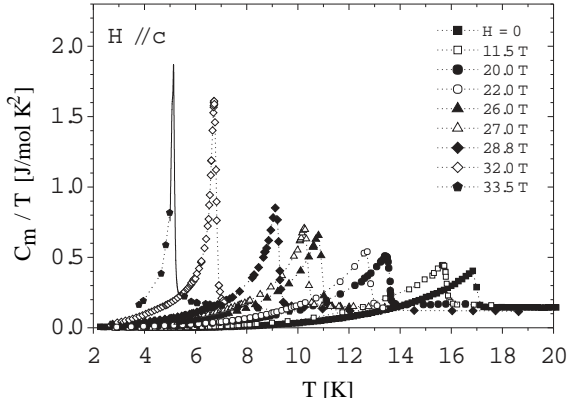}
\caption{Magnetic specific heat C$_m$/T for fields up to 33.5
T (Jaime et al. 2002). \index{specific heat!URu$_2$Si$_2$}}
\label{FIGURUspec}
\end{figure}

The continuous phase transition at T$_0$ is clearly seen in a large
specific heat anomaly which becomes more pronounced in a magnetic
field (fig.~\ref{FIGURUspec}). The magnetic entropy contained in
this peak is orders of magnitude larger compared to other magnetic
U-compounds and alloys if one scales it with the AF ordered
moment \cite{Amitsuka02} which proves that it must be connected to a hidden
order parameter. A similar behaviour is seen in thermal
expansion \cite{Mentink97} along a, however, the tetragonal symmetry
below T$_0$ is preserved and no superstructure
evolves \cite{Kernavanois99}. The shape of specific heat and thermal
expansion anomalies and their sharpening in an external field is
reminiscent of antiferroquadrupolar (AFQ) phase transitions due to
localised f-electrons known in the 4f-Ce-hexaborides and
Tm-intermetallics. A quadrupolar order parameter is even under time
reversal and is expected to be in competition with the dipolar AF order.

\begin{figure}
\includegraphics[width=7.5cm]{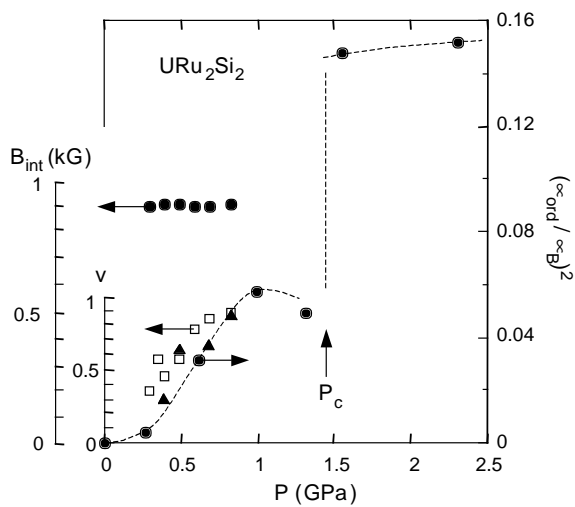}
\caption{Evolution of AF Bragg intensity (full circles, right scale)
with hydrostatic pressure. Full circles (leftmost scale) denote
internal staggered field B$_{int}$ from $^{29}$Si NMR. Triangles and
squares denote AF volume fraction v from the
same experiment (Amitsuka et al. 2002). 
\index{hydrostatic pressure!URu$_2$Si$_2$}}
\label{FIGURUmoment}
\end{figure}

Pressure experiments have indeed found this competition. It is
evident from the comparison of the pressure dependent ordered AF
moment, and the position and intensity of $^{29}$Si-NMR
satellites. While the former increases continuously with pressure, the
splitting which is proportional to the local 5f-moment is pressure
independent, however, the split satellite intensity which is a measure
of the AF volume fraction also increases with pressure
(fig.~\ref{FIGURUmoment}). Analysis of these results show
that the AF moment per magnetic U-site $\mu_{AF}\simeq$ 0.25$\mu_B$/U
is not small. It is the small AF volume fraction at ambient pressure
which leads to the small overall
ordered moment of 0.03$\mu_B$. As pressure increases the ordered
moment also increases due to the increase in AF volume which
continuously replaces the volume fraction with the hidden order. The
replacement is complete at p $\simeq$ 1 GPa. Somewhat above at p$_c$
another phase transition takes place in the complete AF-phase leading
to a sudden increase of the on-site U-moment, however we note that this has not
been observed in magnetic x-ray diffraction experiments. The observed moderate
increase of T$_m$ with pressure \cite{Amitsuka99} is compatible with
the observation of a pressure independent U-moment.
Subsequent uniaxial pressure experiments \cite{Yokoyama02} have shown
that only the [100] or [110] uniaxial pressure in the tetragonal plane
leads to the increase of the AF volume fraction or destruction of
the hidden order, whereas the [001] uniaxial pressure has little
effect. This anisotropic behaviour under pressure suggests that the
hidden order parameter is associated with the tetragonal
fourfold symmetry plane.

\subsubsection{Theoretical models: localised vs. itinerant}
\label{sub3URU}

We shall not recount the many attempts to explain the broken symmetry
states of \URU but rather concentrate on two typical examples that
seem to be compatible with the recent important results of the
pressure investigations just described.\\

{\em AFQ order of local induced quadrupole moments}
\index{antiferroquadrupolar order}

The competitive behaviour of AF and hidden order points to even time
reversal symmetry of the latter. If one assumes that the localised
5f electrons are involved in the hidden order then the already
mentioned AFQ order in the singlet CEF level scheme is the most
obvious choice. This model was proposed by \citeasnoun{Santini94} and
\citeasnoun{Santini98} and is also the most well studied one. It
is an entirely localised 5f-model and is defined by the Hamiltonian
\begin{eqnarray}
\label{QUADRU}
H&=&\sum_i\sum_{k,q}B^q_kO^q_k(i)+ K\sum_{i\in A,j\in B}Q(i)Q(j)
-\sum_ig\mu_BJ_z(i)H
\end{eqnarray}
whose terms represent CEF potential in Stevens operator representation,
quadrupolar interactions and Zeeman energy
respectively. K is an effective quadrupolar inter-site
exchange mediated by conduction electrons, similar as in 4f-systems
\cite{Thalmeier91a}. The CEF parameters B$^q_k$ are obtained from
fitting to the susceptibility which leads to the CEF states in
table~\ref{tab:URUCEF}. Since the ground state is a singlet, quadrupole
order can only appear as {\em induced} order, i.e. nondiagonal matrix
elements $\la\Gamma_3|Q|\Gamma^1_1\ra\neq$ 0 must be present. This is
possible for quadrupole operators Q$_{\Gamma_3}$ = (J$^2_x$-J$^2_y$) or   
 Q$_{\Gamma_5}$ = (J$_x$J$_y$+J$_y$J$_x$). Naively this would be
compatible with
uniaxial pressure results which show that the hidden order parameter is
most sensitive to strains within the tetragonal xy-plane.
Mean field calculations were performed with eq.~(\ref{QUADRU}) based
on the above CEF scheme under the assumption of two
sublattice AFQ ordering. They give a qualitatively correct
behaviour of the specific heat, thermal expansion, nonlinear and
anisotropic susceptibility  and magnetisation. For example, the
thermal expansion \index{thermal expansion} exhibts step like
anomalies at T$_0$(H) which increase with field strength. This is reproduced by
the model since in tetragonal symmetry (i = a,c)
$\alpha_i=g_i\partial\la O^0_2\ra/\partial T$
where Q = O$^0_2$ = 3J$_z^2$-J(J+1) and g$_i$ is an effective
magnetoelastic coupling constant. Close to T$_0$ this is proportional
to the growth rate $\partial\la Q\ra^2/\partial T$ which leads to
jumps in thermal expansion coefficient
$\alpha_a$ = (1/a)($\partial$a/$\partial T$) etc. reminiscent of those
in the specific heat.

In principle the AFQ order parameter may also lead to a lattice
superstructure with the AFQ vector \v Q via magnetoelastic coupling
terms. This has not been found \cite{Kernavanois99} but may simply
be too small to observe as it is the case in other AFQ compounds like
CeB$_6$. More disturbing is the fact that in the AFQ state an external
field in the ab-plane should also induce a field dependent staggered
magnetisation which has not been observed either.

In the AFQ hidden order model there is no natural connection to an AF
order parameter. For this purpose a dipolar interaction term has to be
introduced \cite{Santini98} and the resulting magnetic ordering
temperature T$_m$ is
unrelated to T$_0$, if they are equal this has to be interpreted as
accidental. In the light of the new pressure experiments it seems that
they are close at ambient pressure, slightly favoring AFQ order. For
p$>$ 0 T$_0$(p) and T$_m$(p) cross at the critical pressure p$_c$,
stabilizing the AF local moment phase for p $>p_c$. At ambient
pressure local stress around crystal imperfections might already lead
to a small AF volume fraction which then increases upon applying
external pressure. Such a situation may be phenomenologically described
by a Landau free energy \index{free energy} ansatz \cite{Shah00}
\begin{eqnarray}
f_L&=&\alpha_0(T-T_0(p))Q^2+\beta Q^4+\nonumber\\
&&\alpha'_0(T-T_m(p))m^2+\beta'm^4+ \gamma_0 m^2Q^2   
\end{eqnarray}
where the last term is a coupling term for AF (m) and AFQ (Q) order
parameters. For $\gamma_0 >$ 0 the the two order parameters are in
competition. If T$_m$(p) and T$_0$(p) are close the inclusion of
inhomogeneities might then lead to a phase separation where m and Q order
parameters exist in macroscopically separate volume fractions. The
Landau approach has been extended by including the strain coordinates
to analyze neutron diffraction results under uniaxial pressure
\cite{Yokoyama03}. 

Although the AFQ scenario for the hidden order parameter seems most
attractive for the explanation of macroscopic anomalies at T$_0$ there
is no direct experimental proof. This would require the observation of 
either i) the induced AF magnetic order in an external field or ii)
direct signature of 5f-orbital order in resonant x-ray scattering as
e.g. in the AFQ phase of CeB$_6$ \cite{Nakao01}.\\ 

{\em Hidden order as unconventional density wave}
\index{hidden order}\index{unconventional pairing}

A complementary view of hidden order in \URU starts from a completely
itinerant view of 5f-electrons by describing them within a
multiband extended Hubbard model \cite{Ikeda98,Ikeda99}. Naturally this
approach is unable to account for the CEF signatures in specific heat,
susceptiblity and similar quantities. In this model the large jump
$\Delta$C(T$_0$) is due to a condensation of electron-hole pairs
(sect.~\ref{Sect:Theory}). Unlike
conventional CDW or SDW states however the pair states belong to
nontrivial anisotropic representations of the symmetry group, similar
to  Cooper pairs in unconventional superconductors. The symmetry
classification of unconventional electron-hole pair states and their
stability analysis is given in \cite{Gulacsi87,Schulz89,Ozaki92}. In
full generality this \index{nesting property}
has only been done for n.n. tight binding  models with perfect nesting
property of 2D conduction bands $\epsilon_{\v k \pm\v Q}= -\epsilon_{\v
k}$ and \v Q = ($\frac{1}{2},\frac{1}{2}$,0). In the low energy corner
of the U-V phase diagram (U,V= on-site and n.n. Coulomb interaction
respectively) the stable
state is an unconventional triplet particle-hole pair condensate
('d-SDW'). In
contrast to a common SDW where the gap function $\Delta^s$(\v k) = const is
momentum independent, in the d-wave case one has $\Delta_{s_1s_2}^d(\v
k)=\Delta^d(\v k)\sigma^z_{s_1s_2}$ with 
\begin{eqnarray}
\label{dSDW}
\Delta^d(\v k)=i\Delta^d_0(\cos k_x-\cos k_y)
\end{eqnarray}
The order parameter is purely imaginary because it is connected with
persistent commensurate spin currents around lattice plaquettes. This
state breaks spin roational symmetry but not time reversal symmetry
because the latter \index{time reversal symmetry}
inverses both current and spin direction leaving the order parameter invariant.
The staggered magnetic moment in the s-or d-wave case is given by
\begin{eqnarray}
\label{STAGGER}
M_Q^z&=&
\sum_{\v k\sigma}\sigma\la c^\dagger_{\v k\sigma}c_{\v k +\v Q\sigma}\ra
=\sum_{\v k}\frac{\Delta^{d,s}(\v k)}{2E(\v k)}th\frac{E(\v k)}{2T}
\end{eqnarray}
where E$_{\v k}$ = [$\epsilon_{\v k}^2+\Delta^2_{\v k}$]$^\frac{1}{2}$ is
the quasiparticle energy which is fully symmetric in both s,d
cases. Therefore M$_Q^z>$ 0 for conventional s-SDW but M$_Q^z\equiv$ 0
for d-SDW case since the sign change in eq.~(\ref{dSDW}) as function
of \v k makes the integral in eq.~(\ref{STAGGER}) vanish for d-SDW
(this notation convention is a misnomer since M$_Q^z\equiv$ 0 means that
there is precisely {\em no} `spin density' in the d-wave case).
 
Despite the vanishing staggered moment in the d-SDW state there is a
large BCS-like specific heat anomaly similar to the conventional s-SDW
case. Therefore the d-SDW state would naturally explain the most
prominent property of the hidden order phase in \URU. The tiny moments
would then immediately appear at the hidden order transition
(T$_m\equiv$ T$_0$) around impurities, either through the creation of
a s-SDW component by a proximity effect \cite{Ikeda98} or through anisotropic
exchange interactions of the condensate with magnetic
impurities \cite{Virosztek02}. In the latter case the temperature
dependence of the staggered moment can be explained. The phase
separation of hidden order and staggered micromagnetism is then again
seen as a consequence of the sample defects and inhomogenous impurity
concentration. 

\subsubsection{High field phase diagram and metamagnetism}
\label{sub4URU} \index{metamagnetism}

\begin{figure}
\includegraphics[width=7.5cm]{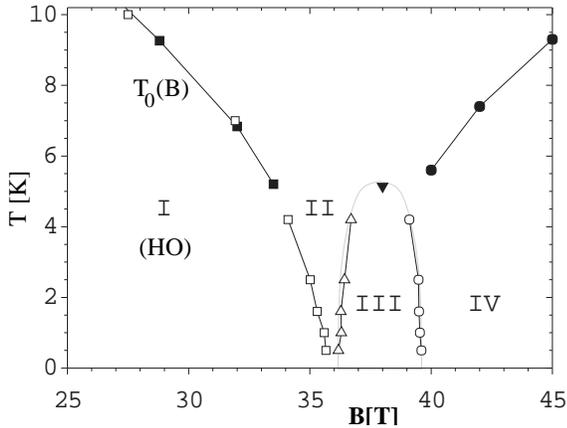}
\caption{B-T phase diagram of hidden order (HO) phase (I), metamagnetic
phase (III) and high field phase (IV). Open symbols correspond to
resistivity anomalies and filled symbols to specific heat maxima. For
zero field T$_0$ = 17.5 K(HO) (Jaime et al. 2002). 
\index{B-T phase diagram!URu$_2$Si$_2$}}
\label{FIGURUmeta}
\end{figure}

Besides pressure, magnetic fields also destroys the hidden order as can
be seen from the progressive T$_0$(B)-reduction in
fig.~\ref{FIGURUspec}. The corresponding critical field line
forms the boundary to the hidden order phase (I) in fig.~\ref{FIGURUmeta}.
Assuming the AFQ hidden order model of \URU there is an important
difference to other known AFQ
compounds which have a {\em degenerate} CEF ground state like cubic
$\Gamma_8$ in CeB$_6$. There T$_0$(B) first {\em increases} with field
strength due to the lifting of orbital degeneracy. Since in \URU T$_0$(B)
decreases monotonously this is a strong argument for the induced AFQ
order in a singlet-singlet system.  At low temperatures the hidden
order phase is destroyed at a field B$_0$(0) = 36.1 T. For B $>$ B$_0$(0)
M(B) exhibits multistep metamagnetic behaviour \cite{Sugiyama99} which
may in principle be explained by the AFQ model as a field induced crossing of
singlet CEF levels \cite{Santini98} complemented by the effect of
inter-site exchange interactions. Originally it was thought that the
metamagnetic transition extends to temperatures much higher than
T$_0$ \cite{Sugiyama99}. Recent pulsed field experiments
\cite{Jaime02} have shown however that the metamagnetic phase is much
more confined (region III in fig.~\ref{FIGURUmeta}). Alternatively
this region has also been interpreted as reentrant hidden order phase
\cite{Harrison03}. Finally above 40 T one reaches a moment
of 1.5$\mu_B$/U which is much larger than the zero field value at high
pressure, hence at high field not only the AF volume fraction is
maximal but also the magnetic ground state of U$^{4+}$ has a larger
moment. The structure of magnetic phases in the high field regime is
still unknown. 

\subsubsection{Collective excitations in the ordered phase}
\label{sub5URU}

Inelastic neutron scattering 
\index{inelastic neutron scattering!URu$_2$Si$_2$} has perhaps
provided the most convincing
evidence that the localised 5f electron picture is a good starting
point for \URU. In this case the dipolar CEF transition from the
singlet ground state $|0'\ra$ to excited singlet $|2'\ra$ with dipolar
matrix element $\alpha$ = $\la 0'|J_z |2'\ra$  should disperse into a
\index{magnetic exciton} collective `magnetic exciton'
band. Here the primes indicate that the corresponding singlet states
in  table~\ref{tab:URUCEF} are modified due to the presence of AF order
which mixes the unprimed states with higher CEF states, without this
mixing
obviously $\alpha$ would vanish. These magnetic excitons have indeed been
found \cite{Broholm91} as well propagating modes in a large part of the
tetragonal BZ. They may be described by a
similar expression as magnetic excitons for UPd$_2$Al$_3$. The
relatively large matrix element $\alpha$ = 1.2 obtained from fitting
to experimental results \cite{Broholm91} is not in contradiction to the
small overall AF moment, since the latter is only an effect of the
small AF volume fraction. The $|0'\ra$ to $|1'\ra$ singlet excitation
has no dipolar matrix element, it is the excitation whose softening 
is governed by the quadrupolar matrix element $\alpha_Q=\la
0'|Q|1'\ra$ leading to the induced AFQ (hidden) order at T$_0$. This
quadrupolar mode does not directly appear in the dipolar response
function and the INS cross section. It has been suggested that its
presence can be seen indirectly through an influcence on the dipolar
excitation \cite{Santini00}.

\subsubsection{The superconducting state}
\label{sub6URU}
As in other U-HF compounds the superconducting state in \URU  with
T$_c$ = 1.4 K is embedded in the AF phase and here in addition in the
hidden order phase with T $_0\gg$ T$_c$. Although there are a number of
signatures for an unconventional pair state, it has attracted much
less attention than the hidden order phase, possibly because there is
no direct evidence for a multicomponent SC order
parameter \cite{Thalmeier91b}. The evidence for gap anisotropy is
obtained from various low temperature power laws for specific
heat \cite{Hasselbach93}  ($\sim$T$^2$) and NMR relaxation
rate \cite{Matsuda96} ($\sim$T$^3$) which suggests the
presence of line nodes. A plot of C/T vs. T is shown in
fig.~\ref{FIGOPDspec}
in comparison with theoretical curves \cite{Hasselbach93} obtained 
for gap functions allowed in the Landau theory of tetragonal
superconductors \cite{Volovik85}. The region of experimental linear T
behaviour in C/T is surprisingly large. This cannot be
explained within a Landau approach which is restricted to the vicinity
of T$_c$. In addition the reduced specific heat jump $\Delta$C(T$_c$) as
compared to a constant gap supports the existence of a gap anisotropy
with line nodes. This is also the conclusion from point contact
spectroscopy \cite{Hasselbach92}. 
The upper critical field curves of \URU show two anomalies \cite{Keller95}:
First the anisotropy ratio B$^a_{c2}$/B$^c_{c2}$ = 4 is large, it 
cannot be fully explained by the anisotropy of Pauli limiting fields
given by $\sqrt{\chi_c/\chi_a}\simeq\sqrt{5}$, orbital effects due to
effective mass antisotropy must be involved. Furthermore, a distinct upward
curvature of B$^a_{c2}$ is observed, similar to critical fields in the
borocarbide superconductors where it was associated with two band
effects or alternatively anisotropic Fermi velocities. The effect of
AF on B$_{c2}$ should be negligible due to its small volume fraction.

As discussed in sect.~\ref{Sect:Theory} the observation of the dHvA effect
far in the vortex phase (B $\ll$ B$_{c2}$) is a sure signature of nodal
superconductivity. Although oscillations of the three Fermi surface
sheets in \URU have been seen below B$_{c2}$ \cite{Ohkuni99}  the
amplitude falls of
quite rapidly with B, especially for field along c. Therefore, these
experiments are not able to confirm the existence of nodes in $\Delta$(\v k).

\subsection{Superconductivity in the non-Fermi liquid state of \UBE
and \UTB} \label{Sect:UBe13}
\index{UBe$_{13}$}\index{U$_{1-x}$Th$_x$Be$_{13}$}

This cubic compound was discovered rather early
\cite{Ott83,Ott84} as a superconducting HF system. The U atom in this
structure is embedded in an icosahedral cage of 12 Be-atoms.
A global understanding of the normal state and
symmetry breaking in both superconducting and magnetic state is still elusive.
Firstly `pure' \UBE crystals do not have the highest quality as
compared to e.g. \UPT so
that the symmetry of the anisotropic SC gap functions has not been
identified, furthermore the Th-doped crystals \UTB show a perplexing
variety of SC and possibly also magnetic phases whose microscopic origin and 
order parameter symmetries are not understood. The T-x phase diagram
of \UTB has been investigated with a variety of methods and the most
detailed results have been obtained using the thermal expansion method
 \cite{Kromer98,Kromer00b,Kromer02} (fig.~\ref{FIGUBTH13phases}). The most
important question concerns the nature of the low temperature phase
(C) at intermediate doping (0.02 $<$ x $<$ 0.045). This phase may
either be a nonunitary SC phase with condensate magnetic moments due
to unconventional Cooper pairs \cite{Sigrist89} or a phase with coexisting
anisotropic SC and a SDW type phase suggested by \citeasnoun{Kromer98}
on experimental grounds and already proposed in a theory by
\citeasnoun{Kato87}. In the former case one has to assume a
crossing of two SC phases with two independent unconventional order
parameters as function of the Th-concentration \cite{Sigrist91} and in
the latter a crossing of a SC and SDW phase line. Already in the normal
state \UBE is a rather anomalous metal, e.g. non-Fermi-liqud (nFl)
behaviour has been observed and attributed to a multichannel Kondo effect.

\subsubsection{Normal state and nFl properties of \UBE}
\index{non-Fermi liquid state}

The 5f-electron level in \UBE is close to the Fermi level as seen
from photoemission results. Therefore the 
HF-behaviour which sets in below a fluctuation temperature
of T$^*\simeq$ 8 - 25 K does not correspond well to the Kondo picture which
requires 5f states to lie sufficiently removed from the Fermi
energy. Nevertheless the Kondo model \index{Kondo resonance}
in its single and multichannel version has been employed for this HF compound.
The specific heat coefficient is strongly enhanced with
$\gamma\simeq$ 1J/molK$^2$. Furthermore resistivity, specific heat
and thermal expansion reveal the presence of a second low energy scale
with T$_m\simeq$ 2 K where these quantities exhibit an additional
maximum anomaly. This proves that at T$_m$ a \index{Fermi liquid}
coherent Landau FL 
state has not yet evolved and that the superconducting transition at
T$_c\simeq$ 0.9 K in \UBE happens within a strongly anomalous 
nFl state. In fact if superconductivity is suppressed by a strong magnetic
field the $\gamma$ value shows a roughly logarithmic increase with
decreasing temperature typical for a nFl state. No saturation of
$\gamma$(T) is observed down to lowest temperatures
\cite{Helfrich98}. In most other U-HF compounds superconductivty is
embedded in a weakly
AF state. Despite a number of attempts no long range AF order has been found in
UBe$_{13}$, $\mu$SR measurements set an upper limit of 10$^{-3}\mu_B$ for
the moments on U-sites.

\subsubsection{The 5f-ground state of U}

The observed normal state nFl behaviour raises the still controversial
question of the magnetic ground state of U-atoms in \UBE. The high
temperature susceptiblities do not allow to distinguish between a
5f$^3$ and 5f$^2$ state of U. In the former case proposed in
\citeasnoun{Felten86}
a Kramers degenerate magnetic $\Gamma_6$ CEF ground state and and
two excited  $\Gamma_8$ quartets at 180 K and still higher energy were
infered from analysis of specific heat which exhibits a CEF Schottky
anomaly around 80 K. In this case the HF state would be due to a
conventional Kondo mechanism involving the magnetic $\Gamma_6$
localised doublet. However this picture is in conflict with the
pronounced nFl anomalies mentioned above.
Indeed the nFl behaviour was taken as direct evidence for a 5f$^2$
configuration with a nonmagnetic $\Gamma_3$ doublet ground state which has a
nonvanishing quadrupolar moment \cite{Cox87,Cox88} whose fluctuations scatter
conduction electrons. The effective Hamiltonian is of the two-channel
Kondo type which leads to an overcompensation of quadrupolar
moments and therefore to typical logarithmic nFl anomalies in specific
heat and susceptibility at low temperatures \cite{Schlottmann93,Zvyagin00}. The
$\Gamma_3$ quadrupolar model would predict a positive nonlinear magnetic
susceptibility in $\chi_3$ with considerable anisotropy at low temperatures,
instead the opposite, namely almost isotropic, negative $\chi_3(T)$
was found \cite{Ramirez94}. This suggests that the ground
state is a magnetic Kramers doublet $\Gamma_6$. To explain nFl
behaviour one either has to invoke a magnetic multichannel Kondo
effect or closeness to a quantum critical point in \UBE.

\subsubsection{The superconducting state in \UBE}

The T$_c$ values for superconductivity which occurs in the nFl state
depend considerably on the type of sample. There are two classes with
`high' T$_c$ ($\simeq$ 0.9 K) and `low' T$_c$ ($\simeq$ 0.75) which,
however, are not much different in their impurity content. Low
temperature `power law' dependencies give conflicting information
on the node structure of the gap and no firm conclusion on whether spin
singlet or triplet pairing is realized in \UBE can be drawn. Therefore,
these results will not be discussed further here although an axial
order parameter with point nodes seems to be favored. The most direct
evidence obtained sofar for unconventional superconductivity is
connected with the giant ultrasonic absorption anomaly observed
directly below T$_c$ \cite{Golding85,Mueller86} which was attributed to
collective modes or domain-wall damping due to a multicomponent order
parameter \cite{Sigrist91}.

\begin{figure}
\includegraphics[width=75mm]{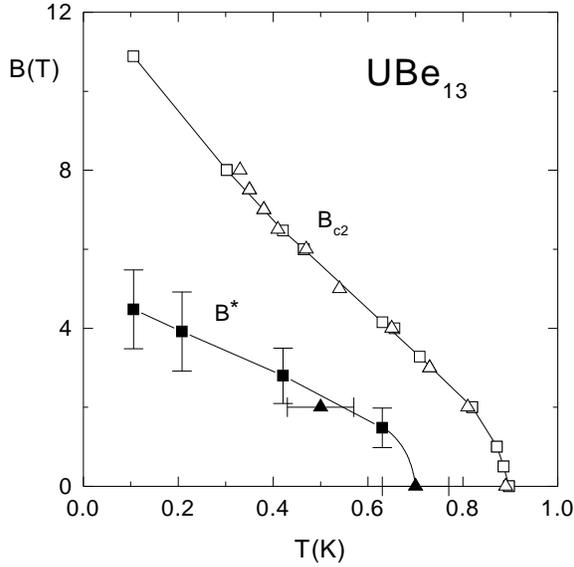}
\caption{Upper critical field B$_{c2}$ and magnetic anomaly line B$^*$
obtained from specific heat (squares) and thermal expansion (triangles)
experiments (Helfrich 1998, Kromer et al. 1998) for a `high' T$_c$
sample. \index{upper critical field!UBe$_{13}$}}
\label{FIGUBE13bccrit}
\end{figure}

The B-T phase diagram of \UBE presented in fig.~\ref{FIGUBE13bccrit}
has quite 
anomalous appearance: The upper critical field curve B$_{c2}$(T) shows an
inflection point around 0.45 K, furthermore deep in the
superconducting regime another anomaly line B$^*$(T) starting at
T$_L$= 0.7 K has been
identified both by specific heat \cite{Helfrich98} and thermal
expansion measurements \cite{Kromer98,Kromer00b}. 
\index{thermal expansion!U$_{1-x}$Th$_x$Be$_{13}$} This line might be 
connected to the onset of magnetic correlations which, however, do not
lead to long range order. This interpretation is supported by
recent thermal expansion results \cite{Kromer00b} who have shown that the
line can be followed as function of Th-doping and eventually,
according to this picture, long range SDW order appears above a Th
concentration of x$_{c1}$ = 0.02.

\subsubsection{Superconducting phase diagram of Th-doped crystals}

\begin{figure}
\includegraphics[width=75mm]{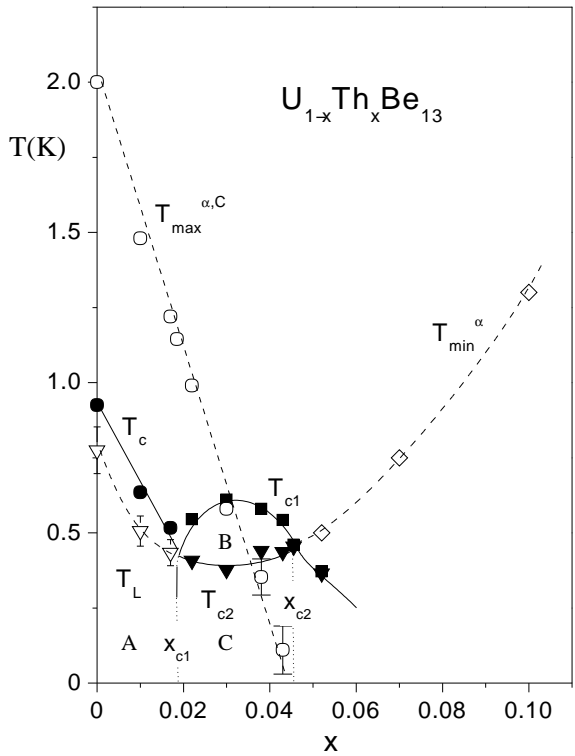}
\caption{Superconducting phase diagram of \UTB for x $<$ 0.10 (Kromer
et al. 2002).
full lines and symbols: thermodynamic phase boundaries;
broken lines and open symbols: lines of anomalies. T$_L$ and
T$^{\alpha}_{min}$ denote line of anomalies from minimum in the thermal 
expansion coefficient $\alpha$(T), T$^{\alpha ,C}_{max}$ is the line of
anomalies from the maximum in C(T) and $\alpha$(T). A, B and C denote
distinct SC phases in the T$_c$- crossing model. 
\index{thermal expansion!U$_{1-x}$Th$_x$Be$_{13}$}
\index{specific heat!U$_{1-x}$Th$_x$Be$_{13}$}}
\label{FIGUBTH13phases}
\end{figure}

The T-x superconducting phase diagram of the thorated \UTB-crystals
(x $\leq$ 0.10) whose most recent version \cite{Kromer02}, is shown in
fig.~\ref{FIGUBTH13phases} has attracted enormous interest because it
was taken as strong evidence for unconventional superconductivity. Mainly two
observations favored an interpretation in terms of exclusively
SC phase transitions into SC phases denoted A,B and C in
fig.~\ref{FIGUBTH13phases}:
i) at x$_{c1}$ = 0.02 a cusp-like increase of
T$_c$ into T$_{c1}$ occurs with seemingly different T$_c$ pressure coefficients
below and above x$_{c1}$, suggesting that SC phases below T$_c$
(A, x $<$ x$_{c1}$) and T$_{c1}$ (B, x $>$ x$_{c1}$) are different.
ii) Below T$_{c2}$ for x$_{c1}<$ x $<$ x$_{c2}$ with x$_{c2}$ = 0.045 a second
superconducting transition into phase C takes place which was
infered from the sudden increase of slope in
H$_{c1}$ \cite{Rauchschwalbe87}. Furthermore in
phase C magnetic moments $\mu\simeq$ 10$^{-3}$ per U site appear
according to $\mu$SR results \cite{Heffner90}. This lead to the
suggestion that the SC order parameter in the C phase is nonunitary,
i.e. Cooper pairs carry a magnetic moment which appears around
impurity sites.

These observations lead to a scenario within the Ginzburg-Landau theory
approach based on the basic assumption of a crossing of T$_c(\Gamma_1,x)$
and T$_c(\Gamma_5,x)$ of two different superconducting order
parameters at the critical concentration x$_{c1}$ \cite{Sigrist91}.
They belong to cubic O$_h$ representations $\Gamma_1$ (fully symmetric)
and $\Gamma_5$ (threefold orbital degeneracy), though different pairs
of representations are also possible. Below x$_{c1}$ (and hence also for
stoichiometric \UBE) T$_c(\Gamma_5,x)>$ T$_c(\Gamma_1,x)$ and the
unconventional $\Gamma_5$ (A) SC state is stable. Above x$_{c1}$ the
opposite inequality holds and the conventional $\Gamma_1$ (B) SC phase
is stable immediately below T$_{c1}$. For lower temperature when the
size of the SC order parameter increases the fourth order mixing terms in
the GL functional favor a mixed $\Gamma_1\oplus\Gamma_5$ SC state (C)
which becomes stable below the transition at T$_{c2}$. For a specific
region in the parameter space of the GL functional the mixed state may
be nonunitary and therefore lead to Cooper pairs with magnetic
moments. For further increase of Th-concentration enhanced pair
breaking eventually leads again to a decreasing T$_c(\Gamma_1,x)$ and
a second crossing at x$_{c2}$ appears.

However, the results of recent detailed thermal expansion experiments
\index{thermal expansion!U$_{1-x}$Th$_x$Be$_{13}$}
including additional Th-concentrations have challenged this
 T$_c(\Gamma_5,x)$- T$_c(\Gamma_1,x)$ crossing
interpretation \cite{Kromer00a,Kromer00b,Kromer02}. Firstly,
investigation of additional Th dopings has shown that pressure
dependence of T$_c$ and T$_{c1}$ is not really different outside the
critical region around x$_{c1}$. More importantly, however, a new line
of anomalies T$_L$ (dashed line in fig.~\ref{FIGUBTH13phases}) has
been found that starts at the aforementioned anomaly for \UBE in
fig.~\ref{FIGUBE13bccrit} (B$^*$=0) and
continues until it merges with T$_{c2}$ for x $>$ x$_{c1}$ and then even
beyond x$_{c2}$. For x = 0 T$_L$ was interpreted as onset
temperature of
magnetic correlations. The thermal expansion anomaly at T$_L$ becomes
ever sharper when x approaches x$_{c1}$ which suggests increasing
magnetic correlation length until finally at x$_{c1}$ at the real phase
transition line T$_{c2}$ a true SDW state has evolved which
continues beyond x$_{c2}$. In this picture then the phases A and B
are the same superconducting phases and C is characterised by
coexistence of the superconducting B-phase with the SDW state
which also is responsible for the observed small moments
(10$^{-3}\mu_B$), there is no need to assume an exoctic nonunitary SC
order parameter in the C-phase. The evolution of a SDW state within
the SC phase has already been proposed by \citeasnoun{Kato87}
However in this picture the
increased slope of H$_{c1}$ below T$_{c2}$ has no obvious explanation.
Furthermore the sudden drop of flux creep rate by several orders of
magnitude below T$_{c2}$ in phase C \cite{Dumont02} can be
explained in the crossing scenario  where the nonunitary nature of C
provides an efficient mechanism for flux pinning
\cite{Sigrist99}. Indeed the drop in the flux creep rate is not
observed for pure \UBE below T$_L$ \cite{Mota03}. 

In thermal expansion investigations it also became clear that the additional
temperature scale T$_{max}$ is continuously reduced with increasing
Th-concentration. Amazingly it hits T$_{c1}$ exactly at its maximum
and (observed for B $>$ B$_{c2}$) vanishes at x$_{c2}$ as shown by the
dashed T$_{max}$ line in fig.~\ref{FIGUBTH13phases}. If this energy
scale is due to  magnetic excitations it suggests a close connection
to the mechanism of Cooper pairing.

The discussion in this section has necessarily been rather qualitative
as there is no developed microscopic theory for this complex behaviour
where indeed SC and perhaps SDW order evolve in an incoherent
nonstoichiometric HF metal with pronounced nFl behaviour.

\section{Rare Earth Borocarbide superconductors}
\label{Sect:Boro} \index{borocarbides} \index{RNi$_2$B$_2$C}

\begin{figure}
\begin{minipage}[t b]{\columnwidth}
\includegraphics[width=.30\columnwidth]{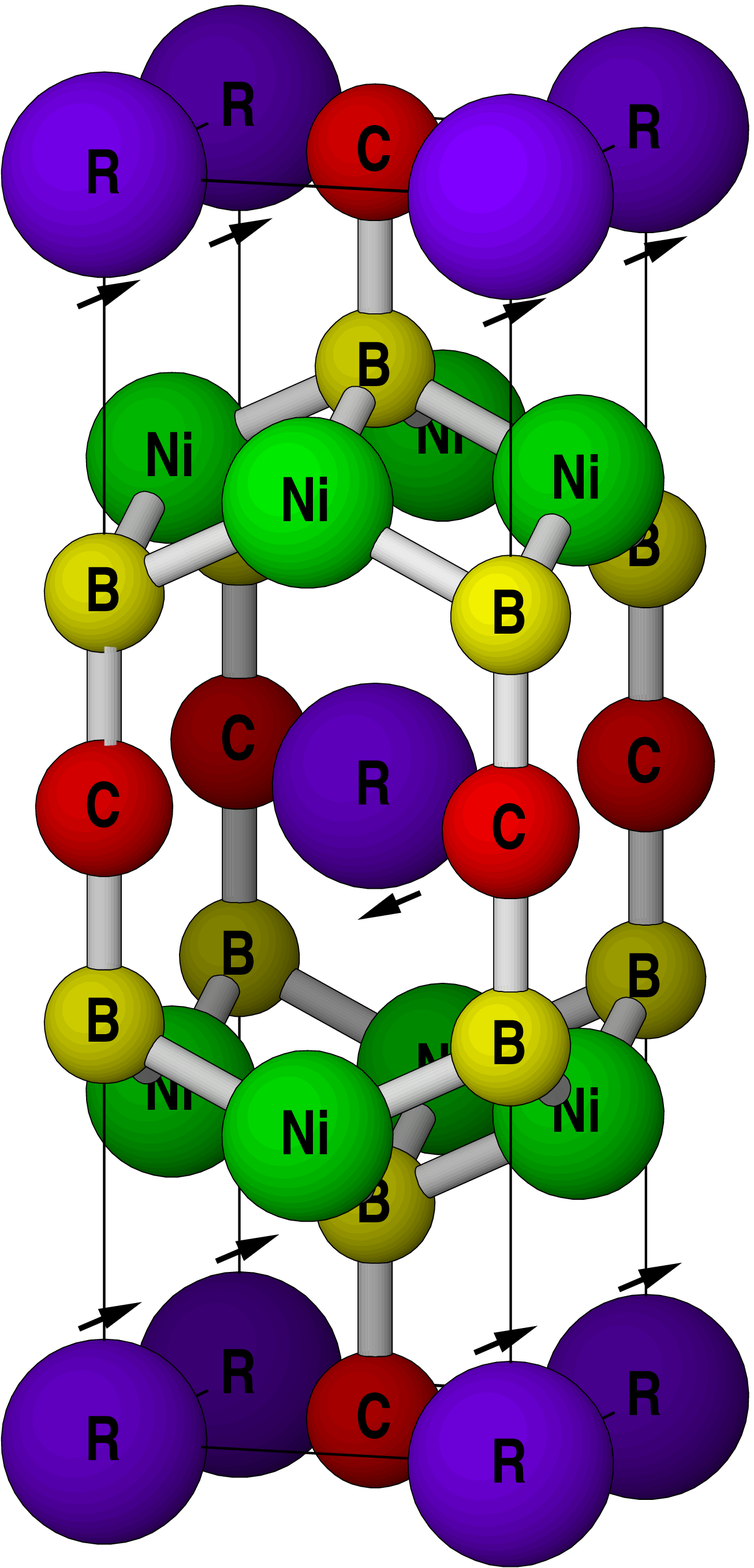} \hfill
\includegraphics[width=90mm]{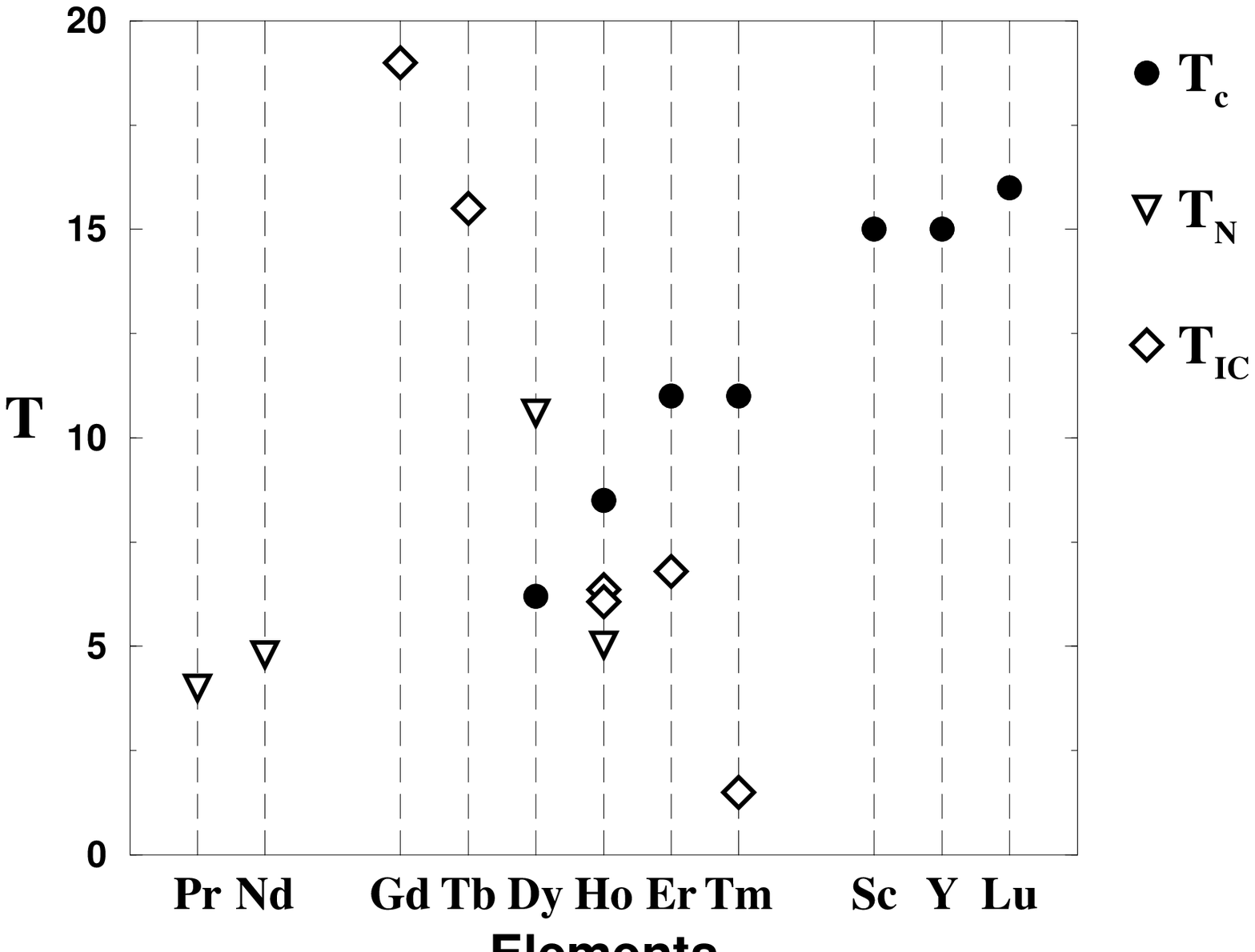}
\end{minipage}
\caption{Left: tetragonal crystal structure (I4/mmm) of RNi$_2$B$_2$C.
Low temperature AF2 magnetic structure with [110] easy axis is
indicated by arrows. a= b= 3.52\AA~and c= 10.53\AA~ for R =
Ho. Reciprocal lattice vectors are given by 2\v a$^*$, 2\v b$^*$ and
2\v c$^*$ where a$^* =\frac{2\pi}{a}$ etc.
Right: magnetic (T$_{IC}$: incommensurate magnetic structure, T$_N$: simple
AF structure) and superconducting (T$_c$) transition temperatures in
Kelvin for the RNi$_2$B$_2$C series.}
\label{FIGcryststrc}
\end{figure}

The superconducting class of layered transition metal borocarbides
\RBC was discovered \cite{Cava94,Nagarajan94} and investigated
rather recently. Here R stands either for nonmagnetic Y, Lu and Sc or
for lanthanide elements in a magnetic R$^{3+}$ state. Several
excellent reviews are available already \cite{Hilscher99,Mueller01},
mostly focusing on the material physics and chemistry of these
compounds. The crystal structure shown in fig.~\ref{FIGcryststrc} is body
centered tetragonal (space group {\it I4/mmm}). It consists of R-C
rock salt type planes separated by Ni$_2$B$_2$ layers built from
NiB$_4$ tetrahedra and stacked along the c-axis. More general
structures with more than one R-C layer are possible \cite{Hilscher99} but will
not be discussed further.  
The nonmagnetic borocarbides have relatively high T$_c's$ around 15 K 
as seen in fig.~\ref{FIGcryststrc}. There is evidence that the
superconducting mechanism is primarily of the electron-phonon (e-p)
type although this cannot explain the large anisotropy of the SC gap.
At first sight the layered structure is similar to the
high-T$_c$ cuprates. However, unlike the copper oxide planes the
NiB$_2$ planes show buckling (fig.~\ref{FIGcryststrc}, left panel), as
a consequence the electronic states
at the Fermi level in the borocarbides do not have quasi-2D
d$_{x^2-y^2}$ character and, therefore, have much weaker correlations
excluding the possibility of AF spin-fluctuation mediated superconductivity. 
The nonmagnetic borocarbides serve as a kind of reference point to
separate the fascinating effects of AF and SC order parameter coupling in the
magnetic \RBC. However, the former have their own peculiarities which
are not yet completely understood. Foremost, despite their alleged
electron-phonon nature, \LuBC and \YBC have strongly anisotropic
gap functions and low energy quasiparticle states as evident from
specific heat and thermal conductivity. Furthermore an
anomalous upturn in H$_{c2}$ has been observed. 
The magnetic \RBC are an excellent class of materials to study the
effects of competition of magnetic order and superconductivity for the
following reasons: The T$_c$'s are relatively high and their size
relative to T$_N$ varies systematically across the R-series. 
Especially interesting are the cases of \RBC with R = Dy, Ho and Er
where T$_c$ and T$_N$ (or T$_{IC}$) are not too different, leading to strong
competition of the magnetic and SC order parameters.
Furthermore the superconducting condensate and magnetic moments are carried by
different types of electrons, namely itinerant 3d-electrons for the
N$_2$B$_2$ layers and localized R$^{3+}$ 4f-electrons for the R-C
layers respectively. Finally they are well separated and their coupling which
is of the local exchange type can be treated in a controlled
perturbative way somewhat akin to the situation in the well
known classes of Chevrel phase \cite{Fischer90} and
ternary compound \cite{Fischer82} magnetic superconductors. 

The antiferromagnetic molecular field establishes a periodic
perturbation characterized by a length scale of the order of the Fermi
wavelength k$_{F}^{-1}\ll\xi _{0}$.
This implies that the spatial extent of the Cooper pairs extends over many
periods of the alternating molecular field. The latter is therefore
effectively averaged to zero and does not suppress superconductivity
via an orbital effect. The system is invariant under the combined operation
of time inversion followed by a translation with a lattice vector
which allows to form Cooper pairs in a spin singlet state with vanishing
(crystal) momentum in the antiferromagnetic lattice. This pair-state
can be considered as a natural generalization of the pairing in time-reversed
states encountered in usual non-magnetic superconductors to which
it reduces in the limit of vanishing staggered magnetization. The
magnetic order does not lead to depairing \cite{Fulde82a}. 

For detailed investigation 
it is necessary to gain a clear understanding of the magnetic
phases of \RBC. A theory of the rich metamagnetic phases
in the B-T phase diagram of \HoBC and \DyBC is a necessary prerequisite to
comprehend the interaction effects of the two order parameters in
these compounds. 

The investigation of alloy series of magnetic R$^{3+}$ disolved in the
nonmagnetic borocarbides allows one to study the pair breaking effects
due to incoherent exchange scattering and its associated de Gennes
scaling of T$_c$. One also observes `inverse' de Gennes
scaling \cite{Mueller01}, i.e. the suppression of T$_c$ by {\em
nonmagnetic} impurities like Lu and Y in the {\em magnetic} \RBC
superconductors which shows quite similar behaviour as function of the
impurity concentration although its underlying physics is very
different.

\subsection{Physical properties of the nonmagnetic borocarbides}
\label{Sect:Borophys} \index{YNi$_2$B$_2$C} \index{LuNi$_2$B$_2$C}

The nonmagnetic \YBC and \LuBC compounds with comparatively high 
T$_c$ of 16.5 K and 15.5 K serve as reference systems
for the more difficult systems \RBC with both magnetic and
superconducting phases. The electron-phonon nature of
superconductivity in \YBC and \LuBC is infered from a substantial s-wave
character of the order parameter as witnessed by the appearance of a
moderate Hebel-Slichter peak in the $^{13}$C NMR relaxation rate
\cite{Mizuno99}. On the other hand the gap function is strongly
anisotropic as can be seen both from temperature and field dependence
of thermodynamic and transport quantities. At low temperatures one
observes a specific heat C$_{s}\sim$T$^3$ \cite{Hilscher99} and thermal
conductivity $\kappa_{xx}\sim$ T \cite{Boaknin01} and
$\kappa_{zz}\sim$ T$^\alpha$ ($\alpha\sim 2-3)$ 
indicating the presence of gap nodes \cite{Izawa02a}. More precisely,
within experimental accuracy, there must be
at least a gap anisotropy of $\Delta_{min}/\Delta_{max}\leq$
10$^{-2}$ \cite{Izawa02a}. For an electron-phonon
superconductor this would be the largest anisotropy ever
observed. This conjecture is also supported by a $\sqrt{H}$ field
dependence \cite{Nohara97} of the low temperature
specific heat and the linear H dependence of the thermal conductivity
$\kappa_{xx}$ for $\vH$ along [001] \cite{Boaknin01}. Since in the
latter case the heat current is perpendicular to the vortices this proves
that quasiparticles must be present in the inter-vortex region. This
is also required to explain the observation of dHvA oscillations far
in the vortex phase. Experimental evidence
therefore demands a nodal gap function for borocarbides and the
s+g wave model proposed in \cite{Maki02} fulfils the requirements. In
addition it explains recent results on ultrasonic attenuation which
also confirmed the existence of gap nodes in the cubic plane
\cite{Watanabe03}. In the following we will address the question of
electron-phonon coupling strength,
isotope effect and positive curvature in the upper critical field. 
Furthermore power law behaviour of the electronic specific heat
and thermal conductivity 
as function of temperature and in addition their field-angle  dependence 
C$_{s}$(T,\v H) and $\kappa_{ij}(T,\v H)$ for T$\ll$T$_c$ in the vortex phase
allows to discuss the nodal structure of the SC gap.

\subsubsection{Evidence for electron-phonon superconductivity}
\index{electron-phonon superconductivity}

The classical argument in favor of the e-p mechanism is
the observation of an isotope effect characterized by the isotope
exponent for a specific atom with mass M as given by $\alpha_M=-d\ln
T_c/d\ln M$. However, this cannot be applied easily to
complex layered superconductors, as evident from the existence of an
isotope effect in the nonphononic cuprate superconductors with
nonopitimal doping. A boron isotope effect has been found in
\YBC ($\alpha_B$ = 0.2) and \LuBC ($\alpha_B$ = 0.11) much smaller than
the BCS value $\alpha_M$ = 0.5 and a nonphononic origin for $\alpha_B$
originating in the influence of boron on the charge density in the
B$_2$Ni$_2$ layers has, therefore, been suggested \cite{Drechsler01}.

The most direct support for the phonon mediated Cooper pairing is
due to scanning tunneling spectroscopy \cite{Martinez02} 
\index{quasiparticle tunneling} which has
shown the existence of a \index{strong coupling effects} strong coupling
signature in the tunneling
DOS due to a soft optical phonon close to the FS nesting wave vector
\vQ. The e-p coupling constant was derived as $\lambda$ = 0.5-0.8 which
is compatible with the value $\lambda$ = 0.53 obtained from resistivity data.

Additional support for the electron phonon mechanism comes from
the comparison of thermodynamic ratios $\Delta(T_c)/\gamma T_c$ and
$\Delta(0)/kT_c$  calculated in strong coupling
theory and the phonon spectra measured by inelastic neutron
scattering \cite{Hilscher99}. The strong coupling corrections of the
above ratios to their BCS values are related to the 
logarithmic moments $\bar{\omega}$ of the phonon DOS F($\omega$) or
more precisely of the $\alpha^2$F($\omega$) Eliashberg function. These
moments may be directly obtained from the INS phonon DOS or indirectly
from the above ratios. A comparison shows a reasonable
agreement between the moments with a T$_c$/$\bar{\omega}$ ratio
characteristic for moderately strong coupling electron phonon
superconductivity. The moments are, however, much smaller than the Debye
energies obtained from the phonon specific heat which again indicates that
certain low lying optical phonon modes play a special role for the
coupling. By varying temperature it has been found that these are soft
optical phonon modes along the [q$_x$,0,0] direction which involve
primarily the vibration of the heavy Y or Lu atoms. The softening
occurs at q$_x\sim$ 0.55 (in r.l.u of 2a$^*$) which corresponds to a
nesting vector of the Fermi surface \cite{Dugdale99} where the \v
q-dependent electronic susceptibility which
determines the renormalized phonon frequency becomes strongly
enhanced. The special role of these optical phonons for the strong
coupling effects is underlined by the
moment ratio ($\bar{\omega}^{\alpha^2F}_{YNi_2B_2C}$/
$\bar{\omega}^{\alpha^2F}_{LuNi_2B_2C}\sim$1.5$\sim (M_{Lu}/M_Y)^\frac{1}{2}$
rather than being equal to the square root of the unit cell masses
which would only be 1.16.  

\subsubsection{Anomalous H$_{c2}$-behaviour}
\index{upper critical field}

The upper critical field in nonmagnetic \LuBC and \YBC compounds shows a
peculiar positive (upward) curvature below T$_c$. This phenomenon is also
observed in various other layered superconductors. Within the standard
Ginzburg Landau
description of H$_{c2}$ for isotropic single band superconductors
the slope is determined by the Fermi velocity and only a
negative curvature is possible. A positive curvature can be obtained
by assuming a strong anisotropy of the Fermi velocity
\cite{Drechsler01} which may be simplified to a two band model with
two Fermi velocities. The upper critical field for the two band model  may
be calculated within the
linearized version of Eliashberg theory \cite{Shulga98}. In this way a fit
for the experimental H$_{c2}$-curves using a two band
Fermi velocity ratio of v$_{F1}$ : v$_{F2}$ = 0.97 : 3.7  (\LuBC) and
0.85 : 3.8 (\YBC) can be obtained. It also has to be assumed that there is
strong e-p coupling in the v$_{F1}$ band and sizable coupling between
v$_{F1}$ and v$_{F2}$ bands. It should be noted that the presence of magnetic
impurity scattering may change the picture \cite{Shulga01} because an
increasing scattering rate decreases the positive curvature. Thus the
extraction of a unique set of parameters for the two band model from the
positive H$_{c2}$ curvature seems difficult. 
As expected for uniaxial crystal structures, there is an a-c
anisotropy of H$_{c2}$ but surprisingly there is
also a strong fourfold anisotropy within the tetragonal ab-plane
\cite{Metlushko97} which cannot be due to FS effects on the level of the
effective mass approximation, it rather should be taken as another
indication of the large gap anisotropy in the ab-plane.

\subsubsection{Specific heat and thermal conductivity results}
\index{specific heat} \index{thermal conductivity}

The low temperature dependence of the specific heat is apparently
described by a power law behaviour C$_s\sim$T$^n$ with n between
2 and 3 \cite{Hilscher99}. Thermal
conductivity $\kappa_{xx}$ \cite{Boaknin01} clearly exhibits T-linear
behaviour for T $\ll$ T$_c$ suggesting the presence of nodal lines or
second order node points as introduced below which would be compatible
with n = 2 for the specific heat.
Furthermore the investigation of field (and field-angle) dependence of
C$_{s}$(T,\vH) and $\kappa_{ij}$(T,\vH) (i,j = x,y,z) is a powerful
method to obtain information on the quasiparticle spectrum and hence
on the anisotropy properties of the gap function
(sect.~\ref{Sect:Theory}). In a conventional
superconductor with isotropic gap the quasiparticles at low
temperature are confined to the vortex core where
they form closely spaced bound states with an energy 
difference $\Delta^2$/E$_F$ much
smaller than kT. Therefore, they can be taken as a cylinder of normal
state electrons of diameter $\xi_0$ (coherence length)  which gives a
field independent contribution
$\xi_0^2\gamma$T per vortex to the linear specific heat. Then in
the vortex phase $\gamma$(H) is predicted to scale with the number of
vortex lines which is $\sim$ H. However in superconductors with
nodes the quasiparticles with momenta in the node direction can tunnel into
the inter-vortex region and then their energy is Doppler shifted by the
supercurrents around the vortex by an amount $\delta$E = m\bf
v\rm$_F\cdot$\bf v\rm$_s$. As shown by \citeasnoun{Volovik93} this
leads to a finite residual quasiparticle DOS at zero energy given by
\index{quasiparticle DOS}
\begin{eqnarray}
\label{VOLDOS}
N_s(0)/N_n=K\sqrt{\frac{H}{H_{c2}}}
\end{eqnarray}
Where K is of order unity and N$_n$ the normal state DOS. As found
later the DOS also depends on the relative angle of the field with
respect to the nodal positions of the gap function. The DOS in
eq.(\ref{VOLDOS}) leads then to a $\sqrt{H}$ behaviour of the specific
heat \cite{Nohara97} and also for $\kappa_{zz}$(H) for temperatures
larger than the quasiparticle scattering rate.
Thermal conduction perpendicular to the vortex lines (\v H $\parallel$
c) as given by $\kappa_{xx}$(H) that starts immediately above H$_{c1}$ for
T $\ll$ T$_c$ can only result from
the presence of extended quasiparticle states outside the vortex cores
and hence is a direct proof for the presence of nodal regions in the
gap. To draw similar conclusions from the specific heat it is
necessary to observe the $\sqrt{H}$ behaviour for H $\ll$ H$_{c2}$ and
in connection the infinite slope of $\gamma$(H) for H$\rightarrow$0.
The Doppler shift picture of magnetothermal properties is only an
approximation. A comparison with exact quasiclassical calculations has
been given in \citeasnoun{Dahm02}

\subsection{Theoretical analysis of nonmagnetic borocarbides} 
\label{Sect:Borotheor}

We briefly discuss the electronic structure and associated FS topology
which is also important for the magnetic
borocarbides. Our main topic here is the explanation of the
quasi-unconventional SC low temperature behaviour of
nonmagnetic Lu- and Y- which may be understood within a a hybrid s+g wave
model for the SC gap function.

\subsubsection{Electronic Structure of the Borocarbides}
\index{band structure}

The crystal structure of the borocarbides is of a layered type which
might one lead to expect quasi 2D features in the electronic
bands. Calculations by various groups
\cite{Mattheis94,Pickett94,Rosner01} have shown however
that they have definitely 3D character which is also suggested by the
rather isotropic resistivity \cite{Fisher97}. Conduction states are 
composed of wide Ni-B-N-sp bands and narrow ($\sim$ 3 eV) Ni-3d bands
centered about 2eV below the Fermi level which is close to a local
peak of the DOS in \LuBC and \YBC. The DOS peak generally
decreases when Lu is replaced by progressively lighter R atoms which
leads to a correspondingly smaller
T$_c$. This also agrees with the observation that
substitution of Ni with Co (hole-doping) or with Cu (electron-doping)
decreases T$_c$ \cite{Gango96,Schmidt97} because the Fermi level moves
away from the local peak position. Unlike in the cuprates all
Ni 3d-states in the borocarbides contribute to conduction bands at the
Fermi surface which therefore consists of many sheets. In \LuBC the main sheet
exhibits an important \index{nesting property} nesting feature
\cite{Dugdale99}. FS lobes along [110] direction in the
tetragonal plane are connected by a nesting vector $\v Q$
=(0.55,0,0). This in turn leads to a peak structure in the staggered
susceptibility $\chi(\v q)$ \cite{Rhee95} at $\v q=\v Q$. In the \RBC
the effective
magnetic RKKY interaction J($\v q$) is proportional to $\chi(\v q)$ and
therefore a-axis incommensurate magnetic order with modulation vector
$\v Q$ is seen in Gd, Tb, Ho, Er and Tm-borocarbides which coexists
with c-axis spiral order in \HoBC (sect.\ref{Sect:Boromag}). Further
evidence of the importance
of the nesting at $\v Q$ comes from the presence of a pronounced Kohn
anomaly at wave vector $\v Q$ in the phonon dispersion of \LuBC
\cite{Dervenagas95}. Finally the previous conjecture of strongly
different Fermi velocities in the two band model has
been confirmed in electronic structure calculations for \YBC \cite{Rosner01}.

\subsubsection{Nodal structure of the superconducting gap and impurity
effects} \index{gap function} \index{gap nodes}

\begin{figure}
\includegraphics[width=75mm]{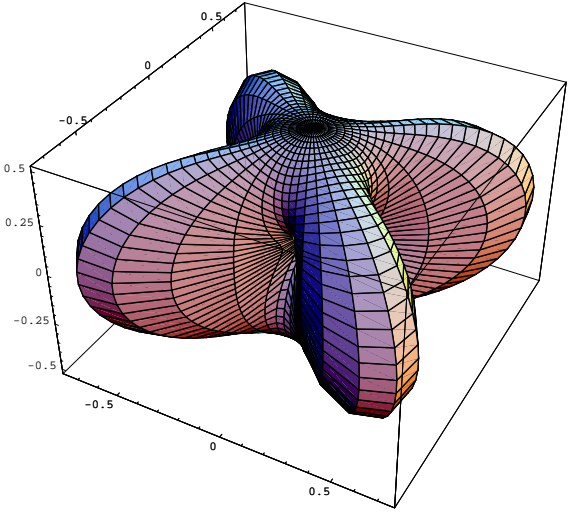}
\caption{Polar plot of the anisotropic SC gap function $\Delta(\v k)$
of the s+g model. Nodal points are along [100] and [010] directions.
(Maki et al. 2002) \index{s+g wave gap}}
\label{FIGgapfun}
\end{figure}

As explained before, thermodynamics and transport behaviour points to
an extremely anisotropic or nodal gap function in the nonmagnetic borocarbides.
A gap function compatible with reported experiments was proposed in
\citeasnoun{Maki02}. It is a hybrid s+g wave gap which is fully
symmetric (A$_{1g}$) under the tetragonal group D$_{4h}$ and has the
form \index{s+g wave gap}
\begin{equation}
\label{GAP}
\Delta(\v k)=\frac{1}{2}\Delta(1-\sin^4\vartheta\cos(4\phi))
\end{equation}
where $\vartheta,\phi$ are the polar and azimuthal angle in \vk-space
respectively. This gap function has four second order node
points at $(\vartheta,\phi)$ = ($\frac{\pi}{2}$,0),
($\frac{\pi}{2},\pi$) and ($\frac{\pi}{2},\pm\frac{\pi}{2}$)
(fig.~\ref{FIGgapfun}) which dominate the quasiparticle DOS for
E $\ll\Delta$ where
\begin{equation}
\frac{N_s(E)}{N_n}=\frac{\pi}{4}\frac{|E|}{\Delta}
\end{equation}
%
%
The form of $\Delta(\v k)$ implies a fine tuning of s and g
amplitudes. At its {\em second order} node points there is no
sign change of the gap function and therefore the derivative is also
zero. The resulting linear quasiparticle DOS then leads to the
observed low temperature specific heat
\begin{eqnarray}
\frac{C_s}{\gamma T}=\frac{27}{4\pi}\zeta(3)(\frac{T}{\Delta})
\end{eqnarray}
where $\gamma$ is the Sommerfeld constant. 
The presence of nodal points in the hybrid s+g wave gap function has no
intrinsic symmetry reason but is due to a `fine-tuning' of s- and
g-wave amplitudes in eq.~(\ref{GAP}). For \YBC this is realised to an
astonishing degree of $\Delta_{min}/\Delta_{max}\leq$ 0.01. 
There is presently no microscopic explanation for this fine tuning but a
phenomenological justification for the stability of an s+g wave order
parameter for a wide range of pair potentials has been given
\cite{Yuan03}. 

The nodal s+g wave gap function has a surprising
behaviour when normal impurity scattering is taken into
account \cite{Yuan03a}. For nodal gap functions belonging to a single
nontrivial representation, like a d-wave gap there is a strong
difference between scattering in the Born limit (small phase shift)
and unitary limit (phase shift $\frac{\pi}{2}$). In the latter
resonance scattering leads to the appearance of a residual zero field DOS 
resulting in a finite specific heat coefficient for T $<\Gamma$
($\Gamma$ = quasiparticle scattering rate). In the hybrid
s+g wave case, quite the opposite behaviour, i.e. gap opening by
impurity scattering, is observed which is almost identical for Born
limit and unitary limit. The \index{quasiparticle DOS} 
quasiparticle DOS including impurity scattering is given by
\index{impurity scattering}
\begin{eqnarray}
\frac{N_s(E)}{N_n}=Im \Bigl\la\frac{i\tilde{\omega}_n}
{\sqrt{\tilde{\omega}^2_n+\tilde{\Delta}^2_{\v k}}}\Bigr\ra
_{i\omega_n\rightarrow E+i\delta}
\end{eqnarray}
%
\begin{figure}
\includegraphics[width=75mm]{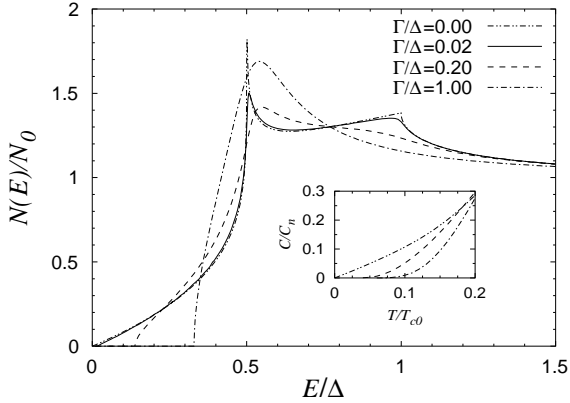}
\caption{Normalised quasiparticle DOS for s+g wave gap in pure \YBC
($\frac{\Gamma}{\Delta}=0$, full line) and for various scattering
strengths $\Gamma$ in the Born limit (for unitary limit results are
quite similar). Induced excitation gap $\omega_g$ increases
monotonically with $\Gamma$. The inset shows the low
temperature specific heat (Yuan et al. 2003). 
\index{quasiparticle DOS!YNi$_2$B$_2$C}}
\label{FIGdensg}
\end{figure}
%
where $\tilde{\omega}_n=\omega_n+i\Sigma_0$ and 
$\tilde{\Delta}_{\v k}=\Delta_{\v k}+\Sigma_1$ are the Matsubara
frequency and gap function renormalised by impurity scattering which
leads to diagonal and nondiagonal self energies $\Sigma_0$ and
$\Sigma_1$ respectively. For $\omega\rightarrow 0$ one finds
$\tilde{\Delta}\rightarrow\frac{1}{2}\Delta +\Gamma$ where $\Gamma$ is
the scattering rate. In other words an energy gap $\omega_g$
immediately opens up for finite $\Gamma$ as can be observed in
fig.~\ref{FIGdensg} which scales approximately as
$\omega_g(\Gamma)=\Gamma/(1+\frac{2\Gamma}{\Delta})$. Thus the s+g wave
fine tuning is destroyed by impurity scattering and nodal
quasiparticles are removed for temperatures T $<\Gamma$.
This would lead to a crossover to exponential low temperature behaviour in the
specific heat. Indeed this gap creation was observed in Pt-substituted
\index{Y(Ni$_{1-x}$Pt$_x$)$_2$B$_2$C}
Y(Ni$_{1-x}$Pt$_x$)$_2$B$_2$C with x = 0.2 by
\citeasnoun{Nohara97} as shown in fig.~\ref{FIGPtsub}. This remarkable
behaviour, which is exactly opposite to the d-wave case where residual
states at the Fermi level are created, can be traced back to the
different character of first order nodes (d-wave) where the gap
function changes sign and second order nodes (s+g wave) where the
angular derivatives of $\Delta(\vartheta,\varphi)$ also vanish and no
sign change of the gap occurs. In the first case the effect of
impurity scattering averages out in the gap equation, in the second it
does not, leading to a finite excitation gap $\omega_g$.

\subsubsection{Thermodynamics and transport in the vortex phase}
\index{vortex phase}

As discussed in sect.~\ref{Sect:Theory} the Volovik effect leads to a momentum
{\em and~} position dependent Doppler shift \index{Doppler shift} of
the quasiparticle
energies caused by the supercurrent flowing around the
vortices. The corresponding DOS change depends both on the field
strength and field direction 
with respect to nodal positions. For the gap model of
eq.(\ref{GAP}) the residual field induced DOS is given by 
\cite{Won03,Thalmeier03}
\begin{eqnarray}
\label{ANGDOS}
\frac{N_s(0)}{N_n}&=&\frac{C_s}{\gamma T}=
\frac{\tilde{v}\sqrt{eH}}{2\Delta}I(\theta,\phi)\nonumber\\
I(\theta,\phi)&=&\frac{1}{2}
\{(1-\sin^2\theta\sin^2\phi)^\frac{1}{2}
+\{(1-\sin^2\theta\cos^2\phi)^\frac{1}{2}\}
\end{eqnarray}
where $\tilde{v}=\sqrt{v_av_c}$ and
I($\frac{\pi}{2},\phi$) = $\max{(|\sin\phi|,|\cos\phi|)}$. This
function has a cusp-like minimum at $\phi$ = n$\frac{\pi}{2}$ when \v H is
sweeping over the node points. Also the residual DOS exhibits naturally the
experimentally observed $\sqrt{H}$ behaviour. For the calculation of
thermal conductivity in the vortex phase of the s+g wave superconductor
one has to be aware that impurity scattering immediately opens a
gap $\omega_g\simeq\Gamma$. To have nodal quasiparticles for transport
available one must fulfil $\omega_g\simeq\Gamma <$ T. On the other
hand to have an appreciable oscillation amplitude one must still be
in the low temperature limit $\Gamma<T\ll \tilde{v}\sqrt{eH}\ll\Delta$. 
The angle dependence of the c-axis thermal conductivity in the leading
order is then given by \index{thermal conductivity}
\begin{eqnarray}
\label{KAPPAZZ}
\frac{\kappa_{zz}}{\kappa_n}&\simeq&
(\frac{2}{\pi})^2\frac{\tilde{v}\sqrt{eH}}{\Delta}I(\theta,\phi)
\end{eqnarray}
i.e. in this limit to leading order it is completely determined by the angular
dependent residual DOS. To determine the proper nodal positions from
the maxima/minima of $\kappa_{zz}(\theta,\phi)$ it is important that
the low temperature limit is reached. 

\begin{figure}
\includegraphics[width=75mm]{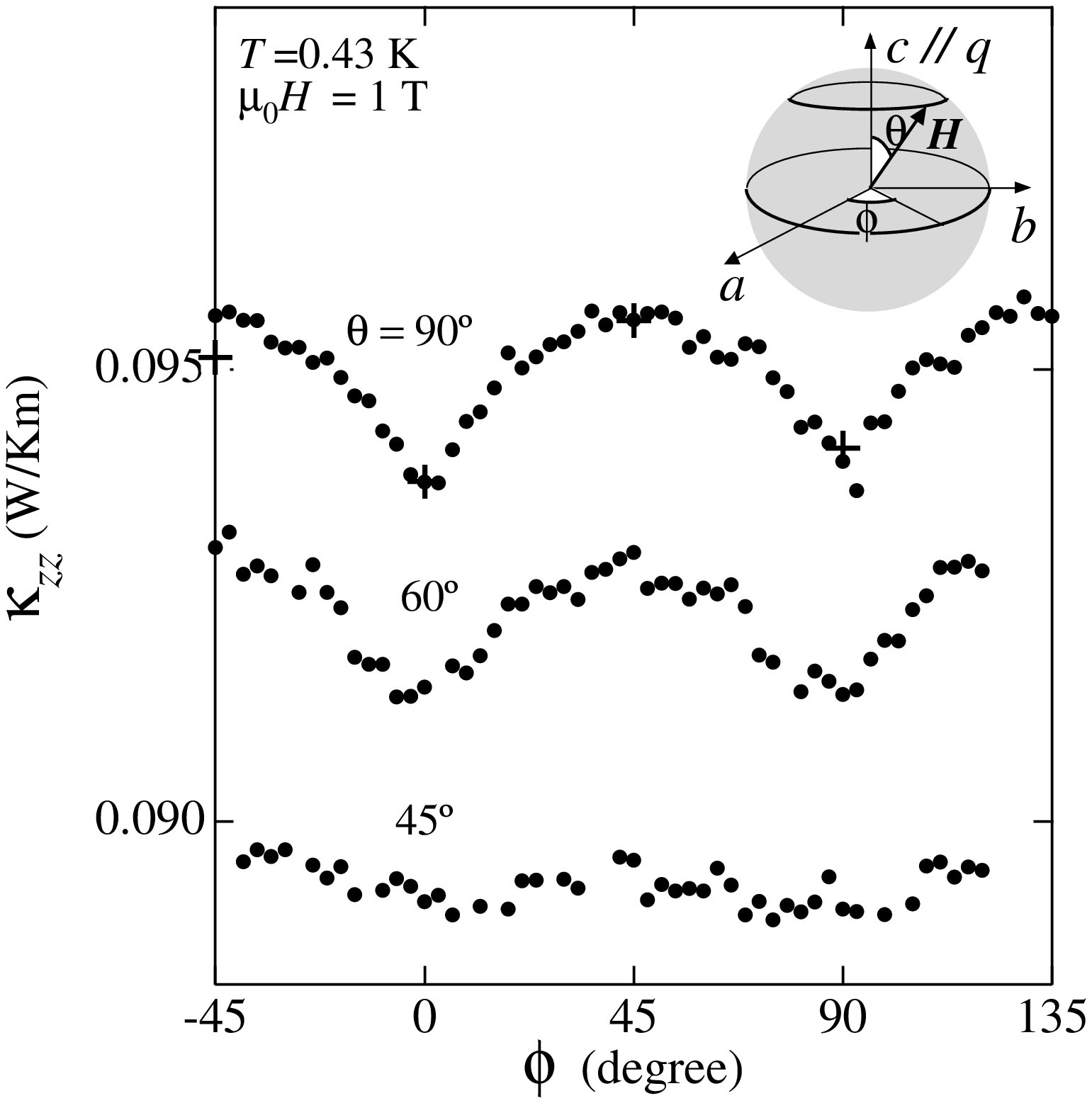}\hfill
\includegraphics[width=75mm]{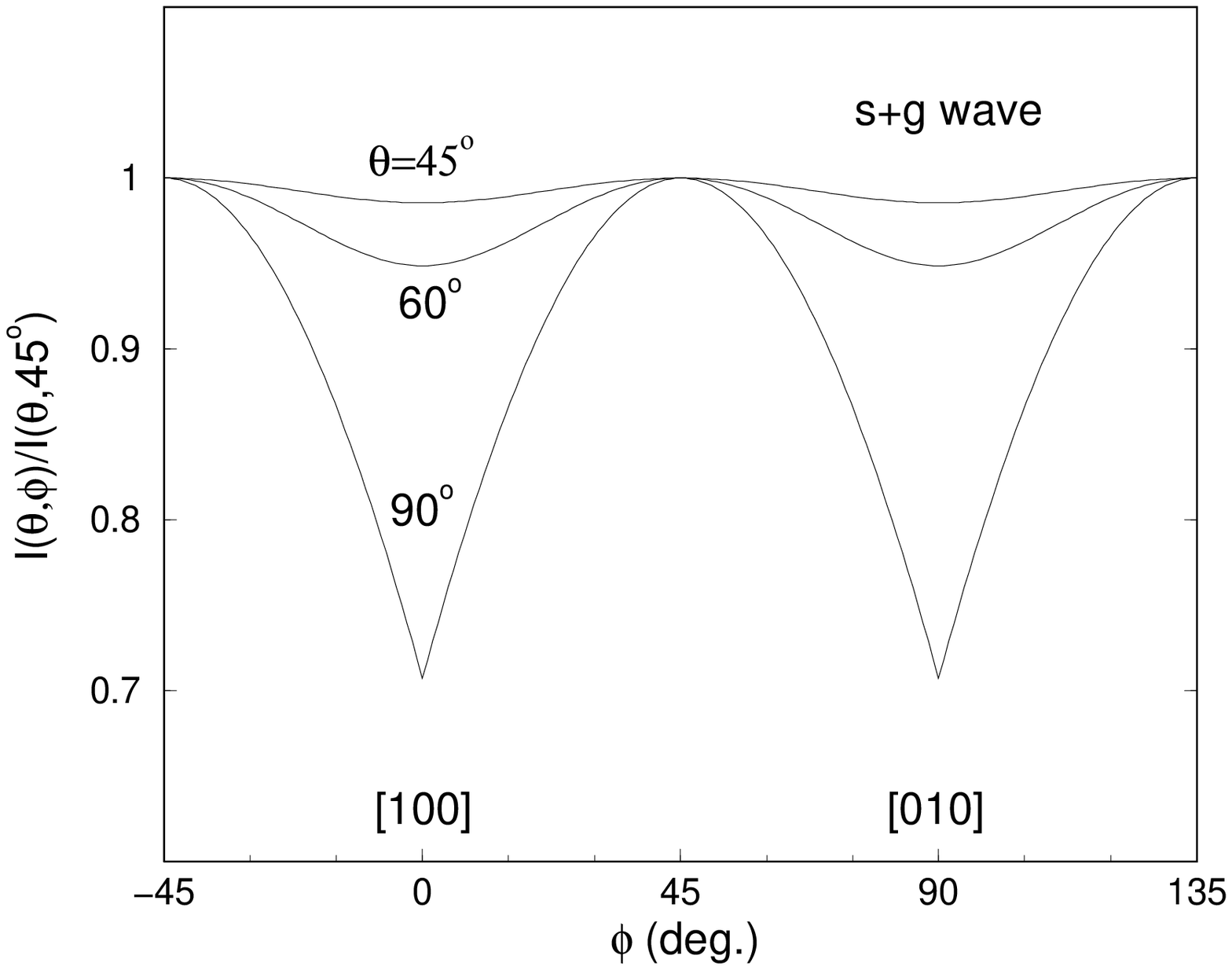}
\caption{Left: experimental c-axis thermal conductivity of \YBC as function of
azimuthal field angle $\phi$ for various polar field angles
$\theta$ (Izawa et al 2002b). The inset shows the field geometry with \v H
being swept around c by varying $\phi$ and keeping $\theta$ constant.
Right: theoretical normalized $\phi$-dependence of angular function
I($\theta$,$\phi$) which determines $\kappa_{zz}$ according to
eqs.~(\ref{ANGDOS}),(\ref{KAPPAZZ}) (Thalmeier and Maki 2003).
\index{thermal conductivity!YNi$_2$B$_2$C}}
\label{FIGkappazz}
\end{figure}

\begin{figure}
\includegraphics[width=75mm]{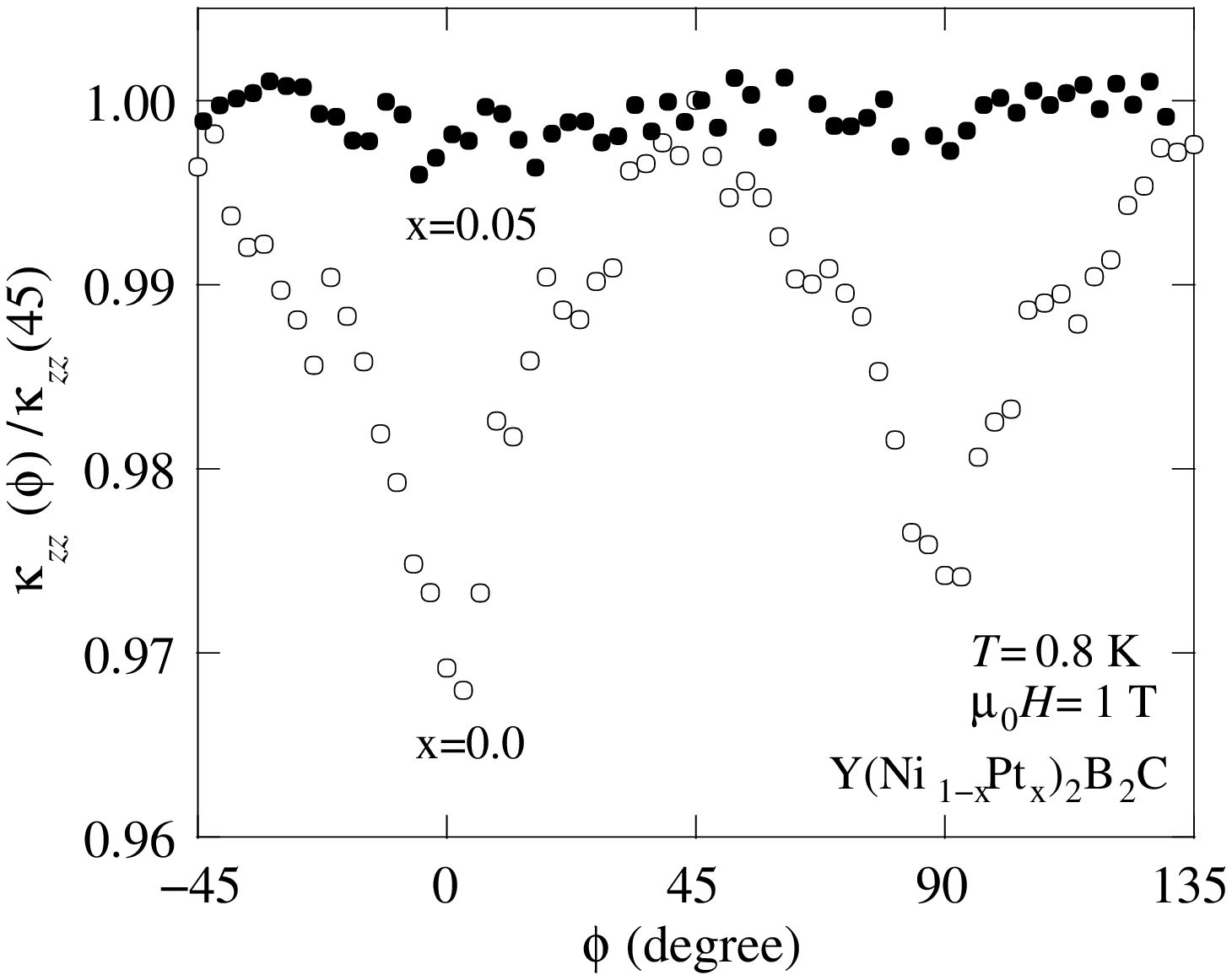}
\includegraphics[width=75mm]{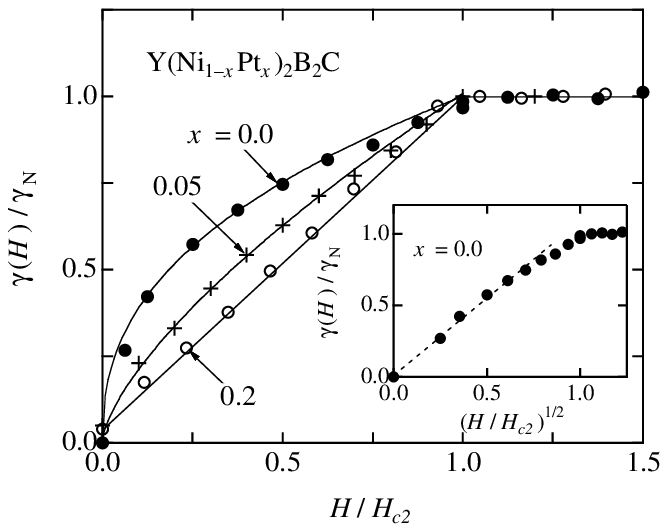}
\caption{Left panel: Cusp-like fourfold oscillations in
$\kappa_{zz}(90^o,\phi)$ of \YBC due to s+g wave point nodes are
completely destroyed by 5\% Pt-doping (Kamata 2003). Right panel:
Transition from $\sqrt{H}$-behaviour of $\gamma$(H) for the
nodal SC at x = 0 to linear in H behaviour for x = 0.2 when a large gap
has opened (Nohara et al. 1997). \index{s+g wave gap}\index{point nodes}}
\label{FIGPtsub}
\end{figure}

The angular dependence of the c-axis
thermal conductivity has been investigated in detail in \cite{Izawa02a}.
The geometry with heat current along c and \v H conically swept around
c is shown in the inset of fig.~\ref{FIGkappazz}. For the in-plane field
($\theta=\frac{\pi}{2}$) pronounced cusps in
$\kappa_{zz}(\frac{\pi}{2},\phi)$ appear for $\phi$ = 0,$\pm\pi$/2 in    
$\kappa_{zz}(\frac{\pi}{2},\phi)$ as is visible
in fig.~\ref{FIGkappazz}.  This is a typical signature for the
existence of point nodes \index{point nodes} in the gap in the
direction of the tetragonal
[100] and [010] axes. When the polar field angle $\theta$ decreases the
oscillations in $\kappa_{zz}(\theta,\phi)$ as function of azimuthal field angle
$\phi$ are rapidly diminished. This behaviour is indeed predicted from
eq.~(\ref{ANGDOS}) as shown in fig.~\ref{FIGkappazz}. A similar
calculation for a d$_{xy}$- gap function $\Delta(\phi)$=
$\Delta(\varphi)=\Delta\sin(2\varphi)$ which has line nodes in the
same directions may be performed \cite{Thalmeier03}. In this case no
cusp appears and the amplitude of $\kappa_{zz}(\theta,\phi)$
oscillation in $\phi$
is almost independent of $\theta$. This speaks strongly in favor of the
s+g wave order parameter as the correct model for \YBC and possibly
\LuBC. It is the first confirmed case of a superconductor with
(second order) node points in the gap functions. The only other known
candidate is UPt$_3$ whose E$_{2u}$ gap function in the B-phase is
supposed to have point nodes at the poles
(sect.~\ref{Sect:UPt3}). Similar experiments as in
fig.~\ref{FIGkappazz} have yet to be performed for this compound. 
The point nodes of \YBC were found to lie  along the tetragonal axis, this
means that the s+g gap function in fig.~\ref{FIGgapfun} is rotated by 45
degrees compared to the one used in \citeasnoun{Maki02}. According to
eq.~(\ref{ANGDOS}) the $\phi$ dependence of the specific heat is
determined by the same angular function and cusps should also appear
there. This was indeed observed \cite{Park02}
(fig.~\ref{FIGOPDspecosc}). In the Pt-substituted compound
\index{Y(Ni$_{1-x}$Pt$_x$)$_2$B$_2$C}
Y(Ni$_{1-x}$Pt$_x$)$_2$B$_2$C already x=0.05 is sufficient to destroy
the angular oscillations in $\kappa_{zz}(\theta,\phi)$
\cite{Kamata03} which is again due to the rapid gap-opening for s+g
wave and destruction of nodal regions by impurity scattering as shown
in fig.~\ref{FIGPtsub}.

In this section we have shown abundant experimental evidence that the
nonmagnetic \YBC and possibly \LuBC have anisotropic gap functions
with point nodes that can be described by a hybrid s+g wave order parameter.
As a consequence the simple electron-phonon type pairing mechanism originally
envisaged for the borocarbides certainly has to be supplemented, for example by
strongly anisotropic Coulomb interactions, to account for the strongly
anisotropic gap function found there.

\subsection{Magnetic borocarbides}
\label{Sect:Boromag} \index{RNi$_2$B$_2$C}

The magnetic 4f-electrons of lanthanides and the superconducting
3d-electrons of Ni are well separated in the RC layers and Ni$_2$B$_2$
layers of \RBC. The coupling between magnetic and superconducting
order parameters is weak enough to be treated as a perturbation but
still strong enough to lead to pronounced effects on the SC properties.
The magnetic and superconducting transition temperatures T$_m$ and
T$_c$ of the series are shown in fig.~\ref{FIGcryststrc}. We will first
discuss some of the basic observations in the series with focus on the
Ho and Er borocarbides where coexistence of both order parameters is
most interesting. The size of T$_m$ and the influence of magnetism on T$_c$ are
both governed by the de Gennes factor dG = (g-1)$^2$J(J+1) where g is
the Land\'{e} factor and J the total angular momentum of the R$^{3+}$ ion.
The former suggests the RKKY interaction via Ni 3d conduction
electrons of intervening Ni$_2$B$_2$ layers as the coupling mechanism
for the 4f moments in RC layers. The magnetic structures exhibit a
great variety which has been described and tabulated in
\citeasnoun{Lynn97} and \citeasnoun{Mueller01} and the following
general features are observed: i) both commensurate (C) and
incommensurate (IC) magnetic order parameters are present. 
ii) For Dy, Ho the commensurate AF structure consists of
ferromagnetically ordered ab-planes stacked along the c-axis.
iii) In IC structures one has two possibly coexisting type of modulations:
helical modulation of stacked FM planes along the c-axis or SDW type
(both longitudinal and transverse) modulation within each ab-plane
with a modulation vector \v Q close to the FS nesting vector. 
iv) The easy axis selected by the CEF potential is [110] (Ho,Dy),
[100] (Er) or [001] (Tm).

As an example the low temperature \HoBC commensurate AF2 structure
is indicated in fig. \ref{FIGcryststrc}. In the RKKY mechanism IC
modulation vectors should result from the maxima of the static
electronic susceptibility $\chi(\v q)$ which
has been approximately calculated by \citeasnoun{Rhee95}. It exhibits a
pronounced peak at the nesting vector \v Q of the Fermi surface which
is indeed quite close to observed a-axis magnetic modulation
vectors \v Q$_a$ = (0.55,0,0) of the Gd, Tb, Ho and Er borocarbide
compounds. On the other hand in \HoBC only a flat maximum is
observed for wave vectors \v q = (0,0,q$_z$) close to the AF vector
\v Q$_{AF}$ = (0,0,0.5) so
that the observed helically modulated structures along \v c$^*$
should not be associated with any FS nesting feature. For theoretical
model calculations of stable magnetic structures it is a more sensible
approach to parametrize the long range RKKY exchange interactions and
compare the results with the experimentally determined field-angle
dependent phase diagram at low temperature. This has been most
sucessfully done for \HoBC \cite{Amici98} and \ErBC \cite{Jensen02}. 
Four of the \RBC (R = Dy, Ho, Er, Tm) show coexistence of
superconductivity and magnetic order. As mentioned before \HoBC and
\DyBC \cite{Winzer99} display the most
spectacular interaction between superconductivity and magnetic order
as already suggested by close transition temperatures T$_m$ and T$_c$
(fig.~\ref{FIGcryststrc}). On the other hand in \ErBC and \TmBC the SC
vortex state exhibits an anomalous behaviour connected with the
magnetic order. In the following we will discuss mainly \HoBC and
\ErBC where most of the work has been performed sofar. 

\subsubsection{Metamagnetism and IC-C lock-in transition in \HoBC}
\index{metamagnetism} \index{incommensurate order} \index{HoNi$_2$B$_2$C}

\begin{figure}
\includegraphics[width=100mm]{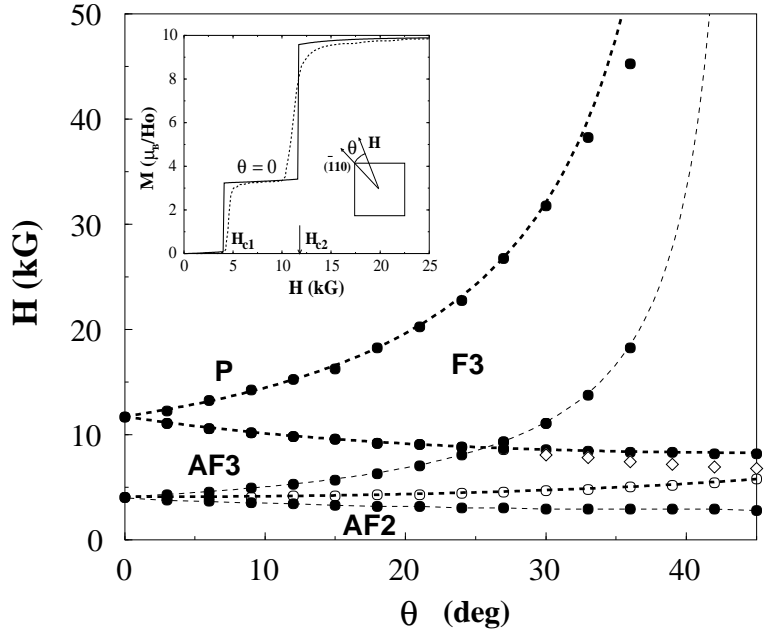}
\caption{Inset: Experimental (dashed line) and
theoretical magnetisation M(H) (full line) for \HoBC with \v H in the
ab-plane at angle $\theta$= 0 with respect to the easy [$\bar{1}$10]
axis. Following the calculated magnetisation steps as function of
$\theta$ leads to the theoretical phase H-$\theta$ phase diagram shown
in the main figure (Amici and Thalmeier 1998). The stacking sequence
of FM planes along c in the unit cell of magnetic phases is given by
AF2($\ua\da$), AF3($\ua\da\ua$) (ferrimagnetic),
F3($\ua\da\rightarrow$) (noncollinear). P($\uparrow$) is the
homogeneously magnetised phase. The weak dashed lines denote 
theoretically possible intermediate phase boundaries not observed
experimentally. \index{HoNi$_2$B$_2$C}}
\label{FIGmetamag}
\end{figure}

The metamagnetism of \HoBC has been investigated theoretically in
considerable detail \cite{Amici98,Amici00}. First we discuss the
4f-magnetism of \HoBC. This problem has two related aspects. For zero
field the first IC magnetic phase with \v Q = (0,0,0.45) appears at
T$_{IC}$ = 6 K which has a helical modulation of the moment of FM
ab-planes along c. At T$_N$ = 5 K a lock-in
transition occurs to the simple AF with wavevector \v Q$_{AF}$ and [110] easy
axis which is the stable low temperature structure. In addition an
a-axis modulation with wave vector \v Q$_a$ exists which will be
neglected here. The application of a magnetic
field \v H at an angle $\theta$ with respect to the easy axis leads to
the appearance of additional phases as witnessed by metamagnetic
transition steps in the magnetisation curve \cite{Canfield97} and shown
in the inset of fig.~\ref{FIGmetamag}. At these steps some of the FM
layers with moments pointing
roughly opposite to the field direction align with the easy axis
closest to the field direction, this leads to a stacking of FM planes
with a resulting net magnetic moment. Following the magnetisation steps
as function of $\theta$ one can construct the metamagnetic phase
diagram (fig.~\ref{FIGmetamag}).  The observed magnetic highfield
phases are listed in the caption of fig.~\ref{FIGmetamag}.
 
\begin{figure}
\includegraphics[width=75mm]{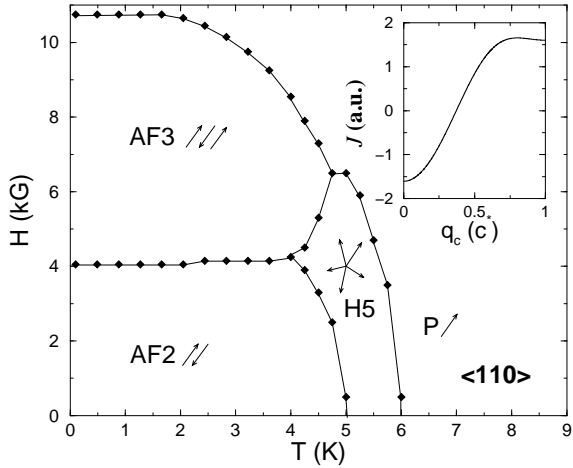}
\caption{Calculated magnetic H-T phase diagram \index{B-T phase
diagram!HoNi$_2$B$_2$C} of \HoBC for field
along the easy direction $\la 110\ra$. Arrows
indicate FM ordered ab-planes (easy axis [110]) with stacking sequence
along c. The inset shows the \index{RKKY interaction} RKKY exchange function obtained from metamagnetic
critical fields with a maximum around 0.8\v c$^*$ that 
corresponds to the H5 helix with T$_{IC}$ = 6 K. The lock-in
transition to the simple AF2 phase appears at T$_N$ = 5 K (Amici et al.
2000). \index{metamagnetism}}
\label{FIGbtphase}
\end{figure}

The metamagnetism of \HoBC is caused by two conflicting interactions,
namely tetragonal CEF potential and the RKKY-interaction. Whereas the
first tries to align the moments along the easy axis as
determined by the CEF potential the second one prefers a helical 
modulation of moments with a wavevector that minimizes the exchange
energy. The favored structure results from a compromise and may
change with temperature and external field. 
The CEF potential is obtained from inelastic neutron scattering experiments
\cite{Gasser96,Cavadini03} by fitting the observed energies and
intensities of CEF transitions to the predictions of an extended point
charge model of the tetragonal CEF which determines the parameters of the model
Hamiltonian H$_{CEF}$. Then
the CEF energies and wave functions may be obtained explicitly. For
Ho$^{3+}$(J=8) the 17 singlet and doublet CEF states split into two
groups: A group of 13 states with energies $>$10 meV  which is
irrelevant in the interesting temperature range and four low lying tetragonal
CEF states consisting of a ground state singlet $\Gamma_4$(0 meV) and
excited doublet $\Gamma_5^*$ (0.15 meV) and another singlet $\Gamma_1$
(0.32 meV). The interacting Ho 4f-states are then described by
\index{crystalline electric field excitations}
\begin{eqnarray}
\label{EXMOD}
H_M=\sum_i[H_{CEF}(\v J_i)-\mu_Bg\v J_i\v B]-
\frac{1}{2}\sum_{ij}J(ij)\v J_i\v J_j
\end{eqnarray}
where the first part describes the CEF potential and Zeeman energy
(g = $\frac{5}{4}$ for Ho$^{3+}$) and \index{RKKY interaction}
the second term the effective RKKY exchange interaction which is to be
fitted empirically to reproduce the critical fields and temperatures
of the metamagnetic phase diagram. In mean field (mf) approximation the
dimensionless magnetisation \v m$_i$ has to fulfil the selfconsistency
relations
\begin{eqnarray}
\label{MFMET}
\v m(\v B^e_i,T)=Tr(e^{\beta H_{mf}(\v J,\v B^e_i)}\v J_i);\;\;\;
\v B_i^e(\{\v m_l\},T)=g\mu_B+\sum_jJ(i,j)\v m_j
\end{eqnarray}
here i runs over all moments \v m$_i$ in the magnetic unit cell in a given
magnetic structure and \v B$_i^e$ is the effective field experienced by
the moment. Because the ab-planes are FM ordered (neglecting
the \v Q$_a$- modulation) these mf-equations may be solved within a
simplified quasi one-dimensional model where only structures are
considered that correspond to a different stacking of FM planes along
c. Structures with unit cell sizes of up to 29 layers have been
considered. Of all structures that are solutions of eq.(\ref{MFMET})
the stable one at a given field and temperature has to minimize the
Helmholtz free energy \index{free energy} per volume
\begin{eqnarray}
f_H=f_G(\v B)+\frac{B^2}{8\pi}-\frac{\v H\cdot\v B}{4\pi}
=f_G(\v B)+2\pi M^2-\frac{H^2}{8\pi}
\end{eqnarray} 
where \v B = \v H+4$\pi$\v M is assumed to be homogeneous throughout the
sample with the homogeneous magnetisation per volume given by 
\v M = (g$\mu_B/v_c)\bar{\v m_i}^2$. The bar indicates the average over
all moments and $v_c$ is the chemical unit cell volume. The effective exchange
parameters between 0$^{th}$ and i$^{th}$ FM layer,
J$_i$=$\sum_{\{j_i\}}$J(0,j) (j$_i$ runs over all the sites of the
i$^{th}$ layer) are considered to be empirical parameters that are
determined from the critical fields of the metamagnetic steps, from
T$_{IC}$ and from the requirement that the 1D Fourier transform J(q$_z$)
of J$_i$ has a maximum close to the IC modulation vector \v
Q. \index{incommensurate order} This
empirically determined exchange function J(q$_z$) is shown
in the inset of fig.~\ref{FIGbtphase}. In this way most of the metamagnetic
phases observed in \HoBC and the field angle have been
explained within the CEF-RKKY-exchange model eq.(\ref{EXMOD}) as shown in
fig.~\ref{FIGmetamag} \cite{Amici98}. The exchange parameters derived
theoretically in this work have been confirmed experimentally by
\citeasnoun{Cavadini03}. However some discrepancies remain,
especially it was found later \cite{Campbell00,Detlefs00} that in the
high field phase
($\uparrow\uparrow\rightarrow$) the ab-planes are no longer FM ordered so that
the simplified one-dimensional model is insufficient for high fields,
but it is completely adequate for studying the 
coexistence behaviour with superconductivity since H$_{c2}(0)\simeq$
0.8 T for \HoBC is below the critical field where the
($\uparrow\uparrow\rightarrow$) phase appears.
 
The inter-layer exchange parameters J$_i$ with i = 0-3, or exchange
function J(q$_z$)
obtained in the analysis of metamagnetism can now be used to predict
the B-T phase diagram for \v H along the easy axis [110] without
invoking any further empirical parameters. The solution of
eq.(\ref{MFMET}) for finite temperature leads to the phase diagram
shown in fig.~\ref{FIGbtphase}. The first phase that appears at T$_{IC}$ is
determined by the maximum gain in exchange energy, it is a
long wavelength commensurate
helix phase H5 (FM moment of ab-plane rotating around the c-axis)
with wave vector \v Q$_{H5}$ = 0.802\v c$^*$ which is not too far from the true
IC wave vector \v Q = 0.9\v c$^*$ observed in neutron diffraction. To
approximate the true IC phase even better with high order C phases, one had
to include still more interlayer interactions, the free energies of these
phases are, however, extremely close. When temperature decreases below
T$_{IC}$ the moments \v m$_i$(T) increase and the CEF contribution to
the free energy which grows with fourth power of the moment size
becomes more important. Because it favours allignment with the easy
axis directions [110] etc. the helix phase becomes less favorable and
finally at T$_N$ = 5 K a lock-in transition to the simple commensurate
AF phase takes place. There are indications from magnetic x-ray
diffraction that an intermediate lock-in to another C-phase
above T$_N$ may exist because a splitting of the IC satellite Bragg
peak was observed. Such behaviour is well known from 1D ANNNI models
that exhibit devil's staircase behaviour \cite{Bak80}. At temperatures
below 4 K application of a magnetic field along the easy axis finally leads to
a metamagnetic transition from the AF2 phase ($\uparrow\downarrow$) to
the ferrimagnetic AF3 phase with unit cell
($\uparrow\downarrow\uparrow$) (fig.\ref{FIGbtphase}).

\subsubsection{Weak Ferromagnetism in \ErBC}
\index{ferromagnetic order}\index{ErNi$_2$B$_2$C}

\begin{figure}
\includegraphics[width=75mm,clip]{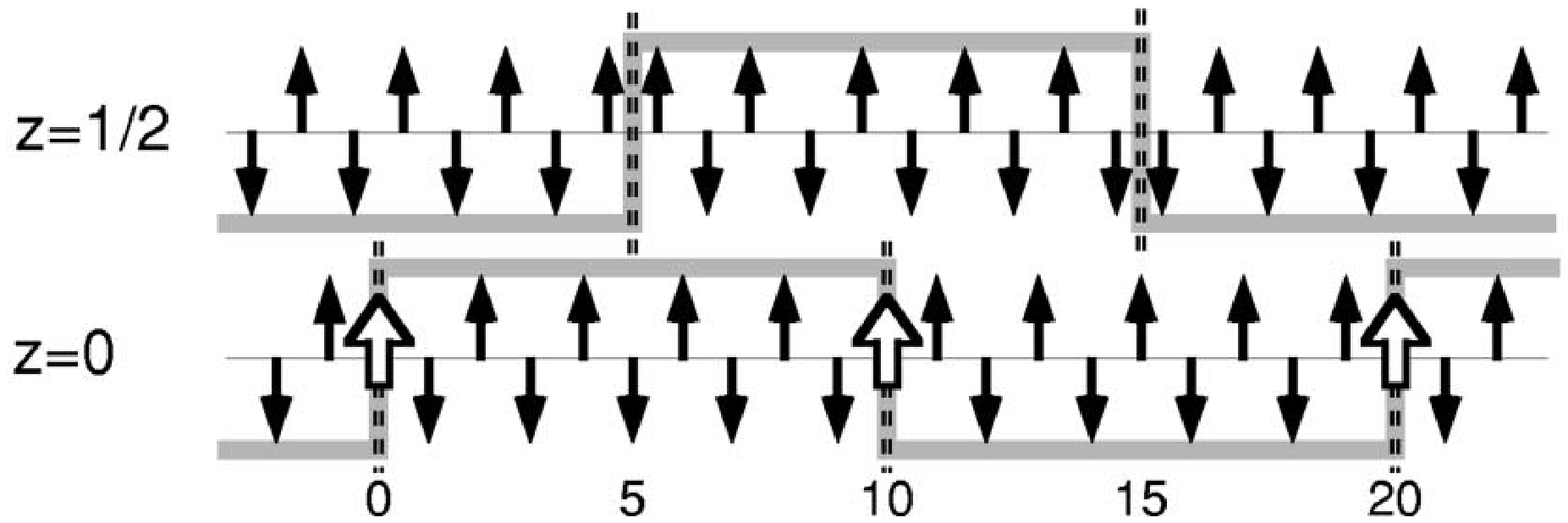}\hfill
\includegraphics[width=75mm,clip]{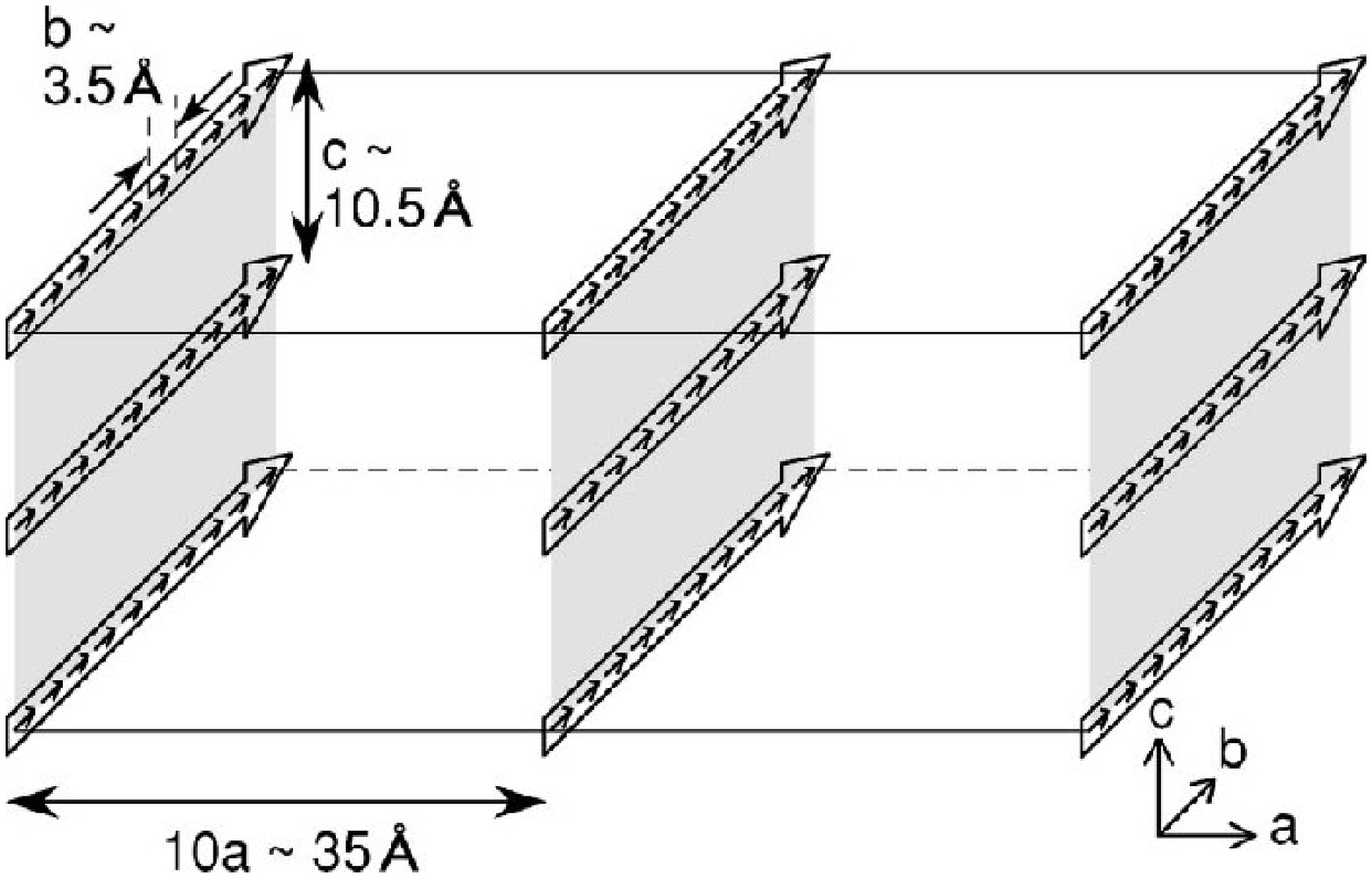}
\caption{Left panel: WFM structure of \ErBC showing the AF moment
ordering (easy axis [010]) along a-direction for two layers (z = 0,1/2)
along c. Moments residing in z=0 antiphase boundaries order
ferromagnetically (open arrows) below T$_{WFM}$. Right panel. Overall
WFM structure is composed of FM sheets in bc-plane stacked along a (shaded in
grey and corresponding to open arrows in left panel) (Kawano-Furukawa
et al. 2002). \index{ferromagnetic order}\index{ErNi$_2$B$_2$C}}
\label{FIGErstruc}
\end{figure}

This compound becomes superconducting at T$_c$ = 11 K  and develops a
transversly polarised SDW of Er moments below T$_{IC}$ = 6 K with a
propagation vector \v Q = (0.533,0,0) corresponding to the FS nesting
vector and a moment $\mu$ = 7.8 $\mu_B$ and an easy axis
[100]. Magnetisation measurements \cite{Canfield96} have shown that below 
T$_{WFM}$ = 2.3 K a remnant magnetisation exists due to the transition
to a weak ferromagnetic (WFM) state with an average moment of
$\mu_{WFM}$ = 0.33$\mu_B$/Er which persist to lowest temperatures. Therefore
\ErBC is the first example of true \index{coexistence} microscopic
SC/FM coexistence for
all temperatures below T$_c$. The origin of the WFM phase has been
investigated with neutron diffraction and its structure
was determined \cite{Kawano99,Kawano02}. On lowering the temperature
from T$_N$ the SDW exhibits a `squaring up' witnessed by magnetic satellite
peaks and finally breaks up into commensurate AF sections separated by
antiphase boundaries as shown in fig.~\ref{FIGErstruc}. In every
second layer along c the boundaries are located between the Er-bonds
creating first disordered moments which finally below T$_{WFM}$
order ferromagnetically. Because only the z=0 and equivalent layers of
the \ErBC bcc structure carry the FM moment one obtains an {\em
average} FM bulk moment of $\mu_{WFM}$ = (2/40)M$_{sat}$ = 0.39$\mu_B$ which
is close to
the moment from magnetisation measurements. The overall WFM structure
(fig.~\ref{FIGErstruc}) of \ErBC can be viewed as FM sheets in the bc-
plane with easy axis b and stacked along a. The large stacking distance
along a is the origin of the bulk WFM moment of 0.33$\mu_B$/Er.  
As in the case of \HoBC metamagnetic transitions appear for applied
field in the ab-plane. They have been explained by a similar
model \cite{Jensen02} including dipolar interactions. The stable zero
field structure has the observed modulation vector with $|\v Q|$ = 11/20.

\subsection{Coexistence of superconductivity and magnetic order}
\label{Sect:Borocoex}

Coexistence behaviour has been studied both for stoichiometric
compounds Er, Tm, Dy and Ho borocarbides as well as for
pseudo-quarternary compounds R$_{1-x}$R'$_x$Ni$_2$B$_2$C where R, R'
are different lanthanide atoms or compounds R(Ni$_{1-x}$M$_x$)B$_2$C
with M denoting 
another transition metal atom. In the stoichiometric compounds the
interaction of both order paramters is most dramatically seen in
the upper critical field H$_{c2}$-anomalies, especially for \HoBC
with an almost reentrant behaviour visible in fig.~\ref{FIGcodope}. Reentrance
can finally be achieved by replacing Ni with Co around x = 0.005. According to
sect.~\ref{Sect:Borotheor} this reduces the DOS at the Fermi level. As seen in
fig.~\ref{FIGcodope}, this results in a lower T$_c$ and destabilizes the
superconducting order parameter against magnetic order, finally for x
$>$ 0.0075 Co reentrance disappears and superconductivity exists only
below the lock-in transition at T$_N$ which remains almost unchanged. 
The main effect of M substitution for Ni is therefore a T$_c$ tuning by 
variation of the DOS. Lanthanide substitution like R'= Ho or Dy
instead of R= Lu or Y has a more subtle effect. For relatively small 
concentrations x of the magnetic R ions T$_c$ is reduced as expected from
the Abrikosov Gor'kov pair breaking theory by paramagnetic impurities in a
nonmagnetic host superconductor, in this case \LuBC. The pair breaking,
like T$_m$ in the magnetic borocarbides, is then controlled by the de
Gennes factor dG. This leads to a reduction of T$_c$ linear in
x$\cdot$dG. For large concentrations (x $>$ 0.5) of magnetic ions
however this concept breaks down dramatically. In fact for x = 1 \DyBC
is superconducting and one observes a kind of `inverse' de Gennes
\index{DyNi$_2$B$_2$C} scaling of the transition temperature with (1-x)dG corresponding to
the concentration (1-x) of {\em nonmagnetic} Lu impurities in the {\em
antiferromagnetic} host superconductor \DyBC which also act like pair
breakers.

\subsubsection{Coexistence of helical SDW, antiferromagnetism and
superconductivity in \HoBC}
\index{coexistence}\index{HoNi$_2$B$_2$C}

In this subsection we discuss a microscopic model \cite{Amici99a,Amici00} which
sucessfully describes the most interesting coexistence behaviour found
in \HoBC. A more phenomenological approach based on GL theory has also been
presented for Ho$_{1-x}$Dy$_x$Ni$_2$B$_2$C in \citeasnoun{Doh99}.
Firstly these compounds should more appropriately be called
superconducting magnets because the magnetic and superconducting
energy scales given by E$_m$ = kT$_m$ and E$_{SC}$ = k$^2$T$_c^2$/E$_F$
\begin{figure}
\includegraphics[width=75mm]{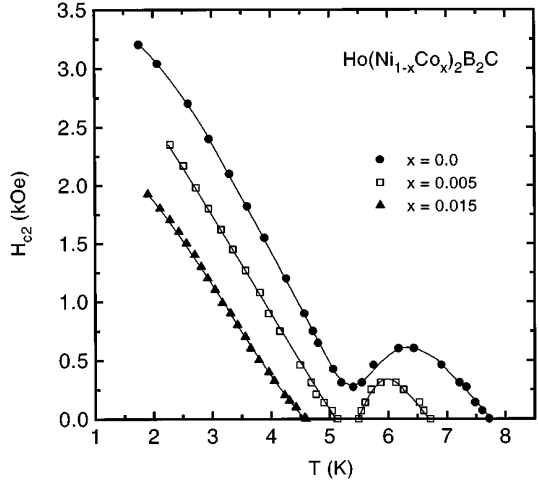}
\caption{Upper critical field curves for \HoBC (x = 0) and Co-substituted
compounds. For x = 0 one observes near reentrance around
T$_{N}\simeq$ 5.3 K. For x = 0.005 the reduced T$_c$ leads to real
reeantrance and for x = 0.015 finally T$_c$ falls below T$_N$ and
ordinary H$_{c2}$ behaviour sets in (Schmidt and Braun 1997).
\index{upper critical field!HoNi$_2$B$_2$C}\index{reentrance behaviour}}
\label{FIGcodope}
\end{figure}
differ strongly. If T$_m\simeq$ T$_c$ as in the present case then
E$_{SC}$/E$_m$ = kT$_c$/E$_F$ $\simeq$ 10$^{-2}$. This is due to the
fact that only a fraction kT$_c$/E$_F$ of conduction electrons
participates in the pair formation whereas the exchange energy of all
localised spins is involved in the magnetic ordering. Thus {\em local
moment} magnetic order energetically completely dominates
superconductivity whose influence on the former will therefore be
neglected \cite{Amici01}. In fact in our previous discussion this was
implicitly assumed by neglecting the effect of superconductivity on the
RKKY interaction via the conduction electron susceptibility . This is
justified because the
latter would only be affected for $|\v q|\leq\xi_0^{-1}$ ($\xi_0$ =
SC coherence length), and we discuss the competition with AF order or
IC phases with modulation vectors close to \v Q$_{AF}$.

The really important microscopic coupling is then caused by the
appearance of the ordered 4f local moments. According to their
magnetic structure they exert an additional periodic potential on the
conduction electrons. For the AF or helix structure the local 4f
moments are described by
\begin{eqnarray}
\la S_{\v R_i}\ra=S(T)(\hat{\v a}\cos(\v Q\cdot\v R_i)+
\hat{\v b}\sin(\v Q\cdot\v R_i))
\end{eqnarray}
where S(T) is the size of the ordered moments and \v Q the helix wave
vector with \v Q = \v Q$_{AF}$ in the AF case. The exchange interaction 
between the conduction and 4f-electrons then leads to an additional
spin dependent periodic potential 
\begin{eqnarray}
\label{HCF}
H_{cf}=\frac{1}{2}IS\sum_{\v k}
(c^{\dagger}_{\v k+\v Q\da}c_{\v k\ua}+
 c^{\dagger}_{\v k-\v Q\ua}c_{\v k\da})
\end{eqnarray}
The exchange constant I may be estimated from the RKKY expression 
I$^2$N(E$_F$)dG$\simeq$ kT$_m$ by using
N(E$_F$) = 4.8 eV$^{-1}$ resulting in I $\simeq$ 5 meV. The periodic
exchange potential of eq.~(\ref{HCF}) has important effects on the
conduction electron states, the original band energies $\epsilon_{\v
k}$ and Bloch states c$_{\v k}$ are
strongly modified close to the magnetic Bragg planes at $\pm \v Q$ and
at \v c$^*$ $\pm$\v Q. The modified magnetic bands and Bloch states
are obtained by a unitary transformation \cite{Herring66} which mixes
states (\v k,$\ua$) and (\v k+\v Q,$\da$) where each pair is decoupled from the
others. This leads to magnetic Bloch states given by
\begin{eqnarray}
c_{\v k+}^\dagger= 
u_{\v k}c_{\v k\ua}^\dagger+v_{\v k}c_{\v k+\v Q\da}^\dagger ;\;\;\;\;
c_{\v k-}^\dagger= 
v_{-\v k}c_{\v k-\v Q\ua}^\dagger+u_{-\v k}c_{\v k\da}^\dagger 
\end{eqnarray}
with the dispersion of the magnetic band  given by 
\begin{eqnarray}
\label{MDISP}
\tilde{\epsilon}_{\v k\pm}=
\frac{1}{2}(\epsilon_{\v k}+\epsilon_{\v k\pm\v Q})+
\frac{1}{2}(\epsilon_{\v k}-\epsilon_{\v k\pm\v Q})
\Bigl(1+\frac{I^2S^2}
{(\epsilon_{\v k}-\epsilon_{\v k\pm\v Q})^2}\Bigr)^\frac{1}{2}
\end{eqnarray}
The (u$_{\v k}$,v$_{\v k}$) satisfy appropriate orthonormality
conditions and fulfil the relation

\begin{eqnarray}
\label{MFACT}
u_{\v k}^2-v_{\v k}^2=\Bigl(\frac
{(\epsilon_{\v k}-\epsilon_{\v k+\v Q})^2}
{(\epsilon_{\v k}-\epsilon_{\v k+\v Q})^2+4I^2S^2}\Bigr)^\frac{1}{2}
\end{eqnarray}

Far away from magnetic Bragg planes $|\epsilon_{\v k}-\epsilon_{\v
k+\v Q}| \gg$ IS this leads to u$_{\v k}\sim 1$ and v$_{\v k}\sim 0$ and
 $\tilde{\epsilon}_{\v k\pm}\sim\epsilon_{\v k\pm}$ so that the
original band states and energies are recovered with (\v k,$\pm$) 
reducing to (\v k,$\ua\da$) Bloch states. An illustration of the
reconstructed FS is shown in fig.\ref{FIGborofermi}. Although the effect
of the exchange potential eq.~(\ref{HCF}) is confined to a small
region of \v k-vectors with 
$\delta k_\perp\leq$ IS/v$_F$ counted from the magnetic Bragg planes,
this modification proves important for superconducting
properties. From the previous consideration on superconducting and
magnetic energy scales it is clear that the Cooper pairs have to be
constructed from the {\em magnetic} Bloch
states. This implies that the BCS pair Hamiltonian which is expressed
in terms of nonmagnetic band states has to be transformed to the new 
magnetic basis states. For the AF structure this has been done in
\cite{Zwicknagl81} and for the more general helix phase in
\cite{Morosov96,Morosov96a}. Therefore, in addition to the modification
eq.~(\ref{MDISP}) of band energies, the nonmagnetic BCS pair potential
V$_{\v k,\v k'}$ will be replaced by an effective pair potential
\index{effective interaction} between magnetic Bloch states:
%
\begin{figure}
\includegraphics[width=75mm]{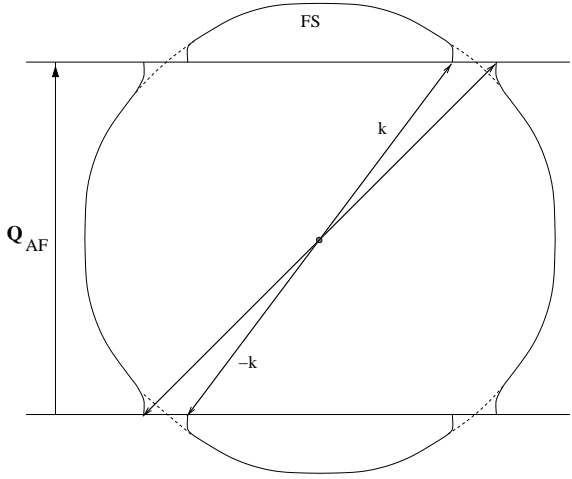}
\caption{
Schematic view of Fermi surface (FS) reconstruction around magnetic
Bragg planes connected by \v Q according to eq.~(\ref{MDISP}). This leads to
strong modification of SC pair potential for
conduction electron states with momenta \v k, -\v k as indicated.}
\label{FIGborofermi}
\end{figure}
%
\begin{eqnarray}
\tilde{V}_{\v k,\v k'}=
(u_{\v k}^2-v_{\v k}^2)V_{\v k,\v k'}(u_{\v k'}^2-v_{\v k'}^2)
\end{eqnarray}
According to eq.(\ref{MFACT}) $\tilde{V}_{\v k,\v k'}$ vanishes if \v
k or \v k' are located on a
Bragg plane. Naturally the BCS gap equations then lead to a modified
gap function. To keep the discussion of magnetic effects simple we
neglect the g-wave part in eq.(\ref{GAP}) and assume an originally isotropic
s-wave gap $\Delta$. This should not influence the qualitative
coexistence behaviour. Then the above equation leads to an additional
\v k-dependence originating in the exchange potential eq.~(\ref{HCF}) which is
confined to the immediate vicinity of the Bragg planes. It is given by
\begin{eqnarray}
\label{MODGAP}
\Delta(\v k,T)&=&(u_{\v k}^2-v_{\v k}^2)\Delta(T)\nonumber\\
\Delta(T)&=&\int_0^{\omega_D}d\epsilon
\Bigl(V\int_{MFS}\frac{dS'}{(2\pi)^3}
\frac{(u_{\v k'}^2-v_{\v k'}^2)^2}{|\nabla_{\v k'}
\tilde{\epsilon}_{\v k'}|}\Bigr)
\frac{\Delta(T)F(T)}{\sqrt{\epsilon^2+\Delta^2(T)}}
\end{eqnarray}
where F$_{\v k}(T)$=[1-2f$_{\v k}(T)$] and the inner integration is
taken over the reconstructed magnetic Fermi surface (MFS)
corresponding to $\tilde{\epsilon}_{\v k}$. Obviously the modified 
gap function $\Delta(\v k,T)$ vanishes on node lines $\perp$ c lying on
the Bragg planes and accordingly the self consistent gap equation for
$\Delta$(T) will lead to a reduced magnitude of the gap due to the
factor in the numerator of the integral in brackets. However, note that
no  sign change in the modified gap function $\Delta(\v k,T)$ occurs
when \v k crosses a node line, it still belongs to a fully symmetric A$_{1g}$
representation of the underlying AF magnetic structure.
Since (u$_{\v k}^2$-v$_{\v k}^2$) $\simeq$ 1 almost
everywhere except in the immediate vicinity of Bragg planes 
($\delta k_\perp$/k$_F\leq$ ISN(E$_F$)) the reduction of the gap amplitude
is not large and
helix order may coexist with superconductivity. For the special AF
case this is known already from \citeasnoun{Baltensperger63} and the helix
case with general modulation vector \v Q is not very different
because it only has two more Bragg planes than the AF case.
The second equation has the formal appearance of the BCS gap equation,
however, the expression in parentheses has to be interpreted as an
effective electron-phonon interaction V$_e$(T) or dimensionless
$\lambda_e$(T) = N(E$_F$)V$_e$(T) which itself depends on the
temperature via S(T). Its reduction from the nonmagnetic
background value $\lambda$ = N(E$_F$)V is due to the reconstruction
of electronic states confined close to the Bragg planes. Therefore it
may be treated as a perturbation linear in the small parameter
IS(T)/E$_F$. One obtains
\begin{eqnarray}
\label{DELAM}
\delta\lambda(T)=\lambda-\lambda_e(T)=-\frac{V}{8\pi\hbar^2}
\frac{k_r}{v_rv_z}IS(T)
\end{eqnarray}
for a FS piece without nesting features which is cut by a
pair of Bragg planes with a FS radius k$_r$ at the intersection and
radial and parallel velocities v$_r$ and v$_z$ respectively. For a
helical structure with two pairs of Bragg planes at $\pm\frac{1}{2}\v
Q$ and  $\pm\frac{1}{2}(\v c^*-\v Q)$ one has two add two corrections
as in eq.~(\ref{DELAM}) with different FS parameters. The main effect of the
interaction of magnetic order and superconductivity
described by eq.~(\ref{HCF}) has now 
been condensed into eq.~(\ref{DELAM}). The physical consequences of the
reduction of $\lambda$ due to magnetic order are a reduced
superconducting T$_c$ and condensation energy which results in an
anomaly in H$_{c2}$(T) that is directly correlated with the
appearance of the local moment order parameter S(T). The upper
critical field anomaly may approximately be calculated by using an
appropriately modified BCS expression and including the magnetisation response
of the local moments:
\begin{eqnarray}
\label{UPCRIT}
H_{c2}^{[110]}(T)&=&B_{c2}[\lambda_e(T),T]-M^{[110]}(T)\nonumber\\
H_{c2}^{[001]}(T)&=&B_{c2}[\lambda_e(T),T]
\end{eqnarray}
%
\begin{figure}
\includegraphics[width=75mm]{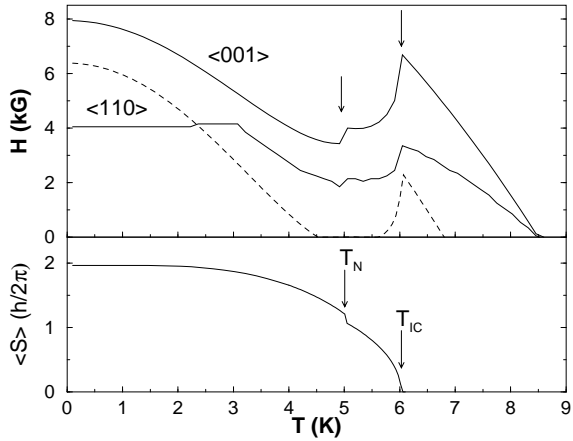}
\caption{Lower panel: size of the magnetic order parameter (local
moment amplitude) $\la S\ra$ vs temperature for the helix (H5) phase
below T$_{IC}$ and the AF2 phase below T$_N$. Upper panel: H$_{c2}$
for two symmetry directions. For $\la 110\ra$ the plateau is caused by
the transition to the ferrimagnetic AF3 phase in
fig.~\ref{FIGmetamag}. The dashed line is  H$_{c2}$ along $\la 001\ra$
with a reduced background T$_c$ that simulates the effect of doping on
the reentrance behaviour (Amici et al. 2000). 
\index{upper critical field!HoNi$_2$B$_2$C}\index{reentrance behaviour}}
\label{FIGcritfield}
\end{figure}
where B$_{c2}(\lambda_e$(T),T) = B$^0_{c2}(\lambda_e$(T))$\cdot$
[1-(T/T$_c(\lambda_e$(T))$^2$] is the BCS expression modified due to
the T-dependence of the effective e-p coupling. The magnetisation for
the [001] direction can be neglected because the 4f moments of CEF
states are well constrained within the easy ab- plane. Using
eqs.~(\ref{DELAM}),(\ref{UPCRIT}) with an appropriate set of parameters
for \HoBC the calculated upper critical field is shown in
fig.~\ref{FIGcritfield}. It clearly shows a pronounced depression of
H$_{c2}$(T) when magnetic order sets in. As discussed before there is
no big difference in the effect of helical order (below T$_{IC}$) and
commensurate AF order (below T$_N$), the overall depression is simply
controlled by the increase of S(T) when T decreases. At this point we
mention again that in our discussion of coexistence behaviour the
a$^*$-incommensurate magnetic modulation has been neglected. This is
fully justified according to hydrostatic pressure 
\index{hydrostatic pressure} experiments by
\citeasnoun{Dertinger01} where it has been shown that: (i) while the a$^*$ 
satellite intensity in ND decreases with pressure and vanishes at
p $>$ 0.7 GPa, (ii) the reentrance minumum in H$_{c2}$ becomes even more
pronounced. This
proves that the a$^*$- modulation of moments plays no important role
in the reentrance behaviour of SC, contrary to what has been claimed
by many previous investigations.
The [110], [001] anisotropy of H$_{c2}$ is due to the
different magnetisation response in both directions. For field lying
in the easy plane H$_{c2}$(T) exhibits a plateau which is due to the
appearance of the ferrimagnetic AF2 phase above the first critical
field in fig.(\ref{FIGbtphase}). This phase has a net magnetic moment that
cannot coexist with superconductivity. If the nonmagnetic background
e-p coupling $\lambda$ is reduced by about 10\% the resulting
reduction in T$_c$ and condensation energy leads to a reentrance
behaviour of H$_{c2}$(T) as shown by the dashed curve. These features
described by the theory of \citeasnoun{Amici00}, namely
depression, anisotropy (plateau) and reentrance behaviour have all been
observed experimentally in \HoBC, the latter under Co-substitution of Ni which
reduces T$_c$ as shown in fig.~\ref{FIGcodope}. 

\subsubsection{Coexistence of superconductivity and weak
ferromagnetism in \ErBC}
\index{coexistence}\index{ferromagnetic order}\index{ErNi$_2$B$_2$C}

At the SDW transition of \ErBC (T$_{IC}$ = 6 K) a similar but
much less pronounced dip in H$_{c2}$ as in \HoBC is observed
indicating the coexistence of both types of order. Even more exciting
is the observation of WFM order below T$_{WFM}$ = 2.3 K which coexists
with SC (T$_c$ = 11 K) to the lowest temperatures. This first
confirmed example of microscopic FM/SC coexistence has 
therefore attracted a lot of attention. Established theories
\cite{Ginzburg57} predict that thermodynamic coexistence in type II
superconductors is only possible when the internal FM field
H$_{int}\sim 4\pi$M is smaller than H$_{c2}$ \cite{Fischer90}. In
\ErBC due to the
large FM layer spacing (fig.~\ref{FIGErstruc}) one estimates from
$\mu$ a small H$_{int}\sim$ 0.5 kG which is of the same order as
H$_{c1}$ and much smaller than H$_{c2}\sim$ 15 kG \cite{Canfield96}.
Despite its smallness H$_{int}$ does have a very peculiar effect: For
applied fields H
with orientation $\theta_a$ close to the c-axis the perpendicular
internal field along the a direction rotates the effective field of the
vortex phase \index{vortex phase} to an
angle $\theta_v$ towards the a-axis. Thus the vortex lattice will show
a misalignment with the applied field by a small angle
$\theta_v-\theta_a$. This effect has been found experimentally by
\citeasnoun{Yaron96}. Theoretically this effect was considered in
the context of a Ginzburg-Landau theory including the terms
corresponding WFM order parameter and the coupling to SC \cite{Ng97}
. Since the London penetration depth $\lambda$ is 26 times larger than
the FM layer spacing in fig.~\ref{FIGErstruc} \cite{Kawano01} it can still be
justified to use an averaged FM moment density of saturation value
0.34$\mu_B$. The tilt angle $\theta_v-\theta_a$ of vortices becomes larger at
lower temperature in agreement with theoretical predictions.
If the internal field H$_{int}$ is at least somewhat larger than
H$_{c1}$ another exotic transition to a spontaneous vortex phase 
in zero applied field may take place. In this phase the internal field
spontaneously forms a vortex lattice oriented along the a direction. Low
field magnetisation measurements and small angle neutron scattering
experiments \cite{Kawano01} seem to support such a possibility but the
issue is not settled.

\section{Rare Earth Skutterudite Superconductors}
\label{Sect:PRSK}\index{skutterudites}

\begin{figure}
\includegraphics[width=75mm,clip]{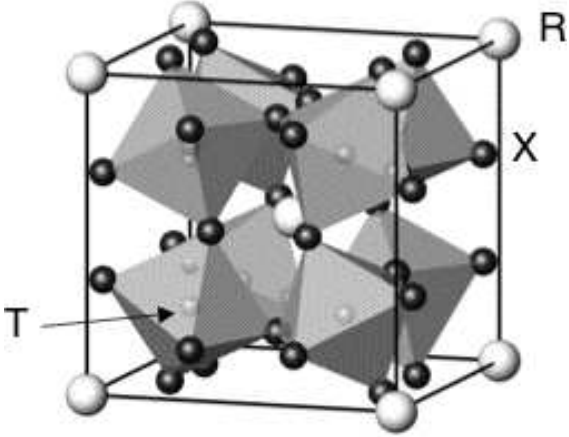}
\caption{Cubic crystal structure of filled skutterudite \RES. R atoms:
large circles, X atoms: middle size circles, T atoms: small 
circles located in the center of TX$_{8}$ octahedra (grey). For \PRS the
lattice constant is a = 9.3017\AA.}
\label{FIGstrc}
\end{figure}

The new heavy fermion \index{PrOs$_4$Sb$_{12}$} superconductor \PRS
which has been discovered
recently \cite{Bauer02} is potentially of similar interest as UPt$_3$
because it represents the second example of multiphase
superconductivity \cite{Izawa03} with a critical temperature
T$_c$ = 1.85 K. This material belongs to the large
class of \RES skutterudite cage compounds (R = alkaline earth, rare
earth or actinide; T = Fe, Ru or Os and X=P, As or Sb) which have a
bcc-filled skutterudite structure with tetrahedral space group
T$_h^4$ (fig.~\ref{FIGstrc}). In this structure large voids formed by tilted
T$_4$X$_{12}$ octahedrons can be filled with R atoms. They are however
rather loosely bound and therefore may have large anharmonic
oscillations (`rattling') in the cage. In addition if equivalent
equilibrium positions are present tunneling split states may
exist. Both effects may lead to interesting low temperature elastic and
transport phenomena i.e. thermoelectric effects
\cite{Sales96,Sales03}. Depending on the cage-filling atom this large
class of compounds displays also
a great variety of interesting effects of strong electron
correlation. Mixed valent and heavy fermion behaviour, magnetic and
quadrupolar order, non-Fermi liquid and Kondo insulating behaviour
has been found, see \citeasnoun {Bauer02},
\citeasnoun{Sales03} and references cited therein. Recently the SC in
non-stoichiometric skutterudites Pr(Os$_{1-x}$Ru$_x$)Sb$_{12}$ has been
investigated throughout the whole concentration range of 0 $\leq$ x
$\leq$ 1 \cite{Frederick03}. While for x = 0 one has an unconventional
HF superconductor the x = 1 compound PrRu$_4$Os$_{12}$ on the other
hand is a conventional SC with T$_c\simeq$ 1 K. The type of SC
changes at x$\simeq$ 0.6
where the transition temperature T$_c$(x) has a minimum value of 0.75 K. 
Here we will focus, however, exclusively on the HF multiphase 
superconductivity in \PRS (x=0).

\subsection{Electronic structure and HF behaviour of \PRS}

The LDA+U band structure \index{band structure} and Fermi surface of
\PRS has been investigated
by \citeasnoun {Sugawara02} and compared with dHvA results. Three FS sheets,
two of them closed and approximately spherical  were identified both
experimentally and in the calculation. Their dHvA masses are m$^*$
= 2.4 - 7.6 m, this is considerably higher than the calculated LDA+U
band masses ranging from m$_b$ = 0.65 - 2.5 m. The observed dHvA
masses however are
still much too small compared to the thermal effective mass m$^*$ estimated
from the the extrapolated $\gamma$-value of the linear specific
heat. One obtains \cite{Vollmer02} $\gamma$ = 313 mJ/mole K$^2$
leading to an estimate of m$^*$ = 50 m. This places \index{heavy fermions}
\PRS among the heavy fermion metals, the first one ever observed for a
Pr-compound with its 4f$^2$ electronic configuration, the estimated
quasiparticle bandwith is T$^*\simeq$ 10 K \cite{Bauer02} which is in the
same range as the lowest CEF splittings. Therefore, the low
temperature behaviour of C(T), $\chi(T)$ and $\rho$(T) is determined
by both CEF effects and heavy quasiparticle formation and it is
difficult to separate them \cite{Maple02}. The latter is rather
anomalous for a HF system since the A-coefficient in $\rho$(T)
 = $\rho_0$+AT$^2$ is smaller by two orders of magnitude compared to
other HF systems. The susceptibility is largely dominated by localised
4f$^2$ CEF states and from the high temperature behaviour an effective
Pr$^{3+}$ moment $\mu$ = 2.97$\mu_B$ is obtained. 

\subsection{Pr-CEF states and antiferroquadrupolar order}
\index{antiferroquadrupolar order}

The CEF level scheme of Pr$^{3+}$ is determined by a CEF potential
with tetrahedral T$_h$ symmetry. For a long time it was thought that it is
equivalent to that of other cubic point groups like T$_d$ or
O$_h$. However, it was shown recently \cite{Takegahara01} that this is
incorrect. Due to the absence of two types of symmetry operations the
CEF potential \index{crystalline electric field excitations} is rather given by
\begin{eqnarray}
H_{CEF}=W\Bigl[x\Bigl(\frac{O_4}{F(4)}\Bigr)+
(1-|x|)\Bigl(\frac{O_6^c}{F(6)}\Bigr) 
+y\Bigl(\frac{O_6^t}{F^t(6)}\Bigr)\Bigr]
\end{eqnarray} 
where O$_4$=O$_4^0$+5O$_4^4$, O$_6^c$=O$_6^0$-21O$_6^4$ and
O$_6^t$=O$_6^2$-O$_6^6$ are Steven' s operators and for Pr$^{3+}$
F(4)=60, F(6)=1260 and F$^t$(6)=30. For the cubic groups O$_h$ and
T$_d$ we have y=0 and there is only one CEF parameter x, aside from
the overall scale W, then H$_{CEF}$ reduces to the well known
form. However in tetrahedral symmetry T$_h$ in general a second CEF
parameter y$\neq$0 appears. The consequences were analysed in detail
in \citeasnoun{Takegahara01}, specifically the $\Gamma_1$, $\Gamma_2$ states
of the cubic CEF case will be mixed into two inequivalent $\Gamma_1$
singlets of T$_h$ and likewise  $\Gamma_4$, $\Gamma_5$ of O$_h$ will
be mixed to become two inequivalent $\Gamma_4$ triplets of T$_h$ where the
degree of mixing depends on y. The experimental determination of the Pr$^{3+}$
CEF level scheme has been performed under the restriction y=0
(i.e. assuming the cubic CEF potential), therefore, we keep the original
notation of states. With this restriction the level scheme has been
determined by fitting $\chi(T)$ and also directly by inelastic neutron
scattering \cite{Maple02}. The latter leads to a level scheme
$\Gamma_3$(0K), $\Gamma_5$(8.2K), $\Gamma_4$(133.5K), $\Gamma_1$(320K),
which is close to the one obtained from $\chi(T)$. The ground state is a
nonmagnetic (non-Kramers) doublet with a quadrupole moment and
quadrupole matrix elements connecting to the first excited triplet
state. An alternative CEF level scheme has been proposed
\cite{Aoki02,Tayama03} from specific heat and magnetisation
experiments. It has a singlet ground state without
magnetic nor quadrupole moment, namely $\Gamma_1(0K)$, $\Gamma_5$(6K),
$\Gamma_4$(65K) and $\Gamma_3$(111K).
The splitting energy $\delta$ ($\simeq$ 6 - 8.2 K) of ground state to
first excited state is also seen as a pronounced zero field Schottky anomaly
sitting on top of the superconducting transition. Application of a
field above the SC H$_{c2}$(0) = 2.2 T suppresses the Schottky anomaly.

\begin{figure}
\includegraphics[width=75mm]{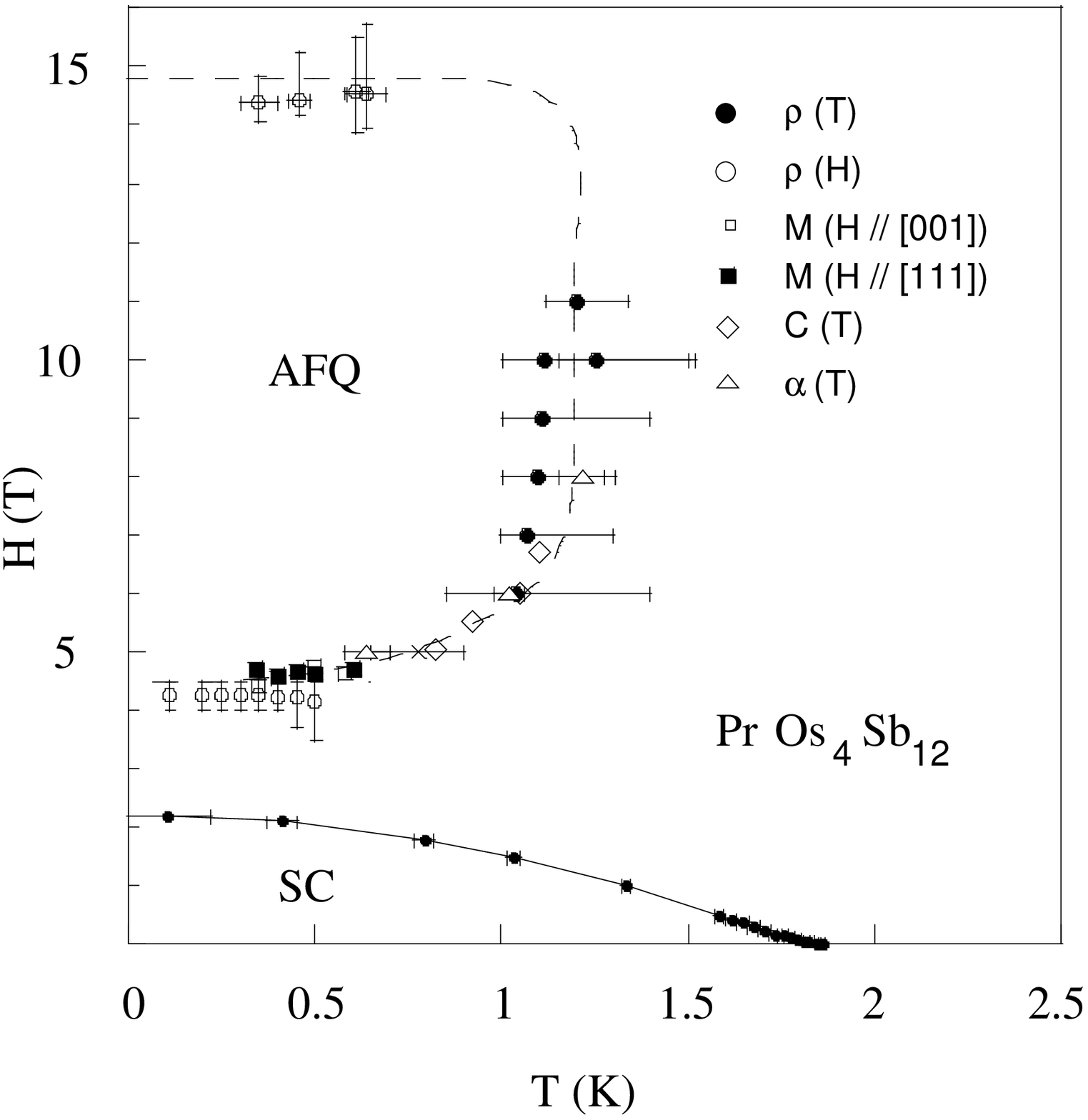}\hfill
\includegraphics[width=75mm]{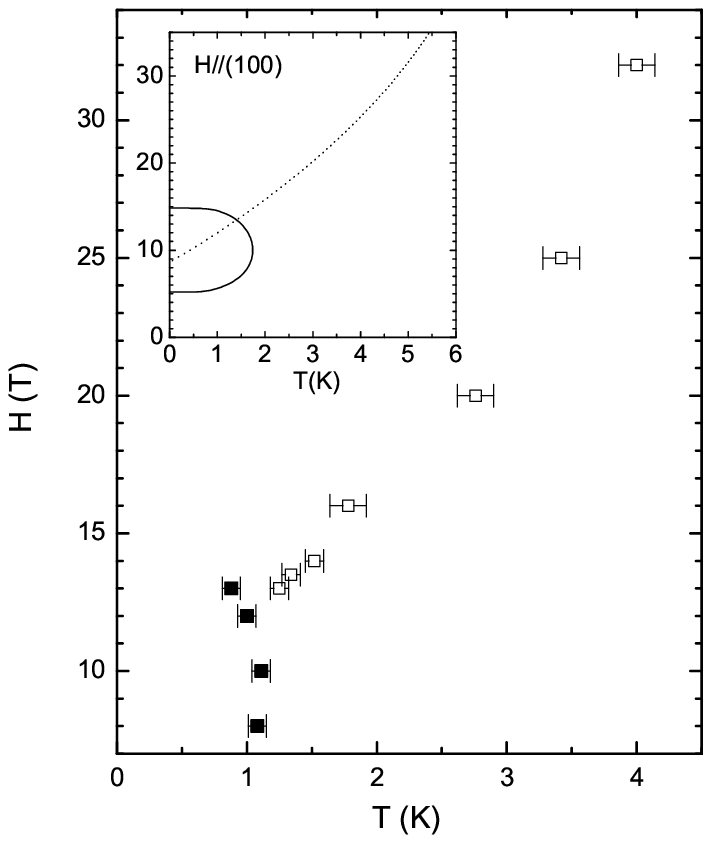}
\caption{Left panel: complete B-T phase diagram
(Maple et al. 2002) with SC regime and region of 
field induced order parameter which is presumably of
antiferroquadrupolar (AFQ) type. Data are obtained from resistivity
($\rho$), magnetisation (M), specific heat (C) and thermal expansion
($\alpha$) measurements. Right panel: high magnetic field phase diagram (Aoki
et al. 2003) with upper part of AFQ phase (full squares) and line of
high field Schottky anomaly from the $\Gamma_1$-$\Gamma_5$
crossing. The inset shows calculation for tetrahedral CEF model.
\index{B-T phase diagram!PrOs$_4$Sb$_{12}$} 
\index{antiferromagnetic order!PrOs$_4$Sb$_{12}$}}
\label{FIGafq}
\end{figure}

However, at larger fields of H $\simeq$ 4 T a new and sharper one
appears at lower temperatures possibly signifying the stabilization of
a long range antiferroquadrupolar ordered phase
\cite{Aoki01,Aoki02,Vollmer02,Maple03}. This is summarized in the phase diagram
shown in the left part of fig.~\ref{FIGafq}. A detailed analysis of
AFQ long range order in the
high field region is still lacking, therefore it is helpful to compare
with other known AFQ ordered 4f systems, where Ce$_{1-x}$La$_x$B$_6$ is
certainly the most well studied one \cite{Shiina97}. It has a
$\Gamma_8$ quartet as CEF ground state which carries both quadrupole
and magnetic moment. Both spontaneous AFQ and AF order are therefore
observed with T$_Q>$T$_N$ and we consider only the former. For x = 0.5
the AFQ phase diagram \cite{Nakamura97} is quite similar to that of
fig.~\ref{FIGafq}, namely the AFQ phase is only induced in finite field and
T$_Q$(H) increases with field. In the Ce$_{1-x}$La$_x$B$_6$ system the
AFQ order has been identified by observing the induced secondary AF
order which is induced by a homogeneous field H in the ordered AFQ
background. This can be done by neutron diffraction and NMR
experiments. Recent neutron diffraction results \cite{Kohgi03} suggest that
the CEF ground state is singlet $\Gamma_1$  and the AFQ order parameter
is of O$_{yz}$ = J$_y$J$_z$+J$_z$J$_y$ type. Additional high field
measurements on C(T,H) for different field directions \cite{Rotundu03}
also give strong evidence for the $\Gamma_1$ ground state scenario and
for a $\Gamma_1$ - $\Gamma_5$ level crossing as origin of the field
induced AFQ phase (fig.~\ref{FIGafq}, right panel).  

The existence of a nonmagnetic ($\Gamma_1$ or $\Gamma_3$) ground state
in \PRS leads one to speculate about the origin of the observed HF
behaviour since the usual Kondo lattice mechanism as in Ce-intermetallic
compounds which demands a Kramers degenerate magnetic ground state
cannot be at work here. In the case that $\Gamma_3$ were realised one might
conjecture that instead of the exchange scattering one has aspherical
Coulomb scattering
from the quadrupolar degrees of freedom of the nonmagnetic $\Gamma_3$
ground state. As discussed for the isoelectronic
U$^{4+}$-configuration in UBe$_{13}$ in sect.~\ref{Sect:UBe13} this is
described by a
multichannel Kondo Hamiltonian for the quadrupolar pseudo-spin. It
can lead to a partial screening of the $\Gamma_3$-quadrupole below
a quadrupolar Kondo temperature T$^*$. At even lower temperatures
logarithmic non-Fermi liquid anomalies in thermodynamic quantities
should develop due
to the partial screening. This cannot be confirmed due to the
intervening SC transition. Therefore this quadrupolar HF mechanism for
\PRS is only a conjecture. If a singlet $\Gamma_1$ ground
state is realised as suggested by neutron diffraction and high field
C(T,H) results mentioned before one has to invoke another mechanism
for heavy mass generation which is similar to the one implied in
U-compounds like \UPD \cite{Zwicknagl03}. It is due to mass
renormalisation by the low lying $\Gamma_1$-$\Gamma_5$ excitation,
i.e. contrary to the previous scenario it is caused by off-diagonal
parts of the conduction electron-CEF level interaction. 

\subsection{The superconducting split transition}
\index{T$_c$-splitting}

\begin{figure}
\includegraphics[width=75mm]{PRSK/PRSK_cjump.eps}\hfill
\includegraphics[width=75mm]{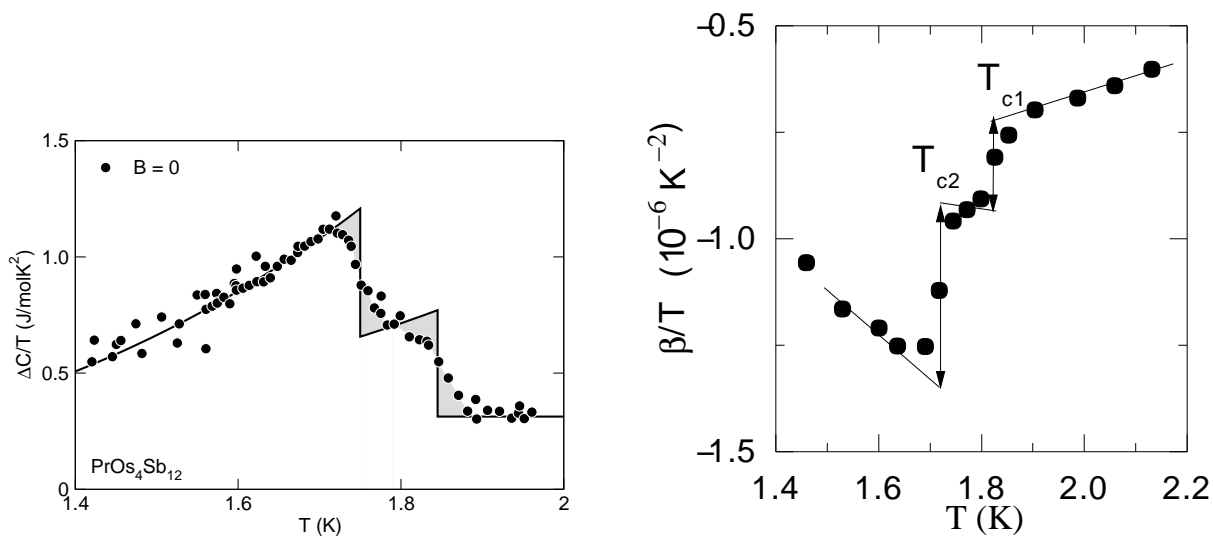}
\caption{Left panel: electronic specific heat $\Delta$C/T
vs. temperature (Vollmer et al. 2002). Solid line is an entropy conserving
construction leading to SC transitions at T$_{c1}$ = 1.85 K and
T$_{c2}$ = 1.75. Total specific heat jump
$\Delta$C$_{SC}/\gamma$T$_c\simeq$ 3. Right panel: corresponding jumps
in the volume thermal expansion at T$_{c1,2}$ (Oeschler et al. 2003,
2003a). \index{specific heat!PrOs$_4$Sb$_{12}$}
\index{thermal expansion!PrOs$_4$Sb$_{12}$}}
\label{FIGcjump}
\end{figure}

The SC specific heat jump is superposed on the Schottky anomaly from
the lowest CEF excitation at $\delta$. Nevertheless its detailed
analysis provides clear evidence for a split SC
transition \cite{Bauer02,Vollmer02} at T$_{c1}$ = 1.85 K and T$_{c2}$
= 1.75 K as shown in fig.~\ref{FIGcjump}. The total jump of both
transitions amounts
to $\Delta_{SC}C/\gamma T_c\simeq$ 3 which exceeds the BCS value 1.43
for a single transition considerably. It also proves that the SC
state is formed from the heavy quasiparticles that cause the enhanced
$\gamma$-value. A T$_c$-splitting of similar size also was clearly
seen in thermal expansion measurements
\citename{Oeschler03} \citeyear{Oeschler03,Oeschler03a}. The two
superconducting transitions in fig.~\ref{FIGcjump} are 
reminiscent of the split transition in UPt$_3$
(sect.~\ref{Sect:UPt3}). There a twofold orbitally degenerate
SC state is split by weak AF order that reduces the hexagonal symmetry
to orthorhombic. This also leads to two critical field curves in the
B-T phase diagram. In \PRS no such symmetry breaking field exists and
the split transition has to belong to two different T$_h$
representations of the SC order parameter or combinations thereof as discussed
below. The critical field curves associated with T$_{c1,2}$ have been
investigated with magnetisation \cite{Tayama03} and specific heat
measurments from which practically parallel curves are obtained. Only the upper
critical field corresponding to T$_{c1}$ with H$_{c2}$(0) $\simeq$ 2.2 T
is shown in fig.~\ref{FIGafq}. 

As usual in the early stage of investigation different experiments
gave inconclusive results on the question of the nature of gap
anisotropy. The low temperature specific heat exhibits a
C$_s$(T) $\sim$ T$^n$ power law in a rather reduced range between 0.65 K
and 1.2 K which points to some kind of nodal state.
In Sb-NQR experiments \cite{Kotegawa02} the nuclear spin lattice
relaxation 1/T$_1$ rate was
determined. It has an itinerant quasiparticle contribution that
contains information on the SC nodal state below T$_c$ and in addition
a localised contribution from broadened CEF excitations which decreases
exponentially for temperatures T $\ll\Delta$. There is no unique way to
separte these contributions, this problem is similar to the two Knight
shift contributions in the case of UPd$_2$Al$_3$ (sect.~\ref{Sect:UPd2Al3})
with its isoelectronic 5f$^2$ localised states. The NQR measurements
did not show any evidence for a coherence peak below T$_c$ which
points to an unconventional SC state, for lower temperatures an
exponential decay of T$^{-1}_1$ in conflict with the existence of gap
nodes was reported. However, this result depends critically on the
subtraction procedure of the localised contribution. 

\subsection{Thermal conductivity in the vortex phase and multiphase
superconductivity in \PRS}
\index{thermal conductivity} \index{multiphase superconductivity}

\begin{figure}
\includegraphics[width=80mm]{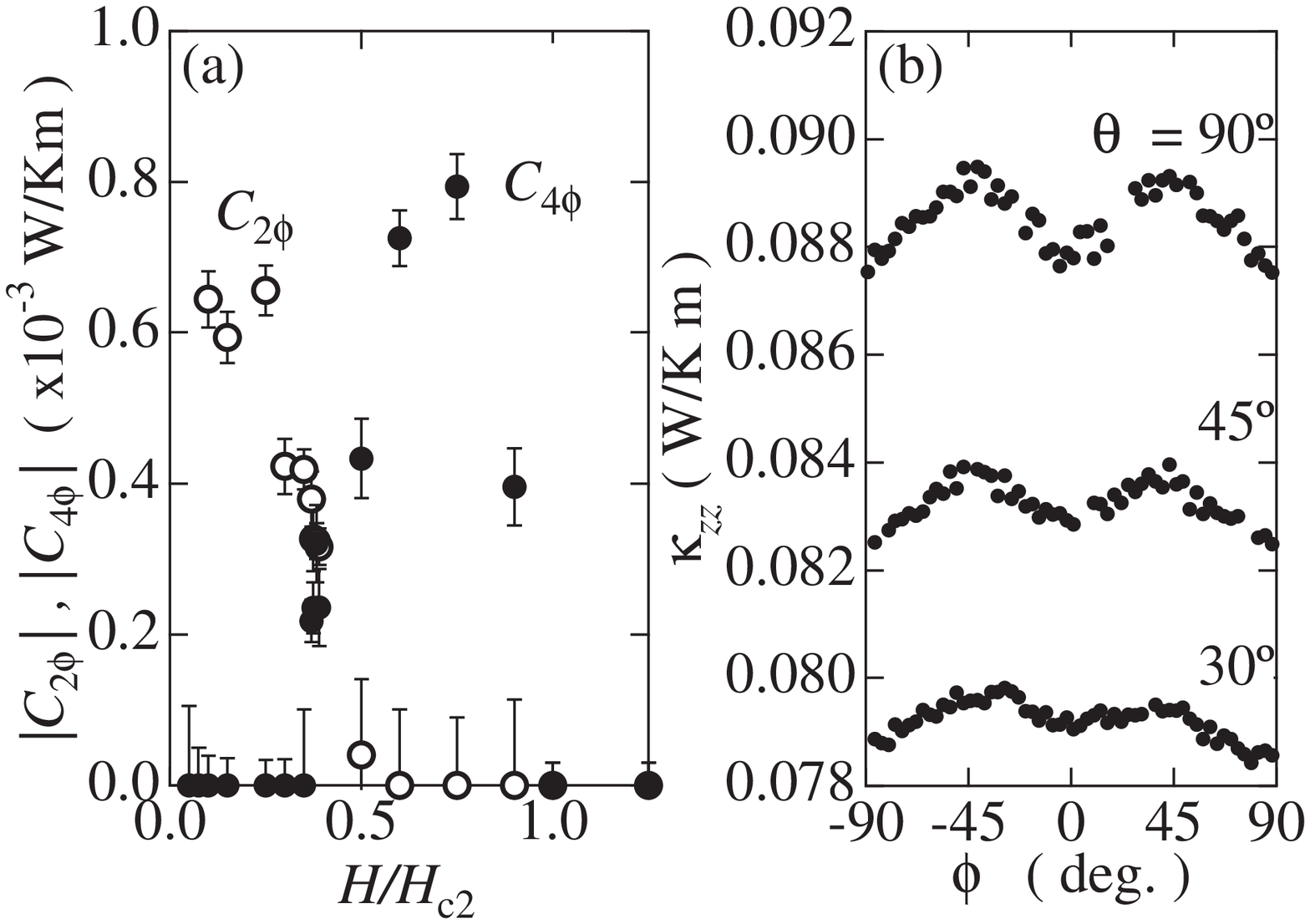}\hfill
\includegraphics[width=70mm]{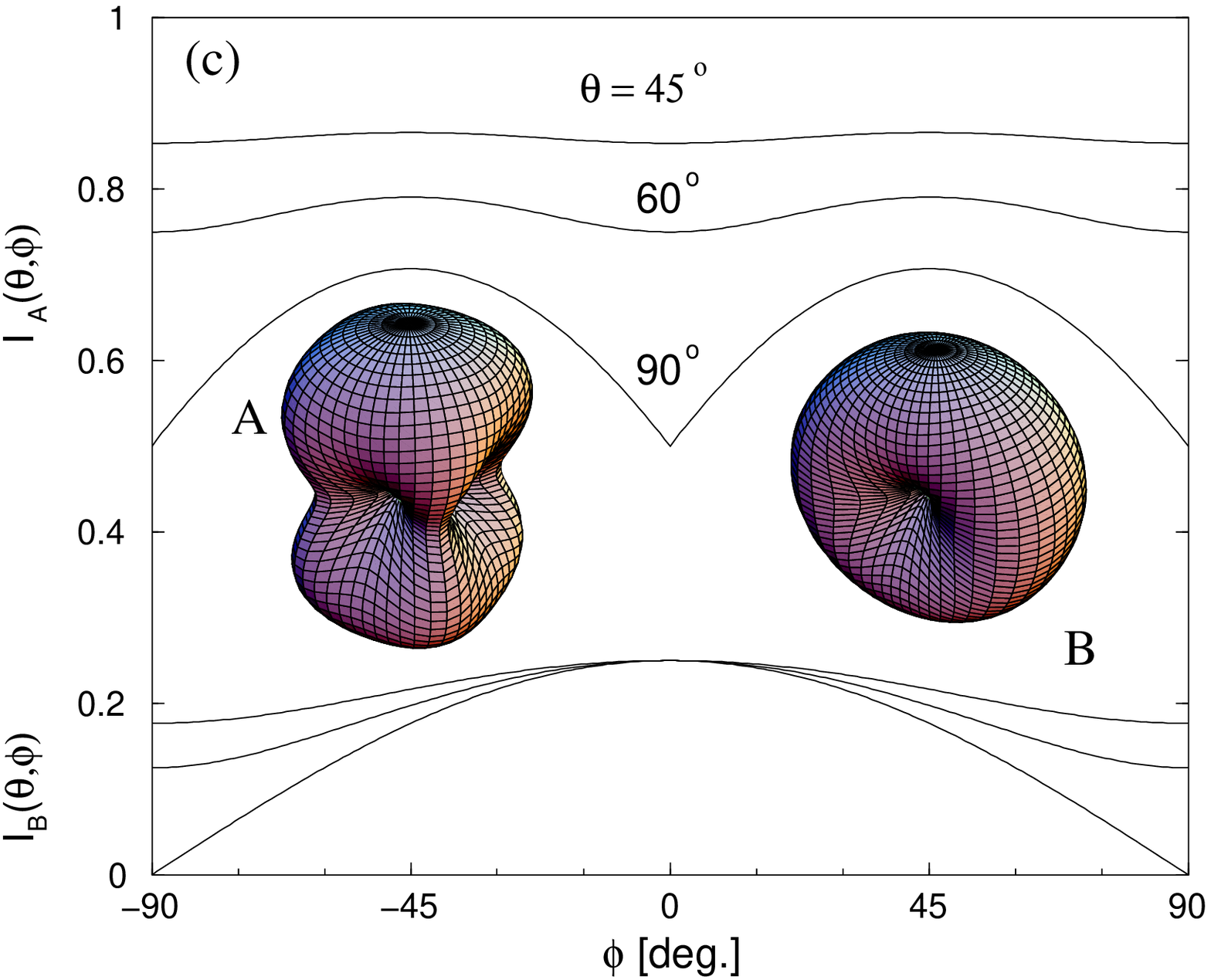}
\caption{(a) Field dependence of the fourfold (C$_{4\phi}$) and twofold
(C$_{2\phi}$) amplitudes of
$\kappa_{zz}(\phi)$ oscillations in (b) as function of field strength at
T = 0.52 K. Sharp transition at H$^*$ = 0.4H$_{c2}$ from twofold to
fourfold oscillation is seen. (b) Fourfold oscillations from
experiment for different polar field angle $\theta$ (Izawa et al. 2003). 
(c) Calculated angular variation of $\kappa_{zz}$ as function of
azimuthal field angle $\phi$ exhibiting fourfold (A) and twofold (B)
oscillations ($\theta$ = polar field angle). The inset shows polar
plots of A- and B-phase gap functions with four and two point nodes
respectively (Maki et al. 2002b). \index{magnetotransport} 
\index{gap function}}
\label{FIGosc}
\end{figure}

The experiments on field-angle dependent thermal conductivity 
described in sect.~\ref{Sect:Theory} are a more powerful method to investigate
the SC state. This method achieved
the determination of critical field curves and the associated B-T
phase diagram in \PRS\cite{Izawa03} and at the same time the nodal
structure of the gap function has at least been partly clarified. 
The same geometry as for borocarbides (sect.~\ref{Sect:Boro}) with conical
field rotation around the heat current direction parallel to a cubic [001] axis
was used and measurements in the whole temperature range from 0.3 K
up to T$_{c1}$ were performed. The azimuthal angle dependence of
$\kappa_{zz}$ was found to be approximately of the empirical form
\begin{equation}
\label{ZZOSC}
\kappa_{zz}(\theta,\phi,H,T)=\kappa_0 + C_{2\phi}(\theta,H,T)\cos 2\phi
+ C_{4\phi}(\theta,H,T)\cos 4\phi
\end{equation}
containing both twofold and fourfold rotations in $\phi$.
The $\phi$ dependence for H = 1.2 T is shown in fig.~\ref{FIGosc} for a few
polar angles $\theta$. It exhibits clearly a dominating fourfold
oscillation. Since the amplitude rapidly decreases with
$\theta$ similar as in the borocarbides this may be interpreted as
evidence for {\em point nodes} in \De along the [100] and [010] cubic axes. 
Surprisingly, when the field is lowered to H$^*$ = 0.8 T $\ll$ H$_{c2}$  a
rather sudden change from a fourfold to a twofold oscillation in
$\phi$ as shown in fig.~\ref{FIGosc} is observed which is interpreted as at
transition to a different SC state with only two point nodes along
[010]. The second critical
field  H$^*$ can be followed to higher temperatures, although with rapidly
decreasing oscillation amplitudes  C$_{2\phi}$ and
C$_{4\phi}$. Apparently the second critical field curve H$^*$(T) ends
at the lower critical temperature T$_{c2}$ observed already in the
second specific heat jump at zero field. 

\begin{figure}
\includegraphics[width=7.5cm]{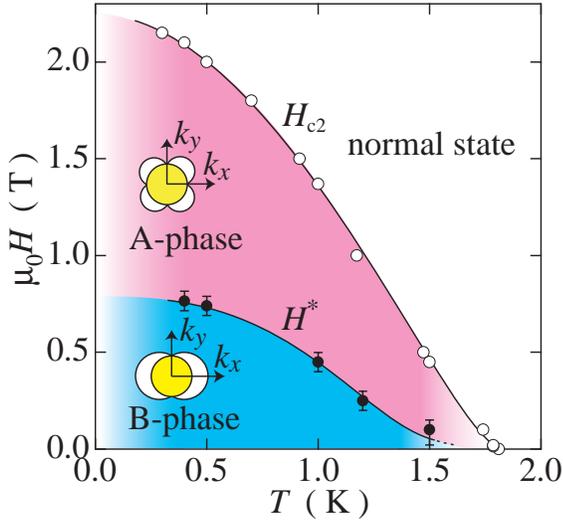}
\caption{B-T phase diagram of \PRS exhibiting B-phase with twofold
symmetry and A-phase with fourfold symmetry in a cubic plane.
The A-phase and B-phase gap symmetries are indicated.
They have four and two point nodes in a cubic plane respectively
(Izawa et al. 2003). \index{B-T phase diagram!PrOs$_4$Sb$_{12}$}}
\label{FIGmulti1}
\end{figure}

This allows one to construct a B-T phase diagram for \PRS shown in
fig.~\ref{FIGmulti1}. Most importantly one has two distinct regions
with A-and B-phase superconductivity characterised by four and two
point nodes along cubic axis [100] and [010]. In the applied geometry
the field is swept around the [001] axis, and therefore possible nodes
along this direction cannot be detected. Additional experiments with
a sweep axis orthogonal to [001] would be necessary.
Furthermore, three different domains (orientation of the
node directions) of A- and B-phase are possible.
Apparently for the sample investigated a specific domain has been 
selected, possibly by internal strain effects. From the B-T phase
diagram fig.~\ref{FIGmulti1} we conclude that \PRS is the second pure
multiphase HF superconductor found after UPt$_3$. There is, however, no
crossing of critical field lines and a corresponding tetracritical
point in \PRS as was observed in fig.~\ref{FIGhcrit} for UPt$_3$. 

\subsection{Gap models for SC A- and B-phases of \PRS}
\index{gap function}

The observation of a different nodal structure in the A- and B-phases
raises the question of the symmetry of the gap function in
\PRS. Various proposals have been made in
\cite{Maki02b,Goryo02,Miyake03,Ichioka03} based
on an empirical approach and compatible with the observed point
nodes. The gap function may be expanded in terms of basis functions
$\psi^\Gamma_i(\vk)$ which transform like representations $\Gamma$ of
the crystal symmetry
group (i=1-d is the degeneracy index, the index l denoting the degree
of $\Gamma$ is suppressed). So far there is no information from NMR
Knight shift or H$_{c2}$- Pauli limiting effects whether PrOs$_4$Sb$_{12}$ has
spin singlet or triplet pairing. In the singlet case the gap function
should then be given by
\begin{eqnarray}
\label{EXPAND}
\Delta(\vk)&=&\sum_{\Gamma,i}
\eta^\Gamma_i\psi^\Gamma_i(\vk)\equiv\Delta f(\vk)
\end{eqnarray}
where the form factor $f(\vk)$ is
normalized to one and $\Delta$ is the temperature dependent maximum gap
value. In the spirit of the Landau theory only a {\em single}
representation with the highest $T_c$ should be realized and for $T\geq T_c$ 
the free energy may then be expanded in terms of possible invariants 
of the order parameter components \cite{Volovik85} $\eta^\Gamma_i$ 
which are determined by Landau parameters $\alpha^\Gamma (T)$ and
$\beta^\Gamma_i$. The node structure is then fixed by the specific
symmetry class of $\Delta$(\vk) defined by the set of $\eta^\Gamma_i$.
However the pure second (l=2) and fourth (l=4) degree representations
in tetragonal T$_h$ symmetry which are given in table~\ref{TABLESKUT}
cannot realize
a gap function with the observed nodal structure, and therefore, one has to
consider the possibility of hybrid order parameter models. A striking example 
of such a hybrid gap function (s+g wave model) has already been
discussed in the previous section for the 
nonmagnetic borocarbide superconductor Y(Lu)Ni$_2$B$_2$C. 
%
\begin{table}
     \caption{\scriptsize Even parity basis functions for SC gap in
       tetrahedral (T$_h$)
       symmetry. (symbols used: $\psi^\Gamma_i(\vk)$ and
       $\psi'^\Gamma_i(\vk)$ (i=1-d) are second (l=2) and fourth
       degree (l=4) basis functions respectively, d
       denotes the degeneracy and $\Gamma_2$ and $\Gamma_3$ are complex
       conjugate combinations of E$_1$ and E$_2$ components)}      
     \vspace{0.5cm} 
     \label{TABLESKUT}
     \begin{tabular}{cccc}
	\hline
	d & $\Gamma$ & $\psi^\Gamma_i(\vk)$ & $\psi'^\Gamma_i(\vk)$\\
	\hline
        1 & A($\Gamma_1$) & 1 & $k_x^4+k_y^4+k_z^4$ \\
        2 & E($\Gamma_2,\Gamma_3$) & ($2k_z^2-k_x^2-k_y^2,k_x^2-k_y^2$) & 
         ($2k_z^4-k_x^4-k_y^4, k_x^4-k_y^4$)\\
        3 & T($\Gamma_5$) & ($k_yk_z, k_zk_x, k_xk_y$) &
         ($k_yk_zk_x^2, k_zk_xk_y^2, k_xk_yk_z^2$)\\
        \hline
     \end{tabular}
\end{table}
%
The symmetry classification of table~\ref{TABLESKUT} suggests the
following simple proposal for a hybrid gap function \De
=$\Delta f(\vk)$ for PrOs$_4$Sb$_{12}$ which has the the observed
nodes along the cubic axis, excluding the still hypothetical ones
along [001]: 

\begin{eqnarray}
\label{GAP2}
\mbox{A-phase~} \Delta(\vk)&=&\Delta(1-k_x^4-k_y^4) \nonumber\\
\mbox{B-phase~} \Delta(\vk)&=&\Delta(1-k_y^4)\ 
\end{eqnarray}

According to table~\ref{TABLESKUT} A- and B-phase are superpositions of
A and E representations with planar and axial symmetry
respectively. They are threefold degenerate and particular domains
with nodes in the $k_yk_z$-plane (A) or along $k_y$ (B) have been
chosen. This selection might be realise due to internal strains in the
SC sample. Polar plots of A,B-phase gap functions are shown in the
inset of fig.~\ref{FIGosc}. The magnetothermal conductivity in the
vortex state of phases A, B in the limit $\Gamma<$ T $\ll\Delta$ can be
calculated as described in \index{thermal conductivity}
\index{vortex phase} sect.~\ref{Sect:Theory} and the result for
$\kappa_{zz}$ is
%
%
\begin{eqnarray}
\frac{\kappa_{zz}}{\kappa_n}=\Bigl(\frac{2}{\pi}\Bigr)^2
\frac{v\sqrt{eH}}{\Delta}I_{A,B}(\theta,\phi) 
\end{eqnarray}
with the angular dependent functions I$_i(\theta,\phi)$ (i=A,B) given by 
\begin{eqnarray}
I_A(\theta,\phi)&=&\frac{1}{2}[(1-\sin^2\theta\sin^2\phi)^\frac{1}{2}
+(1-\sin^2\theta\cos^2\phi)^\frac{1}{2}]\ ,\nonumber\\
I_B(\theta,\phi)&=&\frac{1}{4}(1-\sin^2\theta\sin^2\phi)^\frac{1}{2}
\end{eqnarray}
For in plane field ($\theta$ = $\pi$/2) one has
I$_A(\theta,\phi)$ = $|\sin\phi|+|\cos\phi|$ and I$_B(\theta,\phi)$ =
$|\cos\phi|$, their leading Fourier components vary like $\cos
4\phi$ and $\cos 2\phi$ respectively in accordance with the empirical
expressions in eq.~(\ref{ZZOSC}). However, the above equations really predicts
cusps at $\phi$ = n($\pi$/2) (fig~.\ref{FIGosc}(c)) analogous to the
situation in the borocarbides and not the smooth minima as seen in
fig~.\ref{FIGosc}(b). Therefore, the existence of cusps cannot be safely
infered from the experimental results in fig.~\ref{FIGosc}, this would
necessitate measurements at lower temperatures. Also experiments with
variation of the polar angle $\theta$ have to be performed to check the
possible existence of node points along [001].
Finally $\mu SR$ experiments show indications of the presence of
magnetic moments in the SC phase which is interpreted, similarly as in
\UTB as evidence for a nonunitary triplet SC state \cite{Aoki03}. This
has also been claimed from a recent penetration depth study \cite{Chia03}.

As a preliminary conclusion it seems clear that \PRS is a very
unconventional multiphase HF superconductor of potentially the same
interest as UPt$_3$. Recalling that heavy quasiparticles are
presumably caused by coupling with virtual quadrupolar excitations from the
nonmagnetic 5f
ground state one is lead to speculate that SC in \PRS might also imply
the presence of an unprecedented pairing mechanism based on the
exchange of quadrupolar fluctuations. In addition to the
spin-fluctuation and magnetic-exciton exchange mechanism this would be
the third possibility for Cooper pair formation at work in heavy fermion
compounds. However, at the moment the quadrupolar SC mechanism in \PRS
is still a conjecture. 

\section{Summary and Outlook}

During the past decade a number of exciting and often unanticipated
broken symmetry states and associated physical effects have been discovered in
lanthanide and actinide intermetallic compounds.
Prominent among them are superconductivity characterized by highly
anisotropic unconventional order parameters and superconductivity
coexisting with ferromagnetic order as well as the hidden order of
unconventional density waves. A further example is the inhomogeneous 
superconducting phase appearing in an applied
field which has been predicted theoretically a long time ago. A general
hallmark of these systems is the coexistence and the competition of
various different cooperative phenomena such as superconductivity
and itinerant spin density wave magnetism.

The wealth of experimental data is only partly understood. To describe
the low-temperature ordered phases, the determination of the type and
symmetry of order
parameters is of central importance. The latter restrict the possible
excitations in the ordered phases and hence determine the low-temperature
properties. Order parameters given in terms of expectation values
of physical observables like spin- and charge-densities can be directly
measured e.g. by x-ray and neutron diffraction. The magnetic phases
in lanthanide and actinide compounds are therefore rather well characterized.
This, however, is not the case for hidden order like quadrupolar
ordering or unconventional density waves. 

Superconductivity corresponds to an off-diagonal long range order
parameter which is not
directly observable and it is not surprising that the type of pairing
is still a matter of controversy in many heavy fermion compounds.
Considerable progress has, however, been made recently in detecting
unconventional order parameter symmetries. Angle-resolved studies of
thermodynamic and
transport properties in the vortex phase determine the position of
the order parameter nodes relative to the crystal axes. These results
help to strongly reduce the number of symmetry-allowed superconducting
candidate states.

Important information about the material properties relevant for
superconductivity
is gained from the temperature and frequency dependence of thermodynamic
and transport properties. Experiments indicate severe deviations
from the universal behavior predicted by weak-coupling BCS theory.
The current theoretical analysis of experimental data which relies
on standard weak-coupling theory can yield only qualitative results.
A quantitative treatment must account for strong-coupling effects.
This, however, requires a microscopic picture of the normal state,
i.e., the quasiparticles and their interactions. 

A microscopic picture for the strongly renormalized quasiparticles
has finally emerged for the actinide compounds. The hypothesis of
the dual character of the 5f-electrons is translated into a calculational
scheme which reproduces both the Fermi surfaces and the effective
masses determined by dHvA experiments without adjustable parameter.
The method yields also a model for the residual interaction leading
to the various instabilities of the normal phase. The next step will
be to develop an appropriate Eliashberg-type theory. The dual model approach
should also provide insight into the mysterious hidden order
phases of U-compounds.

In Ce-based compounds, the Cooper pairs are formed by heavy quasiparticles
with predominantly 4f-character which arise through the Kondo
effect in the periodic lattice. This picture has been confirmed in
detail by dHvA and high-resolution
photoemission studies. The strongly renormalised quasiparticles can
be successfully reproduced by a semiphenomenological ansatz. Despite
the efforts to implement modern many-body methods for strong correlations
into realistic electronic structure calculations there is still no
general concept for quantitative microscopic calculations. In particular,
the subtle interplay between local and intersite effects continues
to challenge theorists. The latter may lead to long-range order of
local moments while the former favor the formation of a Fermi liquid
state at low temperatures.

Many heavy fermion materials are on the verge of magnetic instability.
By application of pressure and magnetic field, these materials can
be tuned through a quantum critical point from a metallic antiferromagnet
into a paramagnet. This may also trigger a transition to an unconventional
SC state. Much effort has been devoted to the study of the
behavior in the vicinity of the quantum critical point. Experimental
data exhibit universality with unusual critical exponents. The theoretical
picture, however, is highly controversial at present. 

Superconductivity, magnetism and hidden order
in lanthanide and actinide compounds pose an ongoing challenge. These
compounds serve as model systems to study strong correlations in a
broader context.

\vspace{1.5cm}
\textbf{Acknowledgement}\\
We have benefitted from disucssions and suggestions from many
colleagues, especially we would like to thank for collaboration and
support from
J. W. Allen, A. Amici, N. E. Christensen, T. Dahm, P. Fulde, K. Izawa,
G. G. Lonzarich, K. Maki, Y. Matsuda, E. Runge, N. Sato, R. Shiina,
F. Steglich, G. Varelogiannis, H. Won, A. Yaresko, and Q. Yuan.\\ 
G. Zwicknagl would like to acknowledge hospitality of the
Max-Planck-Institute for the Chemical Physics of Solids, Dresden.

\bibliography{References200503}
\printindex
\end{document}